\documentclass[a4paper,11pt]{article}

\usepackage{jheppub}
\usepackage[table]{xcolor}
\usepackage{slashbox}
\allowdisplaybreaks
\usepackage{graphicx}
\usepackage{epstopdf}

\newcommand{\be}{\begin{equation}}
	\newcommand{\ee}{\end{equation}}
\newcommand{\bea}{\begin{eqnarray}}
	\newcommand{\eea}{\end{eqnarray}}
\newcommand{\nn}{\nonumber}
\newcommand{\bdm}{\begin{displaymath}}
	\newcommand{\edm}{\end{displaymath}}

\def\vv{\vec{v}}

\def\vn{\vec{n}}

\title{N$^3$LO Spin-Orbit Interaction via the
	\\EFT of Spinning Gravitating Objects}

\author[a]{Jung-Wook Kim,}
\author[b]{Mich\`ele Levi,}
\author[c]{Zhewei Yin}

\affiliation[a]{Queen Mary University of London, Mile End Road, 
	London E1 4NS, United Kingdom}

\affiliation[b]{Mathematical Institute, University of Oxford, 
	Oxford OX2 6GG, United Kingdom}

\affiliation[c]{Department of Physics and Astronomy, Uppsala University, 
	75108 Uppsala, Sweden}

\emailAdd{jung-wook.kim@qmul.ac.uk}
\emailAdd{levi@maths.ox.ac.uk}
\emailAdd{zhewei.yin@physics.uu.se}

\abstract{
We present the derivation of the third subleading order (N$^3$LO) spin-orbit 
interaction at the state of the art of post-Newtonian (PN) gravity via the EFT of spinning objects.
The present sector contains the largest and most elaborate collection of Feynman graphs ever 
tackled to date in sectors with spin, and in all PN sectors up to third subleading order. 
Our computations are carried out via advanced multi-loop methods. Their most demanding aspect 
is the imperative transition to a generic dimension across the whole derivation, due to the emergence 
of dimensional-regularization poles across all loop orders as of the N$^3$LO sectors.
At this high order of sectors with spin, it is also critical to extend the formal procedure for the 
reduction of higher-order time derivatives of spin variables beyond linear order for the first time. 
This gives rise to a new unique contribution at the present sector.
The full interaction potential in Lagrangian form and the general Hamiltonian are provided here for 
the first time. 
The consequent gravitational-wave (GW) gauge-invariant observables are also derived, including 
relations among the binding energy, angular momentum, and emitted frequency.
Complete agreement is found between our results, and the binding energy of GW sources, and also with 
the extrapolated scattering angle in the scattering problem, derived via traditional GR. 
In contrast with the latter derivation, our framework is free-standing and generic, and has provided 
theory and results, which have been critical to establish the state of the 
art, and to push the precision frontier for the measurement of GWs.
}

\begin{document}
	

\preprint{QMUL-PH-22-27, UUITP-35/22}
	
\maketitle
	
\flushbottom

\section{Introduction} 
\label{intro}

The successful measurements in ground-based experiments of gravitational waves (GWs) from 
compact-binary inspirals and mergers, since the first breakthrough detection \cite{Abbott:2016blz} by 
the Advanced LIGO \cite{LIGOScientific:2014pky} and Advanced VIRGO \cite{VIRGO:2014yos} collaboration, 
have been already providing abundant and rich data 
\cite{LIGOScientific:2018mvr,LIGOScientific:2020ibl,LIGOScientific:2021djp}. 
Moreover, there is a rapidly-growing worldwide network of GW detectors, including now also KAGRA 
in Japan \cite{KAGRA:2020tym}. 
The sources of these GWs, such as black-hole (BH) binaries, spend most of their evolution time in the 
inspiral phase, where the velocities of orbiting components are non-relativistic. Accordingly they 
have been studied analytically via the post-Newtonian (PN) approximation of General Relativity 
(GR) \cite{Blanchet:2013haa}. 

This PN description forms the basis for the effective-one-body (EOB) approach \cite{Buonanno:1998gg}, 
which in turn enables to model complete theoretical waveforms. These are used to test our knowledge of 
gravity in the strong-field regime \cite{TheLIGOScientific:2016src,TheLIGOScientific:2016wfe}, and also 
of QCD in extreme conditions from neutron-star (NS) binaries \cite{LIGOScientific:2017vwq} or mixed 
NS-BH binaries \cite{LIGOScientific:2021qlt}. 
Since in reality all compact gravitating objects are rotating, it is essential to 
incorporate the effects of spin into any theoretical waveform models  
beyond the basic accuracy of the first PN ($1$PN) order \cite{Abbott:2016izl}. For real 
precessing binaries with compact spinning objects the physics gets dramatically more 
intricate and intriguing.

The experimental breakthrough in the measurements of GWs has ushered in an era of 
booming activity in the analytical studies of the high-precision frontier of PN theory. In the 
conservative sector recent efforts to push the state of the art have peaked at the high 
perturbative order of $5$PN in the point-mass sector via traditional GR  
\cite{Bini:2019nra,Bini:2020wpo,Bini:2020uiq}, and via effective field theory (EFT) 
methods \cite{Goldberger:2004jt,Blumlein:2020pyo}. The point-mass $5$PN results in 
\cite{Bini:2019nra,Bini:2020wpo} have been accomplished via a combined intricate 
deployment of traditional GR methods, with crucial assumptions from the EOB approach, 
and available results from self-force theory, see e.g.~recent review in 
\cite{Barack:2018yvs}. The work in \cite{Antonelli:2020aeb,Antonelli:2020ybz} then 
followed suit, and implemented the same approach of \cite{Bini:2019nra,Bini:2020wpo} to the 
spin-orbit sector, that is linear in the spins, at the $4.5$PN order.

Yet, at the present high orders of the precision frontier it is crucial to deploy 
various independent methodologies in order to carefully push and establish the state 
of the art. The EFT of spinning gravitating objects \cite{Levi:2015msa} constitutes a unique 
free-standing methodology, which has indeed enabled the completion of the state 
of the art at the $4$PN order \cite{Levi:2014gsa,Levi:2015uxa,Levi:2015ixa,Levi:2016ofk}. 
In the present paper we derive the complete next-to-next-to-next-to-leading order (N$^3$LO) spin-orbit 
interaction at the $4.5$PN order via the EFT of spinning gravitating objects, and its automation 
introduced in the public code \texttt{EFTofPNG} \cite{Levi:2017kzq}, also building on 
\cite{Levi:2019kgk,Levi:2020kvb,Levi:2020uwu,Levi:2020lfn}. 
This paper is part of a series, which has recently accomplished the completion all sectors up to the 
$5$PN order \cite{Kim:2021rfj,Kim:2022bwv,Levi:2022dqm,Levi:2022rrq}. 

In sections \ref{theory} and \ref{graphs} of this paper, we present in full detail the formulation and 
evaluation of the Feynman diagrammatic expansion in this sector, which contains the 
largest and most elaborate collection of graphs ever tackled to date in sectors with 
spin, and in all PN sectors up to the third subleading order. Our computations are 
carried out via advanced multi-loop methods, and further development of the \texttt{EFTofPNG} code. 
The most significant computational leap that is taking place in the present sector, is the need to 
switch on a generic dimension and keep track of dimensional-regularization (DimReg) expansions, across 
the whole derivation. This is due to the emergence of DimReg poles across all loop orders 
in sectors as of the N$^3$LO in PN theory.

In section \ref{toreducedaction} we extend the formal procedure introduced in \cite{Levi:2014sba} 
for the reduction of higher-order time derivatives of spin variables via redefinitions, and 
present in full detail the intricate reduction process required in the present sector. 
We find that this sector uniquely requires to apply the redefinition of rotational variables 
beyond linear order for the first time, as previously the linear order has been sufficient for all 
sectors with spin at lower orders.

In sections \ref{finalaction} and \ref{hamilton}, the full interaction potential in Lagrangian form 
and the general Hamiltonian of the sector, are provided here for the first time. 
The Lagrangian potential obtained via our framework enables a direct derivation of the 
physical equations of motion (EOMs) for both the position and spin \cite{Levi:2015msa}. The 
full general Hamiltonian enables to explore possible EOB extensions for this sector, and test the 
performance of such variants of EOB Hamiltonians. The general Hamiltonian also enables to 
study the conserved integrals of motion, which form a representation of the Poincar\'e 
algebra on phase space. Indeed, in \cite{Levi:2022dqm} we find the complete Poincar\'e 
algebra of the present sector, and thus verify the validity of the general Hamiltonian 
derived in the present paper. 

In section \ref{gi}, we proceed to derive the consequent GW gauge-invariant observables, 
namely the relations among binding energy, angular momentum, and emitted frequency. 
We find complete agreement with the binding energy of GW sources, and also with the extrapolated 
scattering angle in the scattering problem, derived via traditional GR \cite{Antonelli:2020ybz}. 
Finally, all of the computations in our framework are also automated as extensions of the public 
\texttt{EFTofPNG} code. 
The independent generic derivation and results presented in this paper, and such development of the 
\texttt{EFTofPNG} code, have been essential to establish the state of the art, and to push the 
precision frontier for the measurement of GWs.

\section{EFT of Spinning Gravitating Objects}
\label{theory}

In order to derive the spin-orbit interaction at this high order,
we should consider the EFT of spinning gravitating objects \cite{Levi:2015msa}.
We start from an effective action of a two-particle system at the orbital scale, that 
captures a compact binary inspiral \cite{Goldberger:2004jt}:
\be \label{2particle}
S_{\text{eff}}=S_{\text{gr}}[g_{\mu\nu}]+\sum_{a=1}^{2}S_{\text{pp}}(\lambda_a),
\ee
where $S_{\text{gr}}$ is some bulk action of the gravitational field, and 
$S_{\text{pp}}$ is a point-particle action for each component of the binary, that is 
localized on their worldline with the parameter $\lambda_a$ for the $a$-th component.

For the bulk gravitational action, which is given in terms of the field modes at the 
orbital scale, $g_{\mu\nu}(x)$, we take the Einstein-Hilbert action of GR, which we 
gauge-fix with the fully-harmonic gauge:
\be \label{Sgr}
S_{\text{gr}}[g_{\mu\nu}]=S_{\text{EH}} + S_{\text{GF}} 
= -\frac{1}{16\pi G_d} \int d^{d+1}x \sqrt{g} \,R 
+ \frac{1}{32\pi G_d} \int d^{d+1}x\sqrt{g}
\,g_{\mu\nu}\Gamma^\mu\Gamma^\nu, 
\ee
where $\Gamma^\mu\equiv\Gamma^\mu_{\rho\sigma}g^{\rho\sigma}$.
Note that at this high non-linear order in the gravitational self-interaction, the 
derivation must be kept in a generic dimension throughout, with $d$ the number of 
spatial dimensions, and a modified minimal subtraction ($\overline{\text{MS}}$), see 
e.g.~\cite{Peskin:1995ev}, for the  $d$-dimensional gravitational constant:
\be
G_d\equiv G_N \left(\sqrt{4\pi e^\gamma} \,R_0 \right)^{d-3},
\ee
with $G_N\equiv G$ Newton's gravitational constant in $3$-dimensional space, 
$\gamma$ 
Euler's constant, and $R_0$ some renormalization scale. 
This is since dimensional regularization (DimReg) is used to evaluate the integrals, and 
poles in the dimensional parameter $\epsilon\equiv d-3$ appear throughout, as shall be 
seen as of the next section.

We make consistent use of a $d+1$ non-relativistic (NR) decomposition of the 
gravitational field, similar to a Kaluza-Klein reduction over the time dimension
\cite{Kol:2007bc,Kol:2010ze}: 
\begin{align}\label{nrkk}
ds^2=g_{\mu\nu}dx^{\mu}dx^{\nu}\equiv 
e^{2\phi}\left(dt-A_idx^i\right)^2-e^{-\frac{2}{d-2}\phi}\gamma_{ij}dx^idx^j,
\end{align}
which defines the NR fields: $\phi$, $A_i$, and 
$\gamma_{ij}\equiv\delta_{ij}+\sigma_{ij}$.
This beneficial parametrization has been implemented first in sectors with spin in 
\cite{Levi:2008nh}, and was incorporated all throughout the \texttt{EFTofPNG} code 
\cite{Levi:2017kzq}.
The propagators of such NR fields contain temporal delta functions, and the momenta 
integrals are then $d$-dimensional. These NR propagators get their relativistic 
corrections order by order via quadratic insertions with two time derivatives. In 
this third subleading sector the propagators get up to three such perturbative 
corrections. The $d$-dimensional propagators and their insertions, 
as well as the higher-point gravitational vertices for our sector are all provided in 
the ancillary files to this publication in 
human visual and machine-readable formats. They were all derived via the 
extension of the \texttt{FeynRul} module of the public code \texttt{EFTofPNG} 
\cite{Levi:2017kzq}.  

Finally and importantly, the point-particle action $S_{\text{pp}}$ should also be considered. 
In the present spin-orbit sector it is the effective action of a spinning 
particle which is minimally coupled to gravity, in the form 
\cite{Levi:2010zu,Levi:2015msa,Levi:2018nxp}:
\begin{align} 
\label{sppspinmc}
S_{\text{pp}}(\lambda)=&\int 
d\lambda\left[-m \sqrt{u^2}-\frac{1}{2} \hat{S}_{\mu\nu} \hat{\Omega}^{\mu\nu}
-\frac{\hat{S}^{\mu\nu} p_{\nu}}{p^2} \frac{D p_{\mu}}{D \lambda}\right],
\end{align}
with $m$ the mass, $u^{\mu}$ the $4$-velocity, $p_{\mu}$ the  
linear-momentum, and $\hat{\Omega}^{\mu\nu}$, $\hat{S}_{\mu\nu}$, the generic angular 
velocity and spin variables of the particle, respectively. These generic rotational 
variables, together with the additional term in eq.~\eqref{sppspinmc} which contains 
the covariant derivative along the worldline, enable to switch the gauge of 
rotational variables. By contrast the action presented in 
\cite{Hanson:1974qy,Bailey:1975fe,Yee:1993ya,Porto:2005ac} does not have generic 
rotational variables, nor does it include the generic last term in eq.~\eqref{sppspinmc}.

Finite-size effects enter only at the $5$PN order in the point-mass sector, and at the 
$6.5$PN order in the spin-orbit sector of the effective action \cite{Levi:2020kvb,Levi:2020uwu}. 
Therefore from the action in eq.~\eqref{sppspinmc} one can extract all the required mass 
and spin worldline-couplings, where both play an important role in the spin-orbit interaction.
Note that the spin couplings are given in 
terms of the local spatial components of the spin tensor in the generalized canonical 
gauge that we formulated in \cite{Levi:2015msa}, so that the indices in all Feynman 
rules are Euclidean. 
All the corresponding Feynman rules of the worldline couplings are provided in the 
ancillary files to this publication in human visual and machine-readable formats.

\section{Diagrammatic Evaluation}
\label{graphs}

We then generated the Feynman graphs that contribute to this sector using the input of Feynman rules 
within the \texttt{EFTofPNG} code. 
The graph distribution among the $4$ orders in $G$ of topologies is shown in table \ref{graphsinG}. 
All in all, there are $1305$ graphs in the present sector, which is to date the largest collection 
of graphs ever tackled in PN sectors with spin, and also of all PN sectors up to third 
subleading order. 
The full list of graphs and their corresponding values can be found in the ancillary 
files to this publication in human visible and machine-readable formats. 
We present only the unique graphs with spin coupling on worldline ``1'', 
where there is another copy of similar graphs from the exchange of worldline labels, 
$1\leftrightarrow2$. 
The graphs are listed in the files according to the enumeration $(n_1, n_2, n_3)$, where $n_1$ 
indicates order in $G$, $n_2$ serial number of topology, $n_3$ serial number of graph within topology, 
see \cite{Levi:2018nxp,Levi:2020kvb} for the list of topologies at each order in $G$. 
The highest-order topologies at $G^4$, and their related integral structure, were analysed in detail 
\cite{Levi:2020kvb}, where the relevant $388$ graphs were evaluated using advanced multi-loop methods.

\begin{table}[t]
\begin{center}
\begin{tabular}{|l|c|c|c|c||c|}
\hline
Order in $G$ & 1 & 2 & 3 & 4 & Total \\
\hline
Number of graphs & 11 & 204 & 702 & 388 & 1305 \\
\hline
\end{tabular}
\caption{The graph distribution 
among topology orders in $G$ in the N$^3$LO spin-orbit sector.}
\label{graphsinG}
\end{center}
\end{table}

Similar to \cite{Levi:2020kvb}, the analysis of the complete present sector also builds on the 
treatment of the N$^2$LO linear-in-spin sectors obtained in \cite{Levi:2011eq,Levi:2015uxa} via EFT 
(also obtained in \cite{Hartung:2011te,Marsat:2012fn,Bohe:2012mr} via traditional GR).
Beyond the exponentiated volume of graphs in this sector, the majority of graphs at order $G^3$ 
are also the most computationally demanding. 
These include $91$ graphs in a $2$-loop topology/integral (see figure 11 graph c3 in 
\cite{Levi:2018nxp}), which requires reduction via integration-by-parts (IBP). 
Together with $219$ nested $2$-loop graphs, and noting that at this order in $G$ the 
propagators also carry time insertions, these graphs were the most time-consuming to 
evaluate. 
Moreover, graphs at order $G^1$ entail the highest load of time derivatives, and require the maximal 
iteration ever implemented for the gauge of rotational variables, see e.g.~\cite{Levi:2014sba} for such 
iteration required at lower orders. 
It was then necessary to further upgrade the automation of the diagrammatic evaluation in order to 
handle the full present sector.
The upgrades that were developed upon the \texttt{EFTofPNG} code involved mainly projection methods 
due to the high-rank numerators of integrals \cite{Passarino:1978jh,Boels:2018nrr,Chen:2019wyb}, and 
IBP methods to reduce the resulting scalar integrals \cite{Smirnov:2006ry}, via our adaption of 
Laporta's algorithm \cite{Laporta:2001dd}. 
We corroborated the values of graphs using parallel evaluations arising from independent development 
and implementation of codes. 

The majority of graphs at each order of $G$ concentrate on the highest-loop ones -- as defined in 
\cite{Levi:2020kvb} in the ``worldline picture''. 
Moreover, similar to what was observed at order $G^4$ in \cite{Levi:2020kvb}, the evaluation of graphs 
yields DimReg poles in the dimensional parameter, $\epsilon_d\equiv d-3$, in conjunction with 
logarithms in $r/R_0$, for individual graphs.
For example, out of the $204$ graphs at order $G^2$, where there are only $2$ possible topologies, 
$165$ are in the single $1$-loop topology, which at this high subleading order yields the 
aforementioned DimReg poles with logarithms. 
Out of the $702$ graphs at order $G^3$, where there are topologies from $0$- to $2$-loop orders, $439$ 
are of $2$-loop order, and the majority of these  $2$-loop topologies yield such DimReg poles and 
logarithms. 
The emergence of DimReg poles across all loop orders in sectors as 
of the N$^3$LO, makes the transition from sectors up to the N$^2$LO -- an especially critical one. 
This necessitates an overall switch to a generic dimension across the whole evaluation, with an 
expansion in the dimensional parameter, $\epsilon_d$, which stands out as the most 
computationally-demanding aspect in the evaluation of this higher-order sector.
By contrast, in sectors up to the NNLO there is only a single $2$-loop topology that yields very few 
graphs overall, which actually requires to keep the dimension generic, while otherwise the dimension 
can be practically specified already at the start of evaluation. Moreover, in sectors up to the NNLO 
even these few graphs that are somewhat more difficult to evaluate, never give rise to DimReg poles in 
their final result.

The systematic zero values in graphs of the $2$ factorizable topologies at order $G^3$, which were 
already observed at the NNLO due to contact interaction terms, are also found here 
\cite{Levi:2011eq,Levi:2020kvb}. 
Moreover, similar to the analysis in \cite{Levi:2020kvb}, all the non-vanishing graphs in these $2$ 
factorizable topologies are proportional to the transcendental number that is a Riemann zeta value, 
$\zeta(2)\equiv\pi^2/6$. The latter feature is also found in graphs of the single $2$-rank topology at 
order $G^3$, that through IBP reduction contains a linear combination of the factorizable and nested 
master integrals \cite{Levi:2020kvb}.

\subsection{Unreduced Action}
\label{rawaction}

Summing up all the graphs in this sector, we obtain the unreduced action. At this stage one gets a very 
bulky action, which contains several large pieces of higher-order time derivatives. The unreduced 
potential in the Lagrangian, $-V_{\text{N}^3\text{LO}}^{\text{SO}}\subset L$, can be expressed 
as:
\bea
V_{\text{N}^3\text{LO}}^{\text{SO}} = \sum_{i=0}^6 \stackrel{(i)}{V}_{3,1} 
+ (1 \leftrightarrow 2),
\eea
where any piece $\stackrel{(i)}{V}$ contains only terms with a total of $i$ 
higher-order time derivatives beyond the velocity and spin variables, 
and the indices $n,l$ in the subscript correspond to the sector $(n,l)$, as in table \ref{overallredef}.
We present the full unreduced potential according to these pieces in 
Appendix \ref{unreduced}, and in the ancillary files to this publication.
Notice that all in all, there are now DimReg poles with logarithms at orders $G^2-G^4$, 
and $\zeta(2)$ factors at orders $G^3$ and $G^4$.

\section{Reduction of Action}
\label{toreducedaction}

\begin{table}[t]
\begin{center}
\begin{tabular}{|l|c|c|c|c|}
\hline
\backslashbox{\quad\boldmath{$l$}}{\boldmath{$n$}} & (N\boldmath{$^{0}$})LO
& N\boldmath{$^{(1)}$}LO & \boldmath{N$^2$LO}
& \boldmath{N$^3$LO} 
\\
\hline
\boldmath{S$^0$} & 
& 
& 
+ 
& 
+ 
\\
\hline
\boldmath{S$^1$} & 
+ 
& 
++ & 
++ 
& 
++ 
\\
\hline
\end{tabular}
\caption{The notation $(n,l)$ and our PN-counting formula for general sectors $(n,l)$ was introduced 
in \cite{Levi:2019kgk}. 
The overall sectors that have to be taken into account for the contributions 
to the present sector through redefinition of variables. 
Sectors with "+" introduce only position shifts, 
whereas sectors with "++" introduce redefinition of both position and rotational variables.}
\label{overallredef}
\end{center}
\end{table}

At this stage the EFT computation is done, and we now need to reduce the raw generalized action, 
which consists of several parts with higher-order time derivatives, to an action 
which contains only the position, velocity, and spin variables. This 
reduction is carried out via a formal procedure of variable redefinitions, which was 
formulated to also handle rotational variables in \cite{Levi:2014sba}.
To that end, it is useful to consider table \ref{overallredef} which summarizes the 
overall redefinitions that have to be taken into account from all the sectors 
relevant to the present one:
We also need to take into account contributions to the present sector, which arise from the application 
of any redefinitions that were required at lower-order sectors. 
As can be seen in table \ref{overallredef}, redefinitions at sectors without spin, 
which obviously consist only of position shifts, start only at the NNLO, that is at 
the $2$PN sector. This is not the case for sectors with spin, in which position shifts 
are already required as of the LO at the $1.5$PN order, and additional redefinitions of rotational 
variables are also required as of the NLO. 

Position shifts, $\Delta \vec{x}$, scale in both PN and spin orders as the sector 
in which they were initially applied \cite{Levi:2014sba}. Thus simple 
power-counting shows that for the present sector position shifts from the sectors of both point-mass, 
i.e.~without spin, and of spin-orbit, 
are only needed to be applied to linear order. 
This is in notable contrast to higher-spin sectors, starting with the NLO quadratic-in-spin 
sector, that already requires to go beyond the linear application of position shifts. 
Rotational redefinitions, $\omega^{ij}$, scale as $v^{-1}S^{-1}$ with respect to the 
sector in which they were initially applied \cite{Levi:2014sba}. Yet since their 
formulation is inherently different than that of position shifts, it is not clear at this 
high order  whether redefinitions of rotational variables are needed to be applied 
beyond the linear order, which has been sufficient for all lower-order sectors. 
To settle this question we extend the formal procedure for redefinition of rotational variables 
from \cite{Levi:2014sba} beyond linear order -- in the following section.

\subsection{Redefinition of Rotational Variables} 
\label{spinredefextend}

We consider a redefinition of the Lorentz rotation matrix which relates local frames 
of the spin. Such a redefinition is also a rotation parametrized in a matrix 
exponential of a small anti-symmetric generator $\omega$:
\bea
\Lambda^{ij}= \Lambda^{ik} \left( e^{ \omega} \right)^{kj},
\eea
and for the shift we get:
\bea
\label{Lorentzshift}
\Delta \Lambda^{ij} =
\Lambda^{ik} \omega^{kj} + \frac{1}{2} \Lambda^{ik} \omega^{kl} \omega^{lj} + 
\mathcal{O} \left( \omega^3 \right),
\eea
with $\Delta_1 \Lambda^{ij}\equiv\Lambda^{ik} \omega^{kj}$ and 
$\Delta_2 \Lambda^{ij} \equiv \frac{1}{2} \Lambda^{ik} \omega^{kl} \omega^{lj}$.
Similar to \cite{Levi:2014sba} we take the redefinition of the spin $\Delta S^{ij}$ 
to be arbitrary, as it is fixed in the end from considering the effect of 
the redefinition of the Lorentz matrix on the rotational kinetic term in the Lagrangian: 
\be
\label{discreteLwspin}
L \supset - \frac{1}{2} S_a^{ij}\Omega_a^{ij} - V(\{S_a\}), \quad a\in\{1,2\},
\ee 
where $\Omega^{ij}\equiv -\Lambda^{ki}\dot{\Lambda}^{kj}$, and we suppressed the 
dependence in position and velocity variables.
Accordingly, in \cite{Levi:2014sba} we fixed $\Delta S$ to linear order in $\omega$:
\bea
\label{linspinshift}
\Delta_1 S^{ij} = S^{ik} \omega^{kj} - S^{jk} \omega^{ki},
\eea
and thus now we would like to fix $\Delta_2 S$ in the spin shift:
\bea
\label{spinshift}
\Delta S^{ij} \equiv \Delta_1 S^{ij} + \Delta_2 S^{ij} 
+ \mathcal{O}  \left( \omega^3 \right).
\eea

We remind that the interaction potentials do not contain any dependence in 
$\dot{\Lambda}$, but only in spin variables, and thus we should only study the effect 
of the Lorentz shift on the rotational kinetic term, and accordingly fix the spin 
shift which affects the potentials.
We find the shift in the kinetic term (up to the prefactor $-1/2$ in the action) as:
\bea \label{spinshifttoquad}
\Delta \left(S^{ij} \Omega^{ij} \right) &=&
\Delta_1 S^{ij} \Omega^{ij} 
- S^{ij}\Delta_1 \Lambda^{ki} \dot{\Lambda}^{kj}
- S^{ij}\Lambda^{ki} \frac{d}{dt}(\Delta_1 \Lambda^{kj}) 
+ \Delta_2 S^{ij} \Omega^{ij} \nn\\
&&
- \Delta_1 S^{ij} \Delta_1 \Lambda^{ki} \dot{\Lambda}^{kj}
- \Delta_1 S^{ij} \Lambda^{ki} \frac{d}{dt}(\Delta_1 \Lambda^{kj}) 
- S^{ij} \Delta_1 \Lambda^{ki} \frac{d}{dt}(\Delta_1 \Lambda^{kj}) \nn\\
&&
- S^{ij}\Delta_2 \Lambda^{ki} \dot{\Lambda}^{kj}
- S^{ij}\Lambda^{ki} \frac{d}{dt}(\Delta_2 \Lambda^{kj}) + \mathcal{O} (\omega^3)
\nn\\
&=&  \dot{S}^{ij} \omega^{ij} + \Delta_2 S^{ij} \Omega^{ij} 
- \left(S^{il} \omega^{lj} - S^{jl} \omega^{li} \right) 
\left[ \Lambda^{km} \omega^{mi} \dot{\Lambda}^{kj} + \Lambda^{ki} \frac{d}{dt} 
\left( \Lambda^{km} \omega^{mj} \right) \right] \nn \\
&& - S^{ij} \Lambda^{kl} \omega^{li} \frac{d}{dt} 
\left( \Lambda^{km} \omega^{mj} \right)
- \frac{1}{2} S^{ij} \left[\Lambda^{kl} \omega^{lm} \omega^{mi}  \dot{\Lambda}^{kj} 
+ \Lambda^{ki} \frac{d}{dt} \left(\Lambda^{kl} \omega^{lm} \omega^{mj} \right) 
\right] \nn \\
&& + \, \mathcal{O} (\omega^3)\nn\\
&=& \dot{S}^{ij} \omega^{ij} + \Delta_2 S^{ij} \Omega^{ij} 
+ \left(S^{il} \omega^{lj} - S^{jl} \omega^{li} \right) 
\left( \Omega^{kj} \omega^{ki} + \Omega^{ik} \omega^{kj} \right) 
\nn \\
&& - \left(S^{ik} \omega^{kj} - S^{jk} \omega^{ki} \right)    \dot{\omega}^{ij}  
+ S^{ij} \Omega^{lm} \omega^{li} \omega^{mj} 
- S^{ij}  \omega^{ki}  \dot{ \omega}^{kj}  
\nn\\
&& + \frac{1}{2} S^{ij} \left( \Omega^{kj} \omega^{kl} \omega^{li} 
+ \Omega^{ik} \omega^{kl} \omega^{lj} \right)- \frac{1}{2} S^{ij} 
\left( \dot{\omega}^{ik} \omega^{kj} + \omega^{ik} \dot{\omega}^{kj}  \right)  
+ \mathcal{O} (\omega^3)
\nn\\
&=& \dot{S}^{ij} \omega^{ij} +   S^{ij} \dot{\omega}^{ik} \omega^{kj}   
+ \left( \Delta_2 S^{ij} + S^{ik} \omega^{jl} \omega^{kl} - S^{kl} \omega^{ik} 
\omega^{jl} \right) \Omega^{ij}   + \mathcal{O} (\omega^3),
\eea
where we used that $\Lambda^{ik} \Lambda^{jk} = \delta^{ij}$, and dropped total time 
derivatives.
Therefore, to eliminate dependence in $\Omega^{ij}$ from the shift of the kinetic term, one should 
fix:
\bea
\label{spinshiftquad}
\Delta_2 S^{ij} = S^{kl} \omega^{ik} \omega^{jl} - \frac{1}{2} \left( S^{ik} 
\omega^{jl}  - S^{jk} \omega^{il} \right)\omega^{kl}. 
\eea
Thus the new spin redefinition to quadratic order is also fixed from the Lorentz 
shift. Note however that here we have a new feature: Beyond the spin shift in 
eq.~\eqref{spinshiftquad} that affects spin-dependent potentials through the spin 
variables, we find in eq.~\eqref{spinshifttoquad} a new addition to the 
spin-dependent potentials, originating from the kinetic term:
\be
\label{addpotfromkin}
\Delta V_{\dot{\omega}\omega} = \frac{1}{4}   S^{ij} 
\big(\dot{\omega}^{ik} \omega^{kj} - \omega^{ik} \dot{\omega}^{kj}\big). 
\ee
To recap, we obtained the generic redefinition of spin variables in 
eq.~\eqref{spinshiftquad}, and we discovered a new addition to the spin potentials in 
eq.~\eqref{addpotfromkin}, once one goes beyond linear order in the application of redefinition 
of rotational variables.

\subsection{Redefinition of Position and Spin}
\label{redefinitions}

As noted we now need to take into account all the contributions that arise via 
redefinitions applied in lower-order sectors, before we fix the new redefinition 
of variables that is needed to reduce our newly computed action from section 
\ref{rawaction}. 
The reduction is carried out consecutively over the sectors in table \ref{overallredef}, according to 
their actual PN order, see \cite{Levi:2019kgk} for our PN-counting formula for general $(n,l)$ 
sectors. 
Tables \ref{los1redef}--\ref{n3los1redef} summarize the build-up of redefinitions 
that are applied at each of the $6$ relevant sectors as noted in table \ref{overallredef}, 
namely the $5$ at lower orders, and the present one. 
Each of the tables \ref{los1redef}--\ref{n3los1redef} specifies from which 
sector the redefinition arose, and to which sector it is applied, so as to yield a 
contribution to the sector under consideration. 
Note that in all these tables only the left column involves new redefinitions that are newly defined 
in the sector under consideration, while all the rest are redefinitions that were already fixed in 
lower-order sectors. 

In sectors where redefinitions of both the position and rotational variables are  required, we present 
here the redefinitions, such that at each sector they were carried out first for the position, and then 
for the rotational variables. 
One can of course carry out the redefinition of rotational variables first, then of the position, in 
which case the redefinitions would be modified, albeit eventually leading to physically equivalent 
results. 
The redefinitions of position or rotational variables are fixed at each sector by iteratively 
eliminating terms with the most higher-order time derivatives at each step, until no terms with 
higher-order time derivatives are left \cite{Levi:2014sba}. 
After all the redefinitions for the sector are fixed, they are summed, and can all be applied  
at once at higher-order sectors. 
To recap, the reduction of higher-order time derivatives from the action 
is intricate. 
Yet once it is streamlined in an automated algorithm, its run-time 
over all sectors is quite rapid. 

Let us then present according to the above 
procedure all redefinitions over the relevant sectors in order. 
\begin{table}[t]
\begin{center}
\begin{tabular}{|l|c|}
\hline
\backslashbox{from}{\boldmath{to}} 
& (0P)N 
\\
\hline
LO S$^1$ & $\Delta \vec{x}$
\\
\hline
\end{tabular}
\caption{Contribution to the LO spin-orbit sector 
from position shifts in a lower-order sector.}
\label{los1redef}
\end{center}
\end{table}
First, for the LO spin-orbit sector at the $1.5$PN order, our unreduced action (our unreduced 
potentials are all computed directly with the \texttt{EFTofPNG} code), and the position shift as noted 
in table \ref{los1redef}, are identical to those we presented in \cite{Levi:2015msa}. 
Next, we proceed to the $2$PN sector, where our unreduced potential  can be expressed as:
\bea
V_{\text{2PN}} = \sum_{i=0}^2 \stackrel{(i)}{V}_{2,0}, 
\eea
with $n,l$ in the subscript as the sector's indices $(n,l)$, introduced in
table \ref{overallredef}, and with: 
\bea
\stackrel{(0)}{V}_{2,0}&=&- 	\frac{1}{16} m_{1} v_{1}^{6} - 	\frac{1}{16} m_{2} v_{2}^{6} \nn\\ && +  	\frac{G m_{1} m_{2}}{8 r} \Big[ 10 v_{1}^2 \vec{v}_{1}\cdot\vec{v}_{2} - 3 v_{1}^2 v_{2}^2 + 10 \vec{v}_{1}\cdot\vec{v}_{2} v_{2}^2 - 2 ( \vec{v}_{1}\cdot\vec{v}_{2})^{2} - 7 v_{1}^{4} - 7 v_{2}^{4} \nn\\ 
&& + 6 \vec{v}_{1}\cdot\vec{n} v_{1}^2 \vec{v}_{2}\cdot\vec{n} - 12 \vec{v}_{1}\cdot\vec{n} \vec{v}_{2}\cdot\vec{n} \vec{v}_{1}\cdot\vec{v}_{2} + 6 \vec{v}_{1}\cdot\vec{n} \vec{v}_{2}\cdot\vec{n} v_{2}^2 + v_{2}^2 ( \vec{v}_{1}\cdot\vec{n})^{2} \nn\\ 
&& + v_{1}^2 ( \vec{v}_{2}\cdot\vec{n})^{2} -3 ( \vec{v}_{1}\cdot\vec{n})^{2} ( \vec{v}_{2}\cdot\vec{n})^{2} \Big]\nn\\ && - 	\frac{G^2 m_{1} m_{2}{}^2}{4 r{}^2} \Big[ 7 v_{1}^2 - 14 \vec{v}_{1}\cdot\vec{v}_{2} + 8 v_{2}^2 + 2 ( \vec{v}_{1}\cdot\vec{n})^{2} \Big] - 	\frac{G^2 m_{1}{}^2 m_{2}}{4 r{}^2} \Big[ 8 v_{1}^2 - 14 \vec{v}_{1}\cdot\vec{v}_{2} \nn\\ 
&& + 7 v_{2}^2 + 2 ( \vec{v}_{2}\cdot\vec{n})^{2} \Big]\nn\\ && - 	\frac{G^3 m_{1}{}^3 m_{2}}{2 r{}^3} - 	\frac{3 G^3 m_{1}{}^2 m_{2}{}^2}{r{}^3} - 	\frac{G^3 m_{1} m_{2}{}^3}{2 r{}^3},
\eea
where the kinetic term is included, and:
\bea
\stackrel{(1)}{V}_{2,0}&=& - 	\frac{1}{8} G m_{1} m_{2} \Big[ 12 \vec{v}_{1}\cdot\vec{a}_{1} \vec{v}_{2}\cdot\vec{n} - 14 \vec{v}_{2}\cdot\vec{n} \vec{a}_{1}\cdot\vec{v}_{2} + \vec{a}_{1}\cdot\vec{n} v_{2}^2 - v_{1}^2 \vec{a}_{2}\cdot\vec{n} + 14 \vec{v}_{1}\cdot\vec{n} \vec{v}_{1}\cdot\vec{a}_{2} \nn\\ 
&& - 12 \vec{v}_{1}\cdot\vec{n} \vec{v}_{2}\cdot\vec{a}_{2} + \vec{a}_{2}\cdot\vec{n} ( \vec{v}_{1}\cdot\vec{n})^{2} - \vec{a}_{1}\cdot\vec{n} ( \vec{v}_{2}\cdot\vec{n})^{2} \Big],
\eea
\bea
\stackrel{(2)}{V}_{2,0}&=&- 	\frac{1}{8} G m_{1} m_{2} r \Big[ 15 \vec{a}_{1}\cdot\vec{a}_{2} - \vec{a}_{1}\cdot\vec{n} \vec{a}_{2}\cdot\vec{n} \Big].
\eea

\begin{table}[t]
\begin{center}
\begin{tabular}{|l|c|}
\hline
\backslashbox{from}{\boldmath{to}} 
& (0P)N 
\\
\hline
2PN & $\Delta \vec{x}$
\\
\hline
\end{tabular}
\caption{Contribution to the $2$PN sector 
from position shifts in a lower-order sector.}
\label{2pnredef}
\end{center}
\end{table}

\noindent The position shift as noted in table \ref{2pnredef} is then fixed as:
\bea
\left(\Delta \vec{x}_1\right)_{\text{2PN}} &=&  \frac{1}{8} G m_{2} \Big[ v_{2}^2 \vec{n} + 12 \vec{v}_{2}\cdot\vec{n} \vec{v}_{1} - 14 \vec{v}_{2}\cdot\vec{n} \vec{v}_{2} - ( \vec{v}_{2}\cdot\vec{n})^{2} \vec{n} \Big] \nn\\ &&+  	\frac{7 G^2 m_{1} m_{2}}{8 r} \vec{n}\nn\\ && + 	\frac{1}{16} G m_{2} r \Big[ 15 \vec{a}_{2} - \vec{a}_{2}\cdot\vec{n} \vec{n} \Big].
\eea

\begin{table}[t]
\begin{center}
\begin{tabular}{|l|c|c|}
\hline
\backslashbox{from}{\boldmath{to}} 
& (0P)N & 1PN
\\
\hline
LO S$^1$ &  & $\Delta \vec{x}$
\\
\hline
NLO S$^1$ & $\Delta \vec{x}$, $\Delta \vec{S}$ &
\\
\hline
\end{tabular}
\caption{Contributions to the NLO spin-orbit sector 
from position shifts and spin redefinition in lower-order sectors.}
\label{nlos1redef}
\end{center}
\end{table}

Next, we proceed to the NLO spin-orbit sector at the $2.5$PN order, where our unreduced potential is 
identical to that we presented in \cite{Levi:2015msa}. As noted in table \ref{nlos1redef}, in this 
sector new redefinitions of both position and rotational variables are required. The position shift 
is fixed here as:
\bea
\left(\Delta \vec{x}_1\right)_{\text{NLO}}^{\text{SO}} &=& 	\frac{1}{8 m_{1}} v_{1}^2 \vec{S}_{1}\times\vec{v}_{1} \nn\\ && - 	\frac{G m_{2}}{2 m_{1} r} \Big[ 2 \vec{S}_{1}\times\vec{n} \vec{v}_{2}\cdot\vec{n} - \vec{S}_{1}\times\vec{v}_{1} + 6 \vec{S}_{1}\times\vec{v}_{2} \Big] \nn\\ 
&&- 	\frac{G}{4 r} \Big[ \vec{S}_{2}\times\vec{n}\cdot\vec{v}_{2} \vec{n} + 4 \vec{S}_{2}\times\vec{n} \vec{v}_{2}\cdot\vec{n}  - 11 \vec{S}_{2}\times\vec{v}_{2} \Big]\nn\\ && - 	G \dot{\vec{S}}_{2}\times\vec{n},
\eea
whereas the spin redefinition at this sector is identical to that we presented in \cite{Levi:2015msa}:
\bea
\label{leadingomega}
(\omega^{ij}_1)_{\text{NLO}}^{\text{SO}} &=&- 	\frac{G m_{2}}{r} \Big[ 3 v_{2}^i v_{1}^j + \vec{v}_{2}\cdot\vec{n} n^i v_{1}^j - \vec{v}_{2}\cdot\vec{n} n^i v_{2}^j - \left(i \leftrightarrow j \right) \Big]. 
\eea

We proceed to consider the $3$PN sector. Our unreduced potential can be expressed as:
\bea
V_{\text{3PN}} = \sum_{i=0}^4 \stackrel{(i)}{V}_{3,0}, 
\label{eq:3pnds}
\eea
where the explicit pieces are provided in appendix \ref{n3los1redefapp}.
\begin{table}[t]
\begin{center}
\begin{tabular}{|l|c|c|}
\hline
\backslashbox{from}{\boldmath{to}} 
& (0P)N & 1PN
\\
\hline
2PN & & $\Delta \vec{x}$ 
\\
\hline
3PN & $\Delta \vec{x}$  & 
\\
\hline
\end{tabular}
\caption{Contributions to the $3$PN sector 
from position shifts in lower-order sectors.}
\label{3pnredef}
\end{center}
\end{table}
Notice there, that now DimReg poles with logarithms show up in pieces with and without higher-order 
time derivatives. 
Accordingly, the new redefinitions at this sector, as noted in table \ref{3pnredef}, are expected to 
also include such poles, and they should be applied on the Newtonian potential, including its piece that is linear in the DimReg zero, which reads:
\bea
\Delta V_\text{N} &=& -\epsilon \, \frac{G m_1 m_2}{r} \left[ \frac{1}{2} - \log \left( \frac{ r}{R_0} \right) \right]. 
\eea
Note that this extra piece also contains a logarithm.
Now, we also add to the unreduced $3$PN potential the following total time derivative (TTD):
\bea
\Delta V_{\text{$3$PN}}^{\text{TTD}}=\frac{d}{dt} \left[    \frac{ G^3 m_1^3 m_2  }{3 r^2} \left(   3\vn \cdot \vv_1 -2 \vn \cdot \vv_2 \right) \left( \frac{1}{\epsilon} - 3 \log \frac{r}{R_0} \right) \right] + (1 \leftrightarrow 2),
\label{3pnttd}
\eea
in order to ensure that we land on a reduced $3$PN potential that contains neither poles nor 
logarithms, after the redefinition of position. The new one that we fix at this sector reads:
\bea
\left(\Delta \vec{x}_1\right)_{\text{3PN}} &=& \sum_{i=0}^3  \stackrel{(i)}{ \Delta \vec{x}_1}_{(3,0)},
\eea
with the explicit pieces provided in appendix \ref{n3los1redefapp}.

We proceed to the N$^2$LO spin-orbit sector at the $3.5$PN order. Our unreduced potential can be 
expressed as:
\bea
V_{\text{N}^2\text{LO}}^{\text{SO}} = \sum_{i=0}^4 \stackrel{(i)}{V}_{2,1} + (1 \leftrightarrow 2),
\eea
with the explicit pieces provided in appendix \ref{n3los1redefapp}.
\begin{table}[t]
\begin{center}
\begin{tabular}{|l|c|c|c|c|}
\hline
\backslashbox{from}{\boldmath{to}} 
& (0P)N & 1PN & LO S$^1$ & 2PN 
\\
\hline
LO S$^1$ &  &  & & $\Delta \vec{x}$ 
\\
\hline
2PN &  &  & $\Delta \vec{x}$ &  
\\
\hline
NLO S$^1$ &  & $\Delta \vec{x}$ & $\Delta \vec{S}$ &
\\
\hline
N$^2$LO S$^1$ & $\Delta \vec{x}$, $\Delta \vec{S}$ & & &
\\
\hline
\end{tabular}
\caption{Contributions to the N$^2$LO spin-orbit sector 
from position shifts and spin redefinitions in lower-order sectors.}
\label{n2los1redef}
\end{center}
\end{table}
As noted in table \ref{n2los1redef}, the new redefinitions in the N$^2$LO spin-orbit sector for both 
the position and spin variables can be expressed as:
\bea
\left(\Delta \vec{x}_1\right)_{\text{N}^2\text{LO}}^{\text{SO}} &=& \sum_{i=0}^3  \stackrel{(i)}{ \Delta \vec{x}_1}_{(2,1)},\\
(\omega^{ij})_{\text{N}^2\text{LO}}^{\text{SO}} &=& \sum_{k=0}^1  \stackrel{(k)}{ \omega^{ij}_1}_{(2,1)} - (i \leftrightarrow j) ,
\eea
with the explicit pieces provided in appendix \ref{n3los1redefapp}.

\begin{table}[t]
\begin{center}
\begin{tabular}{|l|c|c|c|c|c|c|}
\hline
\backslashbox{from}{\boldmath{to}} 
& (0P)N & 1PN & LO S$^1$ & 2PN &  NLO S$^1$ & 3PN
\\
\hline
LO S$^1$ & & & &  & & $\Delta \vec{x}$
\\
\hline
2PN & & & & & $\Delta \vec{x}$ &  
\\
\hline
NLO S$^1$ & $(\Delta \Lambda)^2 $ & &  & $\Delta \vec{x}$ & $ \Delta \vec{S}$ & 
\\
\hline
3PN & & & $\Delta \vec{x}$ & & & 
\\
\hline
N$^2$LO S$^1$ & & $\Delta \vec{x}$ & $\Delta \vec{S}$ & & & 
\\
\hline
N$^3$LO S$^1$ & $\Delta \vec{x}$, $\Delta \vec{S}$ & & & & &
\\
\hline
\end{tabular}
\caption{Contributions to the N$^3$LO spin-orbit sector 
from position shifts and spin redefinitions in lower-order sectors.}
\label{n3los1redef}
\end{center}
\end{table}

Finally, we arrive at the present N$^3$LO spin-orbit sector at the $4.5$PN order. As we already noted 
in our unreduced action in section \ref{rawaction}, there are DimReg poles with logarithms that show up 
in our unreduced action in pieces with and without higher-order time derivatives. 
Accordingly, the new redefinitions at the present sector, as noted in table \ref{n3los1redef}, are also 
expected to include such poles, and thus similar to the $3$PN sector (without spins), these 
redefinitions should be applied on the LO spin-orbit potential, including its piece that is linear in 
the DimReg zero, which also contains a logarithm, that reads:
\bea
\Delta V_{\text{LO}}^{\text{SO}} &=& 
\epsilon \Bigg[\frac{2 G m_2 (\hat{n} \times (\vec{v}_1 - \vec{v}_2)) \cdot \vec{S}_1}{r^2} \left( 1 - \log \left( \frac{ r}{R_0} \right) \right) + (1 \leftrightarrow 2) \Bigg].
\eea
Also similar to the $3$PN sector, we also add in advance to the present unreduced potential a TTD:
\bea
\Delta V_{\text{N$^3$LO:SO}}^{\text{TTD}}
&=& \frac{d}{dt} \Bigg[\left(\frac{1}{\epsilon} - 3 \log \frac{r}{R_0} \right) 
\bigg ( -\frac{2 G^3 m_{2}{}^3}{15 r{}^3} \Big[ 3 \vec{S}_{1}\times\vec{n}\cdot\vec{v}_{1} \big( 5 \vec{v}_{1}\cdot\vec{n} - 12 \vec{v}_{2}\cdot\vec{n} \big)  \nn\\ 
&& + 24 \vec{S}_{1}\times\vec{n}\cdot\vec{v}_{2} \big( \vec{v}_{1}\cdot\vec{n} 
+ 7 \vec{v}_{2}\cdot\vec{n} \big) - 20 \vec{S}_{1}\times\vec{v}_{1}\cdot\vec{v}_{2} \Big] \nn\\ 
&& - \frac{G^3 m_{1}{}^2 m_{2}}{30 r{}^3} \Big[ 3 \vec{S}_{1}\times\vec{n}\cdot\vec{v}_{1} \big( 157 \vec{v}_{1}\cdot\vec{n} + 79 \vec{v}_{2}\cdot\vec{n} \big) \nn\\ 
&&  \phantom{\frac{ G^3 }{5 r{}^3}}- 12 \vec{S}_{1}\times\vec{n}\cdot\vec{v}_{2} \big( 19 \vec{v}_{1}\cdot\vec{n} + \vec{v}_{2}\cdot\vec{n} \big) + 155 \vec{S}_{1}\times\vec{v}_{1}\cdot\vec{v}_{2} \Big] \bigg)  \Bigg] \nn\\
&& + (1 \leftrightarrow 2 ),
\label{n3los1ttd}
\eea
which leads -- after the redefinition of variables that we fix at this sector -- to 
a reduced potential that is free of poles and logarithms. 

Now, as noted in table \ref{n3los1redef}, we also need to recall the result of section 
\ref{spinredefextend}, where we extended the redefinition of rotational variables beyond linear order, 
and found a new addition to the spin potentials in eq.~\eqref{addpotfromkin}, which scales as 
$S\dot{\omega}\omega$. 
Examining the leading redefinition with $\omega$, that appears in eq.~\eqref{leadingomega}, we see that 
it scales as $v^4$, and thus we find that it gives rise to a new addition in the present sector, which 
contributes the following piece to the unreduced potential:
\bea
\Delta V_{\text{N$^3$LO:SO}}^{\dot{\omega}\omega} &=&- 	\frac{3 G^2 m_{2}{}^2}{2 r{}^3} \Big[ \vec{S}_{1}\times\vec{n}\cdot\vec{v}_{1} \big( 2 \vec{v}_{1}\cdot\vec{v}_{2} v_{2}^2 - ( \vec{v}_{1}\cdot\vec{v}_{2})^{2} - v_{2}^{4} + 2 \vec{v}_{1}\cdot\vec{n} \vec{v}_{2}\cdot\vec{n} \vec{v}_{1}\cdot\vec{v}_{2} \nn\\ 
&& - 2 \vec{v}_{1}\cdot\vec{n} \vec{v}_{2}\cdot\vec{n} v_{2}^2 - 2 \vec{v}_{1}\cdot\vec{v}_{2} ( \vec{v}_{2}\cdot\vec{n})^{2} + 2 v_{2}^2 ( \vec{v}_{2}\cdot\vec{n})^{2} \big) + \vec{S}_{1}\times\vec{n}\cdot\vec{v}_{2} \big( v_{1}^2 \vec{v}_{1}\cdot\vec{v}_{2}  \nn\\ 
&& - v_{1}^2 v_{2}^2 + \vec{v}_{1}\cdot\vec{v}_{2} v_{2}^2 - ( \vec{v}_{1}\cdot\vec{v}_{2})^{2} -2 \vec{v}_{1}\cdot\vec{n} v_{1}^2 \vec{v}_{2}\cdot\vec{n} + 2 \vec{v}_{1}\cdot\vec{n} \vec{v}_{2}\cdot\vec{n} \vec{v}_{1}\cdot\vec{v}_{2}  \nn\\ 
&& + 2 v_{1}^2 ( \vec{v}_{2}\cdot\vec{n})^{2} - 2 \vec{v}_{1}\cdot\vec{v}_{2} ( \vec{v}_{2}\cdot\vec{n})^{2} \big) - \vec{S}_{1}\times\vec{v}_{1}\cdot\vec{v}_{2} \big( \vec{v}_{1}\cdot\vec{n} \vec{v}_{1}\cdot\vec{v}_{2} - \vec{v}_{2}\cdot\vec{n} \vec{v}_{1}\cdot\vec{v}_{2} \nn\\ 
&&  - \vec{v}_{1}\cdot\vec{n} v_{2}^2 + \vec{v}_{2}\cdot\vec{n} v_{2}^2 -2 \vec{v}_{2}\cdot\vec{n} ( \vec{v}_{1}\cdot\vec{n})^{2} + 4 \vec{v}_{1}\cdot\vec{n} ( \vec{v}_{2}\cdot\vec{n})^{2} - 2 ( \vec{v}_{2}\cdot\vec{n})^{3} \big) \Big]\nn\\
&&+ 	\frac{G^2 m_{2}{}^2}{2 r{}^2} \Big[ \vec{S}_{1}\times\vec{n}\cdot\vec{v}_{1} \big( 3 \vec{v}_{2}\cdot\vec{n} \vec{a}_{1}\cdot\vec{v}_{2} + 3 \vec{v}_{1}\cdot\vec{v}_{2} \vec{a}_{2}\cdot\vec{n} - 3 v_{2}^2 \vec{a}_{2}\cdot\vec{n} - 3 \vec{v}_{2}\cdot\vec{n} \vec{v}_{1}\cdot\vec{a}_{2} \nn\\ 
&& + \vec{a}_{1}\cdot\vec{n} ( \vec{v}_{2}\cdot\vec{n})^{2} - \vec{a}_{2}\cdot\vec{n} ( \vec{v}_{2}\cdot\vec{n})^{2} \big) - \vec{S}_{1}\times\vec{n}\cdot\vec{a}_{1} \big( 3 \vec{v}_{2}\cdot\vec{n} \vec{v}_{1}\cdot\vec{v}_{2} - 3 \vec{v}_{2}\cdot\vec{n} v_{2}^2 \nn\\ 
&& + \vec{v}_{1}\cdot\vec{n} ( \vec{v}_{2}\cdot\vec{n})^{2} - ( \vec{v}_{2}\cdot\vec{n})^{3} \big) + \vec{S}_{1}\times\vec{v}_{1}\cdot\vec{a}_{1} \big( 9 v_{2}^2 + 7 ( \vec{v}_{2}\cdot\vec{n})^{2} \big) \nn\\ 
&& - \vec{S}_{1}\times\vec{n}\cdot\vec{v}_{2} \big( 3 \vec{v}_{2}\cdot\vec{n} \vec{a}_{1}\cdot\vec{v}_{2} + 3 v_{1}^2 \vec{a}_{2}\cdot\vec{n} - 3 \vec{v}_{1}\cdot\vec{v}_{2} \vec{a}_{2}\cdot\vec{n} - 3 \vec{v}_{2}\cdot\vec{n} \vec{v}_{1}\cdot\vec{a}_{2} \nn\\ 
&& + \vec{a}_{1}\cdot\vec{n} ( \vec{v}_{2}\cdot\vec{n})^{2} - \vec{a}_{2}\cdot\vec{n} ( \vec{v}_{2}\cdot\vec{n})^{2} \big) - 3 \vec{S}_{1}\times\vec{v}_{1}\cdot\vec{v}_{2} \big( 3 \vec{a}_{1}\cdot\vec{v}_{2} - 3 \vec{v}_{1}\cdot\vec{a}_{2}  \nn\\ 
&& + \vec{a}_{1}\cdot\vec{n} \vec{v}_{2}\cdot\vec{n} - \vec{v}_{1}\cdot\vec{n} \vec{a}_{2}\cdot\vec{n} \big) + \vec{S}_{1}\times\vec{a}_{1}\cdot\vec{v}_{2} \big( 9 \vec{v}_{1}\cdot\vec{v}_{2} + 3 \vec{v}_{1}\cdot\vec{n} \vec{v}_{2}\cdot\vec{n} + 4 ( \vec{v}_{2}\cdot\vec{n})^{2} \big) \nn\\ 
&& + \vec{S}_{1}\times\vec{n}\cdot\vec{a}_{2} \big( 3 v_{1}^2 \vec{v}_{2}\cdot\vec{n} - 3 \vec{v}_{2}\cdot\vec{n} \vec{v}_{1}\cdot\vec{v}_{2} + \vec{v}_{1}\cdot\vec{n} ( \vec{v}_{2}\cdot\vec{n})^{2} - ( \vec{v}_{2}\cdot\vec{n})^{3} \big) \nn\\ 
&& - \vec{S}_{1}\times\vec{v}_{1}\cdot\vec{a}_{2} \big( 9 \vec{v}_{1}\cdot\vec{v}_{2} + 3 \vec{v}_{1}\cdot\vec{n} \vec{v}_{2}\cdot\vec{n} + 4 ( \vec{v}_{2}\cdot\vec{n})^{2} \big) + \vec{S}_{1}\times\vec{v}_{2}\cdot\vec{a}_{2} \big( 9 v_{1}^2 \nn\\ 
&& + 6 \vec{v}_{1}\cdot\vec{n} \vec{v}_{2}\cdot\vec{n} + ( \vec{v}_{2}\cdot\vec{n})^{2} \big) \Big]
+ \left( 1 \leftrightarrow 2 \right).
\eea
Finally, as noted in table \ref{n3los1redef}, the new redefinitions for both the position and spin 
variables, that are fixed in the present sector, can be written as:
\bea
\label{n3los1xredefs}
\left(\Delta \vec{x}_1\right)_{\text{N}^3\text{LO}}^{\text{SO}} &=& \sum_{i=0}^5  \stackrel{(i)}{ \Delta \vec{x}_1}_{(3,1)},\\
(\omega^{ij})_{\text{N}^3\text{LO}}^{\text{SO}} &=& \sum_{k=0}^2  \stackrel{(k)}{ \omega^{ij}_1}_{(3,1)} - (i \leftrightarrow j) ,
\eea
where the explicit pieces in these redefinitions are provided in appendix 
\ref{n3los1redefapp}. Notice there that both the position shifts and the spin 
redefinitions in the present sector contain pieces that are proportional to the DimReg poles. 
All of the ingredients of the reduction process, that was detailed in this section, are 
also provided in the ancillary files to this publication in machine-readable format.

\subsection{Generic Action}
\label{finalaction}

After we carry out the full procedure of reduction described above, we get a 
generic action, that is an action with no higher-order time derivatives 
beyond velocity and spin. 
It is provided in appendix \ref{explicitstdaction}, 
and in machine-readable format in the ancillary files to this publication.
Clearly this reduced action is much more compact than that which was initially obtained 
from the EFT computation, see section \ref{rawaction} and appendix \ref{unreduced}. 
This final action still displays a new important feature in the sectors with spin: 
The occurrence of a transcendental number -- the 
Riemann zeta value $\zeta(2)\propto \pi^2$ -- at orders $G^3$ and $G^4$. 

In contrast, we reached an action which is free from the DimReg poles and 
logarithms that showed up in intermediate stages. 
This is thanks to the $2$ TTDs in eqs.~\eqref{3pnttd}, \eqref{n3los1ttd}, that we added in 
advance to the unreduced actions of the $2$ relevant sectors at the N$^3$LO, which start 
exhibiting such poles.
Thus the DimReg poles, which are clearly unphysical, can be removed at the level of 
the action by adding TTDs to the unreduced (or reduced) action, as we illustrated here
prior to the reduction. 
Alternatively, one can just apply the full reduction procedure, and then use -- as is -- the 
reduced action with the unphysical DimReg poles (which would have appeared there at orders 
$G^3$ and $G^4$), for the extraction of physical observables. 
The latter would obviously not contain such poles nor 
logarithms, that drop out of physical observables.

As explained in \cite{Levi:2015msa} due to our generalized canonical gauge, the equations of motion 
(EOMs) for both the position and spin are now readily obtained via a variation of the action in 
eq.~\eqref{reducedwopole}. Specifically for the spin, the generic form of the EOMs from the potentials derived via our EFT of spinning objects \cite{Levi:2015msa} was provided there as:
\begin{align}
\label{Seom}
\dot{S}_a^{ij}=&
- 4 S_a^{k[i} \delta^{j]l}\frac{\delta\int{dt\,V}}{\delta S_a^{kl}}
=- 4 S_a^{k[i} \delta^{j]l} \left[\frac{\partial V}
{\partial S_a^{kl}} - \frac{d}{dt} \frac{\partial V}{\partial \dot{S}_a^{kl}} \right],
\end{align}
such that only a variation of the spin potentials with respect to the spin variables 
is required in order to get the precession equations.

\section{General Hamiltonian}
\label{hamilton}

From the generic action in eq.~\eqref{reducedwopole}, one can proceed to derive the 
full general Hamiltonian in a straightforward manner, where one should make a Legendre 
transform only with respect to the position variables. For this Legendre transform, we 
need to take into account all the reduced actions resulting from the reduction procedure 
detailed in section \ref{redefinitions}: In the point-mass sector from the Newtonian up 
to the $3$PN order, and in the spin-orbit sector up to the N$^3$LO. We then obtain the 
general Hamiltonian for the present sector, which is provided in appendix \ref{explicitgenham}, 
and in machine-readable format in the ancillary files to this publication.

\subsection{Simplified Hamiltonian}
\label{spechamilton}

Let us start by recalling some standard conventions useful to express simplified 
Hamiltonians, and subsequently various gauge-invariant relations, see also
\cite{Levi:2014sba} for more details. First we consider some quantities related with  
the masses of the binary, namely the total mass $m \equiv m_1+m_2$, the 
mass ratio $q \equiv m_1/m_2$, the reduced mass $\mu \equiv m_1 m_2 / m$, and the 
symmetric mass ratio $\nu \equiv m_1 m_2/m^2 = \mu/m = q/(1+q)^2$.

To simplify the general Hamiltonian it is customary to first specify to the center-of-mass 
(COM) frame where $\vec{p} \equiv \vec{p}_1 = - \vec{p}_2$. The COM specification 
actually results in the major simplification of general Hamiltonians.
We then rescale all variables to become dimensionless:
\begin{equation} 
\tilde{H} \equiv \frac{H}{\mu} , \qquad
\tilde{r} \equiv \frac{r}{G m} , \qquad
\tilde{L} \equiv \frac{L}{G m \mu}, \qquad 
\tilde{S}_a \equiv \frac{S_a}{G m \mu},
\end{equation}
with the orbital angular momentum $\vec{L} \equiv r \vec{n} \times \vec{p}$.
The simplified Hamiltonian in the COM frame then reads:
\bea
\label{hcom}
\tilde{H}_{\text{N}^3\text{LO}}^{\text{SO}}& = &
\frac{\nu  }{\tilde{r}^6} \tilde{\vec{L}} \cdot \left[
\left( \tilde{\vec{S}}_1 + \tilde{\vec{S}}_2 \right)
\left(\left(-\frac{79867}{1200}-\frac{27 \pi ^2}{2}\right) \nu ^2+\left(\frac{479 \pi ^2}{24}-\frac{752881}{3600}\right) \nu -\frac{1497}{50} \phantom{\frac{\tilde{L}^0}{\tilde{r^0}}}\right. \right.\nn\\
&&+ \frac{\tilde{L}^2}{\tilde{r}} \left(  \left(\frac{199109}{2400}+\frac{27 \pi ^2}{2}\right) \nu ^2+\left(\frac{254077}{1200}-\frac{675 \pi ^2}{64}\right) \nu -\frac{3}{50} \right)\nn\\
&&+\frac{\tilde{L}^4}{\tilde{r}^2} \left(-\frac{29 \nu ^3}{32}-\frac{263 \nu ^2}{16}+\frac{83 \nu }{32} \right)  +\frac{\tilde{L}^6}{\tilde{r}^3} \left( \frac{151 \nu ^3}{128}-\frac{101 \nu ^2}{32}+\frac{109 \nu }{128} \right)\nn\\
&&  +\tilde{p}_r^2 \tilde{r}  \left( \left(-\frac{635261}{2400}-54 \pi ^2\right) \nu ^2+\left(\frac{675 \pi ^2}{16}-\frac{296233}{1200}\right) \nu +\frac{6}{25}\phantom{ \frac{\tilde{L}^2}{\tilde{r}}} \right. \nn\\
&&\left.+ \frac{\tilde{L}^2}{\tilde{r}}\left( -\frac{17 \nu ^3}{4}-\frac{1695 \nu ^2}{32}+\frac{815 \nu }{32}  \right) + \frac{\tilde{L}^4}{\tilde{r}^2}\left(\frac{777 \nu ^3}{128}-\frac{387 \nu ^2}{32}+\frac{471 \nu }{128}  \right) \right)  \nn\\
&&  +\tilde{p}_r^4 \tilde{r}^2 \left(  -\frac{263 \nu ^3}{32}-\frac{705 \nu ^2}{8}+\frac{1091 \nu }{32}+\frac{\tilde{L}^2}{\tilde{r}}\left(  \frac{1671 \nu ^3}{128}-\frac{471 \nu ^2}{32}+\frac{375 \nu }{128} \right) \right) \nn\\
&& +\tilde{p}_r^6 \tilde{r}^3 \left( \frac{2025 \nu ^3}{128}-\frac{325 \nu ^2}{32}+\frac{153 \nu }{128} \right) \Bigg) \nn\\
&&
+\left(\tilde{\vec{S}}_1/q +\tilde{\vec{S}}_2 q \right)
\left( \left(-\frac{79867}{1200}-\frac{27 \pi ^2}{2}\right) \nu ^2+\left(\frac{859 \pi ^2}{64}-\frac{62191}{480}\right) \nu -\frac{303}{8}\phantom{\frac{\tilde{L}^0}{\tilde{r^0}}}\right.\nn\\
&& + \frac{\tilde{L}^2}{\tilde{r}} \left(\left(\frac{188609}{2400}+\frac{27 \pi ^2}{2}\right) \nu ^2+\left(\frac{20887}{160}-\frac{771 \pi ^2}{128}\right) \nu +\frac{63}{16} \right)\nn\\
&&+\frac{\tilde{L}^4}{\tilde{r}^2} \left( -\frac{29 \nu ^3}{32}-\frac{297 \nu ^2}{32}+\frac{39 \nu }{16}-\frac{41}{16} \right)   + \frac{\tilde{L}^6}{\tilde{r}^3}\left( \frac{29 \nu ^3}{32}-\frac{123 \nu ^2}{32}+\frac{151 \nu }{64}-\frac{45}{128} \right) \nn\\
&&  +\tilde{p}_r^2 \tilde{r}  \left( \left(-\frac{20248}{75}-54 \pi ^2\right) \nu ^2+\left(\frac{771 \pi ^2}{32}-\frac{5683}{160}\right) \nu -\frac{1001}{16}  
\phantom{ \frac{\tilde{L}^2}{\tilde{r}}} \right. \nn\\
&& + \frac{\tilde{L}^2}{\tilde{r}}\left( -\frac{17 \nu ^3}{4}-\frac{377 \nu ^2}{16}+\frac{1843 \nu }{32}-\frac{63}{8} \right) 
+ \frac{\tilde{L}^4}{\tilde{r}^2}\left(\frac{153 \nu ^3}{32}-\frac{531 \nu ^2}{32}+\frac{525 \nu }{64}-\frac{135}{128}  \right) \Bigg)  \nn\\
&&  +\tilde{p}_r^4 \tilde{r}^2 \Bigg( -\frac{263 \nu ^3}{32}-\frac{1401 \nu ^2}{16}+\frac{2391 \nu }{32}-\frac{85}{16} \nn \\
&&+\frac{\tilde{L}^2}{\tilde{r}}\left(  \frac{339 \nu ^3}{32}-\frac{1491 \nu ^2}{64}+\frac{597 \nu }{64}-\frac{135}{128} \right) \Bigg)\nn\\
&& +\tilde{p}_r^6 \tilde{r}^3 \left( \frac{425 \nu ^3}{32}-\frac{745 \nu ^2}{64}+\frac{223 \nu }{64}-\frac{45}{128} \right) \Bigg) \Bigg].
\eea
Note that the new addition to the potential from eq.~\eqref{addpotfromkin} vanishes 
in the COM frame, and thus does not affect this simplified Hamiltonian. 

Second, it is also common to make the simplifying assumption that the spins are aligned 
with the orbital angular momentum, namely that the conditions
$\vec{S}_a \cdot \vec{n} = \vec{S}_a \cdot \vec{p}=0$ hold for both spins.
Yet for the spin-orbit sector there is only a single spin in the interaction, which only 
appears in the Hamiltonian in the overall coupling $S_a\cdot L$ as can be seen in eq.~\eqref{hcom}.  
Therefore the aligned-spins constraints do not affect the spin-orbit COM Hamiltonian, and yield no 
further loss of information for generic spin orientations. 
For this reason, in the simple spin-orbit sector, results obtained for the aligned-spins configuration 
can be trivially extended to generic spin orientations.

At this point we remark that a simplified EOB Hamiltonian was reconstructed in 
\cite{Antonelli:2020ybz} via some ansatz, assumptions from the EOB approach, and 
available results from self-force theory. That EOB Hamiltonian is similar to our 
simplified Hamiltonian in eq.~\eqref{hcom} in that they are both specified to the 
COM frame, and trivially satisfy the aligned-spins constraints, by virtue of the 
simplicity of the spin-orbit coupling. 
Yet, the simplified EOB Hamiltonian in \cite{Antonelli:2020ybz} differs from ours:  
It is based on some EOB ansatz, namely it is not fixed from theory, 
and it is specified to the so-called ``quasi-isotropic'' gauge, where dependence 
in factors of $L^2$ is hidden, and which is not clearly well-defined.

Finally, an additional common simplification of the Hamiltonian in eq.~\eqref{hcom} is 
to specify it to circular orbits, with the necessary condition: 
$p_r \equiv \vec{p} \cdot \vec{n}=0 \Rightarrow p^2=p_r^2+L^2/r^2\to L^2/r^2$.
Notice that this assumption of circular orbits, namely of a constant orbital 
separation, makes such a simplified Hamiltonian particularly useful in the 
inspiral phase, but not as useful when the binary is approaching merger, and the PN 
approximation breaks down.

\section{Gauge-Invariant Relations}
\label{gi}

Even when we implement the aforementioned necessary condition for circular orbits, the 
resulting simplified Hamiltonians are still gauge-dependent since they depend on the 
radial coordinate. To eliminate the latter, we solve for the additional 
condition for circular orbits: 
$\dot{p}_r=-\partial\tilde{H}(\tilde{r},\tilde{L})/\partial\tilde{r}=0$, to obtain 
$\tilde{r}(\tilde{L})$. We can then substitute this relation back in the Hamiltonian to get the 
gauge-invariant binding energy $e\equiv\tilde{H}$ in terms of the gauge-invariant 
angular momentum. In this manner, we obtain a gauge-invariant relation between the 
binding energy and the angular momentum, due to the present sector:
\bea
\label{etoL}
(e)^{\text{N}^3\text{LO}}_{\text{SO}} (\tilde{L}) &= &\frac{\nu}{\tilde{L}^{11}} 
\left[\left( \frac{35 \nu ^3}{128}+\frac{369 \nu ^2}{64}
+\left(\frac{3193 \pi ^2}{192}-\frac{1104095}{1152}\right) \nu +1512  \right) 
\left( \tilde{S}_1 + \tilde{S}_2 \right) \right.\nn\\
&&\left.+\left( 6 \nu ^2+\left(\frac{205 \pi ^2}{16}-\frac{71021}{96}\right) \nu^1 
+\frac{123363  }{128} \right) \left( \tilde{S}_1/q + \tilde{S}_2 q\right)\right].
\eea
This relation is a critical tool for comparing and evaluating the performance of 
different analytical and numerical descriptions of the binary dynamics.

We can proceed to compute the angular momentum as a function of the orbital 
frequency. From the latter it is useful to define the gauge-invariant PN 
parameter $x \equiv \tilde{\omega}^{2/3}$,
inferred from Kepler's third law in the Newtonian limit, such that $x^{-1}$ measures
the orbital separation. Yet, since the latter is now expressed in terms of the 
observable emitted frequency, it is gauge-invariant. We can compute this
frequency from Hamilton's equation for the orbital phase: 
$d\phi/d\tilde{t}\equiv\tilde{\omega}
=\partial\tilde{H}(\tilde{r},\tilde{L})/\partial\tilde{L}=0$. 
Then, using the definition of the parameter $x$, and the relation we already found, 
$\tilde{r}(\tilde{L})$, we obtain $x(\tilde{L})$, and inverting it we get the 
relation between the angular momentum and the frequency, due to the 
contribution of the present sector:
\bea
\label{Ltox}
\frac{1}{\tilde{L}^2} &\supset& x^{11/2} \nu \left[ \left( 
-\frac{12439 \nu ^3}{15552}+\frac{17735 \nu ^2}{288}
+\left(\frac{416569}{576}-\frac{3599 \pi ^2}{96}\right) \nu  \right)  
\left( \tilde{S}_1 + \tilde{S}_2 \right)\right.\nn\\
&&\left.+\left(-\frac{70 \nu ^3}{81}+\frac{521 \nu ^2}{12}
+\left(\frac{13891}{24}-\frac{205 \pi ^2}{8}\right) \nu +\frac{243  }{64} \right) 
\left( \tilde{S}_1/q + \tilde{S}_2 q\right)\right].
\eea

The final gauge-invariant relation that we derive here is the energy as a function
of the frequency. This binding energy, in conjunction with the energy flux as 
functions of the frequency of the GWs, is used to derive the change in the 
frequency of the GWs over time, namely the phasing of GWs. Plugging the relation 
$\tilde{L}(x)$ in eq.~\eqref{Ltox} to the relation $e(\tilde{L})$ in 
eq.~\eqref{etoL}, where the point-mass contributions up to $3$PN should also be taken 
into account, leads to the gauge-invariant relation, commonly referred to as the binding 
energy, for the present sector:
\bea
(e)^{\text{N}^3\text{LO}}_{\text{SO}} (\tilde{x}) &=&  x^{11/2} \nu 
\left[ \left(-\frac{265 \nu ^3}{3888}-\frac{1979 \nu ^2}{36}
+\left(\frac{19679}{144}+\frac{29 \pi ^2}{24}\right) \nu -45  \right) 
\left( \tilde{S}_1 + \tilde{S}_2 \right) \right.\nn\\
&&\left.+\left( -\frac{25 \nu ^3}{324}-\frac{1109 \nu ^2}{24}+\frac{565 \nu }{8}
-\frac{135  }{16} \right) \left( \tilde{S}_1/q + \tilde{S}_2 q\right)\right].
\eea
This binding energy is in complete agreement with that derived via traditional GR methods in 
\cite{Antonelli:2020ybz}, which assumed input and results from the EOB approach, and from self-force 
theory. 

As noted in section \ref{spechamilton}, due to the simplicity of the spin-orbit coupling, one can just 
take the COM Hamiltonian in eq.~\eqref{hcom}, assume that the binding energy that it stands for can 
be extended to a kinetic energy of scattering, and compute the scattering angle, which 
assumes aligned spins \cite{Damour:1988mr,Antonelli:2020ybz}. 
As detailed in \cite{Kim:2022bwv}, we carried out this computation, and our consequent scattering angle 
is in complete agreement with that derived in \cite{Antonelli:2020ybz} via traditional GR methods, 
that assumed input and results from the EOB approach, and from self-force theory.

\section{Conclusions}

In this paper we derived the complete N$^3$LO spin-orbit interaction at the $4.5$PN order
via the EFT of spinning gravitating objects \cite{Levi:2015msa}. At the present high 
orders of loop and of spin at the precision frontier, it is crucial to deploy various independent 
methodologies, in order to push and carefully establish the state of the art of PN theory. The EFT of 
spinning gravitating objects uniquely constitutes such a free-standing framework, including its 
automation in the public code \texttt{EFTofPNG} \cite{Levi:2017kzq}, which has enabled the 
completion of the state of the art at the $4$PN order 
\cite{Levi:2014gsa,Levi:2015uxa,Levi:2015ixa,Levi:2016ofk}. 
Together with the present paper, the EFT of spinning gravitating objects has also uniquely enabled the 
recent completion of the new precision frontier at the $5$PN order in
\cite{Kim:2021rfj,Kim:2022bwv,Levi:2022dqm,Levi:2022rrq}. 

The sector studied in this paper contains the largest and most elaborate collection of Feynman 
graphs ever tackled to date in sectors with spins, and in all PN sectors up to the third subleading 
order. Our computations were carried out via advanced multi-loop methods, and required further 
code development mainly for the projection and IBP methods. 
The most significant computational leap, that took place in the present sector, is the need to 
switch on a generic dimension, and keep track of DimReg expansions, across the whole derivation. 
This is due to the emergence of DimReg poles across all loop orders in sectors as of the N$^3$LO in PN 
theory. This transition constituted the most computationally-demanding aspect of the EFT computation in 
this work. 

At this high order of sectors with spins, it was also critical to extend the formal procedure, which was 
introduced in \cite{Levi:2014sba} for the reduction of higher-order time derivatives from spin variables 
via redefinitions. We found that the present sector uniquely requires to apply 
the redefinition of rotational variables beyond linear order, which has been sufficient for all other 
sectors with spins at lower-orders, and even up to the $5$PN order. As a consequence, we found a new 
unique addition to the spin potentials, originating from the rotational kinetic term. Though the 
reduction process for the present high-order sector with spins is intricate, we found that once we 
streamlined it in an automated algorithm, it was executed efficiently and rapidly.

In this paper we provided the full interaction potential in 
Lagrangian form, and the general Hamiltonian of the sector for the first time. 
The Lagrangian potential obtained via our framework enables a direct derivation of the 
physical EOMs for both the position and spin \cite{Levi:2015msa}. The general Hamiltonian enables 
to explore possible EOB applications extended to this sector, and test the performance of such 
variants of EOB Hamiltonians. The general Hamiltonian also enables to study the conserved integrals of 
motion, which form a representation of the Poincar\'e algebra on phase space.  Indeed, in 
\cite{Levi:2022dqm} we found the complete Poincar\'e algebra of the present sector, and thus verified 
the validity of the general Hamiltonian derived in the present paper. 

We also derived consequent GW gauge-invariant observables, namely relations among the binding energy, 
angular momentum, and orbital frequency. We found complete agreement with the binding 
energy of GW sources, as well as with the extrapolated scattering angle for aligned spins in the 
scattering problem, derived via traditional GR methods \cite{Antonelli:2020ybz}. 
Yet, the approach implemented in \cite{Antonelli:2020ybz} for the spin-orbit sector, following 
\cite{Bini:2019nra,Bini:2020wpo}, is an ad-hoc approach, which is limited to this specific sector. It 
relied on some EOB assumptions, and available high-order results taken from self-force theory, and even 
required results at the $4$PN order (including the tail), which go beyond the N$^3$LO, namely the 
subleading order that was being targeted \cite{Antonelli:2020ybz}. 

Moreover, the EOB Hamiltonian reconstruction that was carried out in \cite{Antonelli:2020ybz} cannot be 
extended beyond the spin-orbit sector, namely to higher orders in spin, not even to the simple sector 
that is bilinear in the spins of the binary. This is since the derivation in \cite{Antonelli:2020ybz} 
assumes the aligned-spins simplification, whose results can be trivially extended to generic spin 
orientations -- only in the spin-orbit sector, due to the unique simplicity of the spin-orbit coupling. 
Such an extension to generic spin orientations, which give rise to rich physics that is 
present for real precessing binaries, and manifests as modulations of the gravitational waveform, is 
impossible for all other sectors with spin.

By contrast, our framework is completely independent and generic: It provides a conceptual methodology 
to tackle the various generic sectors that are needed to complete a certain PN accuracy. It thus 
constitutes a unique elementary methodology to study PN theory. Our framework is free-standing, and 
provides self-contained theory, derivations, results, and technology, which are critical to 
push the precision frontier at the present high orders.
Finally, all of the computations in our framework are also automated as extensions of the public 
\texttt{EFTofPNG} code. 
The independent generic derivation and results presented in this paper, and such development of the 
\texttt{EFTofPNG} code, have been essential to establish the state of 
the art, and to push the precision frontier for the measurement of GWs.

\acknowledgments

J-WK was supported by the Science and Technology Facilities Council (STFC) 
Consolidated Grant ST/T000686/1 ``\textit{Amplitudes, Strings and Duality}''.
ML has been supported by the STFC Rutherford Grant ST/V003895
``\textit{Harnessing QFT for Gravity}'' 
and by the Mathematical Institute University of Oxford. 
ZY is supported by the Knut and Alice Wallenberg Foundation under grants 
KAW 2018.0116 and KAW 2018.0162.

\appendix

\section{Unreduced Action} 
\label{unreduced}

As noted in section \ref{rawaction} the unreduced potential can be written as:
\bea
V_{\text{N}^3\text{LO}}^{\text{SO}} = \sum_{i=0}^6 \stackrel{(i)}{V}_{3,1} + (1 
\leftrightarrow 2),
\eea
where any piece $\stackrel{(i)}{V}$ contains only terms with a total of $i$ 
higher-order time derivatives, namely beyond the velocity and spin 
variables, and the indices $n,l$ in the subscript correspond to the sector $(n,l)$ 
as indicated in table \ref{overallredef}.

First, we have a piece without higher-order time derivatives:
\bea
\stackrel{(0)}{V}_{3,1}  &=& - 	\frac{G m_{2}}{8 r{}^2} \Big[ \vec{S}_{1}\times\vec{n}\cdot\vec{v}_{1} \big( 5 v_{1}^2 \vec{v}_{1}\cdot\vec{v}_{2} v_{2}^2 - v_{1}^2 ( \vec{v}_{1}\cdot\vec{v}_{2})^{2} - 5 v_{1}^{6} + 8 \vec{v}_{1}\cdot\vec{v}_{2} v_{1}^{4} - 4 v_{2}^2 v_{1}^{4} \nn\\ 
&& - 4 v_{1}^2 v_{2}^{4} + 6 \vec{v}_{1}\cdot\vec{v}_{2} v_{2}^{4} - 5 v_{2}^{6} -9 \vec{v}_{1}\cdot\vec{n} v_{1}^2 \vec{v}_{2}\cdot\vec{n} \vec{v}_{1}\cdot\vec{v}_{2} + 15 \vec{v}_{1}\cdot\vec{n} v_{1}^2 \vec{v}_{2}\cdot\vec{n} v_{2}^2 \nn\\ 
&& - 6 \vec{v}_{1}\cdot\vec{n} \vec{v}_{2}\cdot\vec{n} \vec{v}_{1}\cdot\vec{v}_{2} v_{2}^2 + 3 v_{1}^2 v_{2}^2 ( \vec{v}_{1}\cdot\vec{n})^{2} - 6 \vec{v}_{1}\cdot\vec{n} \vec{v}_{2}\cdot\vec{n} ( \vec{v}_{1}\cdot\vec{v}_{2})^{2} + 3 v_{1}^2 v_{2}^2 ( \vec{v}_{2}\cdot\vec{n})^{2} \nn\\ 
&& + 9 \vec{v}_{1}\cdot\vec{n} \vec{v}_{2}\cdot\vec{n} v_{1}^{4} + 3 ( \vec{v}_{2}\cdot\vec{n})^{2} v_{1}^{4} + 3 ( \vec{v}_{1}\cdot\vec{n})^{2} v_{2}^{4} \nn\\ 
&& + 9 \vec{v}_{1}\cdot\vec{n} \vec{v}_{2}\cdot\vec{n} v_{2}^{4} -15 \vec{v}_{2}\cdot\vec{n} v_{2}^2 ( \vec{v}_{1}\cdot\vec{n})^{3} - 15 v_{1}^2 ( \vec{v}_{1}\cdot\vec{n})^{2} ( \vec{v}_{2}\cdot\vec{n})^{2} \nn\\ 
&& - 15 \vec{v}_{1}\cdot\vec{n} v_{1}^2 ( \vec{v}_{2}\cdot\vec{n})^{3} - 15 v_{2}^2 ( \vec{v}_{1}\cdot\vec{n})^{2} ( \vec{v}_{2}\cdot\vec{n})^{2} + 35 ( \vec{v}_{1}\cdot\vec{n})^{3} ( \vec{v}_{2}\cdot\vec{n})^{3} \big) \nn\\ 
&& + \vec{S}_{1}\times\vec{n}\cdot\vec{v}_{2} \big( v_{1}^2 \vec{v}_{1}\cdot\vec{v}_{2} v_{2}^2 - 2 ( \vec{v}_{1}\cdot\vec{v}_{2})^{3} - 2 v_{2}^2 ( \vec{v}_{1}\cdot\vec{v}_{2})^{2} + v_{1}^2 v_{2}^{4} - 3 \vec{v}_{1}\cdot\vec{v}_{2} v_{2}^{4} \nn\\ 
&& + 5 v_{2}^{6} -9 \vec{v}_{1}\cdot\vec{n} v_{1}^2 \vec{v}_{2}\cdot\vec{n} v_{2}^2 - 3 \vec{v}_{1}\cdot\vec{v}_{2} v_{2}^2 ( \vec{v}_{1}\cdot\vec{n})^{2} - 3 v_{1}^2 \vec{v}_{1}\cdot\vec{v}_{2} ( \vec{v}_{2}\cdot\vec{n})^{2} \nn\\ 
&& + 6 \vec{v}_{1}\cdot\vec{n} \vec{v}_{2}\cdot\vec{n} ( \vec{v}_{1}\cdot\vec{v}_{2})^{2} - 3 v_{1}^2 v_{2}^2 ( \vec{v}_{2}\cdot\vec{n})^{2} - 3 ( \vec{v}_{1}\cdot\vec{n})^{2} v_{2}^{4} - 9 \vec{v}_{1}\cdot\vec{n} \vec{v}_{2}\cdot\vec{n} v_{2}^{4} \nn\\ 
&& + 15 \vec{v}_{2}\cdot\vec{n} v_{2}^2 ( \vec{v}_{1}\cdot\vec{n})^{3} + 15 \vec{v}_{1}\cdot\vec{n} v_{1}^2 ( \vec{v}_{2}\cdot\vec{n})^{3} + 15 \vec{v}_{1}\cdot\vec{v}_{2} ( \vec{v}_{1}\cdot\vec{n})^{2} ( \vec{v}_{2}\cdot\vec{n})^{2} \nn\\ 
&& + 15 v_{2}^2 ( \vec{v}_{1}\cdot\vec{n})^{2} ( \vec{v}_{2}\cdot\vec{n})^{2} -35 ( \vec{v}_{1}\cdot\vec{n})^{3} ( \vec{v}_{2}\cdot\vec{n})^{3} \big) - \vec{S}_{1}\times\vec{v}_{1}\cdot\vec{v}_{2} \big( 3 \vec{v}_{1}\cdot\vec{n} v_{1}^2 \vec{v}_{1}\cdot\vec{v}_{2} \nn\\ 
&& + 2 v_{1}^2 \vec{v}_{2}\cdot\vec{n} \vec{v}_{1}\cdot\vec{v}_{2} - \vec{v}_{1}\cdot\vec{n} v_{1}^2 v_{2}^2 - 3 v_{1}^2 \vec{v}_{2}\cdot\vec{n} v_{2}^2 + 2 \vec{v}_{2}\cdot\vec{n} ( \vec{v}_{1}\cdot\vec{v}_{2})^{2} - 2 \vec{v}_{1}\cdot\vec{n} v_{1}^{4} \nn\\ 
&& - \vec{v}_{2}\cdot\vec{n} v_{1}^{4} - 2 \vec{v}_{1}\cdot\vec{n} v_{2}^{4} - 3 \vec{v}_{2}\cdot\vec{n} v_{2}^{4} + 3 v_{1}^2 \vec{v}_{2}\cdot\vec{n} ( \vec{v}_{1}\cdot\vec{n})^{2} - 6 \vec{v}_{2}\cdot\vec{n} \vec{v}_{1}\cdot\vec{v}_{2} ( \vec{v}_{1}\cdot\vec{n})^{2} \nn\\ 
&& + 9 \vec{v}_{2}\cdot\vec{n} v_{2}^2 ( \vec{v}_{1}\cdot\vec{n})^{2} + 3 v_{1}^2 ( \vec{v}_{2}\cdot\vec{n})^{3} + 6 \vec{v}_{1}\cdot\vec{n} \vec{v}_{1}\cdot\vec{v}_{2} ( \vec{v}_{2}\cdot\vec{n})^{2} \nn\\ 
&& + 6 \vec{v}_{1}\cdot\vec{n} v_{2}^2 ( \vec{v}_{2}\cdot\vec{n})^{2} -15 ( \vec{v}_{1}\cdot\vec{n})^{2} ( \vec{v}_{2}\cdot\vec{n})^{3} \big) \Big]\nn\\ && +  	\frac{G^2 m_{1} m_{2}}{3 r{}^3} \Big[ \vec{S}_{1}\times\vec{n}\cdot\vec{v}_{1} \big( 36 v_{1}^2 \vec{v}_{1}\cdot\vec{v}_{2} - 4 v_{1}^2 v_{2}^2 - 3 \vec{v}_{1}\cdot\vec{v}_{2} v_{2}^2 - 20 ( \vec{v}_{1}\cdot\vec{v}_{2})^{2} - 30 v_{1}^{4} \nn\\ 
&& + 21 v_{2}^{4} + 184 \vec{v}_{1}\cdot\vec{n} \vec{v}_{2}\cdot\vec{n} \vec{v}_{1}\cdot\vec{v}_{2} - 252 \vec{v}_{1}\cdot\vec{n} \vec{v}_{2}\cdot\vec{n} v_{2}^2 + 144 v_{1}^2 ( \vec{v}_{1}\cdot\vec{n})^{2} \nn\\ 
&& - 132 \vec{v}_{1}\cdot\vec{v}_{2} ( \vec{v}_{1}\cdot\vec{n})^{2} + 148 v_{2}^2 ( \vec{v}_{1}\cdot\vec{n})^{2} - 80 v_{1}^2 ( \vec{v}_{2}\cdot\vec{n})^{2} - 12 v_{2}^2 ( \vec{v}_{2}\cdot\vec{n})^{2} \nn\\ 
&& + 96 ( \vec{v}_{1}\cdot\vec{n})^{4} - 456 \vec{v}_{2}\cdot\vec{n} ( \vec{v}_{1}\cdot\vec{n})^{3} + 264 ( \vec{v}_{1}\cdot\vec{n})^{2} ( \vec{v}_{2}\cdot\vec{n})^{2} + 96 \vec{v}_{1}\cdot\vec{n} ( \vec{v}_{2}\cdot\vec{n})^{3} \big) \nn\\ 
&& + 2 \vec{S}_{1}\times\vec{n}\cdot\vec{v}_{2} \big( 21 v_{1}^2 \vec{v}_{1}\cdot\vec{v}_{2} - 10 v_{1}^2 v_{2}^2 + 15 \vec{v}_{1}\cdot\vec{v}_{2} v_{2}^2 - 14 ( \vec{v}_{1}\cdot\vec{v}_{2})^{2} - 3 v_{1}^{4} \nn\\ 
&& - 9 v_{2}^{4} -36 \vec{v}_{1}\cdot\vec{n} v_{1}^2 \vec{v}_{2}\cdot\vec{n} + 16 \vec{v}_{1}\cdot\vec{n} \vec{v}_{2}\cdot\vec{n} \vec{v}_{1}\cdot\vec{v}_{2} + 36 \vec{v}_{1}\cdot\vec{n} \vec{v}_{2}\cdot\vec{n} v_{2}^2 \nn\\ 
&& - 36 \vec{v}_{1}\cdot\vec{v}_{2} ( \vec{v}_{1}\cdot\vec{n})^{2} - 8 v_{2}^2 ( \vec{v}_{1}\cdot\vec{n})^{2} + 28 v_{1}^2 ( \vec{v}_{2}\cdot\vec{n})^{2} -24 ( \vec{v}_{1}\cdot\vec{n})^{4} \nn\\ 
&& + 120 \vec{v}_{2}\cdot\vec{n} ( \vec{v}_{1}\cdot\vec{n})^{3} - 96 ( \vec{v}_{1}\cdot\vec{n})^{2} ( \vec{v}_{2}\cdot\vec{n})^{2} \big) - \vec{S}_{1}\times\vec{v}_{1}\cdot\vec{v}_{2} \big( 6 \vec{v}_{1}\cdot\vec{n} v_{1}^2 + 22 v_{1}^2 \vec{v}_{2}\cdot\vec{n} \nn\\ 
&& - 106 \vec{v}_{1}\cdot\vec{n} \vec{v}_{1}\cdot\vec{v}_{2} + 26 \vec{v}_{2}\cdot\vec{n} \vec{v}_{1}\cdot\vec{v}_{2} + 52 \vec{v}_{1}\cdot\vec{n} v_{2}^2 + 27 \vec{v}_{2}\cdot\vec{n} v_{2}^2 + 8 ( \vec{v}_{1}\cdot\vec{n})^{3} \nn\\ 
&& + 92 \vec{v}_{2}\cdot\vec{n} ( \vec{v}_{1}\cdot\vec{n})^{2} - 136 \vec{v}_{1}\cdot\vec{n} ( \vec{v}_{2}\cdot\vec{n})^{2} \big) \Big] - 	\frac{G^2 m_{2}{}^2}{48 r{}^3} \Big[ \vec{S}_{1}\times\vec{n}\cdot\vec{v}_{1} \big( 903 v_{1}^2 \vec{v}_{1}\cdot\vec{v}_{2} \nn\\ 
&& - 472 v_{1}^2 v_{2}^2 - 1488 \vec{v}_{1}\cdot\vec{v}_{2} v_{2}^2 - 140 ( \vec{v}_{1}\cdot\vec{v}_{2})^{2} - 267 v_{1}^{4} + 1464 v_{2}^{4} + 672 \vec{v}_{1}\cdot\vec{n} v_{1}^2 \vec{v}_{2}\cdot\vec{n} \nn\\ 
&& - 3680 \vec{v}_{1}\cdot\vec{n} \vec{v}_{2}\cdot\vec{n} \vec{v}_{1}\cdot\vec{v}_{2} + 8256 \vec{v}_{1}\cdot\vec{n} \vec{v}_{2}\cdot\vec{n} v_{2}^2 - 48 v_{1}^2 ( \vec{v}_{1}\cdot\vec{n})^{2} - 336 \vec{v}_{1}\cdot\vec{v}_{2} ( \vec{v}_{1}\cdot\vec{n})^{2} \nn\\ 
&& + 448 v_{2}^2 ( \vec{v}_{1}\cdot\vec{n})^{2} - 296 v_{1}^2 ( \vec{v}_{2}\cdot\vec{n})^{2} + 3000 \vec{v}_{1}\cdot\vec{v}_{2} ( \vec{v}_{2}\cdot\vec{n})^{2} - 8064 v_{2}^2 ( \vec{v}_{2}\cdot\vec{n})^{2} \nn\\ 
&& + 384 \vec{v}_{2}\cdot\vec{n} ( \vec{v}_{1}\cdot\vec{n})^{3} - 960 ( \vec{v}_{1}\cdot\vec{n})^{2} ( \vec{v}_{2}\cdot\vec{n})^{2} + 768 \vec{v}_{1}\cdot\vec{n} ( \vec{v}_{2}\cdot\vec{n})^{3} - 120 ( \vec{v}_{2}\cdot\vec{n})^{4} \big) \nn\\ 
&& - 8 \vec{S}_{1}\times\vec{n}\cdot\vec{v}_{2} \big( 9 v_{1}^2 \vec{v}_{1}\cdot\vec{v}_{2} - 34 v_{1}^2 v_{2}^2 + 78 \vec{v}_{1}\cdot\vec{v}_{2} v_{2}^2 - 122 ( \vec{v}_{1}\cdot\vec{v}_{2})^{2} + 69 v_{2}^{4} \nn\\ 
&& + 24 \vec{v}_{1}\cdot\vec{n} v_{1}^2 \vec{v}_{2}\cdot\vec{n} + 568 \vec{v}_{1}\cdot\vec{n} \vec{v}_{2}\cdot\vec{n} \vec{v}_{1}\cdot\vec{v}_{2} + 72 \vec{v}_{1}\cdot\vec{n} \vec{v}_{2}\cdot\vec{n} v_{2}^2 + 12 \vec{v}_{1}\cdot\vec{v}_{2} ( \vec{v}_{1}\cdot\vec{n})^{2} \nn\\ 
&& + 136 v_{2}^2 ( \vec{v}_{1}\cdot\vec{n})^{2} + 112 v_{1}^2 ( \vec{v}_{2}\cdot\vec{n})^{2} - 606 \vec{v}_{1}\cdot\vec{v}_{2} ( \vec{v}_{2}\cdot\vec{n})^{2} \nn\\ 
&& - 324 v_{2}^2 ( \vec{v}_{2}\cdot\vec{n})^{2} -48 \vec{v}_{2}\cdot\vec{n} ( \vec{v}_{1}\cdot\vec{n})^{3} - 672 ( \vec{v}_{1}\cdot\vec{n})^{2} ( \vec{v}_{2}\cdot\vec{n})^{2} + 912 \vec{v}_{1}\cdot\vec{n} ( \vec{v}_{2}\cdot\vec{n})^{3} \nn\\ 
&& - 183 ( \vec{v}_{2}\cdot\vec{n})^{4} \big) + \vec{S}_{1}\times\vec{v}_{1}\cdot\vec{v}_{2} \big( 423 \vec{v}_{1}\cdot\vec{n} v_{1}^2 - 748 v_{1}^2 \vec{v}_{2}\cdot\vec{n} - 2372 \vec{v}_{1}\cdot\vec{n} \vec{v}_{1}\cdot\vec{v}_{2} \nn\\ 
&& + 4448 \vec{v}_{2}\cdot\vec{n} \vec{v}_{1}\cdot\vec{v}_{2} + 2320 \vec{v}_{1}\cdot\vec{n} v_{2}^2 - 3504 \vec{v}_{2}\cdot\vec{n} v_{2}^2 + 112 ( \vec{v}_{1}\cdot\vec{n})^{3} \nn\\ 
&& + 928 \vec{v}_{2}\cdot\vec{n} ( \vec{v}_{1}\cdot\vec{n})^{2} - 3304 \vec{v}_{1}\cdot\vec{n} ( \vec{v}_{2}\cdot\vec{n})^{2} + 1520 ( \vec{v}_{2}\cdot\vec{n})^{3} \big) \Big]\nn\\ && +  	\frac{G^3 m_{1}{}^2 m_{2}}{150 r{}^4} \Big[ \vec{S}_{1}\times\vec{n}\cdot\vec{v}_{1} \big( 2917 v_{1}^2 - 2438 \vec{v}_{1}\cdot\vec{v}_{2} - 479 v_{2}^2 -1175 \vec{v}_{1}\cdot\vec{n} \vec{v}_{2}\cdot\vec{n} \nn\\ 
&& - 13400 ( \vec{v}_{1}\cdot\vec{n})^{2} + 14200 ( \vec{v}_{2}\cdot\vec{n})^{2} \big) - 3 \vec{S}_{1}\times\vec{n}\cdot\vec{v}_{2} \big( 907 v_{1}^2 - 1279 \vec{v}_{1}\cdot\vec{v}_{2} + 372 v_{2}^2 \nn\\ 
&& + 3625 \vec{v}_{1}\cdot\vec{n} \vec{v}_{2}\cdot\vec{n} - 3875 ( \vec{v}_{1}\cdot\vec{n})^{2} + 125 ( \vec{v}_{2}\cdot\vec{n})^{2} \big) - \vec{S}_{1}\times\vec{v}_{1}\cdot\vec{v}_{2} \big( 2404 \vec{v}_{1}\cdot\vec{n} \nn\\ 
&& - 4045 \vec{v}_{2}\cdot\vec{n} \big) \Big] + 	\frac{G^3 m_{2}{}^3}{150 r{}^4} \Big[ 3 \vec{S}_{1}\times\vec{n}\cdot\vec{v}_{1} \big( 725 v_{1}^2 - 1916 \vec{v}_{1}\cdot\vec{v}_{2} + 1191 v_{2}^2 \nn\\ 
&& + 5500 \vec{v}_{1}\cdot\vec{n} \vec{v}_{2}\cdot\vec{n} - 2200 ( \vec{v}_{1}\cdot\vec{n})^{2} - 3400 ( \vec{v}_{2}\cdot\vec{n})^{2} \big) + \vec{S}_{1}\times\vec{n}\cdot\vec{v}_{2} \big( 827 v_{1}^2 \nn\\ 
&& + 3752 \vec{v}_{1}\cdot\vec{v}_{2} - 4579 v_{2}^2 -21700 \vec{v}_{1}\cdot\vec{n} \vec{v}_{2}\cdot\vec{n} - 5800 ( \vec{v}_{1}\cdot\vec{n})^{2} + 27800 ( \vec{v}_{2}\cdot\vec{n})^{2} \big) \nn\\ 
&& + \vec{S}_{1}\times\vec{v}_{1}\cdot\vec{v}_{2} \big( 5977 \vec{v}_{1}\cdot\vec{n} - 2044 \vec{v}_{2}\cdot\vec{n} \big) \Big] \nn\\ 
&& + 	\frac{G^3 m_{1}{}^2 m_{2}}{5 r{}^4} {\Big( \frac{1}{\epsilon} - 3\log \frac{r}{R_0} \Big)} \Big[ \vec{S}_{1}\times\vec{n}\cdot\vec{v}_{1} \big( 51 v_{1}^2 + 16 \vec{v}_{1}\cdot\vec{v}_{2} - 67 v_{2}^2 -80 \vec{v}_{1}\cdot\vec{n} \vec{v}_{2}\cdot\vec{n} \nn\\ 
&& - 255 ( \vec{v}_{1}\cdot\vec{n})^{2} + 335 ( \vec{v}_{2}\cdot\vec{n})^{2} \big) - 2 \vec{S}_{1}\times\vec{n}\cdot\vec{v}_{2} \big( 19 v_{1}^2 - 18 \vec{v}_{1}\cdot\vec{v}_{2} - v_{2}^2 \nn\\ 
&& + 90 \vec{v}_{1}\cdot\vec{n} \vec{v}_{2}\cdot\vec{n} - 95 ( \vec{v}_{1}\cdot\vec{n})^{2} + 5 ( \vec{v}_{2}\cdot\vec{n})^{2} \big) - 2 \vec{S}_{1}\times\vec{v}_{1}\cdot\vec{v}_{2} \big( 46 \vec{v}_{1}\cdot\vec{n} - 85 \vec{v}_{2}\cdot\vec{n} \big) \Big] \nn\\ 
&& + 	\frac{2 G^3 m_{2}{}^3}{5 r{}^4} {\Big( \frac{1}{\epsilon} - 3\log \frac{r}{R_0} \Big)} \Big[ \vec{S}_{1}\times\vec{n}\cdot\vec{v}_{1} \big( 5 v_{1}^2 - 17 \vec{v}_{1}\cdot\vec{v}_{2} + 12 v_{2}^2 + 85 \vec{v}_{1}\cdot\vec{n} \vec{v}_{2}\cdot\vec{n} \nn\\ 
&& - 25 ( \vec{v}_{1}\cdot\vec{n})^{2} - 60 ( \vec{v}_{2}\cdot\vec{n})^{2} \big) + 8 \vec{S}_{1}\times\vec{n}\cdot\vec{v}_{2} \big( v_{1}^2 + 6 \vec{v}_{1}\cdot\vec{v}_{2} - 7 v_{2}^2 -30 \vec{v}_{1}\cdot\vec{n} \vec{v}_{2}\cdot\vec{n} \nn\\ 
&& - 5 ( \vec{v}_{1}\cdot\vec{n})^{2} + 35 ( \vec{v}_{2}\cdot\vec{n})^{2} \big) + 3 \vec{S}_{1}\times\vec{v}_{1}\cdot\vec{v}_{2} \big( 11 \vec{v}_{1}\cdot\vec{n} + 8 \vec{v}_{2}\cdot\vec{n} \big) \Big] \nn\\ 
&& + 	\frac{G^3 m_{1} m_{2}{}^2}{192 r{}^4} \Big[ \vec{S}_{1}\times\vec{n}\cdot\vec{v}_{1} \big( {(10208 - 1341 \pi^2)} v_{1}^2 - {(28544 - 1368 \pi^2)} \vec{v}_{1}\cdot\vec{v}_{2} \nn\\ 
&& + {(18336 - 27 \pi^2)} v_{2}^2 + {(115360 - 6840 \pi^2)} \vec{v}_{1}\cdot\vec{n} \vec{v}_{2}\cdot\vec{n} - {(41440 - 6705 \pi^2)} ( \vec{v}_{1}\cdot\vec{n})^{2} \nn\\ 
&& - {(77088 - 135 \pi^2)} ( \vec{v}_{2}\cdot\vec{n})^{2} \big) - \vec{S}_{1}\times\vec{n}\cdot\vec{v}_{2} \big( {(5952 - 747 \pi^2)} v_{1}^2 \nn\\ 
&& - {(17056 - 2772 \pi^2)} \vec{v}_{1}\cdot\vec{v}_{2} + {(11104 - 2025 \pi^2)} v_{2}^2 + {(71072 - 13860 \pi^2)} \vec{v}_{1}\cdot\vec{n} \vec{v}_{2}\cdot\vec{n} \nn\\ 
&& - {(28320 - 3735 \pi^2)} ( \vec{v}_{1}\cdot\vec{n})^{2} - {(45920 - 10125 \pi^2)} ( \vec{v}_{2}\cdot\vec{n})^{2} \big) \nn\\ 
&& + 2 \vec{S}_{1}\times\vec{v}_{1}\cdot\vec{v}_{2} \big( {(9472 + 63 \pi^2)} \vec{v}_{1}\cdot\vec{n} - {(11824 + 1359 \pi^2)} \vec{v}_{2}\cdot\vec{n} \big) \Big]\nn\\ && - 	\frac{31 G^4 m_{1} m_{2}{}^3}{3 r{}^5} \Big[ \vec{S}_{1}\times\vec{n}\cdot\vec{v}_{1} - \vec{S}_{1}\times\vec{n}\cdot\vec{v}_{2} \Big] - 	\frac{3 G^4 m_{2}{}^4}{8 r{}^5} \Big[ \vec{S}_{1}\times\vec{n}\cdot\vec{v}_{1} - \vec{S}_{1}\times\vec{n}\cdot\vec{v}_{2} \Big] \nn\\ 
&& - 	\frac{5 G^4 m_{1}{}^2 m_{2}{}^2}{3 r{}^5} {\Big( \frac{1}{\epsilon} - 4\log \frac{r}{R_0} \Big)} \Big[ \vec{S}_{1}\times\vec{n}\cdot\vec{v}_{1} - \vec{S}_{1}\times\vec{n}\cdot\vec{v}_{2} \Big] \nn\\ 
&& + 	\frac{G^4 m_{1}{}^2 m_{2}{}^2}{6 r{}^5} \Big[ {(40 - 13 \pi^2)} \vec{S}_{1}\times\vec{n}\cdot\vec{v}_{1} - {(40 - 13 \pi^2)} \vec{S}_{1}\times\vec{n}\cdot\vec{v}_{2} \Big]  + (1 \leftrightarrow 2),
\eea
where the piece at order $G^4$, namely on the last 3 lines, was computed in 
\cite{Levi:2020kvb}.

Then, we separate the piece with one higher-order time derivative into:
\bea
\stackrel{(1)}{V}_{3,1}  = (V_a)_{3,1} + (V_{\dot{S}})_{3,1},
\eea
where
\bea
(V_a)_{3,1}  &=& \frac{G m_{2}}{8 r} \Big[ \vec{S}_{1}\times\vec{n}\cdot\vec{v}_{1} \big( 12 v_{1}^2 \vec{v}_{1}\cdot\vec{a}_{1} \vec{v}_{2}\cdot\vec{n} - 6 \vec{v}_{1}\cdot\vec{a}_{1} \vec{v}_{2}\cdot\vec{n} \vec{v}_{1}\cdot\vec{v}_{2} - 5 v_{1}^2 \vec{v}_{2}\cdot\vec{n} \vec{a}_{1}\cdot\vec{v}_{2} \nn\\ 
&& - 4 \vec{v}_{2}\cdot\vec{n} \vec{v}_{1}\cdot\vec{v}_{2} \vec{a}_{1}\cdot\vec{v}_{2} + v_{1}^2 \vec{a}_{1}\cdot\vec{n} v_{2}^2 + 4 \vec{v}_{1}\cdot\vec{n} \vec{v}_{1}\cdot\vec{a}_{1} v_{2}^2 + 7 \vec{v}_{1}\cdot\vec{a}_{1} \vec{v}_{2}\cdot\vec{n} v_{2}^2 \nn\\ 
&& - 3 \vec{v}_{1}\cdot\vec{n} \vec{a}_{1}\cdot\vec{v}_{2} v_{2}^2 - 4 \vec{v}_{2}\cdot\vec{n} \vec{a}_{1}\cdot\vec{v}_{2} v_{2}^2 - v_{1}^2 v_{2}^2 \vec{a}_{2}\cdot\vec{n} + 5 \vec{v}_{1}\cdot\vec{n} v_{1}^2 \vec{v}_{1}\cdot\vec{a}_{2} \nn\\ 
&& + 3 v_{1}^2 \vec{v}_{2}\cdot\vec{n} \vec{v}_{1}\cdot\vec{a}_{2} + 4 \vec{v}_{1}\cdot\vec{n} \vec{v}_{1}\cdot\vec{v}_{2} \vec{v}_{1}\cdot\vec{a}_{2} + 4 \vec{v}_{1}\cdot\vec{n} v_{2}^2 \vec{v}_{1}\cdot\vec{a}_{2} - 7 \vec{v}_{1}\cdot\vec{n} v_{1}^2 \vec{v}_{2}\cdot\vec{a}_{2} \nn\\ 
&& - 4 v_{1}^2 \vec{v}_{2}\cdot\vec{n} \vec{v}_{2}\cdot\vec{a}_{2} + 4 \vec{v}_{1}\cdot\vec{n} \vec{v}_{1}\cdot\vec{v}_{2} \vec{v}_{2}\cdot\vec{a}_{2} - 12 \vec{v}_{1}\cdot\vec{n} v_{2}^2 \vec{v}_{2}\cdot\vec{a}_{2} - \vec{a}_{2}\cdot\vec{n} v_{1}^{4} \nn\\ 
&& + \vec{a}_{1}\cdot\vec{n} v_{2}^{4} -9 \vec{v}_{1}\cdot\vec{n} \vec{a}_{1}\cdot\vec{n} \vec{v}_{2}\cdot\vec{n} v_{2}^2 + 9 \vec{v}_{1}\cdot\vec{n} v_{1}^2 \vec{v}_{2}\cdot\vec{n} \vec{a}_{2}\cdot\vec{n} + 3 v_{1}^2 \vec{a}_{2}\cdot\vec{n} ( \vec{v}_{1}\cdot\vec{n})^{2} \nn\\ 
&& + 3 v_{2}^2 \vec{a}_{2}\cdot\vec{n} ( \vec{v}_{1}\cdot\vec{n})^{2} - 9 \vec{v}_{2}\cdot\vec{n} \vec{v}_{1}\cdot\vec{a}_{2} ( \vec{v}_{1}\cdot\vec{n})^{2} + 3 \vec{v}_{2}\cdot\vec{a}_{2} ( \vec{v}_{1}\cdot\vec{n})^{3} \nn\\ 
&& + 12 \vec{v}_{2}\cdot\vec{n} \vec{v}_{2}\cdot\vec{a}_{2} ( \vec{v}_{1}\cdot\vec{n})^{2} - 3 v_{1}^2 \vec{a}_{1}\cdot\vec{n} ( \vec{v}_{2}\cdot\vec{n})^{2} - 12 \vec{v}_{1}\cdot\vec{n} \vec{v}_{1}\cdot\vec{a}_{1} ( \vec{v}_{2}\cdot\vec{n})^{2} \nn\\ 
&& - 3 \vec{v}_{1}\cdot\vec{a}_{1} ( \vec{v}_{2}\cdot\vec{n})^{3} + 9 \vec{v}_{1}\cdot\vec{n} \vec{a}_{1}\cdot\vec{v}_{2} ( \vec{v}_{2}\cdot\vec{n})^{2} \nn\\ 
&& - 3 \vec{a}_{1}\cdot\vec{n} v_{2}^2 ( \vec{v}_{2}\cdot\vec{n})^{2} -15 \vec{v}_{2}\cdot\vec{n} \vec{a}_{2}\cdot\vec{n} ( \vec{v}_{1}\cdot\vec{n})^{3} + 15 \vec{v}_{1}\cdot\vec{n} \vec{a}_{1}\cdot\vec{n} ( \vec{v}_{2}\cdot\vec{n})^{3} \big) \nn\\ 
&& - \vec{S}_{1}\times\vec{n}\cdot\vec{a}_{1} \big( 3 v_{1}^2 \vec{v}_{2}\cdot\vec{n} \vec{v}_{1}\cdot\vec{v}_{2} - 2 \vec{v}_{1}\cdot\vec{n} v_{1}^2 v_{2}^2 - 5 v_{1}^2 \vec{v}_{2}\cdot\vec{n} v_{2}^2 + 2 \vec{v}_{2}\cdot\vec{n} \vec{v}_{1}\cdot\vec{v}_{2} v_{2}^2 \nn\\ 
&& + 2 \vec{v}_{2}\cdot\vec{n} ( \vec{v}_{1}\cdot\vec{v}_{2})^{2} - 3 \vec{v}_{2}\cdot\vec{n} v_{1}^{4} - 2 \vec{v}_{1}\cdot\vec{n} v_{2}^{4} - 3 \vec{v}_{2}\cdot\vec{n} v_{2}^{4} + 9 \vec{v}_{2}\cdot\vec{n} v_{2}^2 ( \vec{v}_{1}\cdot\vec{n})^{2} \nn\\ 
&& + 6 \vec{v}_{1}\cdot\vec{n} v_{1}^2 ( \vec{v}_{2}\cdot\vec{n})^{2} + 3 v_{1}^2 ( \vec{v}_{2}\cdot\vec{n})^{3} + 6 \vec{v}_{1}\cdot\vec{n} v_{2}^2 ( \vec{v}_{2}\cdot\vec{n})^{2} -15 ( \vec{v}_{1}\cdot\vec{n})^{2} ( \vec{v}_{2}\cdot\vec{n})^{3} \big) \nn\\ 
&& + \vec{S}_{1}\times\vec{v}_{1}\cdot\vec{a}_{1} \big( 36 v_{1}^2 \vec{v}_{1}\cdot\vec{v}_{2} - 15 v_{1}^2 v_{2}^2 + 12 \vec{v}_{1}\cdot\vec{v}_{2} v_{2}^2 - 30 v_{1}^{4} - 7 v_{2}^{4} + 12 \vec{v}_{1}\cdot\vec{n} v_{1}^2 \vec{v}_{2}\cdot\vec{n} \nn\\ 
&& - 8 \vec{v}_{1}\cdot\vec{n} \vec{v}_{2}\cdot\vec{n} \vec{v}_{1}\cdot\vec{v}_{2} + 7 \vec{v}_{1}\cdot\vec{n} \vec{v}_{2}\cdot\vec{n} v_{2}^2 + 2 v_{2}^2 ( \vec{v}_{1}\cdot\vec{n})^{2} + 3 v_{1}^2 ( \vec{v}_{2}\cdot\vec{n})^{2} \nn\\ 
&& + v_{2}^2 ( \vec{v}_{2}\cdot\vec{n})^{2} -6 ( \vec{v}_{1}\cdot\vec{n})^{2} ( \vec{v}_{2}\cdot\vec{n})^{2} - 3 \vec{v}_{1}\cdot\vec{n} ( \vec{v}_{2}\cdot\vec{n})^{3} \big) \nn\\ 
&& + \vec{S}_{1}\times\vec{n}\cdot\vec{v}_{2} \big( 6 \vec{v}_{2}\cdot\vec{n} \vec{v}_{1}\cdot\vec{v}_{2} \vec{a}_{1}\cdot\vec{v}_{2} - 3 \vec{v}_{1}\cdot\vec{a}_{1} \vec{v}_{2}\cdot\vec{n} v_{2}^2 - \vec{a}_{1}\cdot\vec{n} \vec{v}_{1}\cdot\vec{v}_{2} v_{2}^2 \nn\\ 
&& + \vec{v}_{1}\cdot\vec{n} \vec{a}_{1}\cdot\vec{v}_{2} v_{2}^2 + 2 \vec{v}_{2}\cdot\vec{n} \vec{a}_{1}\cdot\vec{v}_{2} v_{2}^2 + v_{1}^2 \vec{v}_{1}\cdot\vec{v}_{2} \vec{a}_{2}\cdot\vec{n} + v_{1}^2 v_{2}^2 \vec{a}_{2}\cdot\vec{n} - v_{1}^2 \vec{v}_{2}\cdot\vec{n} \vec{v}_{1}\cdot\vec{a}_{2} \nn\\ 
&& - 6 \vec{v}_{1}\cdot\vec{n} \vec{v}_{1}\cdot\vec{v}_{2} \vec{v}_{1}\cdot\vec{a}_{2} - 2 \vec{v}_{1}\cdot\vec{n} v_{2}^2 \vec{v}_{1}\cdot\vec{a}_{2} + 3 \vec{v}_{1}\cdot\vec{n} v_{1}^2 \vec{v}_{2}\cdot\vec{a}_{2} + 4 v_{1}^2 \vec{v}_{2}\cdot\vec{n} \vec{v}_{2}\cdot\vec{a}_{2} \nn\\ 
&& + 12 \vec{v}_{1}\cdot\vec{n} v_{2}^2 \vec{v}_{2}\cdot\vec{a}_{2} - \vec{a}_{1}\cdot\vec{n} v_{2}^{4} + 9 \vec{v}_{1}\cdot\vec{n} \vec{a}_{1}\cdot\vec{n} \vec{v}_{2}\cdot\vec{n} v_{2}^2 - 9 \vec{v}_{1}\cdot\vec{n} v_{1}^2 \vec{v}_{2}\cdot\vec{n} \vec{a}_{2}\cdot\vec{n} \nn\\ 
&& - 3 \vec{v}_{1}\cdot\vec{v}_{2} \vec{a}_{2}\cdot\vec{n} ( \vec{v}_{1}\cdot\vec{n})^{2} - 3 v_{2}^2 \vec{a}_{2}\cdot\vec{n} ( \vec{v}_{1}\cdot\vec{n})^{2} + 3 \vec{v}_{2}\cdot\vec{n} \vec{v}_{1}\cdot\vec{a}_{2} ( \vec{v}_{1}\cdot\vec{n})^{2} \nn\\ 
&& - 3 \vec{v}_{2}\cdot\vec{a}_{2} ( \vec{v}_{1}\cdot\vec{n})^{3} - 12 \vec{v}_{2}\cdot\vec{n} \vec{v}_{2}\cdot\vec{a}_{2} ( \vec{v}_{1}\cdot\vec{n})^{2} + 3 \vec{v}_{1}\cdot\vec{a}_{1} ( \vec{v}_{2}\cdot\vec{n})^{3} \nn\\ 
&& + 3 \vec{a}_{1}\cdot\vec{n} \vec{v}_{1}\cdot\vec{v}_{2} ( \vec{v}_{2}\cdot\vec{n})^{2} - 3 \vec{v}_{1}\cdot\vec{n} \vec{a}_{1}\cdot\vec{v}_{2} ( \vec{v}_{2}\cdot\vec{n})^{2} + 3 \vec{a}_{1}\cdot\vec{n} v_{2}^2 ( \vec{v}_{2}\cdot\vec{n})^{2} \nn\\ 
&& + 15 \vec{v}_{2}\cdot\vec{n} \vec{a}_{2}\cdot\vec{n} ( \vec{v}_{1}\cdot\vec{n})^{3} - 15 \vec{v}_{1}\cdot\vec{n} \vec{a}_{1}\cdot\vec{n} ( \vec{v}_{2}\cdot\vec{n})^{3} \big) - \vec{S}_{1}\times\vec{v}_{1}\cdot\vec{v}_{2} \big( 36 v_{1}^2 \vec{v}_{1}\cdot\vec{a}_{1} \nn\\ 
&& - 4 \vec{v}_{1}\cdot\vec{a}_{1} \vec{v}_{1}\cdot\vec{v}_{2} - 2 v_{1}^2 \vec{a}_{1}\cdot\vec{v}_{2} - 6 \vec{v}_{1}\cdot\vec{v}_{2} \vec{a}_{1}\cdot\vec{v}_{2} + 12 \vec{v}_{1}\cdot\vec{a}_{1} v_{2}^2 - \vec{a}_{1}\cdot\vec{v}_{2} v_{2}^2 \nn\\ 
&& - 2 v_{1}^2 \vec{v}_{1}\cdot\vec{a}_{2} - 6 \vec{v}_{1}\cdot\vec{v}_{2} \vec{v}_{1}\cdot\vec{a}_{2} - 2 v_{2}^2 \vec{v}_{1}\cdot\vec{a}_{2} + 3 v_{1}^2 \vec{v}_{2}\cdot\vec{a}_{2} + 12 v_{2}^2 \vec{v}_{2}\cdot\vec{a}_{2} \nn\\ 
&& + v_{1}^2 \vec{a}_{1}\cdot\vec{n} \vec{v}_{2}\cdot\vec{n} - 10 \vec{v}_{1}\cdot\vec{n} \vec{v}_{1}\cdot\vec{a}_{1} \vec{v}_{2}\cdot\vec{n} - 2 \vec{a}_{1}\cdot\vec{n} \vec{v}_{2}\cdot\vec{n} \vec{v}_{1}\cdot\vec{v}_{2} + 3 \vec{a}_{1}\cdot\vec{n} \vec{v}_{2}\cdot\vec{n} v_{2}^2 \nn\\ 
&& - 3 v_{1}^2 \vec{v}_{2}\cdot\vec{n} \vec{a}_{2}\cdot\vec{n} - 2 \vec{v}_{1}\cdot\vec{n} \vec{v}_{1}\cdot\vec{v}_{2} \vec{a}_{2}\cdot\vec{n} - 2 \vec{v}_{1}\cdot\vec{n} v_{2}^2 \vec{a}_{2}\cdot\vec{n} + 2 \vec{v}_{1}\cdot\vec{n} \vec{v}_{2}\cdot\vec{n} \vec{v}_{1}\cdot\vec{a}_{2} \nn\\ 
&& - 8 \vec{v}_{1}\cdot\vec{n} \vec{v}_{2}\cdot\vec{n} \vec{v}_{2}\cdot\vec{a}_{2} + 2 \vec{v}_{1}\cdot\vec{a}_{2} ( \vec{v}_{1}\cdot\vec{n})^{2} - 3 \vec{v}_{2}\cdot\vec{a}_{2} ( \vec{v}_{1}\cdot\vec{n})^{2} - \vec{a}_{1}\cdot\vec{v}_{2} ( \vec{v}_{2}\cdot\vec{n})^{2} \nn\\ 
&& + 9 \vec{v}_{2}\cdot\vec{n} \vec{a}_{2}\cdot\vec{n} ( \vec{v}_{1}\cdot\vec{n})^{2} - 3 \vec{a}_{1}\cdot\vec{n} ( \vec{v}_{2}\cdot\vec{n})^{3} \big) + \vec{S}_{1}\times\vec{a}_{1}\cdot\vec{v}_{2} \big( 8 v_{1}^2 \vec{v}_{1}\cdot\vec{v}_{2} - 9 v_{1}^2 v_{2}^2 \nn\\ 
&& + 5 \vec{v}_{1}\cdot\vec{v}_{2} v_{2}^2 + 6 ( \vec{v}_{1}\cdot\vec{v}_{2})^{2} - 9 v_{1}^{4} - 8 v_{2}^{4} + 5 \vec{v}_{1}\cdot\vec{n} v_{1}^2 \vec{v}_{2}\cdot\vec{n} + 4 \vec{v}_{1}\cdot\vec{n} \vec{v}_{2}\cdot\vec{n} \vec{v}_{1}\cdot\vec{v}_{2} \nn\\ 
&& + 3 \vec{v}_{1}\cdot\vec{n} \vec{v}_{2}\cdot\vec{n} v_{2}^2 + 3 v_{2}^2 ( \vec{v}_{1}\cdot\vec{n})^{2} + 3 v_{1}^2 ( \vec{v}_{2}\cdot\vec{n})^{2} - \vec{v}_{1}\cdot\vec{v}_{2} ( \vec{v}_{2}\cdot\vec{n})^{2} \nn\\ 
&& - v_{2}^2 ( \vec{v}_{2}\cdot\vec{n})^{2} -9 ( \vec{v}_{1}\cdot\vec{n})^{2} ( \vec{v}_{2}\cdot\vec{n})^{2} + 3 \vec{v}_{1}\cdot\vec{n} ( \vec{v}_{2}\cdot\vec{n})^{3} \big) + \vec{S}_{1}\times\vec{n}\cdot\vec{a}_{2} \big( 2 v_{1}^2 \vec{v}_{2}\cdot\vec{n} \vec{v}_{1}\cdot\vec{v}_{2} \nn\\ 
&& + 3 \vec{v}_{1}\cdot\vec{n} v_{1}^2 v_{2}^2 + 2 v_{1}^2 \vec{v}_{2}\cdot\vec{n} v_{2}^2 - 2 \vec{v}_{1}\cdot\vec{n} ( \vec{v}_{1}\cdot\vec{v}_{2})^{2} + 3 \vec{v}_{1}\cdot\vec{n} v_{2}^{4} -6 \vec{v}_{2}\cdot\vec{n} \vec{v}_{1}\cdot\vec{v}_{2} ( \vec{v}_{1}\cdot\vec{n})^{2} \nn\\ 
&& - 3 v_{2}^2 ( \vec{v}_{1}\cdot\vec{n})^{3} - 6 \vec{v}_{2}\cdot\vec{n} v_{2}^2 ( \vec{v}_{1}\cdot\vec{n})^{2} - 9 \vec{v}_{1}\cdot\vec{n} v_{1}^2 ( \vec{v}_{2}\cdot\vec{n})^{2} + 15 ( \vec{v}_{1}\cdot\vec{n})^{3} ( \vec{v}_{2}\cdot\vec{n})^{2} \big) \nn\\ 
&& + \vec{S}_{1}\times\vec{v}_{1}\cdot\vec{a}_{2} \big( 2 v_{1}^2 \vec{v}_{1}\cdot\vec{v}_{2} - 2 v_{1}^2 v_{2}^2 + 2 ( \vec{v}_{1}\cdot\vec{v}_{2})^{2} - 3 v_{2}^{4} -3 \vec{v}_{1}\cdot\vec{n} v_{1}^2 \vec{v}_{2}\cdot\vec{n} \nn\\ 
&& + 4 \vec{v}_{1}\cdot\vec{n} \vec{v}_{2}\cdot\vec{n} \vec{v}_{1}\cdot\vec{v}_{2} + 4 \vec{v}_{1}\cdot\vec{n} \vec{v}_{2}\cdot\vec{n} v_{2}^2 - 2 \vec{v}_{1}\cdot\vec{v}_{2} ( \vec{v}_{1}\cdot\vec{n})^{2} + 2 v_{2}^2 ( \vec{v}_{1}\cdot\vec{n})^{2} \nn\\ 
&& + 3 v_{1}^2 ( \vec{v}_{2}\cdot\vec{n})^{2} + 3 \vec{v}_{2}\cdot\vec{n} ( \vec{v}_{1}\cdot\vec{n})^{3} - 9 ( \vec{v}_{1}\cdot\vec{n})^{2} ( \vec{v}_{2}\cdot\vec{n})^{2} \big) - \vec{S}_{1}\times\vec{v}_{2}\cdot\vec{a}_{2} \big( v_{1}^2 \vec{v}_{1}\cdot\vec{v}_{2} \nn\\ 
&& + v_{1}^2 v_{2}^2 -3 \vec{v}_{1}\cdot\vec{n} v_{1}^2 \vec{v}_{2}\cdot\vec{n} - \vec{v}_{1}\cdot\vec{v}_{2} ( \vec{v}_{1}\cdot\vec{n})^{2} - v_{2}^2 ( \vec{v}_{1}\cdot\vec{n})^{2} + 3 \vec{v}_{2}\cdot\vec{n} ( \vec{v}_{1}\cdot\vec{n})^{3} \big) \Big]\nn\\ && - 	\frac{G^2 m_{1} m_{2}}{6 r{}^2} \Big[ 2 \vec{S}_{1}\times\vec{n}\cdot\vec{v}_{1} \big( 4 v_{1}^2 \vec{a}_{1}\cdot\vec{n} + 122 \vec{v}_{1}\cdot\vec{n} \vec{v}_{1}\cdot\vec{a}_{1} + 17 \vec{v}_{1}\cdot\vec{a}_{1} \vec{v}_{2}\cdot\vec{n} \nn\\ 
&& + 14 \vec{a}_{1}\cdot\vec{n} \vec{v}_{1}\cdot\vec{v}_{2} - 85 \vec{v}_{1}\cdot\vec{n} \vec{a}_{1}\cdot\vec{v}_{2} + 17 \vec{v}_{2}\cdot\vec{n} \vec{a}_{1}\cdot\vec{v}_{2} + 10 \vec{a}_{1}\cdot\vec{n} v_{2}^2 + 20 v_{1}^2 \vec{a}_{2}\cdot\vec{n} \nn\\ 
&& + 3 v_{2}^2 \vec{a}_{2}\cdot\vec{n} - 74 \vec{v}_{1}\cdot\vec{n} \vec{v}_{1}\cdot\vec{a}_{2} - 6 \vec{v}_{2}\cdot\vec{n} \vec{v}_{1}\cdot\vec{a}_{2} + 120 \vec{v}_{1}\cdot\vec{n} \vec{v}_{2}\cdot\vec{a}_{2} \nn\\ 
&& + 6 \vec{v}_{2}\cdot\vec{n} \vec{v}_{2}\cdot\vec{a}_{2} -296 \vec{v}_{1}\cdot\vec{n} \vec{a}_{1}\cdot\vec{n} \vec{v}_{2}\cdot\vec{n} - 48 \vec{v}_{1}\cdot\vec{n} \vec{v}_{2}\cdot\vec{n} \vec{a}_{2}\cdot\vec{n} + 128 \vec{a}_{1}\cdot\vec{n} ( \vec{v}_{1}\cdot\vec{n})^{2} \nn\\ 
&& - 44 \vec{a}_{2}\cdot\vec{n} ( \vec{v}_{1}\cdot\vec{n})^{2} + 104 \vec{a}_{1}\cdot\vec{n} ( \vec{v}_{2}\cdot\vec{n})^{2} \big) + 2 \vec{S}_{1}\times\vec{n}\cdot\vec{a}_{1} \big( 90 \vec{v}_{1}\cdot\vec{n} v_{1}^2 - 7 v_{1}^2 \vec{v}_{2}\cdot\vec{n} \nn\\ 
&& - 95 \vec{v}_{1}\cdot\vec{n} \vec{v}_{1}\cdot\vec{v}_{2} + 17 \vec{v}_{2}\cdot\vec{n} \vec{v}_{1}\cdot\vec{v}_{2} + 43 \vec{v}_{1}\cdot\vec{n} v_{2}^2 + 12 \vec{v}_{2}\cdot\vec{n} v_{2}^2 -36 ( \vec{v}_{1}\cdot\vec{n})^{3} \nn\\ 
&& + 16 \vec{v}_{2}\cdot\vec{n} ( \vec{v}_{1}\cdot\vec{n})^{2} - 40 \vec{v}_{1}\cdot\vec{n} ( \vec{v}_{2}\cdot\vec{n})^{2} \big) - \vec{S}_{1}\times\vec{v}_{1}\cdot\vec{a}_{1} \big( 104 v_{1}^2 - 43 \vec{v}_{1}\cdot\vec{v}_{2} - 48 v_{2}^2 \nn\\ 
&& + 206 \vec{v}_{1}\cdot\vec{n} \vec{v}_{2}\cdot\vec{n} - 244 ( \vec{v}_{1}\cdot\vec{n})^{2} + 12 ( \vec{v}_{2}\cdot\vec{n})^{2} \big) + 4 \vec{S}_{1}\times\vec{n}\cdot\vec{v}_{2} \big( 16 v_{1}^2 \vec{a}_{1}\cdot\vec{n} \nn\\ 
&& + 5 \vec{v}_{1}\cdot\vec{n} \vec{v}_{1}\cdot\vec{a}_{1} - 19 \vec{v}_{1}\cdot\vec{a}_{1} \vec{v}_{2}\cdot\vec{n} - 34 \vec{a}_{1}\cdot\vec{n} \vec{v}_{1}\cdot\vec{v}_{2} - 7 \vec{v}_{1}\cdot\vec{n} \vec{a}_{1}\cdot\vec{v}_{2} + 5 \vec{v}_{2}\cdot\vec{n} \vec{a}_{1}\cdot\vec{v}_{2} \nn\\ 
&& + 13 \vec{a}_{1}\cdot\vec{n} v_{2}^2 - 7 v_{1}^2 \vec{a}_{2}\cdot\vec{n} + 4 \vec{v}_{1}\cdot\vec{n} \vec{v}_{1}\cdot\vec{a}_{2} - 18 \vec{v}_{1}\cdot\vec{n} \vec{v}_{2}\cdot\vec{a}_{2} + 112 \vec{v}_{1}\cdot\vec{n} \vec{a}_{1}\cdot\vec{n} \vec{v}_{2}\cdot\vec{n} \nn\\ 
&& - 52 \vec{a}_{1}\cdot\vec{n} ( \vec{v}_{1}\cdot\vec{n})^{2} + 16 \vec{a}_{2}\cdot\vec{n} ( \vec{v}_{1}\cdot\vec{n})^{2} - 52 \vec{a}_{1}\cdot\vec{n} ( \vec{v}_{2}\cdot\vec{n})^{2} \big) + \vec{S}_{1}\times\vec{v}_{1}\cdot\vec{v}_{2} \big( 61 \vec{v}_{1}\cdot\vec{a}_{1} \nn\\ 
&& + 16 \vec{a}_{1}\cdot\vec{v}_{2} - 2 \vec{v}_{1}\cdot\vec{a}_{2} + 48 \vec{v}_{2}\cdot\vec{a}_{2} -26 \vec{v}_{1}\cdot\vec{n} \vec{a}_{1}\cdot\vec{n} - 14 \vec{a}_{1}\cdot\vec{n} \vec{v}_{2}\cdot\vec{n} - 68 \vec{v}_{1}\cdot\vec{n} \vec{a}_{2}\cdot\vec{n} \big) \nn\\ 
&& - \vec{S}_{1}\times\vec{a}_{1}\cdot\vec{v}_{2} \big( 39 v_{1}^2 - 28 \vec{v}_{1}\cdot\vec{v}_{2} - 64 v_{2}^2 + 230 \vec{v}_{1}\cdot\vec{n} \vec{v}_{2}\cdot\vec{n} - 186 ( \vec{v}_{1}\cdot\vec{n})^{2} \nn\\ 
&& - 4 ( \vec{v}_{2}\cdot\vec{n})^{2} \big) + 4 \vec{S}_{1}\times\vec{n}\cdot\vec{a}_{2} \big( 9 \vec{v}_{1}\cdot\vec{n} v_{1}^2 - 14 v_{1}^2 \vec{v}_{2}\cdot\vec{n} - 4 \vec{v}_{1}\cdot\vec{n} \vec{v}_{1}\cdot\vec{v}_{2} \nn\\ 
&& - 9 \vec{v}_{1}\cdot\vec{n} v_{2}^2 -20 ( \vec{v}_{1}\cdot\vec{n})^{3} + 32 \vec{v}_{2}\cdot\vec{n} ( \vec{v}_{1}\cdot\vec{n})^{2} \big) + 2 \vec{S}_{1}\times\vec{v}_{1}\cdot\vec{a}_{2} \big( v_{1}^2 + 13 \vec{v}_{1}\cdot\vec{v}_{2} \nn\\ 
&& + 12 v_{2}^2 -56 \vec{v}_{1}\cdot\vec{n} \vec{v}_{2}\cdot\vec{n} + 34 ( \vec{v}_{1}\cdot\vec{n})^{2} \big) + 2 \vec{S}_{1}\times\vec{v}_{2}\cdot\vec{a}_{2} \big( 7 v_{1}^2 -8 ( \vec{v}_{1}\cdot\vec{n})^{2} \big) \Big] \nn\\ 
&& - 	\frac{G^2 m_{2}{}^2}{24 r{}^2} \Big[ \vec{S}_{1}\times\vec{n}\cdot\vec{v}_{1} \big( 6 v_{1}^2 \vec{a}_{1}\cdot\vec{n} + 21 \vec{v}_{1}\cdot\vec{n} \vec{v}_{1}\cdot\vec{a}_{1} - 192 \vec{v}_{1}\cdot\vec{a}_{1} \vec{v}_{2}\cdot\vec{n} \nn\\ 
&& + 42 \vec{a}_{1}\cdot\vec{n} \vec{v}_{1}\cdot\vec{v}_{2} + 54 \vec{v}_{1}\cdot\vec{n} \vec{a}_{1}\cdot\vec{v}_{2} + 506 \vec{v}_{2}\cdot\vec{n} \vec{a}_{1}\cdot\vec{v}_{2} - 56 \vec{a}_{1}\cdot\vec{n} v_{2}^2 + 156 v_{1}^2 \vec{a}_{2}\cdot\vec{n} \nn\\ 
&& - 56 \vec{v}_{1}\cdot\vec{v}_{2} \vec{a}_{2}\cdot\vec{n} - 752 v_{2}^2 \vec{a}_{2}\cdot\vec{n} + 6 \vec{v}_{1}\cdot\vec{n} \vec{v}_{1}\cdot\vec{a}_{2} + 304 \vec{v}_{2}\cdot\vec{n} \vec{v}_{1}\cdot\vec{a}_{2} + 40 \vec{v}_{1}\cdot\vec{n} \vec{v}_{2}\cdot\vec{a}_{2} \nn\\ 
&& - 1984 \vec{v}_{2}\cdot\vec{n} \vec{v}_{2}\cdot\vec{a}_{2} -96 \vec{v}_{1}\cdot\vec{n} \vec{a}_{1}\cdot\vec{n} \vec{v}_{2}\cdot\vec{n} + 1232 \vec{v}_{1}\cdot\vec{n} \vec{v}_{2}\cdot\vec{n} \vec{a}_{2}\cdot\vec{n} \nn\\ 
&& - 624 \vec{a}_{2}\cdot\vec{n} ( \vec{v}_{1}\cdot\vec{n})^{2} + 80 \vec{a}_{1}\cdot\vec{n} ( \vec{v}_{2}\cdot\vec{n})^{2} - 592 \vec{a}_{2}\cdot\vec{n} ( \vec{v}_{2}\cdot\vec{n})^{2} \big) \nn\\ 
&& + \vec{S}_{1}\times\vec{n}\cdot\vec{a}_{1} \big( 15 \vec{v}_{1}\cdot\vec{n} v_{1}^2 - 84 v_{1}^2 \vec{v}_{2}\cdot\vec{n} + 90 \vec{v}_{1}\cdot\vec{n} \vec{v}_{1}\cdot\vec{v}_{2} + 454 \vec{v}_{2}\cdot\vec{n} \vec{v}_{1}\cdot\vec{v}_{2} \nn\\ 
&& - 112 \vec{v}_{1}\cdot\vec{n} v_{2}^2 - 1032 \vec{v}_{2}\cdot\vec{n} v_{2}^2 -96 \vec{v}_{2}\cdot\vec{n} ( \vec{v}_{1}\cdot\vec{n})^{2} + 160 \vec{v}_{1}\cdot\vec{n} ( \vec{v}_{2}\cdot\vec{n})^{2} - 64 ( \vec{v}_{2}\cdot\vec{n})^{3} \big) \nn\\ 
&& + \vec{S}_{1}\times\vec{v}_{1}\cdot\vec{a}_{1} \big( 270 v_{1}^2 - 480 \vec{v}_{1}\cdot\vec{v}_{2} + 208 v_{2}^2 -192 \vec{v}_{1}\cdot\vec{n} \vec{v}_{2}\cdot\vec{n} + 9 ( \vec{v}_{1}\cdot\vec{n})^{2} \nn\\ 
&& + 94 ( \vec{v}_{2}\cdot\vec{n})^{2} \big) + 4 \vec{S}_{1}\times\vec{n}\cdot\vec{v}_{2} \big( 6 \vec{v}_{1}\cdot\vec{a}_{1} \vec{v}_{2}\cdot\vec{n} + 3 \vec{a}_{1}\cdot\vec{n} \vec{v}_{1}\cdot\vec{v}_{2} + 86 \vec{v}_{2}\cdot\vec{n} \vec{a}_{1}\cdot\vec{v}_{2} \nn\\ 
&& + 34 \vec{a}_{1}\cdot\vec{n} v_{2}^2 - 47 v_{1}^2 \vec{a}_{2}\cdot\vec{n} + 195 \vec{v}_{1}\cdot\vec{v}_{2} \vec{a}_{2}\cdot\vec{n} + 52 v_{2}^2 \vec{a}_{2}\cdot\vec{n} + 29 \vec{v}_{1}\cdot\vec{n} \vec{v}_{1}\cdot\vec{a}_{2} \nn\\ 
&& + 63 \vec{v}_{2}\cdot\vec{n} \vec{v}_{1}\cdot\vec{a}_{2} + 42 \vec{v}_{1}\cdot\vec{n} \vec{v}_{2}\cdot\vec{a}_{2} + 230 \vec{v}_{2}\cdot\vec{n} \vec{v}_{2}\cdot\vec{a}_{2} -24 \vec{v}_{1}\cdot\vec{n} \vec{a}_{1}\cdot\vec{n} \vec{v}_{2}\cdot\vec{n} \nn\\ 
&& - 564 \vec{v}_{1}\cdot\vec{n} \vec{v}_{2}\cdot\vec{n} \vec{a}_{2}\cdot\vec{n} + 188 \vec{a}_{2}\cdot\vec{n} ( \vec{v}_{1}\cdot\vec{n})^{2} - 112 \vec{a}_{1}\cdot\vec{n} ( \vec{v}_{2}\cdot\vec{n})^{2} \nn\\ 
&& + 224 \vec{a}_{2}\cdot\vec{n} ( \vec{v}_{2}\cdot\vec{n})^{2} \big) + 2 \vec{S}_{1}\times\vec{v}_{1}\cdot\vec{v}_{2} \big( 48 \vec{v}_{1}\cdot\vec{a}_{1} + 173 \vec{a}_{1}\cdot\vec{v}_{2} + 140 \vec{v}_{1}\cdot\vec{a}_{2} \nn\\ 
&& - 346 \vec{v}_{2}\cdot\vec{a}_{2} -24 \vec{v}_{1}\cdot\vec{n} \vec{a}_{1}\cdot\vec{n} - 55 \vec{a}_{1}\cdot\vec{n} \vec{v}_{2}\cdot\vec{n} - 220 \vec{v}_{1}\cdot\vec{n} \vec{a}_{2}\cdot\vec{n} + 278 \vec{v}_{2}\cdot\vec{n} \vec{a}_{2}\cdot\vec{n} \big) \nn\\ 
&& + \vec{S}_{1}\times\vec{a}_{1}\cdot\vec{v}_{2} \big( 63 v_{1}^2 + 398 \vec{v}_{1}\cdot\vec{v}_{2} - 440 v_{2}^2 -802 \vec{v}_{1}\cdot\vec{n} \vec{v}_{2}\cdot\vec{n} - 54 ( \vec{v}_{1}\cdot\vec{n})^{2} \nn\\ 
&& + 724 ( \vec{v}_{2}\cdot\vec{n})^{2} \big) + 4 \vec{S}_{1}\times\vec{n}\cdot\vec{a}_{2} \big( 3 \vec{v}_{1}\cdot\vec{n} v_{1}^2 + 9 v_{1}^2 \vec{v}_{2}\cdot\vec{n} + 45 \vec{v}_{1}\cdot\vec{n} \vec{v}_{1}\cdot\vec{v}_{2} \nn\\ 
&& + 7 \vec{v}_{2}\cdot\vec{n} \vec{v}_{1}\cdot\vec{v}_{2} + 17 \vec{v}_{1}\cdot\vec{n} v_{2}^2 + 162 \vec{v}_{2}\cdot\vec{n} v_{2}^2 -4 ( \vec{v}_{1}\cdot\vec{n})^{3} - 36 \vec{v}_{2}\cdot\vec{n} ( \vec{v}_{1}\cdot\vec{n})^{2} \nn\\ 
&& + 22 \vec{v}_{1}\cdot\vec{n} ( \vec{v}_{2}\cdot\vec{n})^{2} - 84 ( \vec{v}_{2}\cdot\vec{n})^{3} \big) - \vec{S}_{1}\times\vec{v}_{1}\cdot\vec{a}_{2} \big( 171 v_{1}^2 - 512 \vec{v}_{1}\cdot\vec{v}_{2} \nn\\ 
&& + 410 v_{2}^2 -272 \vec{v}_{1}\cdot\vec{n} \vec{v}_{2}\cdot\vec{n} + 42 ( \vec{v}_{1}\cdot\vec{n})^{2} + 266 ( \vec{v}_{2}\cdot\vec{n})^{2} \big) - 4 \vec{S}_{1}\times\vec{v}_{2}\cdot\vec{a}_{2} \big( 4 v_{1}^2 \nn\\ 
&& - 34 \vec{v}_{1}\cdot\vec{v}_{2} + 7 v_{2}^2 + 152 \vec{v}_{1}\cdot\vec{n} \vec{v}_{2}\cdot\vec{n} - 8 ( \vec{v}_{1}\cdot\vec{n})^{2} - 143 ( \vec{v}_{2}\cdot\vec{n})^{2} \big) \Big]\nn\\ && +  	\frac{G^3 m_{1}{}^2 m_{2}}{450 r{}^3} \Big[ 12 \vec{S}_{1}\times\vec{n}\cdot\vec{v}_{1} \big( 2398 \vec{a}_{1}\cdot\vec{n} + 401 \vec{a}_{2}\cdot\vec{n} \big) - 3 \vec{S}_{1}\times\vec{n}\cdot\vec{a}_{1} \big( 4358 \vec{v}_{1}\cdot\vec{n} \nn\\ 
&& - 7829 \vec{v}_{2}\cdot\vec{n} \big) - 9 \vec{S}_{1}\times\vec{n}\cdot\vec{v}_{2} \big( 3157 \vec{a}_{1}\cdot\vec{n} + 3 \vec{a}_{2}\cdot\vec{n} \big) - 9 \vec{S}_{1}\times\vec{n}\cdot\vec{a}_{2} \big( 532 \vec{v}_{1}\cdot\vec{n} \nn\\ 
&& + 3 \vec{v}_{2}\cdot\vec{n} \big) + 12000 \vec{S}_{1}\times\vec{v}_{1}\cdot\vec{a}_{1} + 11300 \vec{S}_{1}\times\vec{a}_{1}\cdot\vec{v}_{2} - 700 \vec{S}_{1}\times\vec{v}_{1}\cdot\vec{a}_{2} \Big] \nn\\ 
&& + 	\frac{G^3 m_{1}{}^2 m_{2}}{15 r{}^3} {\Big( \frac{1}{\epsilon} - 3\log \frac{r}{R_0} \Big)} \Big[ 3 \vec{S}_{1}\times\vec{n}\cdot\vec{v}_{1} \big( 161 \vec{a}_{1}\cdot\vec{n} + 67 \vec{a}_{2}\cdot\vec{n} \big) \nn\\ 
&& - 3 \vec{S}_{1}\times\vec{n}\cdot\vec{a}_{1} \big( 39 \vec{v}_{1}\cdot\vec{n} - 157 \vec{v}_{2}\cdot\vec{n} \big) - 6 \vec{S}_{1}\times\vec{n}\cdot\vec{v}_{2} \big( 74 \vec{a}_{1}\cdot\vec{n} + \vec{a}_{2}\cdot\vec{n} \big) \nn\\ 
&& - 6 \vec{S}_{1}\times\vec{n}\cdot\vec{a}_{2} \big( 19 \vec{v}_{1}\cdot\vec{n} + \vec{v}_{2}\cdot\vec{n} \big) + 200 \vec{S}_{1}\times\vec{v}_{1}\cdot\vec{a}_{1} + 305 \vec{S}_{1}\times\vec{a}_{1}\cdot\vec{v}_{2} \nn\\ 
&& + 105 \vec{S}_{1}\times\vec{v}_{1}\cdot\vec{a}_{2} \Big] + 	\frac{2 G^3 m_{2}{}^3}{15 r{}^3} {\Big( \frac{1}{\epsilon} - 3\log \frac{r}{R_0} \Big)} \Big[ 3 \vec{S}_{1}\times\vec{n}\cdot\vec{v}_{1} \big( 5 \vec{a}_{1}\cdot\vec{n} - 67 \vec{a}_{2}\cdot\vec{n} \big) \nn\\ 
&& + 3 \vec{S}_{1}\times\vec{n}\cdot\vec{a}_{1} \big( 10 \vec{v}_{1}\cdot\vec{n} - 17 \vec{v}_{2}\cdot\vec{n} \big) + 3 \vec{S}_{1}\times\vec{n}\cdot\vec{v}_{2} \big( 8 \vec{a}_{1}\cdot\vec{n} + 111 \vec{a}_{2}\cdot\vec{n} \big) \nn\\ 
&& + 3 \vec{S}_{1}\times\vec{n}\cdot\vec{a}_{2} \big( 63 \vec{v}_{1}\cdot\vec{n} + \vec{v}_{2}\cdot\vec{n} \big) - 5 \vec{S}_{1}\times\vec{v}_{1}\cdot\vec{a}_{1} - 25 \vec{S}_{1}\times\vec{a}_{1}\cdot\vec{v}_{2} \nn\\ 
&& - 130 \vec{S}_{1}\times\vec{v}_{1}\cdot\vec{a}_{2} + 110 \vec{S}_{1}\times\vec{v}_{2}\cdot\vec{a}_{2} \Big] + 	\frac{G^3 m_{2}{}^3}{450 r{}^3} \Big[ 36 \vec{S}_{1}\times\vec{n}\cdot\vec{v}_{1} \big( 100 \vec{a}_{1}\cdot\vec{n} \nn\\ 
&& - 879 \vec{a}_{2}\cdot\vec{n} \big) + 9 \vec{S}_{1}\times\vec{n}\cdot\vec{a}_{1} \big( 650 \vec{v}_{1}\cdot\vec{n} - 791 \vec{v}_{2}\cdot\vec{n} \big) + 3 \vec{S}_{1}\times\vec{n}\cdot\vec{v}_{2} \big( 827 \vec{a}_{1}\cdot\vec{n} \nn\\ 
&& + 12454 \vec{a}_{2}\cdot\vec{n} \big) + 18 \vec{S}_{1}\times\vec{n}\cdot\vec{a}_{2} \big( 1642 \vec{v}_{1}\cdot\vec{n} - 891 \vec{v}_{2}\cdot\vec{n} \big) - 2550 \vec{S}_{1}\times\vec{v}_{1}\cdot\vec{a}_{1} \nn\\ 
&& - 6350 \vec{S}_{1}\times\vec{a}_{1}\cdot\vec{v}_{2} - 18600 \vec{S}_{1}\times\vec{v}_{1}\cdot\vec{a}_{2} + 14800 \vec{S}_{1}\times\vec{v}_{2}\cdot\vec{a}_{2} \Big] \nn\\ 
&& + 	\frac{G^3 m_{1} m_{2}{}^2}{288 r{}^3} \Big[ 3 \vec{S}_{1}\times\vec{n}\cdot\vec{v}_{1} \big( {(11680 - 972 \pi^2)} \vec{a}_{1}\cdot\vec{n} - {(16704 - 423 \pi^2)} \vec{a}_{2}\cdot\vec{n} \big) \nn\\ 
&& + 3 \vec{S}_{1}\times\vec{n}\cdot\vec{a}_{1} \big( {(4544 - 828 \pi^2)} \vec{v}_{1}\cdot\vec{n} - {(7792 - 171 \pi^2)} \vec{v}_{2}\cdot\vec{n} \big) \nn\\ 
&& - 3 \vec{S}_{1}\times\vec{n}\cdot\vec{v}_{2} \big( {(10368 - 675 \pi^2)} \vec{a}_{1}\cdot\vec{n} - {(13664 - 1422 \pi^2)} \vec{a}_{2}\cdot\vec{n} \big) \nn\\ 
&& + 3 \vec{S}_{1}\times\vec{n}\cdot\vec{a}_{2} \big( {(4016 + 171 \pi^2)} \vec{v}_{1}\cdot\vec{n} - {(2016 + 810 \pi^2)} \vec{v}_{2}\cdot\vec{n} \big) \nn\\ 
&& + 16 {(374 - 9 \pi^2)} \vec{S}_{1}\times\vec{v}_{1}\cdot\vec{a}_{1} - 8 {(374 + 63 \pi^2)} \vec{S}_{1}\times\vec{a}_{1}\cdot\vec{v}_{2} \nn\\ 
&& - 28 {(908 - 9 \pi^2)} \vec{S}_{1}\times\vec{v}_{1}\cdot\vec{a}_{2} + 4 {(4112 - 153 \pi^2)} \vec{S}_{1}\times\vec{v}_{2}\cdot\vec{a}_{2} \Big]
+ (1 \leftrightarrow 2),
\eea
and
\bea
(V_{\dot{S}})_{3,1} &=& - 	\frac{G m_{2}}{8 r} \Big[ \dot{\vec{S}}_{1}\times\vec{n}\cdot\vec{v}_{1} \big( 3 v_{1}^2 \vec{v}_{2}\cdot\vec{n} \vec{v}_{1}\cdot\vec{v}_{2} - 2 \vec{v}_{1}\cdot\vec{n} v_{1}^2 v_{2}^2 - 5 v_{1}^2 \vec{v}_{2}\cdot\vec{n} v_{2}^2 + 2 \vec{v}_{2}\cdot\vec{n} \vec{v}_{1}\cdot\vec{v}_{2} v_{2}^2 \nn\\ 
&& + 2 \vec{v}_{2}\cdot\vec{n} ( \vec{v}_{1}\cdot\vec{v}_{2})^{2} - 3 \vec{v}_{2}\cdot\vec{n} v_{1}^{4} - 2 \vec{v}_{1}\cdot\vec{n} v_{2}^{4} - 3 \vec{v}_{2}\cdot\vec{n} v_{2}^{4} + 9 \vec{v}_{2}\cdot\vec{n} v_{2}^2 ( \vec{v}_{1}\cdot\vec{n})^{2} \nn\\ 
&& + 6 \vec{v}_{1}\cdot\vec{n} v_{1}^2 ( \vec{v}_{2}\cdot\vec{n})^{2} + 3 v_{1}^2 ( \vec{v}_{2}\cdot\vec{n})^{3} + 6 \vec{v}_{1}\cdot\vec{n} v_{2}^2 ( \vec{v}_{2}\cdot\vec{n})^{2} -15 ( \vec{v}_{1}\cdot\vec{n})^{2} ( \vec{v}_{2}\cdot\vec{n})^{3} \big) \nn\\ 
&& + \dot{\vec{S}}_{1}\times\vec{n}\cdot\vec{v}_{2} \big( 3 v_{1}^2 \vec{v}_{2}\cdot\vec{n} v_{2}^2 + 2 \vec{v}_{1}\cdot\vec{n} \vec{v}_{1}\cdot\vec{v}_{2} v_{2}^2 - 2 \vec{v}_{2}\cdot\vec{n} ( \vec{v}_{1}\cdot\vec{v}_{2})^{2} + 2 \vec{v}_{1}\cdot\vec{n} v_{2}^{4} \nn\\ 
&& + 3 \vec{v}_{2}\cdot\vec{n} v_{2}^{4} -9 \vec{v}_{2}\cdot\vec{n} v_{2}^2 ( \vec{v}_{1}\cdot\vec{n})^{2} - 3 v_{1}^2 ( \vec{v}_{2}\cdot\vec{n})^{3} - 6 \vec{v}_{1}\cdot\vec{n} \vec{v}_{1}\cdot\vec{v}_{2} ( \vec{v}_{2}\cdot\vec{n})^{2} \nn\\ 
&& - 6 \vec{v}_{1}\cdot\vec{n} v_{2}^2 ( \vec{v}_{2}\cdot\vec{n})^{2} + 15 ( \vec{v}_{1}\cdot\vec{n})^{2} ( \vec{v}_{2}\cdot\vec{n})^{3} \big) - \dot{\vec{S}}_{1}\times\vec{v}_{1}\cdot\vec{v}_{2} \big( 2 v_{1}^2 \vec{v}_{1}\cdot\vec{v}_{2} - 9 v_{1}^2 v_{2}^2 \nn\\ 
&& + 2 \vec{v}_{1}\cdot\vec{v}_{2} v_{2}^2 + 2 ( \vec{v}_{1}\cdot\vec{v}_{2})^{2} - 9 v_{1}^{4} - 7 v_{2}^{4} + 5 \vec{v}_{1}\cdot\vec{n} v_{1}^2 \vec{v}_{2}\cdot\vec{n} + 8 \vec{v}_{1}\cdot\vec{n} \vec{v}_{2}\cdot\vec{n} \vec{v}_{1}\cdot\vec{v}_{2} \nn\\ 
&& + 3 v_{2}^2 ( \vec{v}_{1}\cdot\vec{n})^{2} + 3 v_{1}^2 ( \vec{v}_{2}\cdot\vec{n})^{2} - 2 \vec{v}_{1}\cdot\vec{v}_{2} ( \vec{v}_{2}\cdot\vec{n})^{2} \nn\\ 
&& - 2 v_{2}^2 ( \vec{v}_{2}\cdot\vec{n})^{2} -9 ( \vec{v}_{1}\cdot\vec{n})^{2} ( \vec{v}_{2}\cdot\vec{n})^{2} + 6 \vec{v}_{1}\cdot\vec{n} ( \vec{v}_{2}\cdot\vec{n})^{3} \big) \Big]\nn\\ && - 	\frac{G^2 m_{2}{}^2}{48 r{}^2} \Big[ \dot{\vec{S}}_{1}\times\vec{n}\cdot\vec{v}_{1} \big( 21 \vec{v}_{1}\cdot\vec{n} v_{1}^2 - 176 v_{1}^2 \vec{v}_{2}\cdot\vec{n} + 152 \vec{v}_{1}\cdot\vec{n} \vec{v}_{1}\cdot\vec{v}_{2} \nn\\ 
&& + 924 \vec{v}_{2}\cdot\vec{n} \vec{v}_{1}\cdot\vec{v}_{2} - 216 \vec{v}_{1}\cdot\vec{n} v_{2}^2 - 2064 \vec{v}_{2}\cdot\vec{n} v_{2}^2 -160 \vec{v}_{2}\cdot\vec{n} ( \vec{v}_{1}\cdot\vec{n})^{2} \nn\\ 
&& + 312 \vec{v}_{1}\cdot\vec{n} ( \vec{v}_{2}\cdot\vec{n})^{2} - 128 ( \vec{v}_{2}\cdot\vec{n})^{3} \big) + 8 \dot{\vec{S}}_{1}\times\vec{n}\cdot\vec{v}_{2} \big( 4 v_{1}^2 \vec{v}_{2}\cdot\vec{n} + 2 \vec{v}_{1}\cdot\vec{n} \vec{v}_{1}\cdot\vec{v}_{2} \nn\\ 
&& + 147 \vec{v}_{2}\cdot\vec{n} \vec{v}_{1}\cdot\vec{v}_{2} + 69 \vec{v}_{1}\cdot\vec{n} v_{2}^2 + 18 \vec{v}_{2}\cdot\vec{n} v_{2}^2 -16 \vec{v}_{2}\cdot\vec{n} ( \vec{v}_{1}\cdot\vec{n})^{2} - 234 \vec{v}_{1}\cdot\vec{n} ( \vec{v}_{2}\cdot\vec{n})^{2} \nn\\ 
&& + 154 ( \vec{v}_{2}\cdot\vec{n})^{3} \big) + 4 \dot{\vec{S}}_{1}\times\vec{v}_{1}\cdot\vec{v}_{2} \big( 29 v_{1}^2 + 274 \vec{v}_{1}\cdot\vec{v}_{2} - 255 v_{2}^2 -401 \vec{v}_{1}\cdot\vec{n} \vec{v}_{2}\cdot\vec{n} \nn\\ 
&& - 22 ( \vec{v}_{1}\cdot\vec{n})^{2} + 423 ( \vec{v}_{2}\cdot\vec{n})^{2} \big) \Big] - 	\frac{G^2 m_{1} m_{2}}{6 r{}^2} \Big[ \dot{\vec{S}}_{1}\times\vec{n}\cdot\vec{v}_{1} \big( 129 \vec{v}_{1}\cdot\vec{n} v_{1}^2 + 44 v_{1}^2 \vec{v}_{2}\cdot\vec{n} \nn\\ 
&& - 50 \vec{v}_{1}\cdot\vec{n} \vec{v}_{1}\cdot\vec{v}_{2} - 22 \vec{v}_{2}\cdot\vec{n} \vec{v}_{1}\cdot\vec{v}_{2} + 70 \vec{v}_{1}\cdot\vec{n} v_{2}^2 - 30 \vec{v}_{2}\cdot\vec{n} v_{2}^2 + 168 ( \vec{v}_{1}\cdot\vec{n})^{3} \nn\\ 
&& - 584 \vec{v}_{2}\cdot\vec{n} ( \vec{v}_{1}\cdot\vec{n})^{2} + 272 \vec{v}_{1}\cdot\vec{n} ( \vec{v}_{2}\cdot\vec{n})^{2} \big) - 2 \dot{\vec{S}}_{1}\times\vec{n}\cdot\vec{v}_{2} \big( 3 \vec{v}_{1}\cdot\vec{n} v_{1}^2 + 19 v_{1}^2 \vec{v}_{2}\cdot\vec{n} \nn\\ 
&& + 32 \vec{v}_{1}\cdot\vec{n} \vec{v}_{1}\cdot\vec{v}_{2} + 6 \vec{v}_{2}\cdot\vec{n} \vec{v}_{1}\cdot\vec{v}_{2} + 3 \vec{v}_{1}\cdot\vec{n} v_{2}^2 - 27 \vec{v}_{2}\cdot\vec{n} v_{2}^2 + 52 ( \vec{v}_{1}\cdot\vec{n})^{3} \nn\\ 
&& - 196 \vec{v}_{2}\cdot\vec{n} ( \vec{v}_{1}\cdot\vec{n})^{2} + 120 \vec{v}_{1}\cdot\vec{n} ( \vec{v}_{2}\cdot\vec{n})^{2} \big) + \dot{\vec{S}}_{1}\times\vec{v}_{1}\cdot\vec{v}_{2} \big( 34 v_{1}^2 - 18 \vec{v}_{1}\cdot\vec{v}_{2} \nn\\ 
&& + 51 v_{2}^2 -102 \vec{v}_{1}\cdot\vec{n} \vec{v}_{2}\cdot\vec{n} + 4 ( \vec{v}_{1}\cdot\vec{n})^{2} + 24 ( \vec{v}_{2}\cdot\vec{n})^{2} \big) \Big]\nn\\ && +  	\frac{G^3 m_{1}{}^2 m_{2}}{450 r{}^3} \Big[ 3 \dot{\vec{S}}_{1}\times\vec{n}\cdot\vec{v}_{1} \big( 2744 \vec{v}_{1}\cdot\vec{n} + 1027 \vec{v}_{2}\cdot\vec{n} \big) - 9 \dot{\vec{S}}_{1}\times\vec{n}\cdot\vec{v}_{2} \big( 529 \vec{v}_{1}\cdot\vec{n} \nn\\ 
&& + 56 \vec{v}_{2}\cdot\vec{n} \big) - 2306 \dot{\vec{S}}_{1}\times\vec{v}_{1}\cdot\vec{v}_{2} \Big] + 	\frac{G^3 m_{2}{}^3}{150 r{}^3} \Big[ 21 \dot{\vec{S}}_{1}\times\vec{n}\cdot\vec{v}_{1} \big( 100 \vec{v}_{1}\cdot\vec{n} - 113 \vec{v}_{2}\cdot\vec{n} \big) \nn\\ 
&& + 3 \dot{\vec{S}}_{1}\times\vec{n}\cdot\vec{v}_{2} \big( 559 \vec{v}_{1}\cdot\vec{n} + 868 \vec{v}_{2}\cdot\vec{n} \big) - 2300 \dot{\vec{S}}_{1}\times\vec{v}_{1}\cdot\vec{v}_{2} \Big] \nn\\ 
&& + 	\frac{2 G^3 m_{2}{}^3}{5 r{}^3} {\Big( \frac{1}{\epsilon} - 3\log \frac{r}{R_0} \Big)} \Big[ \dot{\vec{S}}_{1}\times\vec{n}\cdot\vec{v}_{1} \big( 10 \vec{v}_{1}\cdot\vec{n} - 17 \vec{v}_{2}\cdot\vec{n} \big) + \dot{\vec{S}}_{1}\times\vec{n}\cdot\vec{v}_{2} \big( 13 \vec{v}_{1}\cdot\vec{n} \nn\\ 
&& + 51 \vec{v}_{2}\cdot\vec{n} \big) - 10 \dot{\vec{S}}_{1}\times\vec{v}_{1}\cdot\vec{v}_{2} \Big] + 	\frac{G^3 m_{1}{}^2 m_{2}}{15 r{}^3} {\Big( \frac{1}{\epsilon} - 3\log \frac{r}{R_0} \Big)} \Big[ 3 \dot{\vec{S}}_{1}\times\vec{n}\cdot\vec{v}_{1} \big( 97 \vec{v}_{1}\cdot\vec{n} \nn\\ 
&& + 21 \vec{v}_{2}\cdot\vec{n} \big) - 12 \dot{\vec{S}}_{1}\times\vec{n}\cdot\vec{v}_{2} \big( 19 \vec{v}_{1}\cdot\vec{n} - 9 \vec{v}_{2}\cdot\vec{n} \big) + 97 \dot{\vec{S}}_{1}\times\vec{v}_{1}\cdot\vec{v}_{2} \Big] \nn\\ 
&& + 	\frac{G^3 m_{1} m_{2}{}^2}{96 r{}^3} \Big[ \dot{\vec{S}}_{1}\times\vec{n}\cdot\vec{v}_{1} \big( {(5520 - 954 \pi^2)} \vec{v}_{1}\cdot\vec{n} - {(9392 - 297 \pi^2)} \vec{v}_{2}\cdot\vec{n} \big) \nn\\ 
&& - \dot{\vec{S}}_{1}\times\vec{n}\cdot\vec{v}_{2} \big( {(512 - 378 \pi^2)} \vec{v}_{1}\cdot\vec{n} - {(2416 - 1017 \pi^2)} \vec{v}_{2}\cdot\vec{n} \big) \nn\\ 
&& - {(4816 + 27 \pi^2)} \dot{\vec{S}}_{1}\times\vec{v}_{1}\cdot\vec{v}_{2} \Big] + (1 \leftrightarrow 2).
\eea

The pieces with further time derivatives are then:
\bea
\stackrel{(2)}{V}_{3,1} &=& - 	\frac{1}{24} G m_{2} \Big[ (3 \vec{S}_{1}\times\vec{n}\cdot\vec{v}_{1} \big( 2 \vec{v}_{1}\cdot\dot{\vec{a}}_{1} v_{2}^2 - 2 \dot{\vec{a}}_{1}\cdot\vec{v}_{2} v_{2}^2 - 2 v_{1}^2 \vec{v}_{1}\cdot\dot{\vec{a}}_{2} + 2 v_{1}^2 \vec{v}_{2}\cdot\dot{\vec{a}}_{2} \nn\\ 
&& - \dot{\vec{a}}_{1}\cdot\vec{n} \vec{v}_{2}\cdot\vec{n} v_{2}^2 - \vec{v}_{1}\cdot\vec{n} v_{1}^2 \dot{\vec{a}}_{2}\cdot\vec{n} + 2 \vec{v}_{1}\cdot\dot{\vec{a}}_{2} ( \vec{v}_{1}\cdot\vec{n})^{2} - 2 \vec{v}_{2}\cdot\dot{\vec{a}}_{2} ( \vec{v}_{1}\cdot\vec{n})^{2} \nn\\ 
&& - 2 \vec{v}_{1}\cdot\dot{\vec{a}}_{1} ( \vec{v}_{2}\cdot\vec{n})^{2} + 2 \dot{\vec{a}}_{1}\cdot\vec{v}_{2} ( \vec{v}_{2}\cdot\vec{n})^{2} + \dot{\vec{a}}_{2}\cdot\vec{n} ( \vec{v}_{1}\cdot\vec{n})^{3} + \dot{\vec{a}}_{1}\cdot\vec{n} ( \vec{v}_{2}\cdot\vec{n})^{3} \big) \nn\\ 
&& + 3 \vec{S}_{1}\times\vec{n}\cdot\dot{\vec{a}}_{1} \big( v_{1}^2 v_{2}^2 + v_{2}^{4} -3 \vec{v}_{1}\cdot\vec{n} \vec{v}_{2}\cdot\vec{n} v_{2}^2 - v_{1}^2 ( \vec{v}_{2}\cdot\vec{n})^{2} - v_{2}^2 ( \vec{v}_{2}\cdot\vec{n})^{2} \nn\\ 
&& + 3 \vec{v}_{1}\cdot\vec{n} ( \vec{v}_{2}\cdot\vec{n})^{3} \big) + 2 \vec{S}_{1}\times\vec{v}_{1}\cdot\dot{\vec{a}}_{1} \big( 18 v_{1}^2 \vec{v}_{2}\cdot\vec{n} - 12 \vec{v}_{2}\cdot\vec{n} \vec{v}_{1}\cdot\vec{v}_{2} + 6 \vec{v}_{1}\cdot\vec{n} v_{2}^2 \nn\\ 
&& + 9 \vec{v}_{2}\cdot\vec{n} v_{2}^2 -6 \vec{v}_{1}\cdot\vec{n} ( \vec{v}_{2}\cdot\vec{n})^{2} - ( \vec{v}_{2}\cdot\vec{n})^{3} \big) + 3 \vec{S}_{1}\times\vec{n}\cdot\vec{v}_{2} \big( \dot{\vec{a}}_{1}\cdot\vec{v}_{2} v_{2}^2 + v_{1}^2 \vec{v}_{1}\cdot\dot{\vec{a}}_{2} \nn\\ 
&& - 2 v_{1}^2 \vec{v}_{2}\cdot\dot{\vec{a}}_{2} + \dot{\vec{a}}_{1}\cdot\vec{n} \vec{v}_{2}\cdot\vec{n} v_{2}^2 + \vec{v}_{1}\cdot\vec{n} v_{1}^2 \dot{\vec{a}}_{2}\cdot\vec{n} - \vec{v}_{1}\cdot\dot{\vec{a}}_{2} ( \vec{v}_{1}\cdot\vec{n})^{2} + 2 \vec{v}_{2}\cdot\dot{\vec{a}}_{2} ( \vec{v}_{1}\cdot\vec{n})^{2} \nn\\ 
&& - \dot{\vec{a}}_{1}\cdot\vec{v}_{2} ( \vec{v}_{2}\cdot\vec{n})^{2} - \dot{\vec{a}}_{2}\cdot\vec{n} ( \vec{v}_{1}\cdot\vec{n})^{3} - \dot{\vec{a}}_{1}\cdot\vec{n} ( \vec{v}_{2}\cdot\vec{n})^{3} \big) + 3 \vec{S}_{1}\times\vec{v}_{1}\cdot\vec{v}_{2} \big( 12 \vec{v}_{1}\cdot\dot{\vec{a}}_{1} \vec{v}_{2}\cdot\vec{n} \nn\\ 
&& - 2 \vec{v}_{2}\cdot\vec{n} \dot{\vec{a}}_{1}\cdot\vec{v}_{2} - v_{1}^2 \dot{\vec{a}}_{2}\cdot\vec{n} + 2 \vec{v}_{1}\cdot\vec{n} \vec{v}_{1}\cdot\dot{\vec{a}}_{2} - 4 \vec{v}_{1}\cdot\vec{n} \vec{v}_{2}\cdot\dot{\vec{a}}_{2} + \dot{\vec{a}}_{2}\cdot\vec{n} ( \vec{v}_{1}\cdot\vec{n})^{2} \big) \nn\\ 
&& + \vec{S}_{1}\times\dot{\vec{a}}_{1}\cdot\vec{v}_{2} \big( 18 v_{1}^2 \vec{v}_{2}\cdot\vec{n} + 6 \vec{v}_{2}\cdot\vec{n} \vec{v}_{1}\cdot\vec{v}_{2} + 18 \vec{v}_{1}\cdot\vec{n} v_{2}^2 \nn\\ 
&& + 15 \vec{v}_{2}\cdot\vec{n} v_{2}^2 -18 \vec{v}_{1}\cdot\vec{n} ( \vec{v}_{2}\cdot\vec{n})^{2} + ( \vec{v}_{2}\cdot\vec{n})^{3} \big) - 3 \vec{S}_{1}\times\vec{n}\cdot\dot{\vec{a}}_{2} \big( v_{1}^2 \vec{v}_{1}\cdot\vec{v}_{2} \nn\\ 
&& + v_{1}^2 v_{2}^2 -3 \vec{v}_{1}\cdot\vec{n} v_{1}^2 \vec{v}_{2}\cdot\vec{n} - \vec{v}_{1}\cdot\vec{v}_{2} ( \vec{v}_{1}\cdot\vec{n})^{2} - v_{2}^2 ( \vec{v}_{1}\cdot\vec{n})^{2} + 3 \vec{v}_{2}\cdot\vec{n} ( \vec{v}_{1}\cdot\vec{n})^{3} \big) \nn\\ 
&& + \vec{S}_{1}\times\vec{v}_{1}\cdot\dot{\vec{a}}_{2} \big( 6 \vec{v}_{1}\cdot\vec{n} v_{1}^2 - 9 v_{1}^2 \vec{v}_{2}\cdot\vec{n} - 6 \vec{v}_{1}\cdot\vec{n} \vec{v}_{1}\cdot\vec{v}_{2} - 6 \vec{v}_{1}\cdot\vec{n} v_{2}^2 -2 ( \vec{v}_{1}\cdot\vec{n})^{3} \nn\\ 
&& + 9 \vec{v}_{2}\cdot\vec{n} ( \vec{v}_{1}\cdot\vec{n})^{2} \big) - 2 \vec{S}_{1}\times\vec{v}_{2}\cdot\dot{\vec{a}}_{2} \big( 3 \vec{v}_{1}\cdot\vec{n} v_{1}^2 - ( \vec{v}_{1}\cdot\vec{n})^{3} \big)) + (3 \ddot{\vec{S}}_{1}\times\vec{n}\cdot\vec{v}_{1} \big( v_{1}^2 v_{2}^2 \nn\\ 
&& + v_{2}^{4} -3 \vec{v}_{1}\cdot\vec{n} \vec{v}_{2}\cdot\vec{n} v_{2}^2 - v_{1}^2 ( \vec{v}_{2}\cdot\vec{n})^{2} - v_{2}^2 ( \vec{v}_{2}\cdot\vec{n})^{2} + 3 \vec{v}_{1}\cdot\vec{n} ( \vec{v}_{2}\cdot\vec{n})^{3} \big) \nn\\ 
&& - 3 \ddot{\vec{S}}_{1}\times\vec{n}\cdot\vec{v}_{2} \big( \vec{v}_{1}\cdot\vec{v}_{2} v_{2}^2 + v_{2}^{4} -3 \vec{v}_{1}\cdot\vec{n} \vec{v}_{2}\cdot\vec{n} v_{2}^2 - \vec{v}_{1}\cdot\vec{v}_{2} ( \vec{v}_{2}\cdot\vec{n})^{2} - v_{2}^2 ( \vec{v}_{2}\cdot\vec{n})^{2} \nn\\ 
&& + 3 \vec{v}_{1}\cdot\vec{n} ( \vec{v}_{2}\cdot\vec{n})^{3} \big) + 3 \ddot{\vec{S}}_{1}\times\vec{v}_{1}\cdot\vec{v}_{2} \big( 6 v_{1}^2 \vec{v}_{2}\cdot\vec{n} + 6 \vec{v}_{2}\cdot\vec{n} \vec{v}_{1}\cdot\vec{v}_{2} + 6 \vec{v}_{1}\cdot\vec{n} v_{2}^2 \nn\\ 
&& + 3 \vec{v}_{2}\cdot\vec{n} v_{2}^2 -6 \vec{v}_{1}\cdot\vec{n} ( \vec{v}_{2}\cdot\vec{n})^{2} + ( \vec{v}_{2}\cdot\vec{n})^{3} \big)) + (3 \vec{S}_{1}\times\vec{n}\cdot\vec{v}_{1} \big( 2 a_{1}^2 v_{2}^2 \nn\\ 
&& + 10 \vec{v}_{1}\cdot\vec{a}_{1} \vec{v}_{1}\cdot\vec{a}_{2} + \vec{a}_{1}\cdot\vec{v}_{2} \vec{v}_{1}\cdot\vec{a}_{2} + 6 v_{1}^2 \vec{a}_{1}\cdot\vec{a}_{2} + 4 \vec{v}_{1}\cdot\vec{v}_{2} \vec{a}_{1}\cdot\vec{a}_{2} + 5 v_{2}^2 \vec{a}_{1}\cdot\vec{a}_{2} \nn\\ 
&& - 11 \vec{v}_{1}\cdot\vec{a}_{1} \vec{v}_{2}\cdot\vec{a}_{2} + 8 \vec{a}_{1}\cdot\vec{v}_{2} \vec{v}_{2}\cdot\vec{a}_{2} + 2 v_{1}^2 a_{2}^2 + v_{1}^2 \vec{a}_{1}\cdot\vec{n} \vec{a}_{2}\cdot\vec{n} + 4 \vec{v}_{1}\cdot\vec{n} \vec{v}_{1}\cdot\vec{a}_{1} \vec{a}_{2}\cdot\vec{n} \nn\\ 
&& + 3 \vec{v}_{1}\cdot\vec{a}_{1} \vec{v}_{2}\cdot\vec{n} \vec{a}_{2}\cdot\vec{n} - 3 \vec{v}_{1}\cdot\vec{n} \vec{a}_{1}\cdot\vec{v}_{2} \vec{a}_{2}\cdot\vec{n} + \vec{a}_{1}\cdot\vec{n} v_{2}^2 \vec{a}_{2}\cdot\vec{n} - 3 \vec{a}_{1}\cdot\vec{n} \vec{v}_{2}\cdot\vec{n} \vec{v}_{1}\cdot\vec{a}_{2} \nn\\ 
&& - 9 \vec{v}_{1}\cdot\vec{n} \vec{v}_{2}\cdot\vec{n} \vec{a}_{1}\cdot\vec{a}_{2} + 3 \vec{v}_{1}\cdot\vec{n} \vec{a}_{1}\cdot\vec{n} \vec{v}_{2}\cdot\vec{a}_{2} + 4 \vec{a}_{1}\cdot\vec{n} \vec{v}_{2}\cdot\vec{n} \vec{v}_{2}\cdot\vec{a}_{2} - 2 a_{2}^2 ( \vec{v}_{1}\cdot\vec{n})^{2} \nn\\ 
&& - 2 a_{1}^2 ( \vec{v}_{2}\cdot\vec{n})^{2} -9 \vec{v}_{1}\cdot\vec{n} \vec{a}_{1}\cdot\vec{n} \vec{v}_{2}\cdot\vec{n} \vec{a}_{2}\cdot\vec{n} \big) + 3 \vec{S}_{1}\times\vec{n}\cdot\vec{a}_{1} \big( 4 \vec{v}_{1}\cdot\vec{a}_{1} v_{2}^2 - 3 \vec{a}_{1}\cdot\vec{v}_{2} v_{2}^2 \nn\\ 
&& + 5 v_{1}^2 \vec{v}_{1}\cdot\vec{a}_{2} + 4 \vec{v}_{1}\cdot\vec{v}_{2} \vec{v}_{1}\cdot\vec{a}_{2} + 4 v_{2}^2 \vec{v}_{1}\cdot\vec{a}_{2} - 7 v_{1}^2 \vec{v}_{2}\cdot\vec{a}_{2} + 4 \vec{v}_{1}\cdot\vec{v}_{2} \vec{v}_{2}\cdot\vec{a}_{2} \nn\\ 
&& - 12 v_{2}^2 \vec{v}_{2}\cdot\vec{a}_{2} -3 \vec{a}_{1}\cdot\vec{n} \vec{v}_{2}\cdot\vec{n} v_{2}^2 + 2 \vec{v}_{1}\cdot\vec{n} v_{1}^2 \vec{a}_{2}\cdot\vec{n} + 3 v_{1}^2 \vec{v}_{2}\cdot\vec{n} \vec{a}_{2}\cdot\vec{n} + 2 \vec{v}_{1}\cdot\vec{n} v_{2}^2 \vec{a}_{2}\cdot\vec{n} \nn\\ 
&& - 6 \vec{v}_{1}\cdot\vec{n} \vec{v}_{2}\cdot\vec{n} \vec{v}_{1}\cdot\vec{a}_{2} + 8 \vec{v}_{1}\cdot\vec{n} \vec{v}_{2}\cdot\vec{n} \vec{v}_{2}\cdot\vec{a}_{2} + 3 \vec{v}_{2}\cdot\vec{a}_{2} ( \vec{v}_{1}\cdot\vec{n})^{2} - 4 \vec{v}_{1}\cdot\vec{a}_{1} ( \vec{v}_{2}\cdot\vec{n})^{2} \nn\\ 
&& + 3 \vec{a}_{1}\cdot\vec{v}_{2} ( \vec{v}_{2}\cdot\vec{n})^{2} -9 \vec{v}_{2}\cdot\vec{n} \vec{a}_{2}\cdot\vec{n} ( \vec{v}_{1}\cdot\vec{n})^{2} + 3 \vec{a}_{1}\cdot\vec{n} ( \vec{v}_{2}\cdot\vec{n})^{3} \big) \nn\\ 
&& + 3 \vec{S}_{1}\times\vec{v}_{1}\cdot\vec{a}_{1} \big( 24 \vec{v}_{1}\cdot\vec{a}_{1} \vec{v}_{2}\cdot\vec{n} - 12 \vec{v}_{2}\cdot\vec{n} \vec{a}_{1}\cdot\vec{v}_{2} + 2 \vec{a}_{1}\cdot\vec{n} v_{2}^2 - 3 v_{1}^2 \vec{a}_{2}\cdot\vec{n} - v_{2}^2 \vec{a}_{2}\cdot\vec{n} \nn\\ 
&& + 12 \vec{v}_{1}\cdot\vec{n} \vec{v}_{1}\cdot\vec{a}_{2} + 3 \vec{v}_{2}\cdot\vec{n} \vec{v}_{1}\cdot\vec{a}_{2} - 11 \vec{v}_{1}\cdot\vec{n} \vec{v}_{2}\cdot\vec{a}_{2} - 4 \vec{v}_{2}\cdot\vec{n} \vec{v}_{2}\cdot\vec{a}_{2} \nn\\ 
&& + 3 \vec{v}_{1}\cdot\vec{n} \vec{v}_{2}\cdot\vec{n} \vec{a}_{2}\cdot\vec{n} + 2 \vec{a}_{2}\cdot\vec{n} ( \vec{v}_{1}\cdot\vec{n})^{2} - 2 \vec{a}_{1}\cdot\vec{n} ( \vec{v}_{2}\cdot\vec{n})^{2} \big) \nn\\ 
&& - 3 \vec{S}_{1}\times\vec{n}\cdot\vec{v}_{2} \big( 5 \vec{a}_{1}\cdot\vec{v}_{2} \vec{v}_{1}\cdot\vec{a}_{2} + 7 \vec{v}_{1}\cdot\vec{v}_{2} \vec{a}_{1}\cdot\vec{a}_{2} + 3 v_{2}^2 \vec{a}_{1}\cdot\vec{a}_{2} - 3 \vec{v}_{1}\cdot\vec{a}_{1} \vec{v}_{2}\cdot\vec{a}_{2} \nn\\ 
&& + 4 \vec{a}_{1}\cdot\vec{v}_{2} \vec{v}_{2}\cdot\vec{a}_{2} + 2 v_{1}^2 a_{2}^2 + 3 \vec{v}_{1}\cdot\vec{a}_{1} \vec{v}_{2}\cdot\vec{n} \vec{a}_{2}\cdot\vec{n} + \vec{a}_{1}\cdot\vec{n} \vec{v}_{1}\cdot\vec{v}_{2} \vec{a}_{2}\cdot\vec{n} \nn\\ 
&& - \vec{v}_{1}\cdot\vec{n} \vec{a}_{1}\cdot\vec{v}_{2} \vec{a}_{2}\cdot\vec{n} + \vec{a}_{1}\cdot\vec{n} v_{2}^2 \vec{a}_{2}\cdot\vec{n} - \vec{a}_{1}\cdot\vec{n} \vec{v}_{2}\cdot\vec{n} \vec{v}_{1}\cdot\vec{a}_{2} - 5 \vec{v}_{1}\cdot\vec{n} \vec{v}_{2}\cdot\vec{n} \vec{a}_{1}\cdot\vec{a}_{2} \nn\\ 
&& + 3 \vec{v}_{1}\cdot\vec{n} \vec{a}_{1}\cdot\vec{n} \vec{v}_{2}\cdot\vec{a}_{2} + 4 \vec{a}_{1}\cdot\vec{n} \vec{v}_{2}\cdot\vec{n} \vec{v}_{2}\cdot\vec{a}_{2} \nn\\ 
&& - 2 a_{2}^2 ( \vec{v}_{1}\cdot\vec{n})^{2} -9 \vec{v}_{1}\cdot\vec{n} \vec{a}_{1}\cdot\vec{n} \vec{v}_{2}\cdot\vec{n} \vec{a}_{2}\cdot\vec{n} \big) + 3 \vec{S}_{1}\times\vec{v}_{1}\cdot\vec{v}_{2} \big( 12 a_{1}^2 \vec{v}_{2}\cdot\vec{n} - \vec{a}_{1}\cdot\vec{v}_{2} \vec{a}_{2}\cdot\vec{n} \nn\\ 
&& - 2 \vec{a}_{1}\cdot\vec{n} \vec{v}_{1}\cdot\vec{a}_{2} - 5 \vec{v}_{2}\cdot\vec{n} \vec{a}_{1}\cdot\vec{a}_{2} + 3 \vec{a}_{1}\cdot\vec{n} \vec{v}_{2}\cdot\vec{a}_{2} - 4 \vec{v}_{1}\cdot\vec{n} a_{2}^2 -3 \vec{a}_{1}\cdot\vec{n} \vec{v}_{2}\cdot\vec{n} \vec{a}_{2}\cdot\vec{n} \big) \nn\\ 
&& + 3 \vec{S}_{1}\times\vec{a}_{1}\cdot\vec{v}_{2} \big( 24 \vec{v}_{1}\cdot\vec{a}_{1} \vec{v}_{2}\cdot\vec{n} - 6 \vec{v}_{2}\cdot\vec{n} \vec{a}_{1}\cdot\vec{v}_{2} + 3 \vec{a}_{1}\cdot\vec{n} v_{2}^2 - 3 v_{1}^2 \vec{a}_{2}\cdot\vec{n} \nn\\ 
&& + \vec{v}_{1}\cdot\vec{v}_{2} \vec{a}_{2}\cdot\vec{n} + v_{2}^2 \vec{a}_{2}\cdot\vec{n} + 2 \vec{v}_{1}\cdot\vec{n} \vec{v}_{1}\cdot\vec{a}_{2} - \vec{v}_{2}\cdot\vec{n} \vec{v}_{1}\cdot\vec{a}_{2} - 9 \vec{v}_{1}\cdot\vec{n} \vec{v}_{2}\cdot\vec{a}_{2} \nn\\ 
&& + 4 \vec{v}_{2}\cdot\vec{n} \vec{v}_{2}\cdot\vec{a}_{2} -3 \vec{v}_{1}\cdot\vec{n} \vec{v}_{2}\cdot\vec{n} \vec{a}_{2}\cdot\vec{n} + 3 \vec{a}_{2}\cdot\vec{n} ( \vec{v}_{1}\cdot\vec{n})^{2} - 3 \vec{a}_{1}\cdot\vec{n} ( \vec{v}_{2}\cdot\vec{n})^{2} \big) \nn\\ 
&& - 3 \vec{S}_{1}\times\vec{n}\cdot\vec{a}_{2} \big( 6 \vec{v}_{1}\cdot\vec{v}_{2} \vec{a}_{1}\cdot\vec{v}_{2} - 3 \vec{v}_{1}\cdot\vec{a}_{1} v_{2}^2 + 2 \vec{a}_{1}\cdot\vec{v}_{2} v_{2}^2 - v_{1}^2 \vec{v}_{1}\cdot\vec{a}_{2} + 4 v_{1}^2 \vec{v}_{2}\cdot\vec{a}_{2} \nn\\ 
&& + 2 \vec{a}_{1}\cdot\vec{n} \vec{v}_{2}\cdot\vec{n} \vec{v}_{1}\cdot\vec{v}_{2} - 2 \vec{v}_{1}\cdot\vec{n} \vec{v}_{2}\cdot\vec{n} \vec{a}_{1}\cdot\vec{v}_{2} + 3 \vec{v}_{1}\cdot\vec{n} \vec{a}_{1}\cdot\vec{n} v_{2}^2 + 2 \vec{a}_{1}\cdot\vec{n} \vec{v}_{2}\cdot\vec{n} v_{2}^2 \nn\\ 
&& - 3 \vec{v}_{1}\cdot\vec{n} v_{1}^2 \vec{a}_{2}\cdot\vec{n} + \vec{v}_{1}\cdot\vec{a}_{2} ( \vec{v}_{1}\cdot\vec{n})^{2} - 4 \vec{v}_{2}\cdot\vec{a}_{2} ( \vec{v}_{1}\cdot\vec{n})^{2} + 3 \vec{v}_{1}\cdot\vec{a}_{1} ( \vec{v}_{2}\cdot\vec{n})^{2} \nn\\ 
&& + 3 \vec{a}_{2}\cdot\vec{n} ( \vec{v}_{1}\cdot\vec{n})^{3} - 9 \vec{v}_{1}\cdot\vec{n} \vec{a}_{1}\cdot\vec{n} ( \vec{v}_{2}\cdot\vec{n})^{2} \big) - 3 \vec{S}_{1}\times\vec{v}_{1}\cdot\vec{a}_{2} \big( 14 \vec{v}_{1}\cdot\vec{n} \vec{v}_{1}\cdot\vec{a}_{1} \nn\\ 
&& + 3 \vec{v}_{1}\cdot\vec{a}_{1} \vec{v}_{2}\cdot\vec{n} + 2 \vec{a}_{1}\cdot\vec{n} \vec{v}_{1}\cdot\vec{v}_{2} - 3 \vec{v}_{1}\cdot\vec{n} \vec{a}_{1}\cdot\vec{v}_{2} + 2 \vec{v}_{2}\cdot\vec{n} \vec{a}_{1}\cdot\vec{v}_{2} - 2 \vec{a}_{1}\cdot\vec{n} v_{2}^2 \nn\\ 
&& + 3 v_{1}^2 \vec{a}_{2}\cdot\vec{n} - 2 \vec{v}_{1}\cdot\vec{n} \vec{v}_{1}\cdot\vec{a}_{2} + 8 \vec{v}_{1}\cdot\vec{n} \vec{v}_{2}\cdot\vec{a}_{2} -3 \vec{v}_{1}\cdot\vec{n} \vec{a}_{1}\cdot\vec{n} \vec{v}_{2}\cdot\vec{n} - 3 \vec{a}_{2}\cdot\vec{n} ( \vec{v}_{1}\cdot\vec{n})^{2} \nn\\ 
&& + 3 \vec{a}_{1}\cdot\vec{n} ( \vec{v}_{2}\cdot\vec{n})^{2} \big) - 3 \vec{S}_{1}\times\vec{a}_{1}\cdot\vec{a}_{2} \big( 7 \vec{v}_{1}\cdot\vec{n} v_{1}^2 + 9 v_{1}^2 \vec{v}_{2}\cdot\vec{n} + 4 \vec{v}_{1}\cdot\vec{n} \vec{v}_{1}\cdot\vec{v}_{2} \nn\\ 
&& - 2 \vec{v}_{2}\cdot\vec{n} \vec{v}_{1}\cdot\vec{v}_{2} + 5 \vec{v}_{1}\cdot\vec{n} v_{2}^2 - 2 \vec{v}_{2}\cdot\vec{n} v_{2}^2 -9 \vec{v}_{2}\cdot\vec{n} ( \vec{v}_{1}\cdot\vec{n})^{2} + 3 \vec{v}_{1}\cdot\vec{n} ( \vec{v}_{2}\cdot\vec{n})^{2} \big) \nn\\ 
&& + 3 \vec{S}_{1}\times\vec{v}_{2}\cdot\vec{a}_{2} \big( 3 \vec{v}_{1}\cdot\vec{a}_{1} \vec{v}_{2}\cdot\vec{n} + \vec{a}_{1}\cdot\vec{n} \vec{v}_{1}\cdot\vec{v}_{2} - \vec{v}_{1}\cdot\vec{n} \vec{a}_{1}\cdot\vec{v}_{2} \nn\\ 
&& + \vec{a}_{1}\cdot\vec{n} v_{2}^2 -3 \vec{v}_{1}\cdot\vec{n} \vec{a}_{1}\cdot\vec{n} \vec{v}_{2}\cdot\vec{n} \big)) + (3 \dot{\vec{S}}_{1}\times\vec{n}\cdot\vec{v}_{1} \big( 4 \vec{v}_{1}\cdot\vec{a}_{1} v_{2}^2 - 3 \vec{a}_{1}\cdot\vec{v}_{2} v_{2}^2 \nn\\ 
&& + 5 v_{1}^2 \vec{v}_{1}\cdot\vec{a}_{2} + 4 \vec{v}_{1}\cdot\vec{v}_{2} \vec{v}_{1}\cdot\vec{a}_{2} + 4 v_{2}^2 \vec{v}_{1}\cdot\vec{a}_{2} - 7 v_{1}^2 \vec{v}_{2}\cdot\vec{a}_{2} + 4 \vec{v}_{1}\cdot\vec{v}_{2} \vec{v}_{2}\cdot\vec{a}_{2} \nn\\ 
&& - 12 v_{2}^2 \vec{v}_{2}\cdot\vec{a}_{2} -3 \vec{a}_{1}\cdot\vec{n} \vec{v}_{2}\cdot\vec{n} v_{2}^2 + 2 \vec{v}_{1}\cdot\vec{n} v_{1}^2 \vec{a}_{2}\cdot\vec{n} + 3 v_{1}^2 \vec{v}_{2}\cdot\vec{n} \vec{a}_{2}\cdot\vec{n} + 2 \vec{v}_{1}\cdot\vec{n} v_{2}^2 \vec{a}_{2}\cdot\vec{n} \nn\\ 
&& - 6 \vec{v}_{1}\cdot\vec{n} \vec{v}_{2}\cdot\vec{n} \vec{v}_{1}\cdot\vec{a}_{2} + 8 \vec{v}_{1}\cdot\vec{n} \vec{v}_{2}\cdot\vec{n} \vec{v}_{2}\cdot\vec{a}_{2} + 3 \vec{v}_{2}\cdot\vec{a}_{2} ( \vec{v}_{1}\cdot\vec{n})^{2} - 4 \vec{v}_{1}\cdot\vec{a}_{1} ( \vec{v}_{2}\cdot\vec{n})^{2} \nn\\ 
&& + 3 \vec{a}_{1}\cdot\vec{v}_{2} ( \vec{v}_{2}\cdot\vec{n})^{2} -9 \vec{v}_{2}\cdot\vec{n} \vec{a}_{2}\cdot\vec{n} ( \vec{v}_{1}\cdot\vec{n})^{2} + 3 \vec{a}_{1}\cdot\vec{n} ( \vec{v}_{2}\cdot\vec{n})^{3} \big) + 6 \dot{\vec{S}}_{1}\times\vec{n}\cdot\vec{a}_{1} \big( v_{1}^2 v_{2}^2 \nn\\ 
&& + v_{2}^{4} -3 \vec{v}_{1}\cdot\vec{n} \vec{v}_{2}\cdot\vec{n} v_{2}^2 - v_{1}^2 ( \vec{v}_{2}\cdot\vec{n})^{2} - v_{2}^2 ( \vec{v}_{2}\cdot\vec{n})^{2} + 3 \vec{v}_{1}\cdot\vec{n} ( \vec{v}_{2}\cdot\vec{n})^{3} \big) \nn\\ 
&& + 3 \dot{\vec{S}}_{1}\times\vec{v}_{1}\cdot\vec{a}_{1} \big( 12 v_{1}^2 \vec{v}_{2}\cdot\vec{n} - 8 \vec{v}_{2}\cdot\vec{n} \vec{v}_{1}\cdot\vec{v}_{2} + 4 \vec{v}_{1}\cdot\vec{n} v_{2}^2 \nn\\ 
&& + 7 \vec{v}_{2}\cdot\vec{n} v_{2}^2 -4 \vec{v}_{1}\cdot\vec{n} ( \vec{v}_{2}\cdot\vec{n})^{2} - ( \vec{v}_{2}\cdot\vec{n})^{3} \big) + 3 \dot{\vec{S}}_{1}\times\vec{n}\cdot\vec{v}_{2} \big( \vec{a}_{1}\cdot\vec{v}_{2} v_{2}^2 - 6 \vec{v}_{1}\cdot\vec{v}_{2} \vec{v}_{1}\cdot\vec{a}_{2} \nn\\ 
&& - 2 v_{2}^2 \vec{v}_{1}\cdot\vec{a}_{2} + 3 v_{1}^2 \vec{v}_{2}\cdot\vec{a}_{2} + 12 v_{2}^2 \vec{v}_{2}\cdot\vec{a}_{2} + 3 \vec{a}_{1}\cdot\vec{n} \vec{v}_{2}\cdot\vec{n} v_{2}^2 - 3 v_{1}^2 \vec{v}_{2}\cdot\vec{n} \vec{a}_{2}\cdot\vec{n} \nn\\ 
&& - 2 \vec{v}_{1}\cdot\vec{n} \vec{v}_{1}\cdot\vec{v}_{2} \vec{a}_{2}\cdot\vec{n} - 2 \vec{v}_{1}\cdot\vec{n} v_{2}^2 \vec{a}_{2}\cdot\vec{n} + 2 \vec{v}_{1}\cdot\vec{n} \vec{v}_{2}\cdot\vec{n} \vec{v}_{1}\cdot\vec{a}_{2} - 8 \vec{v}_{1}\cdot\vec{n} \vec{v}_{2}\cdot\vec{n} \vec{v}_{2}\cdot\vec{a}_{2} \nn\\ 
&& - 3 \vec{v}_{2}\cdot\vec{a}_{2} ( \vec{v}_{1}\cdot\vec{n})^{2} - \vec{a}_{1}\cdot\vec{v}_{2} ( \vec{v}_{2}\cdot\vec{n})^{2} + 9 \vec{v}_{2}\cdot\vec{n} \vec{a}_{2}\cdot\vec{n} ( \vec{v}_{1}\cdot\vec{n})^{2} - 3 \vec{a}_{1}\cdot\vec{n} ( \vec{v}_{2}\cdot\vec{n})^{3} \big) \nn\\ 
&& + 3 \dot{\vec{S}}_{1}\times\vec{v}_{1}\cdot\vec{v}_{2} \big( 24 \vec{v}_{1}\cdot\vec{a}_{1} \vec{v}_{2}\cdot\vec{n} - 2 \vec{v}_{2}\cdot\vec{n} \vec{a}_{1}\cdot\vec{v}_{2} + 3 \vec{a}_{1}\cdot\vec{n} v_{2}^2 - 3 v_{1}^2 \vec{a}_{2}\cdot\vec{n} \nn\\ 
&& + 2 \vec{v}_{1}\cdot\vec{v}_{2} \vec{a}_{2}\cdot\vec{n} + 2 v_{2}^2 \vec{a}_{2}\cdot\vec{n} - 2 \vec{v}_{1}\cdot\vec{n} \vec{v}_{1}\cdot\vec{a}_{2} - 2 \vec{v}_{2}\cdot\vec{n} \vec{v}_{1}\cdot\vec{a}_{2} - 6 \vec{v}_{1}\cdot\vec{n} \vec{v}_{2}\cdot\vec{a}_{2} \nn\\ 
&& + 8 \vec{v}_{2}\cdot\vec{n} \vec{v}_{2}\cdot\vec{a}_{2} -6 \vec{v}_{1}\cdot\vec{n} \vec{v}_{2}\cdot\vec{n} \vec{a}_{2}\cdot\vec{n} + 3 \vec{a}_{2}\cdot\vec{n} ( \vec{v}_{1}\cdot\vec{n})^{2} - 3 \vec{a}_{1}\cdot\vec{n} ( \vec{v}_{2}\cdot\vec{n})^{2} \big) \nn\\ 
&& + 3 \dot{\vec{S}}_{1}\times\vec{a}_{1}\cdot\vec{v}_{2} \big( 12 v_{1}^2 \vec{v}_{2}\cdot\vec{n} + 8 \vec{v}_{2}\cdot\vec{n} \vec{v}_{1}\cdot\vec{v}_{2} + 12 \vec{v}_{1}\cdot\vec{n} v_{2}^2 \nn\\ 
&& + 9 \vec{v}_{2}\cdot\vec{n} v_{2}^2 -12 \vec{v}_{1}\cdot\vec{n} ( \vec{v}_{2}\cdot\vec{n})^{2} + ( \vec{v}_{2}\cdot\vec{n})^{3} \big) + 3 \dot{\vec{S}}_{1}\times\vec{n}\cdot\vec{a}_{2} \big( 3 v_{1}^2 v_{2}^2 - 2 ( \vec{v}_{1}\cdot\vec{v}_{2})^{2} \nn\\ 
&& + 3 v_{2}^{4} -4 \vec{v}_{1}\cdot\vec{n} \vec{v}_{2}\cdot\vec{n} \vec{v}_{1}\cdot\vec{v}_{2} - 4 \vec{v}_{1}\cdot\vec{n} \vec{v}_{2}\cdot\vec{n} v_{2}^2 - 3 v_{2}^2 ( \vec{v}_{1}\cdot\vec{n})^{2} - 3 v_{1}^2 ( \vec{v}_{2}\cdot\vec{n})^{2} \nn\\ 
&& + 9 ( \vec{v}_{1}\cdot\vec{n})^{2} ( \vec{v}_{2}\cdot\vec{n})^{2} \big) - 3 \dot{\vec{S}}_{1}\times\vec{v}_{1}\cdot\vec{a}_{2} \big( 7 \vec{v}_{1}\cdot\vec{n} v_{1}^2 + 9 v_{1}^2 \vec{v}_{2}\cdot\vec{n} + 8 \vec{v}_{1}\cdot\vec{n} \vec{v}_{1}\cdot\vec{v}_{2} \nn\\ 
&& - 4 \vec{v}_{2}\cdot\vec{n} \vec{v}_{1}\cdot\vec{v}_{2} + 2 \vec{v}_{1}\cdot\vec{n} v_{2}^2 - 4 \vec{v}_{2}\cdot\vec{n} v_{2}^2 -9 \vec{v}_{2}\cdot\vec{n} ( \vec{v}_{1}\cdot\vec{n})^{2} + 6 \vec{v}_{1}\cdot\vec{n} ( \vec{v}_{2}\cdot\vec{n})^{2} \big) \nn\\ 
&& + 3 \dot{\vec{S}}_{1}\times\vec{v}_{2}\cdot\vec{a}_{2} \big( 3 v_{1}^2 \vec{v}_{2}\cdot\vec{n} + 2 \vec{v}_{1}\cdot\vec{n} \vec{v}_{1}\cdot\vec{v}_{2} + 2 \vec{v}_{1}\cdot\vec{n} v_{2}^2 -3 \vec{v}_{2}\cdot\vec{n} ( \vec{v}_{1}\cdot\vec{n})^{2} \big)) \Big]\nn\\ && +  	\frac{G^2 m_{1} m_{2}}{6 r} \Big[ (2 \vec{S}_{1}\times\vec{n}\cdot\vec{v}_{1} \big( 18 \vec{v}_{1}\cdot\dot{\vec{a}}_{1} - 10 \dot{\vec{a}}_{1}\cdot\vec{v}_{2} + \vec{v}_{1}\cdot\dot{\vec{a}}_{2} -4 \dot{\vec{a}}_{1}\cdot\vec{n} \vec{v}_{2}\cdot\vec{n} \nn\\ 
&& + 4 \vec{v}_{1}\cdot\vec{n} \dot{\vec{a}}_{2}\cdot\vec{n} \big) + 4 \vec{S}_{1}\times\vec{n}\cdot\dot{\vec{a}}_{1} \big( 6 v_{1}^2 + 2 \vec{v}_{1}\cdot\vec{v}_{2} - 7 v_{2}^2 -28 \vec{v}_{1}\cdot\vec{n} \vec{v}_{2}\cdot\vec{n} + 9 ( \vec{v}_{1}\cdot\vec{n})^{2} \nn\\ 
&& + 14 ( \vec{v}_{2}\cdot\vec{n})^{2} \big) + 4 \vec{S}_{1}\times\vec{v}_{1}\cdot\dot{\vec{a}}_{1} \big( 24 \vec{v}_{1}\cdot\vec{n} + \vec{v}_{2}\cdot\vec{n} \big) - 4 \vec{S}_{1}\times\vec{v}_{1}\cdot\vec{v}_{2} \dot{\vec{a}}_{1}\cdot\vec{n} \nn\\ 
&& + 4 \vec{S}_{1}\times\dot{\vec{a}}_{1}\cdot\vec{v}_{2} \big( 16 \vec{v}_{1}\cdot\vec{n} - \vec{v}_{2}\cdot\vec{n} \big) + 2 \vec{S}_{1}\times\vec{n}\cdot\dot{\vec{a}}_{2} \big( 7 v_{1}^2 -8 ( \vec{v}_{1}\cdot\vec{n})^{2} \big) \nn\\ 
&& + 26 \vec{S}_{1}\times\vec{v}_{1}\cdot\dot{\vec{a}}_{2} \vec{v}_{1}\cdot\vec{n}) + ( \ddot{\vec{S}}_{1}\times\vec{n}\cdot\vec{v}_{1} \big( 15 v_{1}^2 - 2 \vec{v}_{1}\cdot\vec{v}_{2} - 10 v_{2}^2 -68 \vec{v}_{1}\cdot\vec{n} \vec{v}_{2}\cdot\vec{n} \nn\\ 
&& + 24 ( \vec{v}_{1}\cdot\vec{n})^{2} + 20 ( \vec{v}_{2}\cdot\vec{n})^{2} \big) + 2 \ddot{\vec{S}}_{1}\times\vec{n}\cdot\vec{v}_{2} \big( 6 v_{1}^2 - 4 \vec{v}_{1}\cdot\vec{v}_{2} + 5 v_{2}^2 + 20 \vec{v}_{1}\cdot\vec{n} \vec{v}_{2}\cdot\vec{n} \nn\\ 
&& - 6 ( \vec{v}_{1}\cdot\vec{n})^{2} - 10 ( \vec{v}_{2}\cdot\vec{n})^{2} \big) + 2 \ddot{\vec{S}}_{1}\times\vec{v}_{1}\cdot\vec{v}_{2} \big( 25 \vec{v}_{1}\cdot\vec{n} - 18 \vec{v}_{2}\cdot\vec{n} \big)) \nn\\ 
&& + (13 \vec{S}_{1}\times\vec{n}\cdot\vec{v}_{1} \big( a_{1}^2 - 4 \vec{a}_{1}\cdot\vec{a}_{2} -4 \vec{a}_{1}\cdot\vec{n} \vec{a}_{2}\cdot\vec{n} \big) + \vec{S}_{1}\times\vec{n}\cdot\vec{a}_{1} \big( 143 \vec{v}_{1}\cdot\vec{a}_{1} \nn\\ 
&& - 68 \vec{a}_{1}\cdot\vec{v}_{2} - 14 \vec{v}_{1}\cdot\vec{a}_{2} - 14 \vec{v}_{2}\cdot\vec{a}_{2} + 80 \vec{v}_{1}\cdot\vec{n} \vec{a}_{1}\cdot\vec{n} - 128 \vec{a}_{1}\cdot\vec{n} \vec{v}_{2}\cdot\vec{n} \nn\\ 
&& + 16 \vec{v}_{1}\cdot\vec{n} \vec{a}_{2}\cdot\vec{n} + 4 \vec{v}_{2}\cdot\vec{n} \vec{a}_{2}\cdot\vec{n} \big) + \vec{S}_{1}\times\vec{v}_{1}\cdot\vec{a}_{1} \big( 21 \vec{a}_{1}\cdot\vec{n} + 8 \vec{a}_{2}\cdot\vec{n} \big) \nn\\ 
&& + 4 \vec{S}_{1}\times\vec{n}\cdot\vec{v}_{2} \big( 5 a_{1}^2 + 4 \vec{a}_{1}\cdot\vec{a}_{2} + 13 \vec{a}_{1}\cdot\vec{n} \vec{a}_{2}\cdot\vec{n} \big) - 12 \vec{S}_{1}\times\vec{a}_{1}\cdot\vec{v}_{2} \vec{a}_{1}\cdot\vec{n} \nn\\ 
&& + 4 \vec{S}_{1}\times\vec{n}\cdot\vec{a}_{2} \big( 9 \vec{v}_{1}\cdot\vec{a}_{1} - 2 \vec{a}_{1}\cdot\vec{v}_{2} -27 \vec{v}_{1}\cdot\vec{n} \vec{a}_{1}\cdot\vec{n} + 25 \vec{a}_{1}\cdot\vec{n} \vec{v}_{2}\cdot\vec{n} \big) \nn\\ 
&& + 28 \vec{S}_{1}\times\vec{v}_{1}\cdot\vec{a}_{2} \vec{a}_{1}\cdot\vec{n} + 2 \vec{S}_{1}\times\vec{a}_{1}\cdot\vec{a}_{2} \big( 61 \vec{v}_{1}\cdot\vec{n} - 9 \vec{v}_{2}\cdot\vec{n} \big) - 24 \vec{S}_{1}\times\vec{v}_{2}\cdot\vec{a}_{2} \vec{a}_{1}\cdot\vec{n}) \nn\\ 
&& + (2 \dot{\vec{S}}_{1}\times\vec{n}\cdot\vec{v}_{1} \big( 43 \vec{v}_{1}\cdot\vec{a}_{1} - 34 \vec{a}_{1}\cdot\vec{v}_{2} - 16 \vec{v}_{1}\cdot\vec{a}_{2} + 21 \vec{v}_{2}\cdot\vec{a}_{2} + 28 \vec{v}_{1}\cdot\vec{n} \vec{a}_{1}\cdot\vec{n} \nn\\ 
&& - 40 \vec{a}_{1}\cdot\vec{n} \vec{v}_{2}\cdot\vec{n} - 34 \vec{v}_{1}\cdot\vec{n} \vec{a}_{2}\cdot\vec{n} \big) + \dot{\vec{S}}_{1}\times\vec{n}\cdot\vec{a}_{1} \big( 16 v_{1}^2 + 38 \vec{v}_{1}\cdot\vec{v}_{2} \nn\\ 
&& - 53 v_{2}^2 -148 \vec{v}_{1}\cdot\vec{n} \vec{v}_{2}\cdot\vec{n} + 46 ( \vec{v}_{1}\cdot\vec{n})^{2} + 64 ( \vec{v}_{2}\cdot\vec{n})^{2} \big) + 2 \dot{\vec{S}}_{1}\times\vec{v}_{1}\cdot\vec{a}_{1} \big( 69 \vec{v}_{1}\cdot\vec{n} \nn\\ 
&& - 7 \vec{v}_{2}\cdot\vec{n} \big) + 4 \dot{\vec{S}}_{1}\times\vec{n}\cdot\vec{v}_{2} \big( 4 \vec{a}_{1}\cdot\vec{v}_{2} + 9 \vec{v}_{1}\cdot\vec{a}_{2} - 15 \vec{v}_{2}\cdot\vec{a}_{2} -6 \vec{v}_{1}\cdot\vec{n} \vec{a}_{1}\cdot\vec{n} \nn\\ 
&& + 10 \vec{a}_{1}\cdot\vec{n} \vec{v}_{2}\cdot\vec{n} + 15 \vec{v}_{1}\cdot\vec{n} \vec{a}_{2}\cdot\vec{n} \big) - 4 \dot{\vec{S}}_{1}\times\vec{v}_{1}\cdot\vec{v}_{2} \big( \vec{a}_{1}\cdot\vec{n} + 3 \vec{a}_{2}\cdot\vec{n} \big) \nn\\ 
&& + 26 \dot{\vec{S}}_{1}\times\vec{a}_{1}\cdot\vec{v}_{2} \big( 5 \vec{v}_{1}\cdot\vec{n} - 2 \vec{v}_{2}\cdot\vec{n} \big) + 2 \dot{\vec{S}}_{1}\times\vec{n}\cdot\vec{a}_{2} \big( 10 v_{1}^2 + \vec{v}_{1}\cdot\vec{v}_{2} - 15 v_{2}^2 \nn\\ 
&& + 64 \vec{v}_{1}\cdot\vec{n} \vec{v}_{2}\cdot\vec{n} - 50 ( \vec{v}_{1}\cdot\vec{n})^{2} \big) + 2 \dot{\vec{S}}_{1}\times\vec{v}_{1}\cdot\vec{a}_{2} \big( 48 \vec{v}_{1}\cdot\vec{n} - 17 \vec{v}_{2}\cdot\vec{n} \big) \nn\\ 
&& - 34 \dot{\vec{S}}_{1}\times\vec{v}_{2}\cdot\vec{a}_{2} \vec{v}_{1}\cdot\vec{n}) \Big] + 	\frac{G^2 m_{2}{}^2}{24 r} \Big[ (2 \vec{S}_{1}\times\vec{n}\cdot\vec{v}_{1} \big( 5 \dot{\vec{a}}_{1}\cdot\vec{v}_{2} -4 \dot{\vec{a}}_{1}\cdot\vec{n} \vec{v}_{2}\cdot\vec{n} \big) \nn\\ 
&& + 2 \vec{S}_{1}\times\vec{n}\cdot\dot{\vec{a}}_{1} \big( 10 \vec{v}_{1}\cdot\vec{v}_{2} - 12 v_{2}^2 -8 \vec{v}_{1}\cdot\vec{n} \vec{v}_{2}\cdot\vec{n} + 9 ( \vec{v}_{2}\cdot\vec{n})^{2} \big) - 92 \vec{S}_{1}\times\vec{v}_{1}\cdot\dot{\vec{a}}_{1} \vec{v}_{2}\cdot\vec{n} \nn\\ 
&& - 2 \vec{S}_{1}\times\vec{n}\cdot\vec{v}_{2} \big( \dot{\vec{a}}_{1}\cdot\vec{v}_{2} - 36 \vec{v}_{1}\cdot\dot{\vec{a}}_{2} + 8 \vec{v}_{2}\cdot\dot{\vec{a}}_{2} + 4 \dot{\vec{a}}_{1}\cdot\vec{n} \vec{v}_{2}\cdot\vec{n} - 24 \vec{v}_{1}\cdot\vec{n} \dot{\vec{a}}_{2}\cdot\vec{n} \nn\\ 
&& - 4 \vec{v}_{2}\cdot\vec{n} \dot{\vec{a}}_{2}\cdot\vec{n} \big) - 2 \vec{S}_{1}\times\vec{v}_{1}\cdot\vec{v}_{2} \big( 5 \dot{\vec{a}}_{1}\cdot\vec{n} + 12 \dot{\vec{a}}_{2}\cdot\vec{n} \big) - 10 \vec{S}_{1}\times\dot{\vec{a}}_{1}\cdot\vec{v}_{2} \big( 2 \vec{v}_{1}\cdot\vec{n} \nn\\ 
&& + 35 \vec{v}_{2}\cdot\vec{n} \big) + 4 \vec{S}_{1}\times\vec{n}\cdot\dot{\vec{a}}_{2} \big( 16 v_{1}^2 + v_{2}^2 + 48 \vec{v}_{1}\cdot\vec{n} \vec{v}_{2}\cdot\vec{n} - 32 ( \vec{v}_{1}\cdot\vec{n})^{2} + ( \vec{v}_{2}\cdot\vec{n})^{2} \big) \nn\\ 
&& + 32 \vec{S}_{1}\times\vec{v}_{1}\cdot\dot{\vec{a}}_{2} \big( 4 \vec{v}_{1}\cdot\vec{n} - 3 \vec{v}_{2}\cdot\vec{n} \big) - 72 \vec{S}_{1}\times\vec{v}_{2}\cdot\dot{\vec{a}}_{2} \vec{v}_{1}\cdot\vec{n}) + (3 \ddot{\vec{S}}_{1}\times\vec{n}\cdot\vec{v}_{1} \big( 5 \vec{v}_{1}\cdot\vec{v}_{2} \nn\\ 
&& - 8 v_{2}^2 -4 \vec{v}_{1}\cdot\vec{n} \vec{v}_{2}\cdot\vec{n} + 6 ( \vec{v}_{2}\cdot\vec{n})^{2} \big) - 2 \ddot{\vec{S}}_{1}\times\vec{n}\cdot\vec{v}_{2} \big( \vec{v}_{1}\cdot\vec{v}_{2} - 35 v_{2}^2 + 4 \vec{v}_{1}\cdot\vec{n} \vec{v}_{2}\cdot\vec{n} \nn\\ 
&& + 61 ( \vec{v}_{2}\cdot\vec{n})^{2} \big) - 5 \ddot{\vec{S}}_{1}\times\vec{v}_{1}\cdot\vec{v}_{2} \big( 3 \vec{v}_{1}\cdot\vec{n} + 70 \vec{v}_{2}\cdot\vec{n} \big)) - ( \vec{S}_{1}\times\vec{n}\cdot\vec{v}_{1} \big( 123 \vec{a}_{1}\cdot\vec{a}_{2} \nn\\ 
&& + 268 a_{2}^2 + 156 \vec{a}_{1}\cdot\vec{n} \vec{a}_{2}\cdot\vec{n} \big) - \vec{S}_{1}\times\vec{n}\cdot\vec{a}_{1} \big( 30 \vec{a}_{1}\cdot\vec{v}_{2} + 51 \vec{v}_{1}\cdot\vec{a}_{2} \nn\\ 
&& + 20 \vec{v}_{2}\cdot\vec{a}_{2} -24 \vec{a}_{1}\cdot\vec{n} \vec{v}_{2}\cdot\vec{n} - 312 \vec{v}_{1}\cdot\vec{n} \vec{a}_{2}\cdot\vec{n} + 308 \vec{v}_{2}\cdot\vec{n} \vec{a}_{2}\cdot\vec{n} \big) \nn\\ 
&& - 78 \vec{S}_{1}\times\vec{v}_{1}\cdot\vec{a}_{1} \vec{a}_{2}\cdot\vec{n} - 4 \vec{S}_{1}\times\vec{n}\cdot\vec{v}_{2} \big( 38 \vec{a}_{1}\cdot\vec{a}_{2} + 63 a_{2}^2 + 47 \vec{a}_{1}\cdot\vec{n} \vec{a}_{2}\cdot\vec{n} \big) \nn\\ 
&& + 6 \vec{S}_{1}\times\vec{a}_{1}\cdot\vec{v}_{2} \big( 5 \vec{a}_{1}\cdot\vec{n} + 42 \vec{a}_{2}\cdot\vec{n} \big) - 2 \vec{S}_{1}\times\vec{n}\cdot\vec{a}_{2} \big( 3 \vec{v}_{1}\cdot\vec{a}_{1} + 36 \vec{a}_{1}\cdot\vec{v}_{2} \nn\\ 
&& + 32 \vec{v}_{1}\cdot\vec{a}_{2} -6 \vec{v}_{1}\cdot\vec{n} \vec{a}_{1}\cdot\vec{n} - 18 \vec{a}_{1}\cdot\vec{n} \vec{v}_{2}\cdot\vec{n} + 140 \vec{v}_{1}\cdot\vec{n} \vec{a}_{2}\cdot\vec{n} - 12 \vec{v}_{2}\cdot\vec{n} \vec{a}_{2}\cdot\vec{n} \big) \nn\\ 
&& + 3 \vec{S}_{1}\times\vec{v}_{1}\cdot\vec{a}_{2} \big( 23 \vec{a}_{1}\cdot\vec{n} + 40 \vec{a}_{2}\cdot\vec{n} \big) - \vec{S}_{1}\times\vec{a}_{1}\cdot\vec{a}_{2} \big( 153 \vec{v}_{1}\cdot\vec{n} - 188 \vec{v}_{2}\cdot\vec{n} \big) \nn\\ 
&& - 16 \vec{S}_{1}\times\vec{v}_{2}\cdot\vec{a}_{2} \vec{a}_{1}\cdot\vec{n}) + ( \dot{\vec{S}}_{1}\times\vec{n}\cdot\vec{v}_{1} \big( 25 \vec{a}_{1}\cdot\vec{v}_{2} + 3 \vec{v}_{1}\cdot\vec{a}_{2} \nn\\ 
&& + 20 \vec{v}_{2}\cdot\vec{a}_{2} -20 \vec{a}_{1}\cdot\vec{n} \vec{v}_{2}\cdot\vec{n} - 312 \vec{v}_{1}\cdot\vec{n} \vec{a}_{2}\cdot\vec{n} + 308 \vec{v}_{2}\cdot\vec{n} \vec{a}_{2}\cdot\vec{n} \big) \nn\\ 
&& + \dot{\vec{S}}_{1}\times\vec{n}\cdot\vec{a}_{1} \big( 35 \vec{v}_{1}\cdot\vec{v}_{2} - 48 v_{2}^2 -28 \vec{v}_{1}\cdot\vec{n} \vec{v}_{2}\cdot\vec{n} + 36 ( \vec{v}_{2}\cdot\vec{n})^{2} \big) \nn\\ 
&& - 92 \dot{\vec{S}}_{1}\times\vec{v}_{1}\cdot\vec{a}_{1} \vec{v}_{2}\cdot\vec{n} - 4 \dot{\vec{S}}_{1}\times\vec{n}\cdot\vec{v}_{2} \big( \vec{a}_{1}\cdot\vec{v}_{2} - 14 \vec{v}_{1}\cdot\vec{a}_{2} + 3 \vec{v}_{2}\cdot\vec{a}_{2} + 4 \vec{a}_{1}\cdot\vec{n} \vec{v}_{2}\cdot\vec{n} \nn\\ 
&& - 95 \vec{v}_{1}\cdot\vec{n} \vec{a}_{2}\cdot\vec{n} + 141 \vec{v}_{2}\cdot\vec{n} \vec{a}_{2}\cdot\vec{n} \big) - \dot{\vec{S}}_{1}\times\vec{v}_{1}\cdot\vec{v}_{2} \big( 25 \vec{a}_{1}\cdot\vec{n} + 348 \vec{a}_{2}\cdot\vec{n} \big) \nn\\ 
&& - 35 \dot{\vec{S}}_{1}\times\vec{a}_{1}\cdot\vec{v}_{2} \big( \vec{v}_{1}\cdot\vec{n} + 20 \vec{v}_{2}\cdot\vec{n} \big) + 2 \dot{\vec{S}}_{1}\times\vec{n}\cdot\vec{a}_{2} \big( 2 v_{1}^2 + 16 \vec{v}_{1}\cdot\vec{v}_{2} - v_{2}^2 \nn\\ 
&& + 22 \vec{v}_{1}\cdot\vec{n} \vec{v}_{2}\cdot\vec{n} - 4 ( \vec{v}_{1}\cdot\vec{n})^{2} - ( \vec{v}_{2}\cdot\vec{n})^{2} \big) + \dot{\vec{S}}_{1}\times\vec{v}_{1}\cdot\vec{a}_{2} \big( 203 \vec{v}_{1}\cdot\vec{n} - 228 \vec{v}_{2}\cdot\vec{n} \big) \nn\\ 
&& - 8 \dot{\vec{S}}_{1}\times\vec{v}_{2}\cdot\vec{a}_{2} \big( 3 \vec{v}_{1}\cdot\vec{n} + 35 \vec{v}_{2}\cdot\vec{n} \big)) \Big] + 	\frac{2 G^2 m_{1} m_{2}}{3 r} \Big[ \vec{S}_{1}\times\vec{n}\cdot\vec{v}_{1} ( \vec{a}_{1}\cdot\vec{n})^{2} \nn\\ 
&& - \vec{S}_{1}\times\vec{n}\cdot\vec{v}_{2} ( \vec{a}_{1}\cdot\vec{n})^{2} \Big] + 	\frac{G^2 m_{2}{}^2}{6 r} \Big[ 5 \vec{S}_{1}\times\vec{n}\cdot\vec{v}_{1} ( \vec{a}_{2}\cdot\vec{n})^{2} - 3 \vec{S}_{1}\times\vec{n}\cdot\vec{v}_{2} ( \vec{a}_{2}\cdot\vec{n})^{2} \Big]\nn\\ && - 	\frac{G^3 m_{1}{}^2 m_{2}}{18 r{}^2} \Big[ 210 \vec{S}_{1}\times\vec{n}\cdot\dot{\vec{a}}_{1} + 142 \ddot{\vec{S}}_{1}\times\vec{n}\cdot\vec{v}_{1} - (178 \dot{\vec{S}}_{1}\times\vec{n}\cdot\vec{a}_{1} - 87 \dot{\vec{S}}_{1}\times\vec{n}\cdot\vec{a}_{2}) \Big] \nn\\ 
&& - 	\frac{G^3 m_{1}{}^2 m_{2}}{3 r{}^2} {\Big( \frac{1}{\epsilon} - 3\log \frac{r}{R_0} \Big)} \Big[ 14 \vec{S}_{1}\times\vec{n}\cdot\dot{\vec{a}}_{1} + 10 \ddot{\vec{S}}_{1}\times\vec{n}\cdot\vec{v}_{1} - (5 \dot{\vec{S}}_{1}\times\vec{n}\cdot\vec{a}_{1} \nn\\ 
&& + 8 \dot{\vec{S}}_{1}\times\vec{n}\cdot\vec{a}_{2}) \Big] - 	\frac{2 G^3 m_{2}{}^3}{3 r{}^2} {\Big( \frac{1}{\epsilon} - 3\log \frac{r}{R_0} \Big)} \Big[ \vec{S}_{1}\times\vec{n}\cdot\dot{\vec{a}}_{1} + ( \ddot{\vec{S}}_{1}\times\vec{n}\cdot\vec{v}_{1} \nn\\ 
&& + \ddot{\vec{S}}_{1}\times\vec{n}\cdot\vec{v}_{2}) + (2 \dot{\vec{S}}_{1}\times\vec{n}\cdot\vec{a}_{1} + 9 \dot{\vec{S}}_{1}\times\vec{n}\cdot\vec{a}_{2}) \Big] - 	\frac{G^3 m_{2}{}^3}{18 r{}^2} \Big[ 30 \vec{S}_{1}\times\vec{n}\cdot\dot{\vec{a}}_{1} \nn\\ 
&& + (30 \ddot{\vec{S}}_{1}\times\vec{n}\cdot\vec{v}_{1} + 22 \ddot{\vec{S}}_{1}\times\vec{n}\cdot\vec{v}_{2}) + (60 \dot{\vec{S}}_{1}\times\vec{n}\cdot\vec{a}_{1} + 291 \dot{\vec{S}}_{1}\times\vec{n}\cdot\vec{a}_{2}) \Big] \nn\\ 
&& - 	\frac{G^3 m_{1} m_{2}{}^2}{288 r{}^2} \Big[ 576 (16 - \pi^2) \vec{S}_{1}\times\vec{n}\cdot\dot{\vec{a}}_{1} + (2 {(3296 - 225 \pi^2)} \ddot{\vec{S}}_{1}\times\vec{n}\cdot\vec{v}_{1} \nn\\ 
&& - 7 {(160 - 9 \pi^2)} \ddot{\vec{S}}_{1}\times\vec{n}\cdot\vec{v}_{2}) + (18 {(416 - 29 \pi^2)} \dot{\vec{S}}_{1}\times\vec{n}\cdot\vec{a}_{1} \nn\\ 
&& + 3 {(1456 + 33 \pi^2)} \dot{\vec{S}}_{1}\times\vec{n}\cdot\vec{a}_{2}) \Big] 
+ (1 \leftrightarrow 2),
\eea
\bea
\stackrel{(3)}{V}_{3,1} &=& - 	\frac{1}{24} G m_{2} r \Big[ ( \vec{S}_{1}\times\vec{n}\cdot\ddot{\vec{a}}_{1} \big( 3 \vec{v}_{2}\cdot\vec{n} v_{2}^2 - ( \vec{v}_{2}\cdot\vec{n})^{3} \big) + 6 \vec{S}_{1}\times\vec{v}_{1}\cdot\ddot{\vec{a}}_{1} \big( v_{2}^2 + ( \vec{v}_{2}\cdot\vec{n})^{2} \big) \nn\\ 
&& + 9 \vec{S}_{1}\times\ddot{\vec{a}}_{1}\cdot\vec{v}_{2} \big( v_{2}^2 + ( \vec{v}_{2}\cdot\vec{n})^{2} \big) + \vec{S}_{1}\times\vec{n}\cdot\ddot{\vec{a}}_{2} \big( 3 \vec{v}_{1}\cdot\vec{n} v_{1}^2 - ( \vec{v}_{1}\cdot\vec{n})^{3} \big) \nn\\ 
&& + 3 \vec{S}_{1}\times\vec{v}_{1}\cdot\ddot{\vec{a}}_{2} \big( v_{1}^2 + ( \vec{v}_{1}\cdot\vec{n})^{2} \big)) + ( \dddot{\vec{S}}_{1}\times\vec{n}\cdot\vec{v}_{1} \big( 3 \vec{v}_{2}\cdot\vec{n} v_{2}^2 - ( \vec{v}_{2}\cdot\vec{n})^{3} \big) \nn\\ 
&& - \dddot{\vec{S}}_{1}\times\vec{n}\cdot\vec{v}_{2} \big( 3 \vec{v}_{2}\cdot\vec{n} v_{2}^2 - ( \vec{v}_{2}\cdot\vec{n})^{3} \big) + 9 \dddot{\vec{S}}_{1}\times\vec{v}_{1}\cdot\vec{v}_{2} \big( v_{2}^2 + ( \vec{v}_{2}\cdot\vec{n})^{2} \big)) \nn\\ 
&& - (3 \vec{S}_{1}\times\vec{n}\cdot\vec{v}_{1} \big( 2 \vec{v}_{1}\cdot\dot{\vec{a}}_{1} \vec{a}_{2}\cdot\vec{n} - 2 \dot{\vec{a}}_{1}\cdot\vec{v}_{2} \vec{a}_{2}\cdot\vec{n} - 5 \vec{v}_{2}\cdot\vec{n} \dot{\vec{a}}_{1}\cdot\vec{a}_{2} + \dot{\vec{a}}_{1}\cdot\vec{n} \vec{v}_{2}\cdot\vec{a}_{2} \nn\\ 
&& - \vec{v}_{1}\cdot\vec{a}_{1} \dot{\vec{a}}_{2}\cdot\vec{n} + 2 \vec{a}_{1}\cdot\vec{n} \vec{v}_{1}\cdot\dot{\vec{a}}_{2} + 5 \vec{v}_{1}\cdot\vec{n} \vec{a}_{1}\cdot\dot{\vec{a}}_{2} - 2 \vec{a}_{1}\cdot\vec{n} \vec{v}_{2}\cdot\dot{\vec{a}}_{2} - \dot{\vec{a}}_{1}\cdot\vec{n} \vec{v}_{2}\cdot\vec{n} \vec{a}_{2}\cdot\vec{n} \nn\\ 
&& + \vec{v}_{1}\cdot\vec{n} \vec{a}_{1}\cdot\vec{n} \dot{\vec{a}}_{2}\cdot\vec{n} \big) - 3 \vec{S}_{1}\times\vec{n}\cdot\vec{a}_{1} \big( v_{1}^2 \dot{\vec{a}}_{2}\cdot\vec{n} - 4 \vec{v}_{1}\cdot\vec{n} \vec{v}_{1}\cdot\dot{\vec{a}}_{2} + 4 \vec{v}_{1}\cdot\vec{n} \vec{v}_{2}\cdot\dot{\vec{a}}_{2} \nn\\ 
&& - \dot{\vec{a}}_{2}\cdot\vec{n} ( \vec{v}_{1}\cdot\vec{n})^{2} \big) + 3 \vec{S}_{1}\times\vec{v}_{1}\cdot\vec{a}_{1} \big( 2 \vec{v}_{1}\cdot\dot{\vec{a}}_{2} - 2 \vec{v}_{2}\cdot\dot{\vec{a}}_{2} - \vec{v}_{1}\cdot\vec{n} \dot{\vec{a}}_{2}\cdot\vec{n} \big) \nn\\ 
&& + 3 \vec{S}_{1}\times\vec{n}\cdot\dot{\vec{a}}_{1} \big( v_{1}^2 \vec{a}_{2}\cdot\vec{n} + v_{2}^2 \vec{a}_{2}\cdot\vec{n} - 3 \vec{v}_{2}\cdot\vec{n} \vec{v}_{1}\cdot\vec{a}_{2} + 3 \vec{v}_{1}\cdot\vec{n} \vec{v}_{2}\cdot\vec{a}_{2} \nn\\ 
&& + 4 \vec{v}_{2}\cdot\vec{n} \vec{v}_{2}\cdot\vec{a}_{2} -3 \vec{v}_{1}\cdot\vec{n} \vec{v}_{2}\cdot\vec{n} \vec{a}_{2}\cdot\vec{n} \big) - 6 \vec{S}_{1}\times\vec{v}_{1}\cdot\dot{\vec{a}}_{1} \big( 6 \vec{v}_{1}\cdot\vec{a}_{2} \nn\\ 
&& - 5 \vec{v}_{2}\cdot\vec{a}_{2} -2 \vec{v}_{1}\cdot\vec{n} \vec{a}_{2}\cdot\vec{n} - \vec{v}_{2}\cdot\vec{n} \vec{a}_{2}\cdot\vec{n} \big) - 6 \vec{S}_{1}\times\vec{a}_{1}\cdot\dot{\vec{a}}_{1} \big( v_{2}^2 + ( \vec{v}_{2}\cdot\vec{n})^{2} \big) \nn\\ 
&& + 3 \vec{S}_{1}\times\vec{n}\cdot\vec{v}_{2} \big( \dot{\vec{a}}_{1}\cdot\vec{v}_{2} \vec{a}_{2}\cdot\vec{n} + 3 \vec{v}_{2}\cdot\vec{n} \dot{\vec{a}}_{1}\cdot\vec{a}_{2} - \dot{\vec{a}}_{1}\cdot\vec{n} \vec{v}_{2}\cdot\vec{a}_{2} + \vec{v}_{1}\cdot\vec{a}_{1} \dot{\vec{a}}_{2}\cdot\vec{n} \nn\\ 
&& - \vec{a}_{1}\cdot\vec{n} \vec{v}_{1}\cdot\dot{\vec{a}}_{2} - 3 \vec{v}_{1}\cdot\vec{n} \vec{a}_{1}\cdot\dot{\vec{a}}_{2} + 2 \vec{a}_{1}\cdot\vec{n} \vec{v}_{2}\cdot\dot{\vec{a}}_{2} + \dot{\vec{a}}_{1}\cdot\vec{n} \vec{v}_{2}\cdot\vec{n} \vec{a}_{2}\cdot\vec{n} \nn\\ 
&& - \vec{v}_{1}\cdot\vec{n} \vec{a}_{1}\cdot\vec{n} \dot{\vec{a}}_{2}\cdot\vec{n} \big) - 3 \vec{S}_{1}\times\vec{v}_{1}\cdot\vec{v}_{2} \big( 2 \dot{\vec{a}}_{1}\cdot\vec{a}_{2} + 3 \vec{a}_{1}\cdot\dot{\vec{a}}_{2} - \vec{a}_{1}\cdot\vec{n} \dot{\vec{a}}_{2}\cdot\vec{n} \big) \nn\\ 
&& - 3 \vec{S}_{1}\times\vec{a}_{1}\cdot\vec{v}_{2} \big( \vec{v}_{1}\cdot\dot{\vec{a}}_{2} - 2 \vec{v}_{2}\cdot\dot{\vec{a}}_{2} - \vec{v}_{1}\cdot\vec{n} \dot{\vec{a}}_{2}\cdot\vec{n} \big) - 3 \vec{S}_{1}\times\dot{\vec{a}}_{1}\cdot\vec{v}_{2} \big( 4 \vec{v}_{1}\cdot\vec{a}_{2} \nn\\ 
&& - 11 \vec{v}_{2}\cdot\vec{a}_{2} -6 \vec{v}_{1}\cdot\vec{n} \vec{a}_{2}\cdot\vec{n} + \vec{v}_{2}\cdot\vec{n} \vec{a}_{2}\cdot\vec{n} \big) + 3 \vec{S}_{1}\times\vec{n}\cdot\vec{a}_{2} \big( 2 \vec{v}_{2}\cdot\vec{n} \dot{\vec{a}}_{1}\cdot\vec{v}_{2} - \dot{\vec{a}}_{1}\cdot\vec{n} v_{2}^2 \nn\\ 
&& + \dot{\vec{a}}_{1}\cdot\vec{n} ( \vec{v}_{2}\cdot\vec{n})^{2} \big) + 3 \vec{S}_{1}\times\vec{v}_{1}\cdot\vec{a}_{2} \big( 14 \vec{v}_{1}\cdot\dot{\vec{a}}_{1} - 4 \dot{\vec{a}}_{1}\cdot\vec{v}_{2} + \dot{\vec{a}}_{1}\cdot\vec{n} \vec{v}_{2}\cdot\vec{n} \big) \nn\\ 
&& + 3 \vec{S}_{1}\times\dot{\vec{a}}_{1}\cdot\vec{a}_{2} \big( 7 v_{1}^2 + 2 \vec{v}_{1}\cdot\vec{v}_{2} + 6 v_{2}^2 + 15 \vec{v}_{1}\cdot\vec{n} \vec{v}_{2}\cdot\vec{n} - ( \vec{v}_{2}\cdot\vec{n})^{2} \big) \nn\\ 
&& + 3 \vec{S}_{1}\times\vec{v}_{2}\cdot\vec{a}_{2} \big( \dot{\vec{a}}_{1}\cdot\vec{v}_{2} - \dot{\vec{a}}_{1}\cdot\vec{n} \vec{v}_{2}\cdot\vec{n} \big) + 3 \vec{S}_{1}\times\vec{n}\cdot\dot{\vec{a}}_{2} \big( 3 \vec{v}_{1}\cdot\vec{a}_{1} \vec{v}_{2}\cdot\vec{n} + \vec{a}_{1}\cdot\vec{n} \vec{v}_{1}\cdot\vec{v}_{2} \nn\\ 
&& - \vec{v}_{1}\cdot\vec{n} \vec{a}_{1}\cdot\vec{v}_{2} + \vec{a}_{1}\cdot\vec{n} v_{2}^2 -3 \vec{v}_{1}\cdot\vec{n} \vec{a}_{1}\cdot\vec{n} \vec{v}_{2}\cdot\vec{n} \big) - 3 \vec{S}_{1}\times\vec{v}_{1}\cdot\dot{\vec{a}}_{2} \big( 2 \vec{v}_{1}\cdot\vec{a}_{1} + \vec{a}_{1}\cdot\vec{v}_{2} \nn\\ 
&& + 2 \vec{v}_{1}\cdot\vec{n} \vec{a}_{1}\cdot\vec{n} - 3 \vec{a}_{1}\cdot\vec{n} \vec{v}_{2}\cdot\vec{n} \big) - 3 \vec{S}_{1}\times\vec{a}_{1}\cdot\dot{\vec{a}}_{2} \big( 5 v_{1}^2 - \vec{v}_{1}\cdot\vec{v}_{2} - v_{2}^2 -3 \vec{v}_{1}\cdot\vec{n} \vec{v}_{2}\cdot\vec{n} \nn\\ 
&& + 5 ( \vec{v}_{1}\cdot\vec{n})^{2} \big) + 6 \vec{S}_{1}\times\vec{v}_{2}\cdot\dot{\vec{a}}_{2} \big( \vec{v}_{1}\cdot\vec{a}_{1} + \vec{v}_{1}\cdot\vec{n} \vec{a}_{1}\cdot\vec{n} \big)) + (3 \dot{\vec{S}}_{1}\times\vec{n}\cdot\vec{v}_{1} \big( v_{1}^2 \dot{\vec{a}}_{2}\cdot\vec{n} \nn\\ 
&& - 4 \vec{v}_{1}\cdot\vec{n} \vec{v}_{1}\cdot\dot{\vec{a}}_{2} + 4 \vec{v}_{1}\cdot\vec{n} \vec{v}_{2}\cdot\dot{\vec{a}}_{2} - \dot{\vec{a}}_{2}\cdot\vec{n} ( \vec{v}_{1}\cdot\vec{n})^{2} \big) + 3 \dot{\vec{S}}_{1}\times\vec{n}\cdot\dot{\vec{a}}_{1} \big( 3 \vec{v}_{2}\cdot\vec{n} v_{2}^2 \nn\\ 
&& - ( \vec{v}_{2}\cdot\vec{n})^{3} \big) + 12 \dot{\vec{S}}_{1}\times\vec{v}_{1}\cdot\dot{\vec{a}}_{1} \big( v_{2}^2 + ( \vec{v}_{2}\cdot\vec{n})^{2} \big) - 3 \dot{\vec{S}}_{1}\times\vec{n}\cdot\vec{v}_{2} \big( v_{1}^2 \dot{\vec{a}}_{2}\cdot\vec{n} \nn\\ 
&& - 2 \vec{v}_{1}\cdot\vec{n} \vec{v}_{1}\cdot\dot{\vec{a}}_{2} + 4 \vec{v}_{1}\cdot\vec{n} \vec{v}_{2}\cdot\dot{\vec{a}}_{2} - \dot{\vec{a}}_{2}\cdot\vec{n} ( \vec{v}_{1}\cdot\vec{n})^{2} \big) + 6 \dot{\vec{S}}_{1}\times\vec{v}_{1}\cdot\vec{v}_{2} \big( \vec{v}_{1}\cdot\dot{\vec{a}}_{2} - 2 \vec{v}_{2}\cdot\dot{\vec{a}}_{2} \nn\\ 
&& - \vec{v}_{1}\cdot\vec{n} \dot{\vec{a}}_{2}\cdot\vec{n} \big) + 27 \dot{\vec{S}}_{1}\times\dot{\vec{a}}_{1}\cdot\vec{v}_{2} \big( v_{2}^2 + ( \vec{v}_{2}\cdot\vec{n})^{2} \big) - 3 \dot{\vec{S}}_{1}\times\vec{n}\cdot\dot{\vec{a}}_{2} \big( 3 v_{1}^2 \vec{v}_{2}\cdot\vec{n} \nn\\ 
&& + 2 \vec{v}_{1}\cdot\vec{n} \vec{v}_{1}\cdot\vec{v}_{2} + 2 \vec{v}_{1}\cdot\vec{n} v_{2}^2 -3 \vec{v}_{2}\cdot\vec{n} ( \vec{v}_{1}\cdot\vec{n})^{2} \big) + 3 \dot{\vec{S}}_{1}\times\vec{v}_{1}\cdot\dot{\vec{a}}_{2} \big( 5 v_{1}^2 - 2 \vec{v}_{1}\cdot\vec{v}_{2} \nn\\ 
&& - 2 v_{2}^2 -6 \vec{v}_{1}\cdot\vec{n} \vec{v}_{2}\cdot\vec{n} + 5 ( \vec{v}_{1}\cdot\vec{n})^{2} \big) - 6 \dot{\vec{S}}_{1}\times\vec{v}_{2}\cdot\dot{\vec{a}}_{2} \big( v_{1}^2 + ( \vec{v}_{1}\cdot\vec{n})^{2} \big)) \nn\\ 
&& - (3 \ddot{\vec{S}}_{1}\times\vec{n}\cdot\vec{v}_{1} \big( v_{1}^2 \vec{a}_{2}\cdot\vec{n} + v_{2}^2 \vec{a}_{2}\cdot\vec{n} - 3 \vec{v}_{2}\cdot\vec{n} \vec{v}_{1}\cdot\vec{a}_{2} + 3 \vec{v}_{1}\cdot\vec{n} \vec{v}_{2}\cdot\vec{a}_{2} \nn\\ 
&& + 4 \vec{v}_{2}\cdot\vec{n} \vec{v}_{2}\cdot\vec{a}_{2} -3 \vec{v}_{1}\cdot\vec{n} \vec{v}_{2}\cdot\vec{n} \vec{a}_{2}\cdot\vec{n} \big) - 3 \ddot{\vec{S}}_{1}\times\vec{n}\cdot\vec{a}_{1} \big( 3 \vec{v}_{2}\cdot\vec{n} v_{2}^2 - ( \vec{v}_{2}\cdot\vec{n})^{3} \big) \nn\\ 
&& - 6 \ddot{\vec{S}}_{1}\times\vec{v}_{1}\cdot\vec{a}_{1} \big( v_{2}^2 + ( \vec{v}_{2}\cdot\vec{n})^{2} \big) - 3 \ddot{\vec{S}}_{1}\times\vec{n}\cdot\vec{v}_{2} \big( \vec{v}_{1}\cdot\vec{v}_{2} \vec{a}_{2}\cdot\vec{n} + v_{2}^2 \vec{a}_{2}\cdot\vec{n} \nn\\ 
&& - \vec{v}_{2}\cdot\vec{n} \vec{v}_{1}\cdot\vec{a}_{2} + 3 \vec{v}_{1}\cdot\vec{n} \vec{v}_{2}\cdot\vec{a}_{2} + 4 \vec{v}_{2}\cdot\vec{n} \vec{v}_{2}\cdot\vec{a}_{2} -3 \vec{v}_{1}\cdot\vec{n} \vec{v}_{2}\cdot\vec{n} \vec{a}_{2}\cdot\vec{n} \big) \nn\\ 
&& + 9 \ddot{\vec{S}}_{1}\times\vec{v}_{1}\cdot\vec{v}_{2} \big( 3 \vec{v}_{2}\cdot\vec{a}_{2} + 2 \vec{v}_{1}\cdot\vec{n} \vec{a}_{2}\cdot\vec{n} - \vec{v}_{2}\cdot\vec{n} \vec{a}_{2}\cdot\vec{n} \big) - 27 \ddot{\vec{S}}_{1}\times\vec{a}_{1}\cdot\vec{v}_{2} \big( v_{2}^2 \nn\\ 
&& + ( \vec{v}_{2}\cdot\vec{n})^{2} \big) - 3 \ddot{\vec{S}}_{1}\times\vec{n}\cdot\vec{a}_{2} \big( 2 \vec{v}_{2}\cdot\vec{n} \vec{v}_{1}\cdot\vec{v}_{2} + 3 \vec{v}_{1}\cdot\vec{n} v_{2}^2 \nn\\ 
&& + 2 \vec{v}_{2}\cdot\vec{n} v_{2}^2 -3 \vec{v}_{1}\cdot\vec{n} ( \vec{v}_{2}\cdot\vec{n})^{2} \big) + 3 \ddot{\vec{S}}_{1}\times\vec{v}_{1}\cdot\vec{a}_{2} \big( 7 v_{1}^2 + 6 \vec{v}_{1}\cdot\vec{v}_{2} + 4 v_{2}^2 \nn\\ 
&& + 15 \vec{v}_{1}\cdot\vec{n} \vec{v}_{2}\cdot\vec{n} - 3 ( \vec{v}_{2}\cdot\vec{n})^{2} \big) - 3 \ddot{\vec{S}}_{1}\times\vec{v}_{2}\cdot\vec{a}_{2} \big( \vec{v}_{1}\cdot\vec{v}_{2} + v_{2}^2 + 3 \vec{v}_{1}\cdot\vec{n} \vec{v}_{2}\cdot\vec{n} \big)) \nn\\ 
&& - (6 \vec{S}_{1}\times\vec{n}\cdot\vec{v}_{1} \big( a_{1}^2 \vec{a}_{2}\cdot\vec{n} - \vec{a}_{1}\cdot\vec{n} a_{2}^2 \big) + 3 \vec{S}_{1}\times\vec{n}\cdot\vec{a}_{1} \big( 4 \vec{v}_{1}\cdot\vec{a}_{1} \vec{a}_{2}\cdot\vec{n} - 3 \vec{a}_{1}\cdot\vec{v}_{2} \vec{a}_{2}\cdot\vec{n} \nn\\ 
&& - 9 \vec{v}_{2}\cdot\vec{n} \vec{a}_{1}\cdot\vec{a}_{2} + 3 \vec{a}_{1}\cdot\vec{n} \vec{v}_{2}\cdot\vec{a}_{2} - 4 \vec{v}_{1}\cdot\vec{n} a_{2}^2 -3 \vec{a}_{1}\cdot\vec{n} \vec{v}_{2}\cdot\vec{n} \vec{a}_{2}\cdot\vec{n} \big) \nn\\ 
&& - 6 \vec{S}_{1}\times\vec{v}_{1}\cdot\vec{a}_{1} \big( 7 \vec{a}_{1}\cdot\vec{a}_{2} + a_{2}^2 - \vec{a}_{1}\cdot\vec{n} \vec{a}_{2}\cdot\vec{n} \big) + 6 \vec{S}_{1}\times\vec{n}\cdot\vec{v}_{2} \vec{a}_{1}\cdot\vec{n} a_{2}^2 \nn\\ 
&& - 3 \vec{S}_{1}\times\vec{a}_{1}\cdot\vec{v}_{2} \big( 9 \vec{a}_{1}\cdot\vec{a}_{2} - 2 a_{2}^2 -3 \vec{a}_{1}\cdot\vec{n} \vec{a}_{2}\cdot\vec{n} \big) + 3 \vec{S}_{1}\times\vec{n}\cdot\vec{a}_{2} \big( 3 \vec{v}_{1}\cdot\vec{a}_{1} \vec{a}_{2}\cdot\vec{n} \nn\\ 
&& - \vec{a}_{1}\cdot\vec{n} \vec{v}_{1}\cdot\vec{a}_{2} - 5 \vec{v}_{1}\cdot\vec{n} \vec{a}_{1}\cdot\vec{a}_{2} + 4 \vec{a}_{1}\cdot\vec{n} \vec{v}_{2}\cdot\vec{a}_{2} -3 \vec{v}_{1}\cdot\vec{n} \vec{a}_{1}\cdot\vec{n} \vec{a}_{2}\cdot\vec{n} \big) \nn\\ 
&& + 3 \vec{S}_{1}\times\vec{v}_{1}\cdot\vec{a}_{2} \big( 14 a_{1}^2 - 5 \vec{a}_{1}\cdot\vec{a}_{2} + 3 \vec{a}_{1}\cdot\vec{n} \vec{a}_{2}\cdot\vec{n} \big) + 3 \vec{S}_{1}\times\vec{a}_{1}\cdot\vec{a}_{2} \big( 28 \vec{v}_{1}\cdot\vec{a}_{1} \nn\\ 
&& - 9 \vec{a}_{1}\cdot\vec{v}_{2} - \vec{v}_{1}\cdot\vec{a}_{2} + 4 \vec{v}_{2}\cdot\vec{a}_{2} + 9 \vec{a}_{1}\cdot\vec{n} \vec{v}_{2}\cdot\vec{n} + 3 \vec{v}_{1}\cdot\vec{n} \vec{a}_{2}\cdot\vec{n} \big)) \nn\\ 
&& - (3 \dot{\vec{S}}_{1}\times\vec{n}\cdot\vec{v}_{1} \big( 4 \vec{v}_{1}\cdot\vec{a}_{1} \vec{a}_{2}\cdot\vec{n} - 3 \vec{a}_{1}\cdot\vec{v}_{2} \vec{a}_{2}\cdot\vec{n} - 9 \vec{v}_{2}\cdot\vec{n} \vec{a}_{1}\cdot\vec{a}_{2} + 3 \vec{a}_{1}\cdot\vec{n} \vec{v}_{2}\cdot\vec{a}_{2} \nn\\ 
&& - 4 \vec{v}_{1}\cdot\vec{n} a_{2}^2 -3 \vec{a}_{1}\cdot\vec{n} \vec{v}_{2}\cdot\vec{n} \vec{a}_{2}\cdot\vec{n} \big) + 6 \dot{\vec{S}}_{1}\times\vec{n}\cdot\vec{a}_{1} \big( v_{1}^2 \vec{a}_{2}\cdot\vec{n} + v_{2}^2 \vec{a}_{2}\cdot\vec{n} \nn\\ 
&& - 3 \vec{v}_{2}\cdot\vec{n} \vec{v}_{1}\cdot\vec{a}_{2} + 3 \vec{v}_{1}\cdot\vec{n} \vec{v}_{2}\cdot\vec{a}_{2} + 4 \vec{v}_{2}\cdot\vec{n} \vec{v}_{2}\cdot\vec{a}_{2} -3 \vec{v}_{1}\cdot\vec{n} \vec{v}_{2}\cdot\vec{n} \vec{a}_{2}\cdot\vec{n} \big) \nn\\ 
&& - 3 \dot{\vec{S}}_{1}\times\vec{v}_{1}\cdot\vec{a}_{1} \big( 12 \vec{v}_{1}\cdot\vec{a}_{2} - 11 \vec{v}_{2}\cdot\vec{a}_{2} -4 \vec{v}_{1}\cdot\vec{n} \vec{a}_{2}\cdot\vec{n} - 3 \vec{v}_{2}\cdot\vec{n} \vec{a}_{2}\cdot\vec{n} \big) \nn\\ 
&& + 3 \dot{\vec{S}}_{1}\times\vec{n}\cdot\vec{v}_{2} \big( \vec{a}_{1}\cdot\vec{v}_{2} \vec{a}_{2}\cdot\vec{n} + 5 \vec{v}_{2}\cdot\vec{n} \vec{a}_{1}\cdot\vec{a}_{2} - 3 \vec{a}_{1}\cdot\vec{n} \vec{v}_{2}\cdot\vec{a}_{2} + 4 \vec{v}_{1}\cdot\vec{n} a_{2}^2 \nn\\ 
&& + 3 \vec{a}_{1}\cdot\vec{n} \vec{v}_{2}\cdot\vec{n} \vec{a}_{2}\cdot\vec{n} \big) - 3 \dot{\vec{S}}_{1}\times\vec{v}_{1}\cdot\vec{v}_{2} \big( 5 \vec{a}_{1}\cdot\vec{a}_{2} - 4 a_{2}^2 -3 \vec{a}_{1}\cdot\vec{n} \vec{a}_{2}\cdot\vec{n} \big) \nn\\ 
&& - 3 \dot{\vec{S}}_{1}\times\vec{a}_{1}\cdot\vec{v}_{2} \big( 4 \vec{v}_{1}\cdot\vec{a}_{2} - 21 \vec{v}_{2}\cdot\vec{a}_{2} -12 \vec{v}_{1}\cdot\vec{n} \vec{a}_{2}\cdot\vec{n} + 3 \vec{v}_{2}\cdot\vec{n} \vec{a}_{2}\cdot\vec{n} \big) \nn\\ 
&& + 3 \dot{\vec{S}}_{1}\times\vec{n}\cdot\vec{a}_{2} \big( 2 \vec{v}_{2}\cdot\vec{n} \vec{a}_{1}\cdot\vec{v}_{2} - 3 \vec{a}_{1}\cdot\vec{n} v_{2}^2 + 3 v_{1}^2 \vec{a}_{2}\cdot\vec{n} - 2 \vec{v}_{1}\cdot\vec{n} \vec{v}_{1}\cdot\vec{a}_{2} \nn\\ 
&& + 8 \vec{v}_{1}\cdot\vec{n} \vec{v}_{2}\cdot\vec{a}_{2} -3 \vec{a}_{2}\cdot\vec{n} ( \vec{v}_{1}\cdot\vec{n})^{2} + 3 \vec{a}_{1}\cdot\vec{n} ( \vec{v}_{2}\cdot\vec{n})^{2} \big) + 3 \dot{\vec{S}}_{1}\times\vec{v}_{1}\cdot\vec{a}_{2} \big( 28 \vec{v}_{1}\cdot\vec{a}_{1} \nn\\ 
&& - 5 \vec{a}_{1}\cdot\vec{v}_{2} - 2 \vec{v}_{1}\cdot\vec{a}_{2} + 8 \vec{v}_{2}\cdot\vec{a}_{2} + 9 \vec{a}_{1}\cdot\vec{n} \vec{v}_{2}\cdot\vec{n} + 6 \vec{v}_{1}\cdot\vec{n} \vec{a}_{2}\cdot\vec{n} \big) \nn\\ 
&& + 3 \dot{\vec{S}}_{1}\times\vec{a}_{1}\cdot\vec{a}_{2} \big( 14 v_{1}^2 + 8 \vec{v}_{1}\cdot\vec{v}_{2} + 11 v_{2}^2 + 30 \vec{v}_{1}\cdot\vec{n} \vec{v}_{2}\cdot\vec{n} - 3 ( \vec{v}_{2}\cdot\vec{n})^{2} \big) \nn\\ 
&& + 3 \dot{\vec{S}}_{1}\times\vec{v}_{2}\cdot\vec{a}_{2} \big( \vec{a}_{1}\cdot\vec{v}_{2} -3 \vec{a}_{1}\cdot\vec{n} \vec{v}_{2}\cdot\vec{n} \big)) \Big]\nn\\ && +  	\frac{1}{144} G^2 m_{2}{}^2 \Big[ (468 \vec{S}_{1}\times\vec{n}\cdot\dot{\vec{a}}_{1} \vec{a}_{2}\cdot\vec{n} + 12 \vec{S}_{1}\times\vec{n}\cdot\vec{a}_{2} \dot{\vec{a}}_{1}\cdot\vec{n} + 384 \vec{S}_{1}\times\vec{n}\cdot\dot{\vec{a}}_{2} \vec{a}_{1}\cdot\vec{n} \nn\\ 
&& - 347 \vec{S}_{1}\times\dot{\vec{a}}_{1}\cdot\vec{a}_{2} - 253 \vec{S}_{1}\times\vec{a}_{1}\cdot\dot{\vec{a}}_{2} + 648 \vec{S}_{1}\times\vec{a}_{2}\cdot\dot{\vec{a}}_{2}) - (144 \dot{\vec{S}}_{1}\times\vec{n}\cdot\vec{v}_{2} \dot{\vec{a}}_{2}\cdot\vec{n} \nn\\ 
&& - 384 \dot{\vec{S}}_{1}\times\vec{n}\cdot\dot{\vec{a}}_{2} \big( 2 \vec{v}_{1}\cdot\vec{n} - \vec{v}_{2}\cdot\vec{n} \big) + 410 \dot{\vec{S}}_{1}\times\vec{v}_{1}\cdot\dot{\vec{a}}_{2} - 340 \dot{\vec{S}}_{1}\times\vec{v}_{2}\cdot\dot{\vec{a}}_{2}) \nn\\ 
&& + (468 \ddot{\vec{S}}_{1}\times\vec{n}\cdot\vec{v}_{1} \vec{a}_{2}\cdot\vec{n} - 576 \ddot{\vec{S}}_{1}\times\vec{n}\cdot\vec{v}_{2} \vec{a}_{2}\cdot\vec{n} + 12 \ddot{\vec{S}}_{1}\times\vec{n}\cdot\vec{a}_{2} \big( \vec{v}_{1}\cdot\vec{n} - 20 \vec{v}_{2}\cdot\vec{n} \big) \nn\\ 
&& - 347 \ddot{\vec{S}}_{1}\times\vec{v}_{1}\cdot\vec{a}_{2} - 23 \ddot{\vec{S}}_{1}\times\vec{v}_{2}\cdot\vec{a}_{2}) + (936 \dot{\vec{S}}_{1}\times\vec{n}\cdot\vec{a}_{1} \vec{a}_{2}\cdot\vec{n} + 24 \dot{\vec{S}}_{1}\times\vec{n}\cdot\vec{a}_{2} \big( \vec{a}_{1}\cdot\vec{n} \nn\\ 
&& - 26 \vec{a}_{2}\cdot\vec{n} \big) - 694 \dot{\vec{S}}_{1}\times\vec{a}_{1}\cdot\vec{a}_{2}) \Big] + 	\frac{1}{6} G^2 m_{1} m_{2} {\Big( \frac{1}{\epsilon} - 2\log \frac{r}{R_0} \Big)} \Big[ (32 \vec{S}_{1}\times\vec{a}_{1}\cdot\dot{\vec{a}}_{1} \nn\\ 
&& - 40 \vec{S}_{1}\times\dot{\vec{a}}_{1}\cdot\vec{a}_{2} - 8 \vec{S}_{1}\times\vec{a}_{1}\cdot\dot{\vec{a}}_{2}) - (16 \dot{\vec{S}}_{1}\times\vec{v}_{1}\cdot\dot{\vec{a}}_{1} + 5 \dot{\vec{S}}_{1}\times\dot{\vec{a}}_{1}\cdot\vec{v}_{2} \nn\\ 
&& + 11 \dot{\vec{S}}_{1}\times\vec{v}_{1}\cdot\dot{\vec{a}}_{2}) - (32 \ddot{\vec{S}}_{1}\times\vec{v}_{1}\cdot\vec{a}_{1} - 3 \ddot{\vec{S}}_{1}\times\vec{a}_{1}\cdot\vec{v}_{2} + 43 \ddot{\vec{S}}_{1}\times\vec{v}_{1}\cdot\vec{a}_{2} \nn\\ 
&& - 8 \ddot{\vec{S}}_{1}\times\vec{v}_{2}\cdot\vec{a}_{2}) - 80 \dot{\vec{S}}_{1}\times\vec{a}_{1}\cdot\vec{a}_{2} \Big] - 	\frac{1}{6} G^2 m_{2}{}^2 {\Big( \frac{1}{\epsilon} - 2\log \frac{r}{R_0} \Big)} \Big[ (40 \vec{S}_{1}\times\dot{\vec{a}}_{1}\cdot\vec{a}_{2} \nn\\ 
&& + 8 \vec{S}_{1}\times\vec{a}_{1}\cdot\dot{\vec{a}}_{2}) + (16 \dot{\vec{S}}_{1}\times\vec{v}_{1}\cdot\dot{\vec{a}}_{2} - 5 \dot{\vec{S}}_{1}\times\vec{v}_{2}\cdot\dot{\vec{a}}_{2}) + (40 \ddot{\vec{S}}_{1}\times\vec{v}_{1}\cdot\vec{a}_{2} \nn\\ 
&& - 5 \ddot{\vec{S}}_{1}\times\vec{v}_{2}\cdot\vec{a}_{2}) + 80 \dot{\vec{S}}_{1}\times\vec{a}_{1}\cdot\vec{a}_{2} \Big] + 	\frac{1}{36} G^2 m_{1} m_{2} \Big[ (12 \vec{S}_{1}\times\vec{n}\cdot\vec{a}_{1} \dot{\vec{a}}_{2}\cdot\vec{n} \nn\\ 
&& - 24 \vec{S}_{1}\times\vec{n}\cdot\dot{\vec{a}}_{1} \big( 4 \vec{a}_{1}\cdot\vec{n} - 7 \vec{a}_{2}\cdot\vec{n} \big) + 144 \vec{S}_{1}\times\vec{n}\cdot\dot{\vec{a}}_{2} \vec{a}_{1}\cdot\vec{n} + 340 \vec{S}_{1}\times\vec{a}_{1}\cdot\dot{\vec{a}}_{1} \nn\\ 
&& - 152 \vec{S}_{1}\times\dot{\vec{a}}_{1}\cdot\vec{a}_{2} - 40 \vec{S}_{1}\times\vec{a}_{1}\cdot\dot{\vec{a}}_{2}) - (12 \dot{\vec{S}}_{1}\times\vec{n}\cdot\dot{\vec{a}}_{1} \big( 8 \vec{v}_{1}\cdot\vec{n} - 7 \vec{v}_{2}\cdot\vec{n} \big) \nn\\ 
&& - 204 \dot{\vec{S}}_{1}\times\vec{n}\cdot\dot{\vec{a}}_{2} \vec{v}_{1}\cdot\vec{n} + 164 \dot{\vec{S}}_{1}\times\vec{v}_{1}\cdot\dot{\vec{a}}_{1} + 13 \dot{\vec{S}}_{1}\times\dot{\vec{a}}_{1}\cdot\vec{v}_{2} + 25 \dot{\vec{S}}_{1}\times\vec{v}_{1}\cdot\dot{\vec{a}}_{2}) \nn\\ 
&& + (60 \ddot{\vec{S}}_{1}\times\vec{n}\cdot\vec{v}_{1} \vec{a}_{2}\cdot\vec{n} + 12 \ddot{\vec{S}}_{1}\times\vec{n}\cdot\vec{a}_{1} \big( 8 \vec{v}_{1}\cdot\vec{n} - 9 \vec{v}_{2}\cdot\vec{n} \big) - 60 \ddot{\vec{S}}_{1}\times\vec{n}\cdot\vec{v}_{2} \vec{a}_{2}\cdot\vec{n} \nn\\ 
&& + 12 \ddot{\vec{S}}_{1}\times\vec{n}\cdot\vec{a}_{2} \big( 12 \vec{v}_{1}\cdot\vec{n} - 13 \vec{v}_{2}\cdot\vec{n} \big) - 406 \ddot{\vec{S}}_{1}\times\vec{v}_{1}\cdot\vec{a}_{1} + 21 \ddot{\vec{S}}_{1}\times\vec{a}_{1}\cdot\vec{v}_{2} \nn\\ 
&& - 155 \ddot{\vec{S}}_{1}\times\vec{v}_{1}\cdot\vec{a}_{2} + 28 \ddot{\vec{S}}_{1}\times\vec{v}_{2}\cdot\vec{a}_{2}) - (24 \dot{\vec{S}}_{1}\times\vec{n}\cdot\vec{a}_{1} \big( 4 \vec{a}_{1}\cdot\vec{n} - 9 \vec{a}_{2}\cdot\vec{n} \big) \nn\\ 
&& - 96 \dot{\vec{S}}_{1}\times\vec{n}\cdot\vec{a}_{2} \vec{a}_{1}\cdot\vec{n} + 358 \dot{\vec{S}}_{1}\times\vec{a}_{1}\cdot\vec{a}_{2}) \Big] 
+ (1 \leftrightarrow 2),
\eea
\bea
\stackrel{(4)}{V}_{3,1} &=&\frac{1}{24} G m_{2} r{}^2 \Big[ ( \vec{S}_{1}\times\vec{n}\cdot\vec{v}_{1} \big( 8 \dot{\vec{a}}_{1}\cdot\dot{\vec{a}}_{2} - \dot{\vec{a}}_{1}\cdot\vec{n} \dot{\vec{a}}_{2}\cdot\vec{n} \big) + 3 \vec{S}_{1}\times\vec{n}\cdot\dot{\vec{a}}_{1} \big( 2 \vec{v}_{1}\cdot\dot{\vec{a}}_{2} - 2 \vec{v}_{2}\cdot\dot{\vec{a}}_{2} \nn\\ 
&& - \vec{v}_{1}\cdot\vec{n} \dot{\vec{a}}_{2}\cdot\vec{n} \big) - 2 \vec{S}_{1}\times\vec{v}_{1}\cdot\dot{\vec{a}}_{1} \dot{\vec{a}}_{2}\cdot\vec{n} - \vec{S}_{1}\times\vec{n}\cdot\vec{v}_{2} \big( 5 \dot{\vec{a}}_{1}\cdot\dot{\vec{a}}_{2} - \dot{\vec{a}}_{1}\cdot\vec{n} \dot{\vec{a}}_{2}\cdot\vec{n} \big) \nn\\ 
&& + \vec{S}_{1}\times\dot{\vec{a}}_{1}\cdot\vec{v}_{2} \dot{\vec{a}}_{2}\cdot\vec{n} - 3 \vec{S}_{1}\times\vec{n}\cdot\dot{\vec{a}}_{2} \big( \dot{\vec{a}}_{1}\cdot\vec{v}_{2} - \dot{\vec{a}}_{1}\cdot\vec{n} \vec{v}_{2}\cdot\vec{n} \big) - 2 \vec{S}_{1}\times\vec{v}_{1}\cdot\dot{\vec{a}}_{2} \dot{\vec{a}}_{1}\cdot\vec{n} \nn\\ 
&& - 3 \vec{S}_{1}\times\dot{\vec{a}}_{1}\cdot\dot{\vec{a}}_{2} \big( 8 \vec{v}_{1}\cdot\vec{n} - \vec{v}_{2}\cdot\vec{n} \big) + 2 \vec{S}_{1}\times\vec{v}_{2}\cdot\dot{\vec{a}}_{2} \dot{\vec{a}}_{1}\cdot\vec{n}) + (3 \vec{S}_{1}\times\vec{n}\cdot\ddot{\vec{a}}_{1} \big( \vec{v}_{2}\cdot\vec{a}_{2} \nn\\ 
&& + \vec{v}_{2}\cdot\vec{n} \vec{a}_{2}\cdot\vec{n} \big) + 6 \vec{S}_{1}\times\vec{v}_{1}\cdot\ddot{\vec{a}}_{1} \vec{a}_{2}\cdot\vec{n} + 9 \vec{S}_{1}\times\ddot{\vec{a}}_{1}\cdot\vec{v}_{2} \vec{a}_{2}\cdot\vec{n} + 21 \vec{S}_{1}\times\ddot{\vec{a}}_{1}\cdot\vec{a}_{2} \vec{v}_{2}\cdot\vec{n} \nn\\ 
&& - 3 \vec{S}_{1}\times\vec{n}\cdot\ddot{\vec{a}}_{2} \big( \vec{v}_{1}\cdot\vec{a}_{1} + \vec{v}_{1}\cdot\vec{n} \vec{a}_{1}\cdot\vec{n} \big) - 3 \vec{S}_{1}\times\vec{v}_{1}\cdot\ddot{\vec{a}}_{2} \vec{a}_{1}\cdot\vec{n} - 3 \vec{S}_{1}\times\vec{a}_{1}\cdot\ddot{\vec{a}}_{2} \vec{v}_{1}\cdot\vec{n}) \nn\\ 
&& - (3 \dot{\vec{S}}_{1}\times\vec{n}\cdot\ddot{\vec{a}}_{2} \big( v_{1}^2 + ( \vec{v}_{1}\cdot\vec{n})^{2} \big) + 6 \dot{\vec{S}}_{1}\times\vec{v}_{1}\cdot\ddot{\vec{a}}_{2} \vec{v}_{1}\cdot\vec{n}) + (3 \ddot{\vec{S}}_{1}\times\vec{n}\cdot\vec{v}_{1} \big( 2 \vec{v}_{1}\cdot\dot{\vec{a}}_{2} \nn\\ 
&& - 2 \vec{v}_{2}\cdot\dot{\vec{a}}_{2} - \vec{v}_{1}\cdot\vec{n} \dot{\vec{a}}_{2}\cdot\vec{n} \big) - 3 \ddot{\vec{S}}_{1}\times\vec{n}\cdot\vec{v}_{2} \big( \vec{v}_{1}\cdot\dot{\vec{a}}_{2} - 2 \vec{v}_{2}\cdot\dot{\vec{a}}_{2} - \vec{v}_{1}\cdot\vec{n} \dot{\vec{a}}_{2}\cdot\vec{n} \big) \nn\\ 
&& + 3 \ddot{\vec{S}}_{1}\times\vec{v}_{1}\cdot\vec{v}_{2} \dot{\vec{a}}_{2}\cdot\vec{n} + 3 \ddot{\vec{S}}_{1}\times\vec{n}\cdot\dot{\vec{a}}_{2} \big( \vec{v}_{1}\cdot\vec{v}_{2} + v_{2}^2 + 3 \vec{v}_{1}\cdot\vec{n} \vec{v}_{2}\cdot\vec{n} \big) \nn\\ 
&& - 3 \ddot{\vec{S}}_{1}\times\vec{v}_{1}\cdot\dot{\vec{a}}_{2} \big( 8 \vec{v}_{1}\cdot\vec{n} - 3 \vec{v}_{2}\cdot\vec{n} \big) + 6 \ddot{\vec{S}}_{1}\times\vec{v}_{2}\cdot\dot{\vec{a}}_{2} \vec{v}_{1}\cdot\vec{n}) + (3 \dddot{\vec{S}}_{1}\times\vec{n}\cdot\vec{v}_{1} \big( \vec{v}_{2}\cdot\vec{a}_{2} \nn\\ 
&& + \vec{v}_{2}\cdot\vec{n} \vec{a}_{2}\cdot\vec{n} \big) - 3 \dddot{\vec{S}}_{1}\times\vec{n}\cdot\vec{v}_{2} \big( \vec{v}_{2}\cdot\vec{a}_{2} + \vec{v}_{2}\cdot\vec{n} \vec{a}_{2}\cdot\vec{n} \big) + 9 \dddot{\vec{S}}_{1}\times\vec{v}_{1}\cdot\vec{v}_{2} \vec{a}_{2}\cdot\vec{n} \nn\\ 
&& - 3 \dddot{\vec{S}}_{1}\times\vec{n}\cdot\vec{a}_{2} \big( v_{2}^2 + ( \vec{v}_{2}\cdot\vec{n})^{2} \big) + 21 \dddot{\vec{S}}_{1}\times\vec{v}_{1}\cdot\vec{a}_{2} \vec{v}_{2}\cdot\vec{n} - 3 \dddot{\vec{S}}_{1}\times\vec{v}_{2}\cdot\vec{a}_{2} \vec{v}_{2}\cdot\vec{n}) \nn\\ 
&& + (3 \vec{S}_{1}\times\vec{n}\cdot\vec{a}_{1} \big( 5 \vec{a}_{1}\cdot\dot{\vec{a}}_{2} - \vec{a}_{1}\cdot\vec{n} \dot{\vec{a}}_{2}\cdot\vec{n} \big) - 6 \vec{S}_{1}\times\vec{n}\cdot\dot{\vec{a}}_{1} a_{2}^2 + 6 \vec{S}_{1}\times\vec{a}_{1}\cdot\dot{\vec{a}}_{1} \vec{a}_{2}\cdot\vec{n} \nn\\ 
&& - 3 \vec{S}_{1}\times\vec{n}\cdot\vec{a}_{2} \big( 3 \dot{\vec{a}}_{1}\cdot\vec{a}_{2} - \dot{\vec{a}}_{1}\cdot\vec{n} \vec{a}_{2}\cdot\vec{n} \big) + 3 \vec{S}_{1}\times\dot{\vec{a}}_{1}\cdot\vec{a}_{2} \vec{a}_{2}\cdot\vec{n} \nn\\ 
&& - 15 \vec{S}_{1}\times\vec{a}_{1}\cdot\dot{\vec{a}}_{2} \vec{a}_{1}\cdot\vec{n}) + (3 \dot{\vec{S}}_{1}\times\vec{n}\cdot\vec{v}_{1} \big( 5 \vec{a}_{1}\cdot\dot{\vec{a}}_{2} - \vec{a}_{1}\cdot\vec{n} \dot{\vec{a}}_{2}\cdot\vec{n} \big) \nn\\ 
&& + 6 \dot{\vec{S}}_{1}\times\vec{n}\cdot\vec{a}_{1} \big( 2 \vec{v}_{1}\cdot\dot{\vec{a}}_{2} - 2 \vec{v}_{2}\cdot\dot{\vec{a}}_{2} - \vec{v}_{1}\cdot\vec{n} \dot{\vec{a}}_{2}\cdot\vec{n} \big) - 3 \dot{\vec{S}}_{1}\times\vec{v}_{1}\cdot\vec{a}_{1} \dot{\vec{a}}_{2}\cdot\vec{n} \nn\\ 
&& + 9 \dot{\vec{S}}_{1}\times\vec{n}\cdot\dot{\vec{a}}_{1} \big( \vec{v}_{2}\cdot\vec{a}_{2} + \vec{v}_{2}\cdot\vec{n} \vec{a}_{2}\cdot\vec{n} \big) + 12 \dot{\vec{S}}_{1}\times\vec{v}_{1}\cdot\dot{\vec{a}}_{1} \vec{a}_{2}\cdot\vec{n} - 3 \dot{\vec{S}}_{1}\times\vec{n}\cdot\vec{v}_{2} \big( 3 \vec{a}_{1}\cdot\dot{\vec{a}}_{2} \nn\\ 
&& - \vec{a}_{1}\cdot\vec{n} \dot{\vec{a}}_{2}\cdot\vec{n} \big) + 3 \dot{\vec{S}}_{1}\times\vec{a}_{1}\cdot\vec{v}_{2} \dot{\vec{a}}_{2}\cdot\vec{n} + 27 \dot{\vec{S}}_{1}\times\dot{\vec{a}}_{1}\cdot\vec{v}_{2} \vec{a}_{2}\cdot\vec{n} + 63 \dot{\vec{S}}_{1}\times\dot{\vec{a}}_{1}\cdot\vec{a}_{2} \vec{v}_{2}\cdot\vec{n} \nn\\ 
&& - 3 \dot{\vec{S}}_{1}\times\vec{n}\cdot\dot{\vec{a}}_{2} \big( \vec{a}_{1}\cdot\vec{v}_{2} -3 \vec{a}_{1}\cdot\vec{n} \vec{v}_{2}\cdot\vec{n} \big) - 15 \dot{\vec{S}}_{1}\times\vec{v}_{1}\cdot\dot{\vec{a}}_{2} \vec{a}_{1}\cdot\vec{n} \nn\\ 
&& - 3 \dot{\vec{S}}_{1}\times\vec{a}_{1}\cdot\dot{\vec{a}}_{2} \big( 16 \vec{v}_{1}\cdot\vec{n} - 3 \vec{v}_{2}\cdot\vec{n} \big) + 6 \dot{\vec{S}}_{1}\times\vec{v}_{2}\cdot\dot{\vec{a}}_{2} \vec{a}_{1}\cdot\vec{n}) - (6 \ddot{\vec{S}}_{1}\times\vec{n}\cdot\vec{v}_{1} a_{2}^2 \nn\\ 
&& - 9 \ddot{\vec{S}}_{1}\times\vec{n}\cdot\vec{a}_{1} \big( \vec{v}_{2}\cdot\vec{a}_{2} + \vec{v}_{2}\cdot\vec{n} \vec{a}_{2}\cdot\vec{n} \big) - 6 \ddot{\vec{S}}_{1}\times\vec{v}_{1}\cdot\vec{a}_{1} \vec{a}_{2}\cdot\vec{n} - 6 \ddot{\vec{S}}_{1}\times\vec{n}\cdot\vec{v}_{2} a_{2}^2 \nn\\ 
&& - 27 \ddot{\vec{S}}_{1}\times\vec{a}_{1}\cdot\vec{v}_{2} \vec{a}_{2}\cdot\vec{n} + 3 \ddot{\vec{S}}_{1}\times\vec{n}\cdot\vec{a}_{2} \big( \vec{v}_{1}\cdot\vec{a}_{2} - 4 \vec{v}_{2}\cdot\vec{a}_{2} -3 \vec{v}_{1}\cdot\vec{n} \vec{a}_{2}\cdot\vec{n} \big) \nn\\ 
&& - 9 \ddot{\vec{S}}_{1}\times\vec{v}_{1}\cdot\vec{a}_{2} \vec{a}_{2}\cdot\vec{n} - 63 \ddot{\vec{S}}_{1}\times\vec{a}_{1}\cdot\vec{a}_{2} \vec{v}_{2}\cdot\vec{n}) - (12 \dot{\vec{S}}_{1}\times\vec{n}\cdot\vec{a}_{1} a_{2}^2 \nn\\ 
&& + 3 \dot{\vec{S}}_{1}\times\vec{n}\cdot\vec{a}_{2} \big( 5 \vec{a}_{1}\cdot\vec{a}_{2} -3 \vec{a}_{1}\cdot\vec{n} \vec{a}_{2}\cdot\vec{n} \big) - 9 \dot{\vec{S}}_{1}\times\vec{a}_{1}\cdot\vec{a}_{2} \vec{a}_{2}\cdot\vec{n}) \Big]\nn\\ && - 	\frac{1}{18} G^2 m_{1} m_{2} r \Big[ 9 \ddot{\vec{S}}_{1}\times\vec{n}\cdot\dot{\vec{a}}_{1} + 14 \ddot{\vec{S}}_{1}\times\vec{n}\cdot\dot{\vec{a}}_{2} \Big] - 	\frac{157}{144} G^2 m_{2}{}^2 r \ddot{\vec{S}}_{1}\times\vec{n}\cdot\dot{\vec{a}}_{2} \nn\\ 
&& - 	\frac{4}{3} G^2 m_{1} m_{2} r {\Big( \frac{1}{\epsilon} - 2\log \frac{r}{R_0} \Big)} \ddot{\vec{S}}_{1}\times\vec{n}\cdot\dot{\vec{a}}_{2} - 	\frac{4}{3} G^2 m_{2}{}^2 r {\Big( \frac{1}{\epsilon} - 2\log \frac{r}{R_0} \Big)} \ddot{\vec{S}}_{1}\times\vec{n}\cdot\dot{\vec{a}}_{2}  \nn\\
&&+ (1 \leftrightarrow 2),
\eea
\bea
\stackrel{(5)}{V}_{3,1} &=& - 	\frac{1}{72} G m_{2} r{}^3 \Big[ (3 \vec{S}_{1}\times\vec{n}\cdot\ddot{\vec{a}}_{1} \dot{\vec{a}}_{2}\cdot\vec{n} + 3 \vec{S}_{1}\times\vec{n}\cdot\ddot{\vec{a}}_{2} \dot{\vec{a}}_{1}\cdot\vec{n} + 11 \vec{S}_{1}\times\ddot{\vec{a}}_{1}\cdot\dot{\vec{a}}_{2} \nn\\ 
&& + \vec{S}_{1}\times\dot{\vec{a}}_{1}\cdot\ddot{\vec{a}}_{2}) + (9 \ddot{\vec{S}}_{1}\times\vec{n}\cdot\ddot{\vec{a}}_{2} \vec{v}_{1}\cdot\vec{n} + 3 \ddot{\vec{S}}_{1}\times\vec{v}_{1}\cdot\ddot{\vec{a}}_{2}) + (3 \dddot{\vec{S}}_{1}\times\vec{n}\cdot\vec{v}_{1} \dot{\vec{a}}_{2}\cdot\vec{n} \nn\\ 
&& - 3 \dddot{\vec{S}}_{1}\times\vec{n}\cdot\vec{v}_{2} \dot{\vec{a}}_{2}\cdot\vec{n} - 9 \dddot{\vec{S}}_{1}\times\vec{n}\cdot\dot{\vec{a}}_{2} \vec{v}_{2}\cdot\vec{n} + 11 \dddot{\vec{S}}_{1}\times\vec{v}_{1}\cdot\dot{\vec{a}}_{2} - 2 \dddot{\vec{S}}_{1}\times\vec{v}_{2}\cdot\dot{\vec{a}}_{2}) \nn\\ 
&& + (9 \dot{\vec{S}}_{1}\times\vec{n}\cdot\dot{\vec{a}}_{1} \dot{\vec{a}}_{2}\cdot\vec{n} + 33 \dot{\vec{S}}_{1}\times\dot{\vec{a}}_{1}\cdot\dot{\vec{a}}_{2}) + (9 \dot{\vec{S}}_{1}\times\vec{n}\cdot\ddot{\vec{a}}_{2} \vec{a}_{1}\cdot\vec{n} + 3 \dot{\vec{S}}_{1}\times\vec{a}_{1}\cdot\ddot{\vec{a}}_{2}) \nn\\ 
&& + (9 \ddot{\vec{S}}_{1}\times\vec{n}\cdot\vec{a}_{1} \dot{\vec{a}}_{2}\cdot\vec{n} + 33 \ddot{\vec{S}}_{1}\times\vec{a}_{1}\cdot\dot{\vec{a}}_{2}) - 9 \dddot{\vec{S}}_{1}\times\vec{n}\cdot\vec{a}_{2} \vec{a}_{2}\cdot\vec{n} \Big] 
+ (1 \leftrightarrow 2),
\eea
\bea
\stackrel{(6)}{V}_{3,1} =-
\frac{1}{72} G m_{2} r{}^4 \dddot{\vec{S}}_{1}\times\vec{n}\cdot\ddot{\vec{a}}_{2} + (1 \leftrightarrow 2).
\eea

\section{Redefinition of Position and Spin} 
\label{n3los1redefapp}

In this appendix we provide the explicit unreduced potentials, and redefinitions that are fixed in 
order to reduce them, throughout the reduction process that is described in section \ref{redefinitions}.
First we consider the $3$PN sector. Our unreduced potential can be expressed as:
\bea
V_{\text{3PN}} = \sum_{i=0}^4 \stackrel{(i)}{V}_{3,0}, 
\label{eq:3pnds}
\eea
with
\bea
\stackrel{(0)}{V}_{3,0}&=&- 	\frac{5}{128} m_{1} v_{1}^{8} - 	\frac{5}{128} m_{2} v_{2}^{8} \nn\\ && +  	\frac{G m_{1} m_{2}}{16 r} \Big[ 19 v_{1}^2 \vec{v}_{1}\cdot\vec{v}_{2} v_{2}^2 - 6 v_{1}^2 ( \vec{v}_{1}\cdot\vec{v}_{2})^{2} - 2 ( \vec{v}_{1}\cdot\vec{v}_{2})^{3} - 6 v_{2}^2 ( \vec{v}_{1}\cdot\vec{v}_{2})^{2} - 11 v_{1}^{6} \nn\\ 
&& + 17 \vec{v}_{1}\cdot\vec{v}_{2} v_{1}^{4} - 8 v_{2}^2 v_{1}^{4} - 8 v_{1}^2 v_{2}^{4} + 17 \vec{v}_{1}\cdot\vec{v}_{2} v_{2}^{4} - 11 v_{2}^{6} -4 \vec{v}_{1}\cdot\vec{n} v_{1}^2 \vec{v}_{2}\cdot\vec{n} \vec{v}_{1}\cdot\vec{v}_{2} \nn\\ 
&& + 5 \vec{v}_{1}\cdot\vec{n} v_{1}^2 \vec{v}_{2}\cdot\vec{n} v_{2}^2 - 4 \vec{v}_{1}\cdot\vec{n} \vec{v}_{2}\cdot\vec{n} \vec{v}_{1}\cdot\vec{v}_{2} v_{2}^2 + 3 v_{1}^2 v_{2}^2 ( \vec{v}_{1}\cdot\vec{n})^{2} - 5 \vec{v}_{1}\cdot\vec{v}_{2} v_{2}^2 ( \vec{v}_{1}\cdot\vec{n})^{2} \nn\\ 
&& - 5 v_{1}^2 \vec{v}_{1}\cdot\vec{v}_{2} ( \vec{v}_{2}\cdot\vec{n})^{2} - 10 \vec{v}_{1}\cdot\vec{n} \vec{v}_{2}\cdot\vec{n} ( \vec{v}_{1}\cdot\vec{v}_{2})^{2} + 3 v_{1}^2 v_{2}^2 ( \vec{v}_{2}\cdot\vec{n})^{2} + 7 \vec{v}_{1}\cdot\vec{n} \vec{v}_{2}\cdot\vec{n} v_{1}^{4} \nn\\ 
&& + 3 ( \vec{v}_{2}\cdot\vec{n})^{2} v_{1}^{4} + 3 ( \vec{v}_{1}\cdot\vec{n})^{2} v_{2}^{4} + 7 \vec{v}_{1}\cdot\vec{n} \vec{v}_{2}\cdot\vec{n} v_{2}^{4} -3 \vec{v}_{2}\cdot\vec{n} v_{2}^2 ( \vec{v}_{1}\cdot\vec{n})^{3} \nn\\ 
&& - 9 v_{1}^2 ( \vec{v}_{1}\cdot\vec{n})^{2} ( \vec{v}_{2}\cdot\vec{n})^{2} - 3 \vec{v}_{1}\cdot\vec{n} v_{1}^2 ( \vec{v}_{2}\cdot\vec{n})^{3} + 15 \vec{v}_{1}\cdot\vec{v}_{2} ( \vec{v}_{1}\cdot\vec{n})^{2} ( \vec{v}_{2}\cdot\vec{n})^{2} \nn\\ 
&& - 9 v_{2}^2 ( \vec{v}_{1}\cdot\vec{n})^{2} ( \vec{v}_{2}\cdot\vec{n})^{2} + 5 ( \vec{v}_{1}\cdot\vec{n})^{3} ( \vec{v}_{2}\cdot\vec{n})^{3} \Big]\nn\\ && - 	\frac{G^2 m_{1}{}^2 m_{2}}{48 r{}^2} \Big[ 408 v_{1}^2 \vec{v}_{1}\cdot\vec{v}_{2} + 168 v_{1}^2 v_{2}^2 - 396 \vec{v}_{1}\cdot\vec{v}_{2} v_{2}^2 - 48 ( \vec{v}_{1}\cdot\vec{v}_{2})^{2} - 264 v_{1}^{4} \nn\\ 
&& + 135 v_{2}^{4} -816 \vec{v}_{1}\cdot\vec{n} v_{1}^2 \vec{v}_{2}\cdot\vec{n} + 1008 \vec{v}_{1}\cdot\vec{n} \vec{v}_{2}\cdot\vec{n} \vec{v}_{1}\cdot\vec{v}_{2} - 192 \vec{v}_{1}\cdot\vec{n} \vec{v}_{2}\cdot\vec{n} v_{2}^2 \nn\\ 
&& + 816 v_{1}^2 ( \vec{v}_{1}\cdot\vec{n})^{2} - 936 \vec{v}_{1}\cdot\vec{v}_{2} ( \vec{v}_{1}\cdot\vec{n})^{2} + 120 v_{2}^2 ( \vec{v}_{1}\cdot\vec{n})^{2} - 96 v_{1}^2 ( \vec{v}_{2}\cdot\vec{n})^{2} \nn\\ 
&& + 96 \vec{v}_{1}\cdot\vec{v}_{2} ( \vec{v}_{2}\cdot\vec{n})^{2} + 12 v_{2}^2 ( \vec{v}_{2}\cdot\vec{n})^{2} -8 ( \vec{v}_{1}\cdot\vec{n})^{4} + 48 ( \vec{v}_{1}\cdot\vec{n})^{2} ( \vec{v}_{2}\cdot\vec{n})^{2} \nn\\ 
&& - 64 \vec{v}_{1}\cdot\vec{n} ( \vec{v}_{2}\cdot\vec{n})^{3} \Big] + 	\frac{G^2 m_{1} m_{2}{}^2}{48 r{}^2} \Big[ 396 v_{1}^2 \vec{v}_{1}\cdot\vec{v}_{2} - 168 v_{1}^2 v_{2}^2 - 408 \vec{v}_{1}\cdot\vec{v}_{2} v_{2}^2 \nn\\ 
&& + 48 ( \vec{v}_{1}\cdot\vec{v}_{2})^{2} - 135 v_{1}^{4} + 264 v_{2}^{4} + 192 \vec{v}_{1}\cdot\vec{n} v_{1}^2 \vec{v}_{2}\cdot\vec{n} - 1008 \vec{v}_{1}\cdot\vec{n} \vec{v}_{2}\cdot\vec{n} \vec{v}_{1}\cdot\vec{v}_{2} \nn\\ 
&& + 816 \vec{v}_{1}\cdot\vec{n} \vec{v}_{2}\cdot\vec{n} v_{2}^2 - 12 v_{1}^2 ( \vec{v}_{1}\cdot\vec{n})^{2} - 96 \vec{v}_{1}\cdot\vec{v}_{2} ( \vec{v}_{1}\cdot\vec{n})^{2} + 96 v_{2}^2 ( \vec{v}_{1}\cdot\vec{n})^{2} \nn\\ 
&& - 120 v_{1}^2 ( \vec{v}_{2}\cdot\vec{n})^{2} + 936 \vec{v}_{1}\cdot\vec{v}_{2} ( \vec{v}_{2}\cdot\vec{n})^{2} - 816 v_{2}^2 ( \vec{v}_{2}\cdot\vec{n})^{2} + 64 \vec{v}_{2}\cdot\vec{n} ( \vec{v}_{1}\cdot\vec{n})^{3} \nn\\ 
&& - 48 ( \vec{v}_{1}\cdot\vec{n})^{2} ( \vec{v}_{2}\cdot\vec{n})^{2} + 8 ( \vec{v}_{2}\cdot\vec{n})^{4} \Big]\nn\\ && - 	\frac{G^3 m_{1}{}^3 m_{2}}{3 r{}^3} {\Big( \frac{1}{\epsilon} - 3\log \frac{r}{R_0} \Big)} \Big[ 3 v_{1}^2 - 5 \vec{v}_{1}\cdot\vec{v}_{2} + 2 v_{2}^2 + 15 \vec{v}_{1}\cdot\vec{n} \vec{v}_{2}\cdot\vec{n} - 9 ( \vec{v}_{1}\cdot\vec{n})^{2} \nn\\ 
&& - 6 ( \vec{v}_{2}\cdot\vec{n})^{2} \Big] - 	\frac{G^3 m_{1} m_{2}{}^3}{3 r{}^3} {\Big( \frac{1}{\epsilon} - 3\log \frac{r}{R_0} \Big)} \Big[ 2 v_{1}^2 - 5 \vec{v}_{1}\cdot\vec{v}_{2} + 3 v_{2}^2 + 15 \vec{v}_{1}\cdot\vec{n} \vec{v}_{2}\cdot\vec{n} \nn\\ 
&& - 6 ( \vec{v}_{1}\cdot\vec{n})^{2} - 9 ( \vec{v}_{2}\cdot\vec{n})^{2} \Big] - 	\frac{G^3 m_{1}{}^3 m_{2}}{36 r{}^3} \Big[ 127 v_{1}^2 - 241 \vec{v}_{1}\cdot\vec{v}_{2} + 105 v_{2}^2 \nn\\ 
&& + 471 \vec{v}_{1}\cdot\vec{n} \vec{v}_{2}\cdot\vec{n} - 300 ( \vec{v}_{1}\cdot\vec{n})^{2} - 198 ( \vec{v}_{2}\cdot\vec{n})^{2} \Big] - 	\frac{G^3 m_{1} m_{2}{}^3}{36 r{}^3} \Big[ 105 v_{1}^2 - 241 \vec{v}_{1}\cdot\vec{v}_{2} \nn\\ 
&& + 127 v_{2}^2 + 471 \vec{v}_{1}\cdot\vec{n} \vec{v}_{2}\cdot\vec{n} - 198 ( \vec{v}_{1}\cdot\vec{n})^{2} - 300 ( \vec{v}_{2}\cdot\vec{n})^{2} \Big] \nn\\ 
&& - 	\frac{G^3 m_{1}{}^2 m_{2}{}^2}{576 r{}^3} \Big[ {(11200 - 783 \pi^2)} v_{1}^2 - {(23264 - 1566 \pi^2)} \vec{v}_{1}\cdot\vec{v}_{2} \nn\\ 
&& + {(11200 - 783 \pi^2)} v_{2}^2 + {(69216 - 4698 \pi^2)} \vec{v}_{1}\cdot\vec{n} \vec{v}_{2}\cdot\vec{n} - {(35904 - 2349 \pi^2)} ( \vec{v}_{1}\cdot\vec{n})^{2} \nn\\ 
&& - {(35904 - 2349 \pi^2)} ( \vec{v}_{2}\cdot\vec{n})^{2} \Big]\nn\\ && +  	\frac{3 G^4 m_{1}{}^4 m_{2}}{8 r{}^4} + 	\frac{6 G^4 m_{1}{}^3 m_{2}{}^2}{r{}^4} + 	\frac{6 G^4 m_{1}{}^2 m_{2}{}^3}{r{}^4} + 	\frac{3 G^4 m_{1} m_{2}{}^4}{8 r{}^4},
\eea
where the kinetic term is included and:
\bea
\stackrel{(1)}{V}_{3,0}&=& - 	\frac{1}{16} G m_{1} m_{2} \Big[ 28 v_{1}^2 \vec{v}_{1}\cdot\vec{a}_{1} \vec{v}_{2}\cdot\vec{n} - 8 \vec{v}_{1}\cdot\vec{a}_{1} \vec{v}_{2}\cdot\vec{n} \vec{v}_{1}\cdot\vec{v}_{2} - 10 v_{1}^2 \vec{v}_{2}\cdot\vec{n} \vec{a}_{1}\cdot\vec{v}_{2} \nn\\ 
&& - 10 \vec{v}_{2}\cdot\vec{n} \vec{v}_{1}\cdot\vec{v}_{2} \vec{a}_{1}\cdot\vec{v}_{2} + 3 v_{1}^2 \vec{a}_{1}\cdot\vec{n} v_{2}^2 + 12 \vec{v}_{1}\cdot\vec{n} \vec{v}_{1}\cdot\vec{a}_{1} v_{2}^2 + 7 \vec{v}_{1}\cdot\vec{a}_{1} \vec{v}_{2}\cdot\vec{n} v_{2}^2 \nn\\ 
&& - 5 \vec{a}_{1}\cdot\vec{n} \vec{v}_{1}\cdot\vec{v}_{2} v_{2}^2 - 13 \vec{v}_{1}\cdot\vec{n} \vec{a}_{1}\cdot\vec{v}_{2} v_{2}^2 - 10 \vec{v}_{2}\cdot\vec{n} \vec{a}_{1}\cdot\vec{v}_{2} v_{2}^2 + 5 v_{1}^2 \vec{v}_{1}\cdot\vec{v}_{2} \vec{a}_{2}\cdot\vec{n} \nn\\ 
&& - 3 v_{1}^2 v_{2}^2 \vec{a}_{2}\cdot\vec{n} + 10 \vec{v}_{1}\cdot\vec{n} v_{1}^2 \vec{v}_{1}\cdot\vec{a}_{2} + 13 v_{1}^2 \vec{v}_{2}\cdot\vec{n} \vec{v}_{1}\cdot\vec{a}_{2} + 10 \vec{v}_{1}\cdot\vec{n} \vec{v}_{1}\cdot\vec{v}_{2} \vec{v}_{1}\cdot\vec{a}_{2} \nn\\ 
&& + 10 \vec{v}_{1}\cdot\vec{n} v_{2}^2 \vec{v}_{1}\cdot\vec{a}_{2} - 7 \vec{v}_{1}\cdot\vec{n} v_{1}^2 \vec{v}_{2}\cdot\vec{a}_{2} - 12 v_{1}^2 \vec{v}_{2}\cdot\vec{n} \vec{v}_{2}\cdot\vec{a}_{2} + 8 \vec{v}_{1}\cdot\vec{n} \vec{v}_{1}\cdot\vec{v}_{2} \vec{v}_{2}\cdot\vec{a}_{2} \nn\\ 
&& - 28 \vec{v}_{1}\cdot\vec{n} v_{2}^2 \vec{v}_{2}\cdot\vec{a}_{2} - 3 \vec{a}_{2}\cdot\vec{n} v_{1}^{4} + 3 \vec{a}_{1}\cdot\vec{n} v_{2}^{4} -3 \vec{v}_{1}\cdot\vec{n} \vec{a}_{1}\cdot\vec{n} \vec{v}_{2}\cdot\vec{n} v_{2}^2 \nn\\ 
&& + 3 \vec{v}_{1}\cdot\vec{n} v_{1}^2 \vec{v}_{2}\cdot\vec{n} \vec{a}_{2}\cdot\vec{n} + 3 v_{1}^2 \vec{a}_{2}\cdot\vec{n} ( \vec{v}_{1}\cdot\vec{n})^{2} - 5 \vec{v}_{1}\cdot\vec{v}_{2} \vec{a}_{2}\cdot\vec{n} ( \vec{v}_{1}\cdot\vec{n})^{2} \nn\\ 
&& + 3 v_{2}^2 \vec{a}_{2}\cdot\vec{n} ( \vec{v}_{1}\cdot\vec{n})^{2} - 13 \vec{v}_{2}\cdot\vec{n} \vec{v}_{1}\cdot\vec{a}_{2} ( \vec{v}_{1}\cdot\vec{n})^{2} + \vec{v}_{2}\cdot\vec{a}_{2} ( \vec{v}_{1}\cdot\vec{n})^{3} \nn\\ 
&& + 12 \vec{v}_{2}\cdot\vec{n} \vec{v}_{2}\cdot\vec{a}_{2} ( \vec{v}_{1}\cdot\vec{n})^{2} - 3 v_{1}^2 \vec{a}_{1}\cdot\vec{n} ( \vec{v}_{2}\cdot\vec{n})^{2} - 12 \vec{v}_{1}\cdot\vec{n} \vec{v}_{1}\cdot\vec{a}_{1} ( \vec{v}_{2}\cdot\vec{n})^{2} \nn\\ 
&& - \vec{v}_{1}\cdot\vec{a}_{1} ( \vec{v}_{2}\cdot\vec{n})^{3} + 5 \vec{a}_{1}\cdot\vec{n} \vec{v}_{1}\cdot\vec{v}_{2} ( \vec{v}_{2}\cdot\vec{n})^{2} + 13 \vec{v}_{1}\cdot\vec{n} \vec{a}_{1}\cdot\vec{v}_{2} ( \vec{v}_{2}\cdot\vec{n})^{2} \nn\\ 
&& - 3 \vec{a}_{1}\cdot\vec{n} v_{2}^2 ( \vec{v}_{2}\cdot\vec{n})^{2} -3 \vec{v}_{2}\cdot\vec{n} \vec{a}_{2}\cdot\vec{n} ( \vec{v}_{1}\cdot\vec{n})^{3} + 3 \vec{v}_{1}\cdot\vec{n} \vec{a}_{1}\cdot\vec{n} ( \vec{v}_{2}\cdot\vec{n})^{3} \Big]\nn\\ && +  	\frac{G^2 m_{1}{}^2 m_{2}}{6 r} \Big[ 32 v_{1}^2 \vec{a}_{1}\cdot\vec{n} + 124 \vec{v}_{1}\cdot\vec{n} \vec{v}_{1}\cdot\vec{a}_{1} - 24 \vec{v}_{1}\cdot\vec{a}_{1} \vec{v}_{2}\cdot\vec{n} - 21 \vec{a}_{1}\cdot\vec{n} \vec{v}_{1}\cdot\vec{v}_{2} \nn\\ 
&& - 81 \vec{v}_{1}\cdot\vec{n} \vec{a}_{1}\cdot\vec{v}_{2} + 29 \vec{v}_{2}\cdot\vec{n} \vec{a}_{1}\cdot\vec{v}_{2} - 11 \vec{a}_{1}\cdot\vec{n} v_{2}^2 + 5 v_{1}^2 \vec{a}_{2}\cdot\vec{n} - 5 \vec{v}_{1}\cdot\vec{v}_{2} \vec{a}_{2}\cdot\vec{n} \nn\\ 
&& - 68 \vec{v}_{1}\cdot\vec{n} \vec{v}_{1}\cdot\vec{a}_{2} - 5 \vec{v}_{2}\cdot\vec{n} \vec{v}_{1}\cdot\vec{a}_{2} + 25 \vec{v}_{1}\cdot\vec{n} \vec{v}_{2}\cdot\vec{a}_{2} -42 \vec{v}_{1}\cdot\vec{n} \vec{a}_{1}\cdot\vec{n} \vec{v}_{2}\cdot\vec{n} \nn\\ 
&& + 4 \vec{v}_{1}\cdot\vec{n} \vec{v}_{2}\cdot\vec{n} \vec{a}_{2}\cdot\vec{n} + 20 \vec{a}_{1}\cdot\vec{n} ( \vec{v}_{1}\cdot\vec{n})^{2} - \vec{a}_{2}\cdot\vec{n} ( \vec{v}_{1}\cdot\vec{n})^{2} + 22 \vec{a}_{1}\cdot\vec{n} ( \vec{v}_{2}\cdot\vec{n})^{2} \Big] \nn\\ 
&& - 	\frac{G^2 m_{1} m_{2}{}^2}{6 r} \Big[ 25 \vec{v}_{1}\cdot\vec{a}_{1} \vec{v}_{2}\cdot\vec{n} - 5 \vec{a}_{1}\cdot\vec{n} \vec{v}_{1}\cdot\vec{v}_{2} - 5 \vec{v}_{1}\cdot\vec{n} \vec{a}_{1}\cdot\vec{v}_{2} - 68 \vec{v}_{2}\cdot\vec{n} \vec{a}_{1}\cdot\vec{v}_{2} \nn\\ 
&& + 5 \vec{a}_{1}\cdot\vec{n} v_{2}^2 - 11 v_{1}^2 \vec{a}_{2}\cdot\vec{n} - 21 \vec{v}_{1}\cdot\vec{v}_{2} \vec{a}_{2}\cdot\vec{n} + 32 v_{2}^2 \vec{a}_{2}\cdot\vec{n} + 29 \vec{v}_{1}\cdot\vec{n} \vec{v}_{1}\cdot\vec{a}_{2} \nn\\ 
&& - 81 \vec{v}_{2}\cdot\vec{n} \vec{v}_{1}\cdot\vec{a}_{2} - 24 \vec{v}_{1}\cdot\vec{n} \vec{v}_{2}\cdot\vec{a}_{2} + 124 \vec{v}_{2}\cdot\vec{n} \vec{v}_{2}\cdot\vec{a}_{2} + 4 \vec{v}_{1}\cdot\vec{n} \vec{a}_{1}\cdot\vec{n} \vec{v}_{2}\cdot\vec{n} \nn\\ 
&& - 42 \vec{v}_{1}\cdot\vec{n} \vec{v}_{2}\cdot\vec{n} \vec{a}_{2}\cdot\vec{n} + 22 \vec{a}_{2}\cdot\vec{n} ( \vec{v}_{1}\cdot\vec{n})^{2} - \vec{a}_{1}\cdot\vec{n} ( \vec{v}_{2}\cdot\vec{n})^{2} + 20 \vec{a}_{2}\cdot\vec{n} ( \vec{v}_{2}\cdot\vec{n})^{2} \Big]\nn\\ && - 	\frac{5 G^3 m_{1} m_{2}{}^3}{3 r{}^2} \Big[ \vec{a}_{1}\cdot\vec{n} - 7 \vec{a}_{2}\cdot\vec{n} \Big] - 	\frac{2 G^3 m_{1} m_{2}{}^3}{3 r{}^2} {\Big( \frac{1}{\epsilon} - 3\log \frac{r}{R_0} \Big)} \Big[ \vec{a}_{1}\cdot\vec{n} - 7 \vec{a}_{2}\cdot\vec{n} \Big] \nn\\ 
&& - 	\frac{5 G^3 m_{1}{}^3 m_{2}}{3 r{}^2} \Big[ 7 \vec{a}_{1}\cdot\vec{n} - \vec{a}_{2}\cdot\vec{n} \Big] - 	\frac{2 G^3 m_{1}{}^3 m_{2}}{3 r{}^2} {\Big( \frac{1}{\epsilon} - 3\log \frac{r}{R_0} \Big)} \Big[ 7 \vec{a}_{1}\cdot\vec{n} - \vec{a}_{2}\cdot\vec{n} \Big] \nn\\ 
&& - 	\frac{2 G^3 m_{1}{}^2 m_{2}{}^2}{r{}^2} \Big[ 16 \vec{a}_{1}\cdot\vec{n} - \pi^2 \vec{a}_{1}\cdot\vec{n} - {(16 - \pi^2)} \vec{a}_{2}\cdot\vec{n} \Big],
\eea
\bea
\stackrel{(2)}{V}_{3,0}&=& - 	\frac{1}{48} G m_{1} m_{2} r \Big[ \big( 18 \vec{v}_{1}\cdot\dot{\vec{a}}_{1} v_{2}^2 - 21 \dot{\vec{a}}_{1}\cdot\vec{v}_{2} v_{2}^2 - 21 v_{1}^2 \vec{v}_{1}\cdot\dot{\vec{a}}_{2} + 18 v_{1}^2 \vec{v}_{2}\cdot\dot{\vec{a}}_{2} \nn\\ 
&& + 3 \dot{\vec{a}}_{1}\cdot\vec{n} \vec{v}_{2}\cdot\vec{n} v_{2}^2 + 3 \vec{v}_{1}\cdot\vec{n} v_{1}^2 \dot{\vec{a}}_{2}\cdot\vec{n} - 21 \vec{v}_{1}\cdot\dot{\vec{a}}_{2} ( \vec{v}_{1}\cdot\vec{n})^{2} + 18 \vec{v}_{2}\cdot\dot{\vec{a}}_{2} ( \vec{v}_{1}\cdot\vec{n})^{2} \nn\\ 
&& + 18 \vec{v}_{1}\cdot\dot{\vec{a}}_{1} ( \vec{v}_{2}\cdot\vec{n})^{2} - 21 \dot{\vec{a}}_{1}\cdot\vec{v}_{2} ( \vec{v}_{2}\cdot\vec{n})^{2} - \dot{\vec{a}}_{2}\cdot\vec{n} ( \vec{v}_{1}\cdot\vec{n})^{3} - \dot{\vec{a}}_{1}\cdot\vec{n} ( \vec{v}_{2}\cdot\vec{n})^{3} \big) \nn\\ 
&& + 3 \big( 6 a_{1}^2 v_{2}^2 + 20 \vec{v}_{1}\cdot\vec{a}_{1} \vec{v}_{1}\cdot\vec{a}_{2} - 3 \vec{a}_{1}\cdot\vec{v}_{2} \vec{v}_{1}\cdot\vec{a}_{2} + 13 v_{1}^2 \vec{a}_{1}\cdot\vec{a}_{2} + 5 \vec{v}_{1}\cdot\vec{v}_{2} \vec{a}_{1}\cdot\vec{a}_{2} \nn\\ 
&& + 13 v_{2}^2 \vec{a}_{1}\cdot\vec{a}_{2} - 11 \vec{v}_{1}\cdot\vec{a}_{1} \vec{v}_{2}\cdot\vec{a}_{2} + 20 \vec{a}_{1}\cdot\vec{v}_{2} \vec{v}_{2}\cdot\vec{a}_{2} + 6 v_{1}^2 a_{2}^2 -3 v_{1}^2 \vec{a}_{1}\cdot\vec{n} \vec{a}_{2}\cdot\vec{n} \nn\\ 
&& - 12 \vec{v}_{1}\cdot\vec{n} \vec{v}_{1}\cdot\vec{a}_{1} \vec{a}_{2}\cdot\vec{n} - 3 \vec{v}_{1}\cdot\vec{a}_{1} \vec{v}_{2}\cdot\vec{n} \vec{a}_{2}\cdot\vec{n} + 5 \vec{a}_{1}\cdot\vec{n} \vec{v}_{1}\cdot\vec{v}_{2} \vec{a}_{2}\cdot\vec{n} \nn\\ 
&& + 13 \vec{v}_{1}\cdot\vec{n} \vec{a}_{1}\cdot\vec{v}_{2} \vec{a}_{2}\cdot\vec{n} - 3 \vec{a}_{1}\cdot\vec{n} v_{2}^2 \vec{a}_{2}\cdot\vec{n} + 13 \vec{a}_{1}\cdot\vec{n} \vec{v}_{2}\cdot\vec{n} \vec{v}_{1}\cdot\vec{a}_{2} \nn\\ 
&& + 29 \vec{v}_{1}\cdot\vec{n} \vec{v}_{2}\cdot\vec{n} \vec{a}_{1}\cdot\vec{a}_{2} - 3 \vec{v}_{1}\cdot\vec{n} \vec{a}_{1}\cdot\vec{n} \vec{v}_{2}\cdot\vec{a}_{2} - 12 \vec{a}_{1}\cdot\vec{n} \vec{v}_{2}\cdot\vec{n} \vec{v}_{2}\cdot\vec{a}_{2} + 6 a_{2}^2 ( \vec{v}_{1}\cdot\vec{n})^{2} \nn\\ 
&& + 6 a_{1}^2 ( \vec{v}_{2}\cdot\vec{n})^{2} + 3 \vec{v}_{1}\cdot\vec{n} \vec{a}_{1}\cdot\vec{n} \vec{v}_{2}\cdot\vec{n} \vec{a}_{2}\cdot\vec{n} \big) \Big]\nn\\ && +  	\frac{1}{72} G^2 m_{1}{}^2 m_{2} \Big[ 484 a_{1}^2 + 347 \vec{a}_{1}\cdot\vec{a}_{2} + 132 \vec{a}_{1}\cdot\vec{n} \vec{a}_{2}\cdot\vec{n} \Big] + 	\frac{1}{72} G^2 m_{1} m_{2}{}^2 \Big[ 347 \vec{a}_{1}\cdot\vec{a}_{2} \nn\\ 
&& + 484 a_{2}^2 + 132 \vec{a}_{1}\cdot\vec{n} \vec{a}_{2}\cdot\vec{n} \Big] + 	\frac{11}{3} G^2 m_{1}{}^2 m_{2} {\Big( \frac{1}{\epsilon} - 2\log \frac{r}{R_0} \Big)} \Big[ a_{1}^2 + 2 \vec{a}_{1}\cdot\vec{a}_{2} \Big] \nn\\ 
&& + 	\frac{11}{3} G^2 m_{1} m_{2}{}^2 {\Big( \frac{1}{\epsilon} - 2\log \frac{r}{R_0} \Big)} \Big[ 2 \vec{a}_{1}\cdot\vec{a}_{2} + a_{2}^2 \Big] - 	\frac{1}{3} G^2 m_{1}{}^2 m_{2} ( \vec{a}_{1}\cdot\vec{n})^{2} \nn\\ 
&& - 	\frac{1}{3} G^2 m_{1} m_{2}{}^2 ( \vec{a}_{2}\cdot\vec{n})^{2},
\eea
\bea
\stackrel{(3)}{V}_{3,0}&=&\frac{1}{16} G m_{1} m_{2} r{}^2 \Big[ \big( 6 \vec{v}_{1}\cdot\dot{\vec{a}}_{1} \vec{a}_{2}\cdot\vec{n} - 7 \dot{\vec{a}}_{1}\cdot\vec{v}_{2} \vec{a}_{2}\cdot\vec{n} - 15 \vec{v}_{2}\cdot\vec{n} \dot{\vec{a}}_{1}\cdot\vec{a}_{2} + \dot{\vec{a}}_{1}\cdot\vec{n} \vec{v}_{2}\cdot\vec{a}_{2} \nn\\ 
&& - \vec{v}_{1}\cdot\vec{a}_{1} \dot{\vec{a}}_{2}\cdot\vec{n} + 7 \vec{a}_{1}\cdot\vec{n} \vec{v}_{1}\cdot\dot{\vec{a}}_{2} + 15 \vec{v}_{1}\cdot\vec{n} \vec{a}_{1}\cdot\dot{\vec{a}}_{2} - 6 \vec{a}_{1}\cdot\vec{n} \vec{v}_{2}\cdot\dot{\vec{a}}_{2} + \dot{\vec{a}}_{1}\cdot\vec{n} \vec{v}_{2}\cdot\vec{n} \vec{a}_{2}\cdot\vec{n} \nn\\ 
&& - \vec{v}_{1}\cdot\vec{n} \vec{a}_{1}\cdot\vec{n} \dot{\vec{a}}_{2}\cdot\vec{n} \big) + 6 \big( a_{1}^2 \vec{a}_{2}\cdot\vec{n} - \vec{a}_{1}\cdot\vec{n} a_{2}^2 \big) \Big],
\eea
\bea
\stackrel{(4)}{V}_{3,0}&=& 	\frac{1}{144} G m_{1} m_{2} r{}^3 \Big[ 23 \dot{\vec{a}}_{1}\cdot\dot{\vec{a}}_{2} -3 \dot{\vec{a}}_{1}\cdot\vec{n} \dot{\vec{a}}_{2}\cdot\vec{n} \Big].
\eea

The redefinition of position that we fix for the $3$PN sector can be written as:
\bea
\left(\Delta \vec{x}_1\right)_{\text{3PN}} &=& \sum_{i=0}^3  \stackrel{(i)}{ \Delta \vec{x}_1}_{(3,0)},
\eea
with:
\bea
\stackrel{(0)}{ \Delta \vec{x}_1}_{(3,0)} &=&  	\frac{1}{48} G m_{2} \Big[ 6 v_{1}^2 v_{2}^2 \vec{n} - 18 \vec{v}_{1}\cdot\vec{v}_{2} v_{2}^2 \vec{n} + 12 v_{2}^{4} \vec{n} - 24 v_{1}^2 \vec{v}_{2}\cdot\vec{n} \vec{v}_{1} + 24 \vec{v}_{2}\cdot\vec{n} \vec{v}_{1}\cdot\vec{v}_{2} \vec{v}_{1} \nn\\ 
&& + 12 \vec{v}_{1}\cdot\vec{n} v_{2}^2 \vec{v}_{1} + 72 \vec{v}_{2}\cdot\vec{n} v_{2}^2 \vec{v}_{1} + 12 v_{1}^2 \vec{v}_{2}\cdot\vec{n} \vec{v}_{2} + 12 \vec{v}_{2}\cdot\vec{n} \vec{v}_{1}\cdot\vec{v}_{2} \vec{v}_{2} - 18 \vec{v}_{1}\cdot\vec{n} v_{2}^2 \vec{v}_{2} \nn\\ 
&& - 90 \vec{v}_{2}\cdot\vec{n} v_{2}^2 \vec{v}_{2} -6 \vec{v}_{1}\cdot\vec{n} \vec{v}_{2}\cdot\vec{n} v_{2}^2 \vec{n} - 6 v_{1}^2 ( \vec{v}_{2}\cdot\vec{n})^{2} \vec{n} + 18 \vec{v}_{1}\cdot\vec{v}_{2} ( \vec{v}_{2}\cdot\vec{n})^{2} \vec{n} \nn\\ 
&& - 15 v_{2}^2 ( \vec{v}_{2}\cdot\vec{n})^{2} \vec{n} - 12 \vec{v}_{1}\cdot\vec{n} ( \vec{v}_{2}\cdot\vec{n})^{2} \vec{v}_{1} - 20 ( \vec{v}_{2}\cdot\vec{n})^{3} \vec{v}_{1} + 18 \vec{v}_{1}\cdot\vec{n} ( \vec{v}_{2}\cdot\vec{n})^{2} \vec{v}_{2} \nn\\ 
&& + 20 ( \vec{v}_{2}\cdot\vec{n})^{3} \vec{v}_{2} + 6 \vec{v}_{1}\cdot\vec{n} ( \vec{v}_{2}\cdot\vec{n})^{3} \vec{n} + 3 ( \vec{v}_{2}\cdot\vec{n})^{4} \vec{n} \Big]\nn\\ && - 	\frac{G^2 m_{1} m_{2}}{96 r} \Big[ 488 v_{1}^2 \vec{n} - 150 \vec{v}_{1}\cdot\vec{v}_{2} \vec{n} - 254 v_{2}^2 \vec{n} + 1891 \vec{v}_{1}\cdot\vec{n} \vec{v}_{1} - 735 \vec{v}_{2}\cdot\vec{n} \vec{v}_{1} \nn\\ 
&& - 1203 \vec{v}_{1}\cdot\vec{n} \vec{v}_{2} + 740 \vec{v}_{2}\cdot\vec{n} \vec{v}_{2} -108 \vec{v}_{1}\cdot\vec{n} \vec{v}_{2}\cdot\vec{n} \vec{n} - 130 ( \vec{v}_{1}\cdot\vec{n})^{2} \vec{n} + 217 ( \vec{v}_{2}\cdot\vec{n})^{2} \vec{n} \Big] \nn\\ 
&& - 	\frac{G^2 m_{2}{}^2}{12 r} \Big[ 10 \vec{v}_{1}\cdot\vec{v}_{2} \vec{n} - v_{2}^2 \vec{n} + 4 \vec{v}_{2}\cdot\vec{n} \vec{v}_{1} + 10 \vec{v}_{1}\cdot\vec{n} \vec{v}_{2} + 73 \vec{v}_{2}\cdot\vec{n} \vec{v}_{2} -8 \vec{v}_{1}\cdot\vec{n} \vec{v}_{2}\cdot\vec{n} \vec{n} \nn\\ 
&& + 2 ( \vec{v}_{2}\cdot\vec{n})^{2} \vec{n} \Big]\nn\\ && +  	\frac{7 G^3 m_{1}{}^2 m_{2}}{9 r{}^2} \vec{n} + 	\frac{5 G^3 m_{2}{}^3}{3 r{}^2} \vec{n} + 	\frac{G^3 m_{1} m_{2}{}^2}{2 r{}^2} {(47 - 4 \pi^2)} \vec{n},
\eea
\bea
\stackrel{(1)}{ \Delta \vec{x}_1}_{(3,0)} &=&\frac{1}{64} G m_{2} r \Big[ 34 \vec{v}_{1}\cdot\vec{a}_{2} \vec{v}_{1} - 90 \vec{v}_{2}\cdot\vec{a}_{2} \vec{v}_{1} - 34 \vec{v}_{1}\cdot\vec{a}_{2} \vec{v}_{2} + 90 \vec{v}_{2}\cdot\vec{a}_{2} \vec{v}_{2} + 11 v_{1}^2 \vec{a}_{2} \nn\\ 
&& - 50 \vec{v}_{1}\cdot\vec{v}_{2} \vec{a}_{2} + 99 v_{2}^2 \vec{a}_{2} -5 v_{1}^2 \vec{a}_{2}\cdot\vec{n} \vec{n} + 14 \vec{v}_{1}\cdot\vec{v}_{2} \vec{a}_{2}\cdot\vec{n} \vec{n} - 13 v_{2}^2 \vec{a}_{2}\cdot\vec{n} \vec{n} \nn\\ 
&& + 30 \vec{v}_{2}\cdot\vec{n} \vec{v}_{1}\cdot\vec{a}_{2} \vec{n} - 2 \vec{v}_{1}\cdot\vec{n} \vec{v}_{2}\cdot\vec{a}_{2} \vec{n} - 38 \vec{v}_{2}\cdot\vec{n} \vec{v}_{2}\cdot\vec{a}_{2} \vec{n} + 2 \vec{v}_{1}\cdot\vec{n} \vec{a}_{2}\cdot\vec{n} \vec{v}_{1} \nn\\ 
&& - 74 \vec{v}_{2}\cdot\vec{n} \vec{a}_{2}\cdot\vec{n} \vec{v}_{1} - 2 \vec{v}_{1}\cdot\vec{n} \vec{a}_{2}\cdot\vec{n} \vec{v}_{2} + 80 \vec{v}_{2}\cdot\vec{n} \vec{a}_{2}\cdot\vec{n} \vec{v}_{2} - 2 \vec{v}_{1}\cdot\vec{n} \vec{v}_{2}\cdot\vec{n} \vec{a}_{2} \nn\\ 
&& + 88 ( \vec{v}_{2}\cdot\vec{n})^{2} \vec{a}_{2} + 2 \vec{v}_{1}\cdot\vec{n} \vec{v}_{2}\cdot\vec{n} \vec{a}_{2}\cdot\vec{n} \vec{n} + 8 \vec{a}_{2}\cdot\vec{n} ( \vec{v}_{2}\cdot\vec{n})^{2} \vec{n} \Big]\nn\\ && - 	\frac{1}{288} G^2 m_{2}{}^2 \Big[ 1099 \vec{a}_{2} + 201 \vec{a}_{2}\cdot\vec{n} \vec{n} \Big] - 	\frac{1}{144} G^2 m_{1} m_{2} \Big[ 502 \vec{a}_{1} + 561 \vec{a}_{2} -48 \vec{a}_{1}\cdot\vec{n} \vec{n} \nn\\ 
&& + 69 \vec{a}_{2}\cdot\vec{n} \vec{n} \Big] - 	\frac{11}{3} G^2 m_{2}{}^2 {\Big( \frac{1}{\epsilon} - 2\log \frac{r}{R_0} \Big)} \vec{a}_{2} - 	\frac{11}{3} G^2 m_{1} m_{2} {\Big( \frac{1}{\epsilon} - 2\log \frac{r}{R_0} \Big)} \Big[ \vec{a}_{1} \nn\\ 
&& + \vec{a}_{2} \Big],
\eea
\bea
\stackrel{(2)}{ \Delta \vec{x}_1}_{(3,0)} &=& - 	\frac{1}{96} G m_{2} r{}^2 \Big[ \big( \vec{v}_{1}\cdot\dot{\vec{a}}_{2} \vec{n} - 7 \vec{v}_{2}\cdot\dot{\vec{a}}_{2} \vec{n} - 35 \dot{\vec{a}}_{2}\cdot\vec{n} \vec{v}_{1} + 41 \dot{\vec{a}}_{2}\cdot\vec{n} \vec{v}_{2} - 23 \vec{v}_{1}\cdot\vec{n} \dot{\vec{a}}_{2} \nn\\ 
&& + 113 \vec{v}_{2}\cdot\vec{n} \dot{\vec{a}}_{2} + \vec{v}_{1}\cdot\vec{n} \dot{\vec{a}}_{2}\cdot\vec{n} \vec{n} - 7 \vec{v}_{2}\cdot\vec{n} \dot{\vec{a}}_{2}\cdot\vec{n} \vec{n} \big) - 6 \big( 3 a_{2}^2 \vec{n} + 2 \vec{a}_{2}\cdot\vec{n} \vec{a}_{1} \nn\\ 
&& - 22 \vec{a}_{2}\cdot\vec{n} \vec{a}_{2} \big) \Big] + 	\frac{1}{16} G m_{2} r{}^2 \vec{n} ( \vec{a}_{2}\cdot\vec{n})^{2},
\eea
\bea
\stackrel{(3)}{ \Delta \vec{x}_1}_{(3,0)} &=& 	\frac{1}{288} G m_{2} r{}^3 \Big[ 23 \ddot{\vec{a}}_{2} -3 \ddot{\vec{a}}_{2}\cdot\vec{n} \vec{n} \Big].
\eea

We proceed to consider the N$^2$LO spin-orbit sector at the $3.5$PN order. Our unreduced potential can 
be expressed as:
\bea
V_{\text{N}^2\text{LO}}^{\text{SO}} = \sum_{i=0}^4 \stackrel{(i)}{V}_{2,1} + (1 \leftrightarrow 2),
\eea
with:
\bea
\stackrel{(0)}{V}_{2,1}&=&- 	\frac{G m_{2}}{4 r{}^2} \Big[ \vec{S}_{1}\times\vec{n}\cdot\vec{v}_{1} \big( 5 v_{1}^2 \vec{v}_{1}\cdot\vec{v}_{2} - 3 v_{1}^2 v_{2}^2 + 4 \vec{v}_{1}\cdot\vec{v}_{2} v_{2}^2 - 3 v_{1}^{4} - 3 v_{2}^{4} + 6 \vec{v}_{1}\cdot\vec{n} v_{1}^2 \vec{v}_{2}\cdot\vec{n} \nn\\ 
&& - 6 \vec{v}_{1}\cdot\vec{n} \vec{v}_{2}\cdot\vec{n} \vec{v}_{1}\cdot\vec{v}_{2} + 6 \vec{v}_{1}\cdot\vec{n} \vec{v}_{2}\cdot\vec{n} v_{2}^2 + 3 v_{2}^2 ( \vec{v}_{1}\cdot\vec{n})^{2} \nn\\ 
&& + 3 v_{1}^2 ( \vec{v}_{2}\cdot\vec{n})^{2} -15 ( \vec{v}_{1}\cdot\vec{n})^{2} ( \vec{v}_{2}\cdot\vec{n})^{2} \big) + \vec{S}_{1}\times\vec{n}\cdot\vec{v}_{2} \big( v_{1}^2 v_{2}^2 - 2 \vec{v}_{1}\cdot\vec{v}_{2} v_{2}^2 - 2 ( \vec{v}_{1}\cdot\vec{v}_{2})^{2} \nn\\ 
&& + 3 v_{2}^{4} -6 \vec{v}_{1}\cdot\vec{n} \vec{v}_{2}\cdot\vec{n} v_{2}^2 - 3 v_{2}^2 ( \vec{v}_{1}\cdot\vec{n})^{2} - 3 v_{1}^2 ( \vec{v}_{2}\cdot\vec{n})^{2} + 15 ( \vec{v}_{1}\cdot\vec{n})^{2} ( \vec{v}_{2}\cdot\vec{n})^{2} \big) \nn\\ 
&& + \vec{S}_{1}\times\vec{v}_{1}\cdot\vec{v}_{2} \big( \vec{v}_{1}\cdot\vec{n} v_{1}^2 - 2 \vec{v}_{1}\cdot\vec{n} \vec{v}_{1}\cdot\vec{v}_{2} + 2 \vec{v}_{1}\cdot\vec{n} v_{2}^2 + 2 \vec{v}_{2}\cdot\vec{n} v_{2}^2 -6 \vec{v}_{1}\cdot\vec{n} ( \vec{v}_{2}\cdot\vec{n})^{2} \big) \Big]\nn\\ && +  	\frac{G^2 m_{2}{}^2}{4 r{}^3} \Big[ \vec{S}_{1}\times\vec{n}\cdot\vec{v}_{1} \big( 13 v_{1}^2 - 41 \vec{v}_{1}\cdot\vec{v}_{2} + 28 v_{2}^2 -16 \vec{v}_{1}\cdot\vec{n} \vec{v}_{2}\cdot\vec{n} + 8 ( \vec{v}_{1}\cdot\vec{n})^{2} \nn\\ 
&& + 12 ( \vec{v}_{2}\cdot\vec{n})^{2} \big) - 2 \vec{S}_{1}\times\vec{n}\cdot\vec{v}_{2} \big( v_{1}^2 + 7 \vec{v}_{1}\cdot\vec{v}_{2} - 8 v_{2}^2 -56 \vec{v}_{1}\cdot\vec{n} \vec{v}_{2}\cdot\vec{n} - 4 ( \vec{v}_{1}\cdot\vec{n})^{2} \nn\\ 
&& + 62 ( \vec{v}_{2}\cdot\vec{n})^{2} \big) + \vec{S}_{1}\times\vec{v}_{1}\cdot\vec{v}_{2} \big( 7 \vec{v}_{1}\cdot\vec{n} - 38 \vec{v}_{2}\cdot\vec{n} \big) \Big] - 	\frac{2 G^2 m_{1} m_{2}}{r{}^3} \Big[ \vec{S}_{1}\times\vec{n}\cdot\vec{v}_{1} \big( 2 v_{1}^2 \nn\\ 
&& + \vec{v}_{1}\cdot\vec{v}_{2} - 3 v_{2}^2 + 12 \vec{v}_{1}\cdot\vec{n} \vec{v}_{2}\cdot\vec{n} - 16 ( \vec{v}_{1}\cdot\vec{n})^{2} + 4 ( \vec{v}_{2}\cdot\vec{n})^{2} \big) - 2 \vec{S}_{1}\times\vec{n}\cdot\vec{v}_{2} \big( \vec{v}_{1}\cdot\vec{v}_{2} \nn\\ 
&& - v_{2}^2 + 4 \vec{v}_{1}\cdot\vec{n} \vec{v}_{2}\cdot\vec{n} - 4 ( \vec{v}_{1}\cdot\vec{n})^{2} \big) - \vec{S}_{1}\times\vec{v}_{1}\cdot\vec{v}_{2} \big( 2 \vec{v}_{1}\cdot\vec{n} - 5 \vec{v}_{2}\cdot\vec{n} \big) \Big]\nn\\ && +  	\frac{G^3 m_{1}{}^2 m_{2}}{r{}^4} \Big[ \vec{S}_{1}\times\vec{n}\cdot\vec{v}_{1} - \vec{S}_{1}\times\vec{n}\cdot\vec{v}_{2} \Big] + 	\frac{5 G^3 m_{1} m_{2}{}^2}{r{}^4} \Big[ \vec{S}_{1}\times\vec{n}\cdot\vec{v}_{1} - \vec{S}_{1}\times\vec{n}\cdot\vec{v}_{2} \Big] \nn\\ 
&& + 	\frac{G^3 m_{2}{}^3}{r{}^4} \Big[ \vec{S}_{1}\times\vec{n}\cdot\vec{v}_{1} - \vec{S}_{1}\times\vec{n}\cdot\vec{v}_{2} \Big],
\eea
and we separate terms with one higher-order time derivative into:
\bea
\stackrel{(1)}{V}_{2,1}  = (V_a)_{2,1} + (V_{\dot{S}})_{2,1} ,
\eea
with:
\bea
(V_a)_{2,1} &=&- 	\frac{5}{16} \vec{S}_{1}\times\vec{v}_{1}\cdot\vec{a}_{1} v_{1}^{4} \nn\\ &&\frac{G m_{2}}{4 r} \Big[ \vec{S}_{1}\times\vec{n}\cdot\vec{v}_{1} \big( 4 \vec{v}_{1}\cdot\vec{a}_{1} \vec{v}_{2}\cdot\vec{n} - 4 \vec{v}_{2}\cdot\vec{n} \vec{a}_{1}\cdot\vec{v}_{2} + \vec{a}_{1}\cdot\vec{n} v_{2}^2 - v_{1}^2 \vec{a}_{2}\cdot\vec{n} \nn\\ 
&& + 4 \vec{v}_{1}\cdot\vec{n} \vec{v}_{1}\cdot\vec{a}_{2} - 4 \vec{v}_{1}\cdot\vec{n} \vec{v}_{2}\cdot\vec{a}_{2} + 3 \vec{a}_{2}\cdot\vec{n} ( \vec{v}_{1}\cdot\vec{n})^{2} - 3 \vec{a}_{1}\cdot\vec{n} ( \vec{v}_{2}\cdot\vec{n})^{2} \big) \nn\\ 
&& + 2 \vec{S}_{1}\times\vec{n}\cdot\vec{a}_{1} \big( v_{1}^2 \vec{v}_{2}\cdot\vec{n} - \vec{v}_{2}\cdot\vec{n} \vec{v}_{1}\cdot\vec{v}_{2} + \vec{v}_{1}\cdot\vec{n} v_{2}^2 + \vec{v}_{2}\cdot\vec{n} v_{2}^2 -3 \vec{v}_{1}\cdot\vec{n} ( \vec{v}_{2}\cdot\vec{n})^{2} \big) \nn\\ 
&& - \vec{S}_{1}\times\vec{v}_{1}\cdot\vec{a}_{1} \big( 12 v_{1}^2 - 12 \vec{v}_{1}\cdot\vec{v}_{2} + 5 v_{2}^2 -4 \vec{v}_{1}\cdot\vec{n} \vec{v}_{2}\cdot\vec{n} - ( \vec{v}_{2}\cdot\vec{n})^{2} \big) \nn\\ 
&& + \vec{S}_{1}\times\vec{n}\cdot\vec{v}_{2} \big( 2 \vec{v}_{2}\cdot\vec{n} \vec{a}_{1}\cdot\vec{v}_{2} - \vec{a}_{1}\cdot\vec{n} v_{2}^2 + v_{1}^2 \vec{a}_{2}\cdot\vec{n} - 2 \vec{v}_{1}\cdot\vec{n} \vec{v}_{1}\cdot\vec{a}_{2} \nn\\ 
&& + 4 \vec{v}_{1}\cdot\vec{n} \vec{v}_{2}\cdot\vec{a}_{2} -3 \vec{a}_{2}\cdot\vec{n} ( \vec{v}_{1}\cdot\vec{n})^{2} + 3 \vec{a}_{1}\cdot\vec{n} ( \vec{v}_{2}\cdot\vec{n})^{2} \big) - 2 \vec{S}_{1}\times\vec{v}_{1}\cdot\vec{v}_{2} \big( 6 \vec{v}_{1}\cdot\vec{a}_{1} \nn\\ 
&& - \vec{a}_{1}\cdot\vec{v}_{2} - \vec{v}_{1}\cdot\vec{a}_{2} + 2 \vec{v}_{2}\cdot\vec{a}_{2} - \vec{v}_{1}\cdot\vec{n} \vec{a}_{2}\cdot\vec{n} \big) - \vec{S}_{1}\times\vec{a}_{1}\cdot\vec{v}_{2} \big( 6 v_{1}^2 - 4 \vec{v}_{1}\cdot\vec{v}_{2} \nn\\ 
&& + 5 v_{2}^2 -6 \vec{v}_{1}\cdot\vec{n} \vec{v}_{2}\cdot\vec{n} + ( \vec{v}_{2}\cdot\vec{n})^{2} \big) + 2 \vec{S}_{1}\times\vec{n}\cdot\vec{a}_{2} \big( v_{1}^2 \vec{v}_{2}\cdot\vec{n} \nn\\ 
&& + \vec{v}_{1}\cdot\vec{n} v_{2}^2 -3 \vec{v}_{2}\cdot\vec{n} ( \vec{v}_{1}\cdot\vec{n})^{2} \big) + \vec{S}_{1}\times\vec{v}_{1}\cdot\vec{a}_{2} \big( v_{1}^2 - 2 v_{2}^2 + 4 \vec{v}_{1}\cdot\vec{n} \vec{v}_{2}\cdot\vec{n} - ( \vec{v}_{1}\cdot\vec{n})^{2} \big) \nn\\ 
&& - \vec{S}_{1}\times\vec{v}_{2}\cdot\vec{a}_{2} \big( v_{1}^2 - ( \vec{v}_{1}\cdot\vec{n})^{2} \big) \Big]\nn\\ && - 	\frac{2 G^2 m_{1} m_{2}}{r{}^2} \Big[ \vec{S}_{1}\times\vec{n}\cdot\vec{v}_{1} \big( 2 \vec{a}_{1}\cdot\vec{n} + \vec{a}_{2}\cdot\vec{n} \big) + \vec{S}_{1}\times\vec{n}\cdot\vec{a}_{1} \big( 3 \vec{v}_{1}\cdot\vec{n} + 2 \vec{v}_{2}\cdot\vec{n} \big) \nn\\ 
&& - 2 \vec{S}_{1}\times\vec{n}\cdot\vec{a}_{2} \vec{v}_{1}\cdot\vec{n} + 2 \vec{S}_{1}\times\vec{a}_{1}\cdot\vec{v}_{2} + 2 \vec{S}_{1}\times\vec{v}_{1}\cdot\vec{a}_{2} \Big] - 	\frac{G^2 m_{2}{}^2}{4 r{}^2} \Big[ 2 \vec{S}_{1}\times\vec{n}\cdot\vec{v}_{1} \vec{a}_{1}\cdot\vec{n} \nn\\ 
&& + 4 \vec{S}_{1}\times\vec{n}\cdot\vec{a}_{1} \big( \vec{v}_{1}\cdot\vec{n} - \vec{v}_{2}\cdot\vec{n} \big) + 2 \vec{S}_{1}\times\vec{n}\cdot\vec{v}_{2} \big( \vec{a}_{1}\cdot\vec{n} + 14 \vec{a}_{2}\cdot\vec{n} \big) \nn\\ 
&& - 2 \vec{S}_{1}\times\vec{n}\cdot\vec{a}_{2} \big( \vec{v}_{1}\cdot\vec{n} - 18 \vec{v}_{2}\cdot\vec{n} \big) + 14 \vec{S}_{1}\times\vec{v}_{1}\cdot\vec{a}_{1} + 27 \vec{S}_{1}\times\vec{a}_{1}\cdot\vec{v}_{2} + \vec{S}_{1}\times\vec{v}_{1}\cdot\vec{a}_{2} \nn\\ 
&& + 12 \vec{S}_{1}\times\vec{v}_{2}\cdot\vec{a}_{2} \Big],
\eea
and
\bea
(V_{\dot{S}})_{2,1}  &=&  	\frac{G m_{2}}{2 r} \Big[ \dot{\vec{S}}_{1}\times\vec{n}\cdot\vec{v}_{1} \big( v_{1}^2 \vec{v}_{2}\cdot\vec{n} - \vec{v}_{2}\cdot\vec{n} \vec{v}_{1}\cdot\vec{v}_{2} + \vec{v}_{1}\cdot\vec{n} v_{2}^2 + \vec{v}_{2}\cdot\vec{n} v_{2}^2 -3 \vec{v}_{1}\cdot\vec{n} ( \vec{v}_{2}\cdot\vec{n})^{2} \big) \nn\\ 
&& - \dot{\vec{S}}_{1}\times\vec{n}\cdot\vec{v}_{2} \big( \vec{v}_{1}\cdot\vec{n} v_{2}^2 + \vec{v}_{2}\cdot\vec{n} v_{2}^2 -3 \vec{v}_{1}\cdot\vec{n} ( \vec{v}_{2}\cdot\vec{n})^{2} \big) - \dot{\vec{S}}_{1}\times\vec{v}_{1}\cdot\vec{v}_{2} \big( 3 v_{1}^2 \nn\\ 
&& + 2 v_{2}^2 -3 \vec{v}_{1}\cdot\vec{n} \vec{v}_{2}\cdot\vec{n} + ( \vec{v}_{2}\cdot\vec{n})^{2} \big) \Big]\nn\\ && - 	\frac{G^2 m_{1} m_{2}}{r{}^2} \Big[ 6 \dot{\vec{S}}_{1}\times\vec{n}\cdot\vec{v}_{1} \big( 3 \vec{v}_{1}\cdot\vec{n} - \vec{v}_{2}\cdot\vec{n} \big) - 2 \dot{\vec{S}}_{1}\times\vec{n}\cdot\vec{v}_{2} \big( 7 \vec{v}_{1}\cdot\vec{n} - 5 \vec{v}_{2}\cdot\vec{n} \big) \nn\\ 
&& + 7 \dot{\vec{S}}_{1}\times\vec{v}_{1}\cdot\vec{v}_{2} \Big] - 	\frac{G^2 m_{2}{}^2}{4 r{}^2} \Big[ \dot{\vec{S}}_{1}\times\vec{n}\cdot\vec{v}_{1} \big( 3 \vec{v}_{1}\cdot\vec{n} - 4 \vec{v}_{2}\cdot\vec{n} \big) + 2 \dot{\vec{S}}_{1}\times\vec{n}\cdot\vec{v}_{2} \big( \vec{v}_{1}\cdot\vec{n} \nn\\ 
&& + 15 \vec{v}_{2}\cdot\vec{n} \big) + 27 \dot{\vec{S}}_{1}\times\vec{v}_{1}\cdot\vec{v}_{2} \Big],
\eea
and the pieces with further time derivatives:
\bea
\stackrel{(2)}{V}_{2,1}&=&- 	\frac{1}{4} G m_{2} \Big[ ( \vec{S}_{1}\times\vec{n}\cdot\dot{\vec{a}}_{1} \big( v_{2}^2 - ( \vec{v}_{2}\cdot\vec{n})^{2} \big) + 4 \vec{S}_{1}\times\vec{v}_{1}\cdot\dot{\vec{a}}_{1} \vec{v}_{2}\cdot\vec{n} + 6 \vec{S}_{1}\times\dot{\vec{a}}_{1}\cdot\vec{v}_{2} \vec{v}_{2}\cdot\vec{n} \nn\\ 
&& - \vec{S}_{1}\times\vec{n}\cdot\dot{\vec{a}}_{2} \big( v_{1}^2 - ( \vec{v}_{1}\cdot\vec{n})^{2} \big) - 2 \vec{S}_{1}\times\vec{v}_{1}\cdot\dot{\vec{a}}_{2} \vec{v}_{1}\cdot\vec{n}) + ( \ddot{\vec{S}}_{1}\times\vec{n}\cdot\vec{v}_{1} \big( v_{2}^2 - ( \vec{v}_{2}\cdot\vec{n})^{2} \big) \nn\\ 
&& - \ddot{\vec{S}}_{1}\times\vec{n}\cdot\vec{v}_{2} \big( v_{2}^2 - ( \vec{v}_{2}\cdot\vec{n})^{2} \big) + 6 \ddot{\vec{S}}_{1}\times\vec{v}_{1}\cdot\vec{v}_{2} \vec{v}_{2}\cdot\vec{n}) + ( \vec{S}_{1}\times\vec{n}\cdot\vec{v}_{1} \big( 5 \vec{a}_{1}\cdot\vec{a}_{2} \nn\\ 
&& + \vec{a}_{1}\cdot\vec{n} \vec{a}_{2}\cdot\vec{n} \big) + 2 \vec{S}_{1}\times\vec{n}\cdot\vec{a}_{1} \big( 2 \vec{v}_{1}\cdot\vec{a}_{2} - 2 \vec{v}_{2}\cdot\vec{a}_{2} + \vec{v}_{1}\cdot\vec{n} \vec{a}_{2}\cdot\vec{n} \big) \nn\\ 
&& - \vec{S}_{1}\times\vec{v}_{1}\cdot\vec{a}_{1} \vec{a}_{2}\cdot\vec{n} - \vec{S}_{1}\times\vec{n}\cdot\vec{v}_{2} \big( 3 \vec{a}_{1}\cdot\vec{a}_{2} + \vec{a}_{1}\cdot\vec{n} \vec{a}_{2}\cdot\vec{n} \big) + \vec{S}_{1}\times\vec{a}_{1}\cdot\vec{v}_{2} \vec{a}_{2}\cdot\vec{n} \nn\\ 
&& - 2 \vec{S}_{1}\times\vec{n}\cdot\vec{a}_{2} \big( \vec{a}_{1}\cdot\vec{v}_{2} + \vec{a}_{1}\cdot\vec{n} \vec{v}_{2}\cdot\vec{n} \big) - \vec{S}_{1}\times\vec{v}_{1}\cdot\vec{a}_{2} \vec{a}_{1}\cdot\vec{n} - 2 \vec{S}_{1}\times\vec{a}_{1}\cdot\vec{a}_{2} \big( 4 \vec{v}_{1}\cdot\vec{n} \nn\\ 
&& - \vec{v}_{2}\cdot\vec{n} \big) + \vec{S}_{1}\times\vec{v}_{2}\cdot\vec{a}_{2} \vec{a}_{1}\cdot\vec{n}) + (2 \dot{\vec{S}}_{1}\times\vec{n}\cdot\vec{v}_{1} \big( 2 \vec{v}_{1}\cdot\vec{a}_{2} - 2 \vec{v}_{2}\cdot\vec{a}_{2} + \vec{v}_{1}\cdot\vec{n} \vec{a}_{2}\cdot\vec{n} \big) \nn\\ 
&& + 2 \dot{\vec{S}}_{1}\times\vec{n}\cdot\vec{a}_{1} \big( v_{2}^2 - ( \vec{v}_{2}\cdot\vec{n})^{2} \big) + 4 \dot{\vec{S}}_{1}\times\vec{v}_{1}\cdot\vec{a}_{1} \vec{v}_{2}\cdot\vec{n} - 2 \dot{\vec{S}}_{1}\times\vec{n}\cdot\vec{v}_{2} \big( \vec{v}_{1}\cdot\vec{a}_{2} \nn\\ 
&& - 2 \vec{v}_{2}\cdot\vec{a}_{2} + \vec{v}_{1}\cdot\vec{n} \vec{a}_{2}\cdot\vec{n} \big) + 2 \dot{\vec{S}}_{1}\times\vec{v}_{1}\cdot\vec{v}_{2} \vec{a}_{2}\cdot\vec{n} + 12 \dot{\vec{S}}_{1}\times\vec{a}_{1}\cdot\vec{v}_{2} \vec{v}_{2}\cdot\vec{n} \nn\\ 
&& + 2 \dot{\vec{S}}_{1}\times\vec{n}\cdot\vec{a}_{2} \big( v_{2}^2 -2 \vec{v}_{1}\cdot\vec{n} \vec{v}_{2}\cdot\vec{n} \big) - 4 \dot{\vec{S}}_{1}\times\vec{v}_{1}\cdot\vec{a}_{2} \big( 2 \vec{v}_{1}\cdot\vec{n} - \vec{v}_{2}\cdot\vec{n} \big) \nn\\ 
&& + 2 \dot{\vec{S}}_{1}\times\vec{v}_{2}\cdot\vec{a}_{2} \vec{v}_{1}\cdot\vec{n}) \Big]\nn\\ && - 	\frac{G^2 m_{1} m_{2}}{r} \Big[ \dot{\vec{S}}_{1}\times\vec{n}\cdot\vec{a}_{1} + 6 \dot{\vec{S}}_{1}\times\vec{n}\cdot\vec{a}_{2} \Big] - 	\frac{17 G^2 m_{2}{}^2}{4 r} \dot{\vec{S}}_{1}\times\vec{n}\cdot\vec{a}_{2},
\eea
\bea
\stackrel{(3)}{V}_{2,1}&=&	\frac{1}{4} G m_{2} r \Big[ ( \vec{S}_{1}\times\vec{n}\cdot\dot{\vec{a}}_{1} \vec{a}_{2}\cdot\vec{n} + \vec{S}_{1}\times\vec{n}\cdot\dot{\vec{a}}_{2} \vec{a}_{1}\cdot\vec{n} + 7 \vec{S}_{1}\times\dot{\vec{a}}_{1}\cdot\vec{a}_{2} + \vec{S}_{1}\times\vec{a}_{1}\cdot\dot{\vec{a}}_{2}) \nn\\ 
&& + (2 \dot{\vec{S}}_{1}\times\vec{n}\cdot\dot{\vec{a}}_{2} \vec{v}_{1}\cdot\vec{n} + 2 \dot{\vec{S}}_{1}\times\vec{v}_{1}\cdot\dot{\vec{a}}_{2}) + ( \ddot{\vec{S}}_{1}\times\vec{n}\cdot\vec{v}_{1} \vec{a}_{2}\cdot\vec{n} - \ddot{\vec{S}}_{1}\times\vec{n}\cdot\vec{v}_{2} \vec{a}_{2}\cdot\vec{n} \nn\\ 
&& - 2 \ddot{\vec{S}}_{1}\times\vec{n}\cdot\vec{a}_{2} \vec{v}_{2}\cdot\vec{n} + 7 \ddot{\vec{S}}_{1}\times\vec{v}_{1}\cdot\vec{a}_{2} - \ddot{\vec{S}}_{1}\times\vec{v}_{2}\cdot\vec{a}_{2}) + (2 \dot{\vec{S}}_{1}\times\vec{n}\cdot\vec{a}_{1} \vec{a}_{2}\cdot\vec{n} \nn\\ 
&& + 14 \dot{\vec{S}}_{1}\times\vec{a}_{1}\cdot\vec{a}_{2}) \Big],
\eea
\bea
\stackrel{(4)}{V}_{2,1}&=& 	\frac{1}{4} G m_{2} r{}^2 \ddot{\vec{S}}_{1}\times\vec{n}\cdot\dot{\vec{a}}_{2}.
\eea

The new redefinitions in the N$^2$LO spin-orbit sector of both the position and spin 
variables can be expressed as:
\bea
\left(\Delta \vec{x}_1\right)_{\text{N}^2\text{LO}}^{\text{SO}} &=& \sum_{i=0}^3  \stackrel{(i)}{ \Delta \vec{x}_1}_{(2,1)},\\
(\omega^{ij})_{\text{N}^2\text{LO}}^{\text{SO}} &=& \sum_{k=0}^1  \stackrel{(k)}{ \omega^{ij}_1}_{(2,1)} - (i \leftrightarrow j) ,
\eea
with:
\bea
\stackrel{(0)}{ \Delta \vec{x}_1}_{(2,1)}&=&	\frac{1}{16 m_{1}} \vec{S}_{1}\times\vec{v}_{1} v_{1}^{4} \nn\\ && - 	\frac{G m_{2}}{8 m_{1} r} \Big[ 2 \vec{S}_{1}\times\vec{n}\cdot\vec{v}_{1} \big( v_{2}^2 \vec{n} - 4 \vec{v}_{2}\cdot\vec{n} \vec{v}_{2} -3 ( \vec{v}_{2}\cdot\vec{n})^{2} \vec{n} \big) - 2 \vec{S}_{1}\times\vec{n}\cdot\vec{v}_{2} \big( v_{2}^2 \vec{n} \nn\\ 
&& - 2 \vec{v}_{2}\cdot\vec{n} \vec{v}_{2} -3 ( \vec{v}_{2}\cdot\vec{n})^{2} \vec{n} \big) - 4 \vec{S}_{1}\times\vec{v}_{1}\cdot\vec{v}_{2} \big( 4 \vec{v}_{1} - 5 \vec{v}_{2} \big) - 2 \vec{S}_{1}\times\vec{n} \big( 4 \vec{v}_{2}\cdot\vec{n} \vec{v}_{1}\cdot\vec{v}_{2} \nn\\ 
&& - \vec{v}_{1}\cdot\vec{n} v_{2}^2 - 5 \vec{v}_{2}\cdot\vec{n} v_{2}^2 + 3 \vec{v}_{1}\cdot\vec{n} ( \vec{v}_{2}\cdot\vec{n})^{2} + 3 ( \vec{v}_{2}\cdot\vec{n})^{3} \big) - \vec{S}_{1}\times\vec{v}_{1} \big( 5 v_{1}^2 - 16 \vec{v}_{1}\cdot\vec{v}_{2} \nn\\ 
&& + 8 v_{2}^2 -2 ( \vec{v}_{2}\cdot\vec{n})^{2} \big) - 4 \vec{S}_{1}\times\vec{v}_{2} \big( 5 \vec{v}_{1}\cdot\vec{v}_{2} - 5 v_{2}^2 + 2 ( \vec{v}_{2}\cdot\vec{n})^{2} \big) \Big] \nn\\ 
&& - 	\frac{G}{16 r} \Big[ 4 \vec{S}_{2}\times\vec{n}\cdot\vec{v}_{1} \big( v_{2}^2 \vec{n} - 2 \vec{v}_{2}\cdot\vec{n} \vec{v}_{2} -3 ( \vec{v}_{2}\cdot\vec{n})^{2} \vec{n} \big) + \vec{S}_{2}\times\vec{n}\cdot\vec{v}_{2} \big( 7 v_{1}^2 \vec{n} \nn\\ 
&& - 14 \vec{v}_{1}\cdot\vec{v}_{2} \vec{n} + 4 v_{2}^2 \vec{n} + 14 \vec{v}_{1}\cdot\vec{n} \vec{v}_{1} - 2 \vec{v}_{2}\cdot\vec{n} \vec{v}_{1} - 14 \vec{v}_{1}\cdot\vec{n} \vec{v}_{2} \nn\\ 
&& + 2 \vec{v}_{2}\cdot\vec{n} \vec{v}_{2} -6 \vec{v}_{1}\cdot\vec{n} \vec{v}_{2}\cdot\vec{n} \vec{n} - 9 ( \vec{v}_{1}\cdot\vec{n})^{2} \vec{n} + 9 ( \vec{v}_{2}\cdot\vec{n})^{2} \vec{n} \big) + 2 \vec{S}_{2}\times\vec{v}_{1}\cdot\vec{v}_{2} \big( 9 \vec{v}_{1} \nn\\ 
&& - 13 \vec{v}_{2} -13 \vec{v}_{1}\cdot\vec{n} \vec{n} - 3 \vec{v}_{2}\cdot\vec{n} \vec{n} \big) - 4 \vec{S}_{2}\times\vec{n} \big( 2 \vec{v}_{2}\cdot\vec{n} \vec{v}_{1}\cdot\vec{v}_{2} - \vec{v}_{1}\cdot\vec{n} v_{2}^2 - 3 \vec{v}_{2}\cdot\vec{n} v_{2}^2 \nn\\ 
&& + 3 \vec{v}_{1}\cdot\vec{n} ( \vec{v}_{2}\cdot\vec{n})^{2} + 3 ( \vec{v}_{2}\cdot\vec{n})^{3} \big) - \vec{S}_{2}\times\vec{v}_{2} \big( 9 v_{1}^2 - 26 \vec{v}_{1}\cdot\vec{v}_{2} + 24 v_{2}^2 -6 \vec{v}_{1}\cdot\vec{n} \vec{v}_{2}\cdot\vec{n} \nn\\ 
&& - 13 ( \vec{v}_{1}\cdot\vec{n})^{2} - 7 ( \vec{v}_{2}\cdot\vec{n})^{2} \big) \Big]\nn\\ && +  	\frac{G^2 m_{2}}{32 r{}^2} \Big[ 25 \vec{S}_{1}\times\vec{n}\cdot\vec{v}_{1} \vec{n} + 164 \vec{S}_{1}\times\vec{n}\cdot\vec{v}_{2} \vec{n} + 4 \vec{S}_{1}\times\vec{n} \big( 73 \vec{v}_{1}\cdot\vec{n} - 13 \vec{v}_{2}\cdot\vec{n} \big) \nn\\ 
&& + 185 \vec{S}_{1}\times\vec{v}_{1} - 20 \vec{S}_{1}\times\vec{v}_{2} \Big] + 	\frac{G^2 m_{2}{}^2}{4 m_{1} r{}^2} \Big[ 4 \vec{S}_{1}\times\vec{n}\cdot\vec{v}_{1} \vec{n} + 2 \vec{S}_{1}\times\vec{n}\cdot\vec{v}_{2} \vec{n} \nn\\ 
&& + 4 \vec{S}_{1}\times\vec{n} \big( \vec{v}_{1}\cdot\vec{n} + 2 \vec{v}_{2}\cdot\vec{n} \big) + \vec{S}_{1}\times\vec{v}_{1} + 9 \vec{S}_{1}\times\vec{v}_{2} \Big] + 	\frac{G^2 m_{2}}{8 r{}^2} \Big[ 5 \vec{S}_{2}\times\vec{n}\cdot\vec{v}_{2} \vec{n} \nn\\ 
&& - 8 \vec{S}_{2}\times\vec{n} \vec{v}_{2}\cdot\vec{n} - 38 \vec{S}_{2}\times\vec{v}_{2} \Big] + 	\frac{G^2 m_{1}}{32 r{}^2} \Big[ 101 \vec{S}_{2}\times\vec{n}\cdot\vec{v}_{1} \vec{n} + 116 \vec{S}_{2}\times\vec{n}\cdot\vec{v}_{2} \vec{n} \nn\\ 
&& + \vec{S}_{2}\times\vec{n} \big( 487 \vec{v}_{1}\cdot\vec{n} - 287 \vec{v}_{2}\cdot\vec{n} \big) - 405 \vec{S}_{2}\times\vec{v}_{1} + 270 \vec{S}_{2}\times\vec{v}_{2} \Big],
\eea
\bea
\stackrel{(1)}{ \Delta \vec{x}_1}_{(2,1)}&=&	\frac{G m_{2}}{8 m_{1}} \Big[ \vec{S}_{1}\times\vec{n}\cdot\vec{v}_{1} \big( 5 \vec{a}_{2} + \vec{a}_{2}\cdot\vec{n} \vec{n} \big) - \vec{S}_{1}\times\vec{n}\cdot\vec{v}_{2} \big( 3 \vec{a}_{2} + \vec{a}_{2}\cdot\vec{n} \vec{n} \big) - 2 \vec{S}_{1}\times\vec{n}\cdot\vec{a}_{2} \big( \vec{v}_{2} \nn\\ 
&& + \vec{v}_{2}\cdot\vec{n} \vec{n} \big) - \vec{S}_{1}\times\vec{v}_{1}\cdot\vec{a}_{2} \vec{n} + \vec{S}_{1}\times\vec{v}_{2}\cdot\vec{a}_{2} \vec{n} + 2 \vec{S}_{1}\times\vec{n} \big( 3 \vec{v}_{1}\cdot\vec{a}_{2} - 5 \vec{v}_{2}\cdot\vec{a}_{2} \nn\\ 
&& + 3 \vec{v}_{2}\cdot\vec{n} \vec{a}_{2}\cdot\vec{n} \big) - 4 \vec{S}_{1}\times\vec{v}_{1} \vec{a}_{2}\cdot\vec{n} - 8 \vec{S}_{1}\times\vec{a}_{1} \vec{v}_{2}\cdot\vec{n} + 9 \vec{S}_{1}\times\vec{v}_{2} \vec{a}_{2}\cdot\vec{n} \nn\\ 
&& - 6 \vec{S}_{1}\times\vec{a}_{2} \big( \vec{v}_{1}\cdot\vec{n} - 4 \vec{v}_{2}\cdot\vec{n} \big) \Big] + 	\frac{1}{16} G \Big[ 2 \vec{S}_{2}\times\vec{n}\cdot\vec{v}_{1} \big( 3 \vec{a}_{2} + \vec{a}_{2}\cdot\vec{n} \vec{n} \big) \nn\\ 
&& + \vec{S}_{2}\times\vec{n}\cdot\vec{v}_{2} \big( 5 \vec{a}_{2} - \vec{a}_{2}\cdot\vec{n} \vec{n} \big) - 8 \vec{S}_{2}\times\vec{a}_{1}\cdot\vec{v}_{2} \vec{n} - 2 \vec{S}_{2}\times\vec{n}\cdot\vec{a}_{2} \big( 3 \vec{v}_{1} - 3 \vec{v}_{2} - \vec{v}_{1}\cdot\vec{n} \vec{n} \nn\\ 
&& + \vec{v}_{2}\cdot\vec{n} \vec{n} \big) + 16 \vec{S}_{2}\times\vec{v}_{1}\cdot\vec{a}_{2} \vec{n} - 5 \vec{S}_{2}\times\vec{v}_{2}\cdot\vec{a}_{2} \vec{n} + 4 \vec{S}_{2}\times\vec{n} \big( 2 \vec{v}_{1}\cdot\vec{a}_{2} - 3 \vec{v}_{2}\cdot\vec{a}_{2} \nn\\ 
&& + 3 \vec{v}_{2}\cdot\vec{n} \vec{a}_{2}\cdot\vec{n} \big) + 2 \vec{S}_{2}\times\vec{v}_{1} \vec{a}_{2}\cdot\vec{n} - \vec{S}_{2}\times\vec{v}_{2} \big( 8 \vec{a}_{1}\cdot\vec{n} + 9 \vec{a}_{2}\cdot\vec{n} \big) - 2 \vec{S}_{2}\times\vec{a}_{2} \big( 7 \vec{v}_{1}\cdot\vec{n} \nn\\ 
&& + 5 \vec{v}_{2}\cdot\vec{n} \big) \Big] + 	\frac{G m_{2}}{4 m_{1}} \Big[ \dot{\vec{S}}_{1}\times\vec{n} \big( v_{2}^2 - ( \vec{v}_{2}\cdot\vec{n})^{2} \big) - 6 \dot{\vec{S}}_{1}\times\vec{v}_{2} \vec{v}_{2}\cdot\vec{n} \Big] \nn\\ 
&& + 	\frac{1}{8} G \Big[ 4 \dot{\vec{S}}_{2}\times\vec{n}\cdot\vec{v}_{1} \big( \vec{v}_{2} + \vec{v}_{2}\cdot\vec{n} \vec{n} \big) + \dot{\vec{S}}_{2}\times\vec{n}\cdot\vec{v}_{2} \big( \vec{v}_{1} - \vec{v}_{2} + \vec{v}_{1}\cdot\vec{n} \vec{n} - 3 \vec{v}_{2}\cdot\vec{n} \vec{n} \big) \nn\\ 
&& + 11 \dot{\vec{S}}_{2}\times\vec{v}_{1}\cdot\vec{v}_{2} \vec{n} + 2 \dot{\vec{S}}_{2}\times\vec{n} \big( 2 \vec{v}_{1}\cdot\vec{v}_{2} - 3 v_{2}^2 + 2 \vec{v}_{1}\cdot\vec{n} \vec{v}_{2}\cdot\vec{n} + 3 ( \vec{v}_{2}\cdot\vec{n})^{2} \big) \nn\\ 
&& - \dot{\vec{S}}_{2}\times\vec{v}_{2} \big( 11 \vec{v}_{1}\cdot\vec{n} + 3 \vec{v}_{2}\cdot\vec{n} \big) \Big]\nn\\ && +  	\frac{33 G^2 m_{2}}{8 r} \dot{\vec{S}}_{1}\times\vec{n} - 	\frac{237 G^2 m_{1}}{32 r} \dot{\vec{S}}_{2}\times\vec{n} - 	\frac{3 G^2 m_{2}}{r} \dot{\vec{S}}_{2}\times\vec{n} ,
\eea
\bea
\stackrel{(2)}{ \Delta \vec{x}_1}_{(2,1)}&=&	\frac{1}{16} G r \Big[ ( \vec{S}_{2}\times\vec{n}\cdot\dot{\vec{a}}_{2} \vec{n} + 4 \vec{S}_{2}\times\vec{n} \dot{\vec{a}}_{2}\cdot\vec{n} + 19 \vec{S}_{2}\times\dot{\vec{a}}_{2}) + (4 \ddot{\vec{S}}_{2}\times\vec{n}\cdot\vec{v}_{1} \vec{n} \nn\\ 
&& - 3 \ddot{\vec{S}}_{2}\times\vec{n}\cdot\vec{v}_{2} \vec{n} + 4 \ddot{\vec{S}}_{2}\times\vec{n} \big( \vec{v}_{1}\cdot\vec{n} + 3 \vec{v}_{2}\cdot\vec{n} \big) - \ddot{\vec{S}}_{2}\times\vec{v}_{2}) - (2 \dot{\vec{S}}_{2}\times\vec{n}\cdot\vec{a}_{2} \vec{n} \nn\\ 
&& - 12 \dot{\vec{S}}_{2}\times\vec{n} \vec{a}_{2}\cdot\vec{n} - 14 \dot{\vec{S}}_{2}\times\vec{a}_{2}) \Big] + 	\frac{G m_{2} r}{4 m_{1}} \Big[ \vec{S}_{1}\times\vec{n} \dot{\vec{a}}_{2}\cdot\vec{n} - 7 \vec{S}_{1}\times\dot{\vec{a}}_{2} \Big],
\eea
\bea
\stackrel{(3)}{ \Delta \vec{x}_1}_{(2,1)}&=&- 	\frac{1}{4} G r{}^2 \dddot{\vec{S}}_{2}\times\vec{n} ,
\eea
and:
\bea
\stackrel{(0)}{ \omega^{ij}_1}_{(2,1)}&=&- 	\frac{G m_{2}}{4 r} \Big[ 6 v_{1}^2 v_{2}^i v_{1}^j - 6 \vec{v}_{1}\cdot\vec{v}_{2} v_{2}^i v_{1}^j + 10 v_{2}^2 v_{2}^i v_{1}^j - 6 v_{1}^2 v_{1}^i v_{2}^j + 6 \vec{v}_{1}\cdot\vec{v}_{2} v_{1}^i v_{2}^j \nn\\ 
&& - 10 v_{2}^2 v_{1}^i v_{2}^j -2 v_{1}^2 \vec{v}_{2}\cdot\vec{n} v_{1}^i n^j + 4 \vec{v}_{2}\cdot\vec{n} \vec{v}_{1}\cdot\vec{v}_{2} v_{1}^i n^j - \vec{v}_{1}\cdot\vec{n} v_{2}^2 v_{1}^i n^j - 5 \vec{v}_{2}\cdot\vec{n} v_{2}^2 v_{1}^i n^j \nn\\ 
&& - 2 \vec{v}_{2}\cdot\vec{n} \vec{v}_{1}\cdot\vec{v}_{2} v_{2}^i n^j + \vec{v}_{1}\cdot\vec{n} v_{2}^2 v_{2}^i n^j + 5 \vec{v}_{2}\cdot\vec{n} v_{2}^2 v_{2}^i n^j + 2 v_{1}^2 \vec{v}_{2}\cdot\vec{n} n^i v_{1}^j \nn\\ 
&& - 4 \vec{v}_{2}\cdot\vec{n} \vec{v}_{1}\cdot\vec{v}_{2} n^i v_{1}^j + \vec{v}_{1}\cdot\vec{n} v_{2}^2 n^i v_{1}^j + 5 \vec{v}_{2}\cdot\vec{n} v_{2}^2 n^i v_{1}^j - 4 ( \vec{v}_{2}\cdot\vec{n})^{2} v_{2}^i v_{1}^j \nn\\ 
&& + 2 \vec{v}_{2}\cdot\vec{n} \vec{v}_{1}\cdot\vec{v}_{2} n^i v_{2}^j - \vec{v}_{1}\cdot\vec{n} v_{2}^2 n^i v_{2}^j - 5 \vec{v}_{2}\cdot\vec{n} v_{2}^2 n^i v_{2}^j + 4 ( \vec{v}_{2}\cdot\vec{n})^{2} v_{1}^i v_{2}^j \nn\\ 
&& + 3 \vec{v}_{1}\cdot\vec{n} ( \vec{v}_{2}\cdot\vec{n})^{2} v_{1}^i n^j + 3 ( \vec{v}_{2}\cdot\vec{n})^{3} v_{1}^i n^j - 3 \vec{v}_{1}\cdot\vec{n} ( \vec{v}_{2}\cdot\vec{n})^{2} v_{2}^i n^j - 3 ( \vec{v}_{2}\cdot\vec{n})^{3} v_{2}^i n^j \nn\\ 
&& - 3 \vec{v}_{1}\cdot\vec{n} ( \vec{v}_{2}\cdot\vec{n})^{2} n^i v_{1}^j - 3 ( \vec{v}_{2}\cdot\vec{n})^{3} n^i v_{1}^j + 3 \vec{v}_{1}\cdot\vec{n} ( \vec{v}_{2}\cdot\vec{n})^{2} n^i v_{2}^j + 3 ( \vec{v}_{2}\cdot\vec{n})^{3} n^i v_{2}^j \Big]\nn\\ && - 	\frac{G^2 m_{2}{}^2}{8 r{}^2} \Big[ 54 v_{2}^i v_{1}^j - 54 v_{1}^i v_{2}^j + 6 \vec{v}_{1}\cdot\vec{n} v_{1}^i n^j + 18 \vec{v}_{2}\cdot\vec{n} v_{1}^i n^j + 4 \vec{v}_{1}\cdot\vec{n} v_{2}^i n^j \nn\\ 
&& + 15 \vec{v}_{2}\cdot\vec{n} v_{2}^i n^j - 6 \vec{v}_{1}\cdot\vec{n} n^i v_{1}^j - 18 \vec{v}_{2}\cdot\vec{n} n^i v_{1}^j - 4 \vec{v}_{1}\cdot\vec{n} n^i v_{2}^j - 15 \vec{v}_{2}\cdot\vec{n} n^i v_{2}^j \Big] \nn\\ 
&& - 	\frac{G^2 m_{1} m_{2}}{2 r{}^2} \Big[ 13 v_{2}^i v_{1}^j - 13 v_{1}^i v_{2}^j + 31 \vec{v}_{1}\cdot\vec{n} v_{1}^i n^j - 10 \vec{v}_{2}\cdot\vec{n} v_{1}^i n^j - 17 \vec{v}_{1}\cdot\vec{n} v_{2}^i n^j \nn\\ 
&& + 13 \vec{v}_{2}\cdot\vec{n} v_{2}^i n^j - 31 \vec{v}_{1}\cdot\vec{n} n^i v_{1}^j + 10 \vec{v}_{2}\cdot\vec{n} n^i v_{1}^j + 17 \vec{v}_{1}\cdot\vec{n} n^i v_{2}^j - 13 \vec{v}_{2}\cdot\vec{n} n^i v_{2}^j \Big],\nn\\ 
\eea
\bea
\stackrel{(1)}{ \omega^{ij}_1}_{(2,1)}&=&	\frac{1}{4} G m_{2} \Big[ 2 \vec{v}_{2}\cdot\vec{a}_{2} v_{1}^i n^j + v_{2}^2 a_{1}^i n^j - 2 \vec{v}_{2}\cdot\vec{a}_{2} v_{2}^i n^j - v_{2}^2 a_{2}^i n^j - 2 \vec{v}_{2}\cdot\vec{a}_{2} n^i v_{1}^j \nn\\ 
&& + 6 \vec{a}_{2}\cdot\vec{n} v_{2}^i v_{1}^j + 6 \vec{v}_{2}\cdot\vec{n} a_{2}^i v_{1}^j - v_{2}^2 n^i a_{1}^j + 6 \vec{v}_{2}\cdot\vec{n} v_{2}^i a_{1}^j + 2 \vec{v}_{2}\cdot\vec{a}_{2} n^i v_{2}^j - 6 \vec{a}_{2}\cdot\vec{n} v_{1}^i v_{2}^j \nn\\ 
&& - 6 \vec{v}_{2}\cdot\vec{n} a_{1}^i v_{2}^j + v_{2}^2 n^i a_{2}^j - 6 \vec{v}_{2}\cdot\vec{n} v_{1}^i a_{2}^j -2 \vec{v}_{2}\cdot\vec{n} \vec{a}_{2}\cdot\vec{n} v_{1}^i n^j - ( \vec{v}_{2}\cdot\vec{n})^{2} a_{1}^i n^j \nn\\ 
&& + 2 \vec{v}_{2}\cdot\vec{n} \vec{a}_{2}\cdot\vec{n} v_{2}^i n^j + ( \vec{v}_{2}\cdot\vec{n})^{2} a_{2}^i n^j + 2 \vec{v}_{2}\cdot\vec{n} \vec{a}_{2}\cdot\vec{n} n^i v_{1}^j + ( \vec{v}_{2}\cdot\vec{n})^{2} n^i a_{1}^j \nn\\ 
&& - 2 \vec{v}_{2}\cdot\vec{n} \vec{a}_{2}\cdot\vec{n} n^i v_{2}^j - ( \vec{v}_{2}\cdot\vec{n})^{2} n^i a_{2}^j \Big]\nn\\ && +  	\frac{G^2 m_{1} m_{2}}{4 r} \Big[ 5 a_{1}^i n^j + a_{2}^i n^j - 5 n^i a_{1}^j - n^i a_{2}^j \Big].
\eea

Finally, we consider the N$^3$LO spin-orbit sector at the $4.5$PN order. 
Following the discussion in section \ref{redefinitions}, and in particular table 
\ref{n3los1redef}, we implemented the following new redefinition of the position and 
spin variables in the present sector:
\bea
\left(\Delta \vec{x}_1\right)_{\text{N}^3\text{LO}}^{\text{SO}} &=& \sum_{i=0}^5  \stackrel{(i)}{ \Delta \vec{x}_1}_{(3,1)},\\
(\omega^{ij})_{\text{N}^3\text{LO}}^{\text{SO}} &=& \sum_{k=0}^2  \stackrel{(k)}{ \omega^{ij}_1}_{(3,1)} - (i \leftrightarrow j) ,
\eea
where
\bea
\stackrel{(0)}{ \Delta \vec{x}_1}_{(3,1)}&=& \frac{5}{128 m_{1}} \vec{S}_{1}\times\vec{v}_{1} v_{1}^{6} \nn\\ && - 	\frac{G m_{2}}{32 m_{1} r} \Big[ \vec{S}_{1}\times\vec{n}\cdot\vec{v}_{1} \big( v_{1}^2 v_{2}^2 \vec{n} - 4 \vec{v}_{1}\cdot\vec{v}_{2} v_{2}^2 \vec{n} + 8 v_{2}^{4} \vec{n} + 2 v_{1}^2 \vec{v}_{2}\cdot\vec{n} \vec{v}_{1} \nn\\ 
&& + 26 \vec{v}_{2}\cdot\vec{n} \vec{v}_{1}\cdot\vec{v}_{2} \vec{v}_{1} - 7 \vec{v}_{1}\cdot\vec{n} v_{2}^2 \vec{v}_{1} + 5 \vec{v}_{2}\cdot\vec{n} v_{2}^2 \vec{v}_{1} - 6 v_{1}^2 \vec{v}_{2}\cdot\vec{n} \vec{v}_{2} - 4 \vec{v}_{1}\cdot\vec{n} v_{2}^2 \vec{v}_{2} \nn\\ 
&& - 36 \vec{v}_{2}\cdot\vec{n} v_{2}^2 \vec{v}_{2} -24 \vec{v}_{1}\cdot\vec{n} \vec{v}_{2}\cdot\vec{n} v_{2}^2 \vec{n} - 3 v_{1}^2 ( \vec{v}_{2}\cdot\vec{n})^{2} \vec{n} + 12 \vec{v}_{1}\cdot\vec{v}_{2} ( \vec{v}_{2}\cdot\vec{n})^{2} \vec{n} \nn\\ 
&& - 36 v_{2}^2 ( \vec{v}_{2}\cdot\vec{n})^{2} \vec{n} + 21 \vec{v}_{1}\cdot\vec{n} ( \vec{v}_{2}\cdot\vec{n})^{2} \vec{v}_{1} - 5 ( \vec{v}_{2}\cdot\vec{n})^{3} \vec{v}_{1} + 12 \vec{v}_{1}\cdot\vec{n} ( \vec{v}_{2}\cdot\vec{n})^{2} \vec{v}_{2} \nn\\ 
&& + 20 ( \vec{v}_{2}\cdot\vec{n})^{3} \vec{v}_{2} + 40 \vec{v}_{1}\cdot\vec{n} ( \vec{v}_{2}\cdot\vec{n})^{3} \vec{n} + 20 ( \vec{v}_{2}\cdot\vec{n})^{4} \vec{n} \big) + 4 \vec{S}_{1}\times\vec{n}\cdot\vec{v}_{2} \big( v_{1}^2 v_{2}^2 \vec{n} - 2 v_{2}^{4} \vec{n} \nn\\ 
&& - 4 \vec{v}_{2}\cdot\vec{n} \vec{v}_{1}\cdot\vec{v}_{2} \vec{v}_{1} + 2 \vec{v}_{1}\cdot\vec{n} v_{2}^2 \vec{v}_{1} - 2 \vec{v}_{2}\cdot\vec{n} v_{2}^2 \vec{v}_{1} - 2 v_{1}^2 \vec{v}_{2}\cdot\vec{n} \vec{v}_{2} + 4 \vec{v}_{2}\cdot\vec{n} \vec{v}_{1}\cdot\vec{v}_{2} \vec{v}_{2} \nn\\ 
&& + 4 \vec{v}_{2}\cdot\vec{n} v_{2}^2 \vec{v}_{2} + 6 \vec{v}_{1}\cdot\vec{n} \vec{v}_{2}\cdot\vec{n} v_{2}^2 \vec{n} - 3 v_{1}^2 ( \vec{v}_{2}\cdot\vec{n})^{2} \vec{n} + 9 v_{2}^2 ( \vec{v}_{2}\cdot\vec{n})^{2} \vec{n} \nn\\ 
&& - 6 \vec{v}_{1}\cdot\vec{n} ( \vec{v}_{2}\cdot\vec{n})^{2} \vec{v}_{1} + 2 ( \vec{v}_{2}\cdot\vec{n})^{3} \vec{v}_{1} - 2 ( \vec{v}_{2}\cdot\vec{n})^{3} \vec{v}_{2} -10 \vec{v}_{1}\cdot\vec{n} ( \vec{v}_{2}\cdot\vec{n})^{3} \vec{n} \nn\\ 
&& - 5 ( \vec{v}_{2}\cdot\vec{n})^{4} \vec{n} \big) + \vec{S}_{1}\times\vec{v}_{1}\cdot\vec{v}_{2} \big( 2 v_{1}^2 \vec{v}_{1} - 46 \vec{v}_{1}\cdot\vec{v}_{2} \vec{v}_{1} - 59 v_{2}^2 \vec{v}_{1} + 2 v_{1}^2 \vec{v}_{2} + 16 \vec{v}_{1}\cdot\vec{v}_{2} \vec{v}_{2} \nn\\ 
&& + 72 v_{2}^2 \vec{v}_{2} + 6 v_{1}^2 \vec{v}_{2}\cdot\vec{n} \vec{n} + 8 \vec{v}_{2}\cdot\vec{n} \vec{v}_{1}\cdot\vec{v}_{2} \vec{n} - 12 \vec{v}_{2}\cdot\vec{n} v_{2}^2 \vec{n} + 6 \vec{v}_{1}\cdot\vec{n} \vec{v}_{2}\cdot\vec{n} \vec{v}_{1} \nn\\ 
&& + 27 ( \vec{v}_{2}\cdot\vec{n})^{2} \vec{v}_{1} + 8 \vec{v}_{1}\cdot\vec{n} \vec{v}_{2}\cdot\vec{n} \vec{v}_{2} - 32 ( \vec{v}_{2}\cdot\vec{n})^{2} \vec{v}_{2} + 12 ( \vec{v}_{2}\cdot\vec{n})^{3} \vec{n} \big) \nn\\ 
&& - \vec{S}_{1}\times\vec{n} \big( 6 v_{1}^2 \vec{v}_{2}\cdot\vec{n} \vec{v}_{1}\cdot\vec{v}_{2} - \vec{v}_{1}\cdot\vec{n} v_{1}^2 v_{2}^2 - v_{1}^2 \vec{v}_{2}\cdot\vec{n} v_{2}^2 + 4 \vec{v}_{1}\cdot\vec{n} \vec{v}_{1}\cdot\vec{v}_{2} v_{2}^2 \nn\\ 
&& + 36 \vec{v}_{2}\cdot\vec{n} \vec{v}_{1}\cdot\vec{v}_{2} v_{2}^2 - 4 \vec{v}_{2}\cdot\vec{n} v_{1}^{4} - 8 \vec{v}_{1}\cdot\vec{n} v_{2}^{4} - 44 \vec{v}_{2}\cdot\vec{n} v_{2}^{4} + 12 \vec{v}_{2}\cdot\vec{n} v_{2}^2 ( \vec{v}_{1}\cdot\vec{n})^{2} \nn\\ 
&& + 3 \vec{v}_{1}\cdot\vec{n} v_{1}^2 ( \vec{v}_{2}\cdot\vec{n})^{2} + v_{1}^2 ( \vec{v}_{2}\cdot\vec{n})^{3} - 12 \vec{v}_{1}\cdot\vec{n} \vec{v}_{1}\cdot\vec{v}_{2} ( \vec{v}_{2}\cdot\vec{n})^{2} - 20 \vec{v}_{1}\cdot\vec{v}_{2} ( \vec{v}_{2}\cdot\vec{n})^{3} \nn\\ 
&& + 36 \vec{v}_{1}\cdot\vec{n} v_{2}^2 ( \vec{v}_{2}\cdot\vec{n})^{2} + 52 v_{2}^2 ( \vec{v}_{2}\cdot\vec{n})^{3} -20 ( \vec{v}_{1}\cdot\vec{n})^{2} ( \vec{v}_{2}\cdot\vec{n})^{3} - 20 \vec{v}_{1}\cdot\vec{n} ( \vec{v}_{2}\cdot\vec{n})^{4} \nn\\ 
&& - 20 ( \vec{v}_{2}\cdot\vec{n})^{5} \big) + \vec{S}_{1}\times\vec{v}_{1} \big( 18 v_{1}^2 \vec{v}_{1}\cdot\vec{v}_{2} - 7 v_{1}^2 v_{2}^2 + 51 \vec{v}_{1}\cdot\vec{v}_{2} v_{2}^2 - 2 ( \vec{v}_{1}\cdot\vec{v}_{2})^{2} - 18 v_{1}^{4} \nn\\ 
&& - 32 v_{2}^{4} + 14 \vec{v}_{1}\cdot\vec{n} v_{1}^2 \vec{v}_{2}\cdot\vec{n} - 12 \vec{v}_{1}\cdot\vec{n} \vec{v}_{2}\cdot\vec{n} \vec{v}_{1}\cdot\vec{v}_{2} - 3 \vec{v}_{1}\cdot\vec{n} \vec{v}_{2}\cdot\vec{n} v_{2}^2 + v_{2}^2 ( \vec{v}_{1}\cdot\vec{n})^{2} \nn\\ 
&& - 7 v_{1}^2 ( \vec{v}_{2}\cdot\vec{n})^{2} - 19 \vec{v}_{1}\cdot\vec{v}_{2} ( \vec{v}_{2}\cdot\vec{n})^{2} + 24 v_{2}^2 ( \vec{v}_{2}\cdot\vec{n})^{2} -3 ( \vec{v}_{1}\cdot\vec{n})^{2} ( \vec{v}_{2}\cdot\vec{n})^{2} \nn\\ 
&& + 3 \vec{v}_{1}\cdot\vec{n} ( \vec{v}_{2}\cdot\vec{n})^{3} - 6 ( \vec{v}_{2}\cdot\vec{n})^{4} \big) - \vec{S}_{1}\times\vec{v}_{2} \big( 2 v_{1}^2 \vec{v}_{1}\cdot\vec{v}_{2} + 9 v_{1}^2 v_{2}^2 + 72 \vec{v}_{1}\cdot\vec{v}_{2} v_{2}^2 \nn\\ 
&& + 8 ( \vec{v}_{1}\cdot\vec{v}_{2})^{2} - 12 v_{1}^{4} - 76 v_{2}^{4} + 6 \vec{v}_{1}\cdot\vec{n} v_{1}^2 \vec{v}_{2}\cdot\vec{n} + 8 \vec{v}_{1}\cdot\vec{n} \vec{v}_{2}\cdot\vec{n} \vec{v}_{1}\cdot\vec{v}_{2} \nn\\ 
&& - 12 \vec{v}_{1}\cdot\vec{n} \vec{v}_{2}\cdot\vec{n} v_{2}^2 - 9 v_{1}^2 ( \vec{v}_{2}\cdot\vec{n})^{2} - 32 \vec{v}_{1}\cdot\vec{v}_{2} ( \vec{v}_{2}\cdot\vec{n})^{2} + 64 v_{2}^2 ( \vec{v}_{2}\cdot\vec{n})^{2} \nn\\ 
&& + 12 \vec{v}_{1}\cdot\vec{n} ( \vec{v}_{2}\cdot\vec{n})^{3} - 24 ( \vec{v}_{2}\cdot\vec{n})^{4} \big) \Big] - 	\frac{G}{32 r} \Big[ 4 \vec{S}_{2}\times\vec{n}\cdot\vec{v}_{1} \big( v_{2}^{4} \vec{n} + 4 \vec{v}_{2}\cdot\vec{n} \vec{v}_{1}\cdot\vec{v}_{2} \vec{v}_{1} \nn\\ 
&& - 2 \vec{v}_{1}\cdot\vec{n} v_{2}^2 \vec{v}_{1} + 2 \vec{v}_{2}\cdot\vec{n} v_{2}^2 \vec{v}_{1} - 4 \vec{v}_{2}\cdot\vec{n} \vec{v}_{1}\cdot\vec{v}_{2} \vec{v}_{2} - 2 \vec{v}_{2}\cdot\vec{n} v_{2}^2 \vec{v}_{2} -6 \vec{v}_{1}\cdot\vec{n} \vec{v}_{2}\cdot\vec{n} v_{2}^2 \vec{n} \nn\\ 
&& - 6 v_{2}^2 ( \vec{v}_{2}\cdot\vec{n})^{2} \vec{n} + 6 \vec{v}_{1}\cdot\vec{n} ( \vec{v}_{2}\cdot\vec{n})^{2} \vec{v}_{1} - 2 ( \vec{v}_{2}\cdot\vec{n})^{3} \vec{v}_{1} + 2 ( \vec{v}_{2}\cdot\vec{n})^{3} \vec{v}_{2} \nn\\ 
&& + 10 \vec{v}_{1}\cdot\vec{n} ( \vec{v}_{2}\cdot\vec{n})^{3} \vec{n} + 5 ( \vec{v}_{2}\cdot\vec{n})^{4} \vec{n} \big) - \vec{S}_{2}\times\vec{n}\cdot\vec{v}_{2} \big( 16 v_{1}^2 \vec{v}_{1}\cdot\vec{v}_{2} \vec{n} - 10 v_{1}^2 v_{2}^2 \vec{n} \nn\\ 
&& + 18 \vec{v}_{1}\cdot\vec{v}_{2} v_{2}^2 \vec{n} - 2 ( \vec{v}_{1}\cdot\vec{v}_{2})^{2} \vec{n} - 11 v_{1}^{4} \vec{n} - 7 v_{2}^{4} \vec{n} - 16 \vec{v}_{1}\cdot\vec{n} v_{1}^2 \vec{v}_{1} - 2 v_{1}^2 \vec{v}_{2}\cdot\vec{n} \vec{v}_{1} \nn\\ 
&& + 4 \vec{v}_{1}\cdot\vec{n} \vec{v}_{1}\cdot\vec{v}_{2} \vec{v}_{1} + 4 \vec{v}_{2}\cdot\vec{n} \vec{v}_{1}\cdot\vec{v}_{2} \vec{v}_{1} - 20 \vec{v}_{1}\cdot\vec{n} v_{2}^2 \vec{v}_{1} + 16 \vec{v}_{2}\cdot\vec{n} v_{2}^2 \vec{v}_{1} + 16 \vec{v}_{1}\cdot\vec{n} v_{1}^2 \vec{v}_{2} \nn\\ 
&& + 2 v_{1}^2 \vec{v}_{2}\cdot\vec{n} \vec{v}_{2} - 4 \vec{v}_{1}\cdot\vec{n} \vec{v}_{1}\cdot\vec{v}_{2} \vec{v}_{2} - 4 \vec{v}_{2}\cdot\vec{n} \vec{v}_{1}\cdot\vec{v}_{2} \vec{v}_{2} + 18 \vec{v}_{1}\cdot\vec{n} v_{2}^2 \vec{v}_{2} - 10 \vec{v}_{2}\cdot\vec{n} v_{2}^2 \vec{v}_{2} \nn\\ 
&& + 24 \vec{v}_{1}\cdot\vec{n} v_{1}^2 \vec{v}_{2}\cdot\vec{n} \vec{n} - 36 \vec{v}_{1}\cdot\vec{n} \vec{v}_{2}\cdot\vec{n} \vec{v}_{1}\cdot\vec{v}_{2} \vec{n} + 6 \vec{v}_{1}\cdot\vec{n} \vec{v}_{2}\cdot\vec{n} v_{2}^2 \vec{n} + 48 v_{1}^2 ( \vec{v}_{1}\cdot\vec{n})^{2} \vec{n} \nn\\ 
&& - 60 \vec{v}_{1}\cdot\vec{v}_{2} ( \vec{v}_{1}\cdot\vec{n})^{2} \vec{n} + 30 v_{2}^2 ( \vec{v}_{1}\cdot\vec{n})^{2} \vec{n} + 9 v_{1}^2 ( \vec{v}_{2}\cdot\vec{n})^{2} \vec{n} - 6 \vec{v}_{1}\cdot\vec{v}_{2} ( \vec{v}_{2}\cdot\vec{n})^{2} \vec{n} \nn\\ 
&& - 21 v_{2}^2 ( \vec{v}_{2}\cdot\vec{n})^{2} \vec{n} + 20 ( \vec{v}_{1}\cdot\vec{n})^{3} \vec{v}_{1} + 18 \vec{v}_{2}\cdot\vec{n} ( \vec{v}_{1}\cdot\vec{n})^{2} \vec{v}_{1} + 18 \vec{v}_{1}\cdot\vec{n} ( \vec{v}_{2}\cdot\vec{n})^{2} \vec{v}_{1} \nn\\ 
&& - 12 ( \vec{v}_{2}\cdot\vec{n})^{3} \vec{v}_{1} - 20 ( \vec{v}_{1}\cdot\vec{n})^{3} \vec{v}_{2} - 18 \vec{v}_{2}\cdot\vec{n} ( \vec{v}_{1}\cdot\vec{n})^{2} \vec{v}_{2} \nn\\ 
&& - 6 \vec{v}_{1}\cdot\vec{n} ( \vec{v}_{2}\cdot\vec{n})^{2} \vec{v}_{2} -25 ( \vec{v}_{1}\cdot\vec{n})^{4} \vec{n} - 20 \vec{v}_{2}\cdot\vec{n} ( \vec{v}_{1}\cdot\vec{n})^{3} \vec{n} - 15 ( \vec{v}_{1}\cdot\vec{n})^{2} ( \vec{v}_{2}\cdot\vec{n})^{2} \vec{n} \nn\\ 
&& + 30 \vec{v}_{1}\cdot\vec{n} ( \vec{v}_{2}\cdot\vec{n})^{3} \vec{n} + 15 ( \vec{v}_{2}\cdot\vec{n})^{4} \vec{n} \big) + 2 \vec{S}_{2}\times\vec{v}_{1}\cdot\vec{v}_{2} \big( 8 v_{1}^2 \vec{v}_{1} + 10 \vec{v}_{1}\cdot\vec{v}_{2} \vec{v}_{1} + 10 v_{2}^2 \vec{v}_{1} \nn\\ 
&& - 8 v_{1}^2 \vec{v}_{2} - 10 \vec{v}_{1}\cdot\vec{v}_{2} \vec{v}_{2} - 13 v_{2}^2 \vec{v}_{2} -32 \vec{v}_{1}\cdot\vec{n} v_{1}^2 \vec{n} + 4 v_{1}^2 \vec{v}_{2}\cdot\vec{n} \vec{n} + 12 \vec{v}_{1}\cdot\vec{n} \vec{v}_{1}\cdot\vec{v}_{2} \vec{n} \nn\\ 
&& - 6 \vec{v}_{2}\cdot\vec{n} \vec{v}_{1}\cdot\vec{v}_{2} \vec{n} - 10 \vec{v}_{1}\cdot\vec{n} v_{2}^2 \vec{n} - 5 \vec{v}_{2}\cdot\vec{n} v_{2}^2 \vec{n} + 14 \vec{v}_{1}\cdot\vec{n} \vec{v}_{2}\cdot\vec{n} \vec{v}_{1} - 6 ( \vec{v}_{1}\cdot\vec{n})^{2} \vec{v}_{1} \nn\\ 
&& - ( \vec{v}_{2}\cdot\vec{n})^{2} \vec{v}_{1} - 6 \vec{v}_{1}\cdot\vec{n} \vec{v}_{2}\cdot\vec{n} \vec{v}_{2} + 6 ( \vec{v}_{1}\cdot\vec{n})^{2} \vec{v}_{2} + ( \vec{v}_{2}\cdot\vec{n})^{2} \vec{v}_{2} + 38 ( \vec{v}_{1}\cdot\vec{n})^{3} \vec{n} \nn\\ 
&& - 6 \vec{v}_{2}\cdot\vec{n} ( \vec{v}_{1}\cdot\vec{n})^{2} \vec{n} + 9 \vec{v}_{1}\cdot\vec{n} ( \vec{v}_{2}\cdot\vec{n})^{2} \vec{n} + 5 ( \vec{v}_{2}\cdot\vec{n})^{3} \vec{n} \big) + 4 \vec{S}_{2}\times\vec{n} \big( v_{1}^2 \vec{v}_{2}\cdot\vec{n} v_{2}^2 \nn\\ 
&& - 2 \vec{v}_{2}\cdot\vec{n} \vec{v}_{1}\cdot\vec{v}_{2} v_{2}^2 - 2 \vec{v}_{2}\cdot\vec{n} ( \vec{v}_{1}\cdot\vec{v}_{2})^{2} + \vec{v}_{1}\cdot\vec{n} v_{2}^{4} + 5 \vec{v}_{2}\cdot\vec{n} v_{2}^{4} -3 \vec{v}_{2}\cdot\vec{n} v_{2}^2 ( \vec{v}_{1}\cdot\vec{n})^{2} \nn\\ 
&& - v_{1}^2 ( \vec{v}_{2}\cdot\vec{n})^{3} + 2 \vec{v}_{1}\cdot\vec{v}_{2} ( \vec{v}_{2}\cdot\vec{n})^{3} - 6 \vec{v}_{1}\cdot\vec{n} v_{2}^2 ( \vec{v}_{2}\cdot\vec{n})^{2} - 10 v_{2}^2 ( \vec{v}_{2}\cdot\vec{n})^{3} \nn\\ 
&& + 5 ( \vec{v}_{1}\cdot\vec{n})^{2} ( \vec{v}_{2}\cdot\vec{n})^{3} + 5 \vec{v}_{1}\cdot\vec{n} ( \vec{v}_{2}\cdot\vec{n})^{4} + 5 ( \vec{v}_{2}\cdot\vec{n})^{5} \big) + \vec{S}_{2}\times\vec{v}_{2} \big( 16 v_{1}^2 \vec{v}_{1}\cdot\vec{v}_{2} \nn\\ 
&& - 10 v_{1}^2 v_{2}^2 + 26 \vec{v}_{1}\cdot\vec{v}_{2} v_{2}^2 + 10 ( \vec{v}_{1}\cdot\vec{v}_{2})^{2} - 13 v_{1}^{4} - 37 v_{2}^{4} -8 \vec{v}_{1}\cdot\vec{n} v_{1}^2 \vec{v}_{2}\cdot\vec{n} \nn\\ 
&& + 12 \vec{v}_{1}\cdot\vec{n} \vec{v}_{2}\cdot\vec{n} \vec{v}_{1}\cdot\vec{v}_{2} + 10 \vec{v}_{1}\cdot\vec{n} \vec{v}_{2}\cdot\vec{n} v_{2}^2 + 32 v_{1}^2 ( \vec{v}_{1}\cdot\vec{n})^{2} - 12 \vec{v}_{1}\cdot\vec{v}_{2} ( \vec{v}_{1}\cdot\vec{n})^{2} \nn\\ 
&& + 10 v_{2}^2 ( \vec{v}_{1}\cdot\vec{n})^{2} + v_{1}^2 ( \vec{v}_{2}\cdot\vec{n})^{2} - 2 \vec{v}_{1}\cdot\vec{v}_{2} ( \vec{v}_{2}\cdot\vec{n})^{2} + 19 v_{2}^2 ( \vec{v}_{2}\cdot\vec{n})^{2} -19 ( \vec{v}_{1}\cdot\vec{n})^{4} \nn\\ 
&& + 4 \vec{v}_{2}\cdot\vec{n} ( \vec{v}_{1}\cdot\vec{n})^{3} - 9 ( \vec{v}_{1}\cdot\vec{n})^{2} ( \vec{v}_{2}\cdot\vec{n})^{2} - 10 \vec{v}_{1}\cdot\vec{n} ( \vec{v}_{2}\cdot\vec{n})^{3} - 11 ( \vec{v}_{2}\cdot\vec{n})^{4} \big) \Big]\nn\\ && +  	\frac{G^2 m_{2}}{384 r{}^2} \Big[ \vec{S}_{1}\times\vec{n}\cdot\vec{v}_{1} \big( 3047 v_{1}^2 \vec{n} - 4702 \vec{v}_{1}\cdot\vec{v}_{2} \vec{n} + 4159 v_{2}^2 \vec{n} + 11830 \vec{v}_{1}\cdot\vec{n} \vec{v}_{1} \nn\\ 
&& + 3194 \vec{v}_{2}\cdot\vec{n} \vec{v}_{1} - 13618 \vec{v}_{1}\cdot\vec{n} \vec{v}_{2} + 2726 \vec{v}_{2}\cdot\vec{n} \vec{v}_{2} -9086 \vec{v}_{1}\cdot\vec{n} \vec{v}_{2}\cdot\vec{n} \vec{n} \nn\\ 
&& + 2614 ( \vec{v}_{1}\cdot\vec{n})^{2} \vec{n} + 2342 ( \vec{v}_{2}\cdot\vec{n})^{2} \vec{n} \big) + 4 \vec{S}_{1}\times\vec{n}\cdot\vec{v}_{2} \big( 643 v_{1}^2 \vec{n} - 1756 \vec{v}_{1}\cdot\vec{v}_{2} \vec{n} \nn\\ 
&& + 835 v_{2}^2 \vec{n} - 238 \vec{v}_{1}\cdot\vec{n} \vec{v}_{1} - 532 \vec{v}_{2}\cdot\vec{n} \vec{v}_{1} + 98 \vec{v}_{1}\cdot\vec{n} \vec{v}_{2} - 298 \vec{v}_{2}\cdot\vec{n} \vec{v}_{2} \nn\\ 
&& + 2086 \vec{v}_{1}\cdot\vec{n} \vec{v}_{2}\cdot\vec{n} \vec{n} - 496 ( \vec{v}_{1}\cdot\vec{n})^{2} \vec{n} - 2158 ( \vec{v}_{2}\cdot\vec{n})^{2} \vec{n} \big) + 6 \vec{S}_{1}\times\vec{v}_{1}\cdot\vec{v}_{2} \big( 1110 \vec{v}_{1} \nn\\ 
&& - 154 \vec{v}_{2} -1307 \vec{v}_{1}\cdot\vec{n} \vec{n} + 205 \vec{v}_{2}\cdot\vec{n} \vec{n} \big) + 4 \vec{S}_{1}\times\vec{n} \big( 2160 \vec{v}_{1}\cdot\vec{n} v_{1}^2 - 891 v_{1}^2 \vec{v}_{2}\cdot\vec{n} \nn\\ 
&& - 3912 \vec{v}_{1}\cdot\vec{n} \vec{v}_{1}\cdot\vec{v}_{2} + 3028 \vec{v}_{2}\cdot\vec{n} \vec{v}_{1}\cdot\vec{v}_{2} + 2855 \vec{v}_{1}\cdot\vec{n} v_{2}^2 \nn\\ 
&& - 1935 \vec{v}_{2}\cdot\vec{n} v_{2}^2 -437 ( \vec{v}_{1}\cdot\vec{n})^{3} + 2118 \vec{v}_{2}\cdot\vec{n} ( \vec{v}_{1}\cdot\vec{n})^{2} - 5744 \vec{v}_{1}\cdot\vec{n} ( \vec{v}_{2}\cdot\vec{n})^{2} \nn\\ 
&& + 2528 ( \vec{v}_{2}\cdot\vec{n})^{3} \big) - \vec{S}_{1}\times\vec{v}_{1} \big( 155 v_{1}^2 + 1922 \vec{v}_{1}\cdot\vec{v}_{2} - 1261 v_{2}^2 + 7814 \vec{v}_{1}\cdot\vec{n} \vec{v}_{2}\cdot\vec{n} \nn\\ 
&& - 4042 ( \vec{v}_{1}\cdot\vec{n})^{2} - 2140 ( \vec{v}_{2}\cdot\vec{n})^{2} \big) - 4 \vec{S}_{1}\times\vec{v}_{2} \big( 526 v_{1}^2 - 584 \vec{v}_{1}\cdot\vec{v}_{2} \nn\\ 
&& + 565 v_{2}^2 -776 \vec{v}_{1}\cdot\vec{n} \vec{v}_{2}\cdot\vec{n} - 230 ( \vec{v}_{1}\cdot\vec{n})^{2} + 418 ( \vec{v}_{2}\cdot\vec{n})^{2} \big) \Big] \nn\\ 
&& + 	\frac{G^2 m_{2}{}^2}{192 m_{1} r{}^2} \Big[ \vec{S}_{1}\times\vec{n}\cdot\vec{v}_{1} \big( 24 v_{1}^2 \vec{n} + 272 \vec{v}_{1}\cdot\vec{v}_{2} \vec{n} - 141 v_{2}^2 \vec{n} - 120 \vec{v}_{1}\cdot\vec{n} \vec{v}_{1} + 32 \vec{v}_{2}\cdot\vec{n} \vec{v}_{1} \nn\\ 
&& + 272 \vec{v}_{1}\cdot\vec{n} \vec{v}_{2} + 1314 \vec{v}_{2}\cdot\vec{n} \vec{v}_{2} -512 \vec{v}_{1}\cdot\vec{n} \vec{v}_{2}\cdot\vec{n} \vec{n} - 963 ( \vec{v}_{2}\cdot\vec{n})^{2} \vec{n} \big) \nn\\ 
&& - 2 \vec{S}_{1}\times\vec{n}\cdot\vec{v}_{2} \big( 24 v_{1}^2 \vec{n} - 16 \vec{v}_{1}\cdot\vec{v}_{2} \vec{n} - 495 v_{2}^2 \vec{n} + 48 \vec{v}_{1}\cdot\vec{n} \vec{v}_{1} - 19 \vec{v}_{2}\cdot\vec{n} \vec{v}_{1} \nn\\ 
&& - 16 \vec{v}_{1}\cdot\vec{n} \vec{v}_{2} - 2076 \vec{v}_{2}\cdot\vec{n} \vec{v}_{2} + 256 \vec{v}_{1}\cdot\vec{n} \vec{v}_{2}\cdot\vec{n} \vec{n} + 1485 ( \vec{v}_{2}\cdot\vec{n})^{2} \vec{n} \big) \nn\\ 
&& - 4 \vec{S}_{1}\times\vec{v}_{1}\cdot\vec{v}_{2} \big( 572 \vec{v}_{1} - 1196 \vec{v}_{2} + 56 \vec{v}_{1}\cdot\vec{n} \vec{n} + 73 \vec{v}_{2}\cdot\vec{n} \vec{n} \big) + \vec{S}_{1}\times\vec{n} \big( 24 \vec{v}_{1}\cdot\vec{n} v_{1}^2 \nn\\ 
&& - 272 v_{1}^2 \vec{v}_{2}\cdot\vec{n} + 272 \vec{v}_{1}\cdot\vec{n} \vec{v}_{1}\cdot\vec{v}_{2} + 3192 \vec{v}_{2}\cdot\vec{n} \vec{v}_{1}\cdot\vec{v}_{2} - 303 \vec{v}_{1}\cdot\vec{n} v_{2}^2 \nn\\ 
&& - 7896 \vec{v}_{2}\cdot\vec{n} v_{2}^2 -256 \vec{v}_{2}\cdot\vec{n} ( \vec{v}_{1}\cdot\vec{n})^{2} - 609 \vec{v}_{1}\cdot\vec{n} ( \vec{v}_{2}\cdot\vec{n})^{2} - 456 ( \vec{v}_{2}\cdot\vec{n})^{3} \big) \nn\\ 
&& + 3 \vec{S}_{1}\times\vec{v}_{1} \big( 92 v_{1}^2 - 208 \vec{v}_{1}\cdot\vec{v}_{2} - 147 v_{2}^2 + 160 \vec{v}_{1}\cdot\vec{n} \vec{v}_{2}\cdot\vec{n} + 8 ( \vec{v}_{1}\cdot\vec{n})^{2} \nn\\ 
&& - 483 ( \vec{v}_{2}\cdot\vec{n})^{2} \big) + 2 \vec{S}_{1}\times\vec{v}_{2} \big( 584 v_{1}^2 - 1528 \vec{v}_{1}\cdot\vec{v}_{2} + 1554 v_{2}^2 + 245 \vec{v}_{1}\cdot\vec{n} \vec{v}_{2}\cdot\vec{n} \nn\\ 
&& + 56 ( \vec{v}_{1}\cdot\vec{n})^{2} - 162 ( \vec{v}_{2}\cdot\vec{n})^{2} \big) \Big] - 	\frac{G^2 m_{2}}{96 r{}^2} \Big[ 8 \vec{S}_{2}\times\vec{n}\cdot\vec{v}_{1} \big( 35 v_{2}^2 \vec{n} - 24 \vec{v}_{2}\cdot\vec{n} \vec{v}_{1} \nn\\ 
&& + 94 \vec{v}_{2}\cdot\vec{n} \vec{v}_{2} -71 ( \vec{v}_{2}\cdot\vec{n})^{2} \vec{n} \big) + \vec{S}_{2}\times\vec{n}\cdot\vec{v}_{2} \big( 531 v_{1}^2 \vec{n} - 1024 \vec{v}_{1}\cdot\vec{v}_{2} \vec{n} - 3 v_{2}^2 \vec{n} \nn\\ 
&& + 1350 \vec{v}_{1}\cdot\vec{n} \vec{v}_{1} - 4414 \vec{v}_{2}\cdot\vec{n} \vec{v}_{1} - 922 \vec{v}_{1}\cdot\vec{n} \vec{v}_{2} + 3126 \vec{v}_{2}\cdot\vec{n} \vec{v}_{2} + 3682 \vec{v}_{1}\cdot\vec{n} \vec{v}_{2}\cdot\vec{n} \vec{n} \nn\\ 
&& - 1362 ( \vec{v}_{1}\cdot\vec{n})^{2} \vec{n} + 351 ( \vec{v}_{2}\cdot\vec{n})^{2} \vec{n} \big) - 3 \vec{S}_{2}\times\vec{v}_{1}\cdot\vec{v}_{2} \big( 107 \vec{v}_{1} - 156 \vec{v}_{2} + 17 \vec{v}_{1}\cdot\vec{n} \vec{n} \nn\\ 
&& - 194 \vec{v}_{2}\cdot\vec{n} \vec{n} \big) - 4 \vec{S}_{2}\times\vec{n} \big( 24 v_{1}^2 \vec{v}_{2}\cdot\vec{n} - 188 \vec{v}_{2}\cdot\vec{n} \vec{v}_{1}\cdot\vec{v}_{2} - 70 \vec{v}_{1}\cdot\vec{n} v_{2}^2 + 33 \vec{v}_{2}\cdot\vec{n} v_{2}^2 \nn\\ 
&& + 142 \vec{v}_{1}\cdot\vec{n} ( \vec{v}_{2}\cdot\vec{n})^{2} - 115 ( \vec{v}_{2}\cdot\vec{n})^{3} \big) + 3 \vec{S}_{2}\times\vec{v}_{2} \big( 193 v_{1}^2 - 358 \vec{v}_{1}\cdot\vec{v}_{2} - 77 v_{2}^2 \nn\\ 
&& + 344 \vec{v}_{1}\cdot\vec{n} \vec{v}_{2}\cdot\vec{n} - 388 ( \vec{v}_{1}\cdot\vec{n})^{2} + 67 ( \vec{v}_{2}\cdot\vec{n})^{2} \big) \Big] + 	\frac{G^2 m_{1}}{384 r{}^2} \Big[ \vec{S}_{2}\times\vec{n}\cdot\vec{v}_{1} \big( 3281 v_{1}^2 \vec{n} \nn\\ 
&& + 8942 \vec{v}_{1}\cdot\vec{v}_{2} \vec{n} - 917 v_{2}^2 \vec{n} + 6340 \vec{v}_{1}\cdot\vec{n} \vec{v}_{1} + 3644 \vec{v}_{2}\cdot\vec{n} \vec{v}_{1} + 4004 \vec{v}_{1}\cdot\vec{n} \vec{v}_{2} \nn\\ 
&& + 2750 \vec{v}_{2}\cdot\vec{n} \vec{v}_{2} -14408 \vec{v}_{1}\cdot\vec{n} \vec{v}_{2}\cdot\vec{n} \vec{n} + 3886 ( \vec{v}_{1}\cdot\vec{n})^{2} \vec{n} + 4454 ( \vec{v}_{2}\cdot\vec{n})^{2} \vec{n} \big) \nn\\ 
&& - 2 \vec{S}_{2}\times\vec{n}\cdot\vec{v}_{2} \big( 7670 v_{1}^2 \vec{n} - 2293 \vec{v}_{1}\cdot\vec{v}_{2} \vec{n} - 182 v_{2}^2 \vec{n} + 14077 \vec{v}_{1}\cdot\vec{n} \vec{v}_{1} + 1367 \vec{v}_{2}\cdot\vec{n} \vec{v}_{1} \nn\\ 
&& - 2194 \vec{v}_{1}\cdot\vec{n} \vec{v}_{2} - 631 \vec{v}_{2}\cdot\vec{n} \vec{v}_{2} -224 \vec{v}_{1}\cdot\vec{n} \vec{v}_{2}\cdot\vec{n} \vec{n} - 137 ( \vec{v}_{1}\cdot\vec{n})^{2} \vec{n} + 2408 ( \vec{v}_{2}\cdot\vec{n})^{2} \vec{n} \big) \nn\\ 
&& + 2 \vec{S}_{2}\times\vec{v}_{1}\cdot\vec{v}_{2} \big( 6790 \vec{v}_{1} - 2805 \vec{v}_{2} -2237 \vec{v}_{1}\cdot\vec{n} \vec{n} + 2100 \vec{v}_{2}\cdot\vec{n} \vec{n} \big) \nn\\ 
&& + \vec{S}_{2}\times\vec{n} \big( 7983 \vec{v}_{1}\cdot\vec{n} v_{1}^2 - 2619 v_{1}^2 \vec{v}_{2}\cdot\vec{n} + 1986 \vec{v}_{1}\cdot\vec{n} \vec{v}_{1}\cdot\vec{v}_{2} + 10090 \vec{v}_{2}\cdot\vec{n} \vec{v}_{1}\cdot\vec{v}_{2} \nn\\ 
&& + 2045 \vec{v}_{1}\cdot\vec{n} v_{2}^2 - 6009 \vec{v}_{2}\cdot\vec{n} v_{2}^2 + 6330 ( \vec{v}_{1}\cdot\vec{n})^{3} - 19692 \vec{v}_{2}\cdot\vec{n} ( \vec{v}_{1}\cdot\vec{n})^{2} \nn\\ 
&& + 3958 \vec{v}_{1}\cdot\vec{n} ( \vec{v}_{2}\cdot\vec{n})^{2} + 3960 ( \vec{v}_{2}\cdot\vec{n})^{3} \big) - \vec{S}_{2}\times\vec{v}_{1} \big( 2939 v_{1}^2 - 3358 \vec{v}_{1}\cdot\vec{v}_{2} \nn\\ 
&& + 2597 v_{2}^2 -14206 \vec{v}_{1}\cdot\vec{n} \vec{v}_{2}\cdot\vec{n} + 6356 ( \vec{v}_{1}\cdot\vec{n})^{2} + 2570 ( \vec{v}_{2}\cdot\vec{n})^{2} \big) - 2 \vec{S}_{2}\times\vec{v}_{2} \big( 1999 v_{1}^2 \nn\\ 
&& - 2203 \vec{v}_{1}\cdot\vec{v}_{2} - 403 v_{2}^2 + 7361 \vec{v}_{1}\cdot\vec{n} \vec{v}_{2}\cdot\vec{n} - 4913 ( \vec{v}_{1}\cdot\vec{n})^{2} - 856 ( \vec{v}_{2}\cdot\vec{n})^{2} \big) \Big]\nn\\ && - 	\frac{G^3 m_{1} m_{2}}{7200 r{}^3} \Big[ 227766 \vec{S}_{1}\times\vec{n}\cdot\vec{v}_{1} \vec{n} - 41658 \vec{S}_{1}\times\vec{n}\cdot\vec{v}_{2} \vec{n} - 6 \vec{S}_{1}\times\vec{n} \big( 49539 \vec{v}_{1}\cdot\vec{n} \nn\\ 
&& - 76382 \vec{v}_{2}\cdot\vec{n} \big) + 214175 \vec{S}_{1}\times\vec{v}_{1} - 78450 \vec{S}_{1}\times\vec{v}_{2} \Big] \nn\\ 
&& + 	\frac{13 G^3 m_{1} m_{2}}{2 r{}^3} {\Big( \frac{1}{\epsilon} - 3\log \frac{r}{R_0} \Big)} \Big[ 3 \vec{S}_{1}\times\vec{n} \big( \vec{v}_{1}\cdot\vec{n} - \vec{v}_{2}\cdot\vec{n} \big) - \vec{S}_{1}\times\vec{v}_{1} + \vec{S}_{1}\times\vec{v}_{2} \Big] \nn\\ 
&& + 	\frac{29 G^3 m_{2}{}^2}{6 r{}^3} {\Big( \frac{1}{\epsilon} - 3\log \frac{r}{R_0} \Big)} \Big[ 3 \vec{S}_{1}\times\vec{n} \big( \vec{v}_{1}\cdot\vec{n} - \vec{v}_{2}\cdot\vec{n} \big) - \vec{S}_{1}\times\vec{v}_{1} + \vec{S}_{1}\times\vec{v}_{2} \Big] \nn\\ 
&& - 	\frac{G^3 m_{2}{}^3}{1800 m_{1} r{}^3} \Big[ 16200 \vec{S}_{1}\times\vec{n}\cdot\vec{v}_{1} \vec{n} + 19149 \vec{S}_{1}\times\vec{n}\cdot\vec{v}_{2} \vec{n} + 9 \vec{S}_{1}\times\vec{n} \big( 1800 \vec{v}_{1}\cdot\vec{n} \nn\\ 
&& - 1289 \vec{v}_{2}\cdot\vec{n} \big) - 450 \vec{S}_{1}\times\vec{v}_{1} + 14525 \vec{S}_{1}\times\vec{v}_{2} \Big] \nn\\ 
&& - 	\frac{G^3 m_{2}{}^2}{1152 r{}^3} \Big[ 27 \vec{S}_{1}\times\vec{n}\cdot\vec{v}_{1} {(3163 - 350 \pi^2)} \vec{n} - 12 \vec{S}_{1}\times\vec{n}\cdot\vec{v}_{2} {(8213 - 675 \pi^2)} \vec{n} \nn\\ 
&& - 6 \vec{S}_{1}\times\vec{n} \big( {(11267 + 504 \pi^2)} \vec{v}_{1}\cdot\vec{n} - {(5635 - 810 \pi^2)} \vec{v}_{2}\cdot\vec{n} \big) \nn\\ 
&& + {(31805 - 2142 \pi^2)} \vec{S}_{1}\times\vec{v}_{1} + 2 {(13 + 2160 \pi^2)} \vec{S}_{1}\times\vec{v}_{2} \Big] \nn\\ 
&& - 	\frac{29 G^3 m_{1}{}^2}{6 r{}^3} {\Big( \frac{1}{\epsilon} - 3\log \frac{r}{R_0} \Big)} \Big[ 3 \vec{S}_{2}\times\vec{n} \big( \vec{v}_{1}\cdot\vec{n} - \vec{v}_{2}\cdot\vec{n} \big) - \vec{S}_{2}\times\vec{v}_{1} + \vec{S}_{2}\times\vec{v}_{2} \Big] \nn\\ 
&& - 	\frac{29 G^3 m_{1} m_{2}}{6 r{}^3} {\Big( \frac{1}{\epsilon} - 3\log \frac{r}{R_0} \Big)} \Big[ 3 \vec{S}_{2}\times\vec{n} \big( \vec{v}_{1}\cdot\vec{n} - \vec{v}_{2}\cdot\vec{n} \big) - \vec{S}_{2}\times\vec{v}_{1} + \vec{S}_{2}\times\vec{v}_{2} \Big] \nn\\ 
&& - 	\frac{G^3 m_{1}{}^2}{14400 r{}^3} \Big[ 324234 \vec{S}_{2}\times\vec{n}\cdot\vec{v}_{1} \vec{n} - 181158 \vec{S}_{2}\times\vec{n}\cdot\vec{v}_{2} \vec{n} + 3 \vec{S}_{2}\times\vec{n} \big( 339703 \vec{v}_{1}\cdot\vec{n} \nn\\ 
&& - 227636 \vec{v}_{2}\cdot\vec{n} \big) - 104425 \vec{S}_{2}\times\vec{v}_{1} + 13350 \vec{S}_{2}\times\vec{v}_{2} \Big] + 	\frac{G^3 m_{2}{}^2}{14400 r{}^3} \Big[ 864 \vec{S}_{2}\times\vec{n}\cdot\vec{v}_{1} \vec{n} \nn\\ 
&& + 230691 \vec{S}_{2}\times\vec{n}\cdot\vec{v}_{2} \vec{n} + 288 \vec{S}_{2}\times\vec{n} \big( 3 \vec{v}_{1}\cdot\vec{n} + 857 \vec{v}_{2}\cdot\vec{n} \big) + 155875 \vec{S}_{2}\times\vec{v}_{2} \Big] \nn\\ 
&& - 	\frac{G^3 m_{1} m_{2}}{1152 r{}^3} \Big[ 9 \vec{S}_{2}\times\vec{n}\cdot\vec{v}_{1} {(15415 - 1896 \pi^2)} \vec{n} - 54 \vec{S}_{2}\times\vec{n}\cdot\vec{v}_{2} {(2769 - 53 \pi^2)} \vec{n} \nn\\ 
&& + 3 \vec{S}_{2}\times\vec{n} \big( {(29371 - 3240 \pi^2)} \vec{v}_{1}\cdot\vec{n} - {(26609 - 684 \pi^2)} \vec{v}_{2}\cdot\vec{n} \big) \nn\\ 
&& + {(22607 - 2448 \pi^2)} \vec{S}_{2}\times\vec{v}_{1} - 30 {(1483 - 9 \pi^2)} \vec{S}_{2}\times\vec{v}_{2} \Big],
\eea
\bea
\stackrel{(1)}{ \Delta \vec{x}_1}_{(3,1)}&=& 	\frac{G m_{2}}{192 m_{1}} \Big[ 3 \vec{S}_{1}\times\vec{n}\cdot\vec{v}_{1} \big( 3 \vec{v}_{1}\cdot\vec{a}_{2} \vec{v}_{1} - 15 \vec{v}_{2}\cdot\vec{a}_{2} \vec{v}_{1} - 12 \vec{v}_{1}\cdot\vec{a}_{2} \vec{v}_{2} + 52 \vec{v}_{2}\cdot\vec{a}_{2} \vec{v}_{2} + 14 v_{1}^2 \vec{a}_{2} \nn\\ 
&& - 24 \vec{v}_{1}\cdot\vec{v}_{2} \vec{a}_{2} + 66 v_{2}^2 \vec{a}_{2} + 2 v_{1}^2 \vec{a}_{2}\cdot\vec{n} \vec{n} - 8 \vec{v}_{1}\cdot\vec{v}_{2} \vec{a}_{2}\cdot\vec{n} \vec{n} + 18 v_{2}^2 \vec{a}_{2}\cdot\vec{n} \vec{n} \nn\\ 
&& - 20 \vec{v}_{2}\cdot\vec{n} \vec{v}_{1}\cdot\vec{a}_{2} \vec{n} + 4 \vec{v}_{1}\cdot\vec{n} \vec{v}_{2}\cdot\vec{a}_{2} \vec{n} + 44 \vec{v}_{2}\cdot\vec{n} \vec{v}_{2}\cdot\vec{a}_{2} \vec{n} + 5 \vec{v}_{1}\cdot\vec{n} \vec{a}_{2}\cdot\vec{n} \vec{v}_{1} \nn\\ 
&& - \vec{v}_{2}\cdot\vec{n} \vec{a}_{2}\cdot\vec{n} \vec{v}_{1} + 4 \vec{v}_{1}\cdot\vec{n} \vec{a}_{2}\cdot\vec{n} \vec{v}_{2} - 40 \vec{v}_{2}\cdot\vec{n} \vec{a}_{2}\cdot\vec{n} \vec{v}_{2} + 4 \vec{v}_{1}\cdot\vec{n} \vec{v}_{2}\cdot\vec{n} \vec{a}_{2} \nn\\ 
&& - 56 ( \vec{v}_{2}\cdot\vec{n})^{2} \vec{a}_{2} -12 \vec{v}_{1}\cdot\vec{n} \vec{v}_{2}\cdot\vec{n} \vec{a}_{2}\cdot\vec{n} \vec{n} - 48 \vec{a}_{2}\cdot\vec{n} ( \vec{v}_{2}\cdot\vec{n})^{2} \vec{n} \big) - 3 \vec{S}_{1}\times\vec{n}\cdot\vec{a}_{1} \big( v_{2}^2 \vec{v}_{1} \nn\\ 
&& - 4 v_{2}^2 \vec{v}_{2} + 4 \vec{v}_{2}\cdot\vec{n} v_{2}^2 \vec{n} - ( \vec{v}_{2}\cdot\vec{n})^{2} \vec{v}_{1} + 4 ( \vec{v}_{2}\cdot\vec{n})^{2} \vec{v}_{2} -4 ( \vec{v}_{2}\cdot\vec{n})^{3} \vec{n} \big) \nn\\ 
&& + 3 \vec{S}_{1}\times\vec{v}_{1}\cdot\vec{a}_{1} \big( 7 v_{2}^2 \vec{n} - 16 \vec{v}_{2}\cdot\vec{n} \vec{v}_{1} - 34 \vec{v}_{2}\cdot\vec{n} \vec{v}_{2} -7 ( \vec{v}_{2}\cdot\vec{n})^{2} \vec{n} \big) \nn\\ 
&& + 6 \vec{S}_{1}\times\vec{n}\cdot\vec{v}_{2} \big( 6 \vec{v}_{1}\cdot\vec{a}_{2} \vec{v}_{1} + 2 \vec{v}_{2}\cdot\vec{a}_{2} \vec{v}_{1} - 6 \vec{v}_{1}\cdot\vec{a}_{2} \vec{v}_{2} - 6 \vec{v}_{2}\cdot\vec{a}_{2} \vec{v}_{2} + 3 v_{1}^2 \vec{a}_{2} \nn\\ 
&& - 2 \vec{v}_{1}\cdot\vec{v}_{2} \vec{a}_{2} - 19 v_{2}^2 \vec{a}_{2} + v_{1}^2 \vec{a}_{2}\cdot\vec{n} \vec{n} + 2 \vec{v}_{1}\cdot\vec{v}_{2} \vec{a}_{2}\cdot\vec{n} \vec{n} - 9 v_{2}^2 \vec{a}_{2}\cdot\vec{n} \vec{n} + 6 \vec{v}_{2}\cdot\vec{n} \vec{v}_{1}\cdot\vec{a}_{2} \vec{n} \nn\\ 
&& - 2 \vec{v}_{1}\cdot\vec{n} \vec{v}_{2}\cdot\vec{a}_{2} \vec{n} - 18 \vec{v}_{2}\cdot\vec{n} \vec{v}_{2}\cdot\vec{a}_{2} \vec{n} + 2 \vec{v}_{1}\cdot\vec{n} \vec{a}_{2}\cdot\vec{n} \vec{v}_{1} - 2 \vec{v}_{2}\cdot\vec{n} \vec{a}_{2}\cdot\vec{n} \vec{v}_{1} \nn\\ 
&& - 2 \vec{v}_{1}\cdot\vec{n} \vec{a}_{2}\cdot\vec{n} \vec{v}_{2} + 8 \vec{v}_{2}\cdot\vec{n} \vec{a}_{2}\cdot\vec{n} \vec{v}_{2} - 2 \vec{v}_{1}\cdot\vec{n} \vec{v}_{2}\cdot\vec{n} \vec{a}_{2} + 16 ( \vec{v}_{2}\cdot\vec{n})^{2} \vec{a}_{2} \nn\\ 
&& + 6 \vec{v}_{1}\cdot\vec{n} \vec{v}_{2}\cdot\vec{n} \vec{a}_{2}\cdot\vec{n} \vec{n} + 24 \vec{a}_{2}\cdot\vec{n} ( \vec{v}_{2}\cdot\vec{n})^{2} \vec{n} \big) - 3 \vec{S}_{1}\times\vec{v}_{1}\cdot\vec{v}_{2} \big( 8 \vec{v}_{1}\cdot\vec{a}_{2} \vec{n} - 16 \vec{v}_{2}\cdot\vec{a}_{2} \vec{n} \nn\\ 
&& + 9 \vec{a}_{2}\cdot\vec{n} \vec{v}_{1} - 36 \vec{a}_{2}\cdot\vec{n} \vec{v}_{2} + 16 \vec{v}_{1}\cdot\vec{n} \vec{a}_{2} - 60 \vec{v}_{2}\cdot\vec{n} \vec{a}_{2} + 12 \vec{v}_{2}\cdot\vec{n} \vec{a}_{2}\cdot\vec{n} \vec{n} \big) \nn\\ 
&& + 6 \vec{S}_{1}\times\vec{a}_{1}\cdot\vec{v}_{2} \big( 6 v_{2}^2 \vec{n} - 7 \vec{v}_{2}\cdot\vec{n} \vec{v}_{1} - 4 \vec{v}_{2}\cdot\vec{n} \vec{v}_{2} -6 ( \vec{v}_{2}\cdot\vec{n})^{2} \vec{n} \big) \nn\\ 
&& + 12 \vec{S}_{1}\times\vec{n}\cdot\vec{a}_{2} \big( 2 \vec{v}_{1}\cdot\vec{v}_{2} \vec{v}_{1} + v_{2}^2 \vec{v}_{1} + v_{1}^2 \vec{v}_{2} - 2 \vec{v}_{1}\cdot\vec{v}_{2} \vec{v}_{2} - 5 v_{2}^2 \vec{v}_{2} + v_{1}^2 \vec{v}_{2}\cdot\vec{n} \vec{n} \nn\\ 
&& + 2 \vec{v}_{2}\cdot\vec{n} \vec{v}_{1}\cdot\vec{v}_{2} \vec{n} - \vec{v}_{1}\cdot\vec{n} v_{2}^2 \vec{n} - 9 \vec{v}_{2}\cdot\vec{n} v_{2}^2 \vec{n} + 2 \vec{v}_{1}\cdot\vec{n} \vec{v}_{2}\cdot\vec{n} \vec{v}_{1} - ( \vec{v}_{2}\cdot\vec{n})^{2} \vec{v}_{1} \nn\\ 
&& - 2 \vec{v}_{1}\cdot\vec{n} \vec{v}_{2}\cdot\vec{n} \vec{v}_{2} + 4 ( \vec{v}_{2}\cdot\vec{n})^{2} \vec{v}_{2} + 3 \vec{v}_{1}\cdot\vec{n} ( \vec{v}_{2}\cdot\vec{n})^{2} \vec{n} + 8 ( \vec{v}_{2}\cdot\vec{n})^{3} \vec{n} \big) \nn\\ 
&& + 3 \vec{S}_{1}\times\vec{v}_{1}\cdot\vec{a}_{2} \big( 2 v_{1}^2 \vec{n} - 6 v_{2}^2 \vec{n} - 35 \vec{v}_{1}\cdot\vec{n} \vec{v}_{1} - 33 \vec{v}_{2}\cdot\vec{n} \vec{v}_{1} - 20 \vec{v}_{1}\cdot\vec{n} \vec{v}_{2} + 112 \vec{v}_{2}\cdot\vec{n} \vec{v}_{2} \nn\\ 
&& + 4 \vec{v}_{1}\cdot\vec{n} \vec{v}_{2}\cdot\vec{n} \vec{n} + 4 ( \vec{v}_{2}\cdot\vec{n})^{2} \vec{n} \big) - 6 \vec{S}_{1}\times\vec{v}_{2}\cdot\vec{a}_{2} \big( v_{1}^2 \vec{n} + 2 \vec{v}_{1}\cdot\vec{v}_{2} \vec{n} - 9 v_{2}^2 \vec{n} + 2 \vec{v}_{1}\cdot\vec{n} \vec{v}_{1} \nn\\ 
&& - 2 \vec{v}_{2}\cdot\vec{n} \vec{v}_{1} - 2 \vec{v}_{1}\cdot\vec{n} \vec{v}_{2} + 8 \vec{v}_{2}\cdot\vec{n} \vec{v}_{2} + 2 \vec{v}_{1}\cdot\vec{n} \vec{v}_{2}\cdot\vec{n} \vec{n} + 8 ( \vec{v}_{2}\cdot\vec{n})^{2} \vec{n} \big) \nn\\ 
&& + 3 \vec{S}_{1}\times\vec{n} \big( \vec{v}_{1}\cdot\vec{a}_{1} v_{2}^2 - 4 \vec{a}_{1}\cdot\vec{v}_{2} v_{2}^2 + 11 v_{1}^2 \vec{v}_{1}\cdot\vec{a}_{2} - 24 \vec{v}_{1}\cdot\vec{v}_{2} \vec{v}_{1}\cdot\vec{a}_{2} + 72 v_{2}^2 \vec{v}_{1}\cdot\vec{a}_{2} \nn\\ 
&& + v_{1}^2 \vec{v}_{2}\cdot\vec{a}_{2} + 72 \vec{v}_{1}\cdot\vec{v}_{2} \vec{v}_{2}\cdot\vec{a}_{2} - 176 v_{2}^2 \vec{v}_{2}\cdot\vec{a}_{2} + 4 \vec{a}_{1}\cdot\vec{n} \vec{v}_{2}\cdot\vec{n} v_{2}^2 + 5 \vec{v}_{1}\cdot\vec{n} v_{1}^2 \vec{a}_{2}\cdot\vec{n} \nn\\ 
&& - 9 v_{1}^2 \vec{v}_{2}\cdot\vec{n} \vec{a}_{2}\cdot\vec{n} - 8 \vec{v}_{1}\cdot\vec{n} \vec{v}_{1}\cdot\vec{v}_{2} \vec{a}_{2}\cdot\vec{n} - 40 \vec{v}_{2}\cdot\vec{n} \vec{v}_{1}\cdot\vec{v}_{2} \vec{a}_{2}\cdot\vec{n} + 12 \vec{v}_{1}\cdot\vec{n} v_{2}^2 \vec{a}_{2}\cdot\vec{n} \nn\\ 
&& + 104 \vec{v}_{2}\cdot\vec{n} v_{2}^2 \vec{a}_{2}\cdot\vec{n} - 8 \vec{v}_{1}\cdot\vec{n} \vec{v}_{2}\cdot\vec{n} \vec{v}_{1}\cdot\vec{a}_{2} + 24 \vec{v}_{1}\cdot\vec{n} \vec{v}_{2}\cdot\vec{n} \vec{v}_{2}\cdot\vec{a}_{2} - 4 \vec{v}_{2}\cdot\vec{a}_{2} ( \vec{v}_{1}\cdot\vec{n})^{2} \nn\\ 
&& - \vec{v}_{1}\cdot\vec{a}_{1} ( \vec{v}_{2}\cdot\vec{n})^{2} + 4 \vec{a}_{1}\cdot\vec{v}_{2} ( \vec{v}_{2}\cdot\vec{n})^{2} - 56 \vec{v}_{1}\cdot\vec{a}_{2} ( \vec{v}_{2}\cdot\vec{n})^{2} + 128 \vec{v}_{2}\cdot\vec{a}_{2} ( \vec{v}_{2}\cdot\vec{n})^{2} \nn\\ 
&& + 12 \vec{v}_{2}\cdot\vec{n} \vec{a}_{2}\cdot\vec{n} ( \vec{v}_{1}\cdot\vec{n})^{2} - 4 \vec{a}_{1}\cdot\vec{n} ( \vec{v}_{2}\cdot\vec{n})^{3} - 48 \vec{v}_{1}\cdot\vec{n} \vec{a}_{2}\cdot\vec{n} ( \vec{v}_{2}\cdot\vec{n})^{2} \nn\\ 
&& - 80 \vec{a}_{2}\cdot\vec{n} ( \vec{v}_{2}\cdot\vec{n})^{3} \big) - 3 \vec{S}_{1}\times\vec{v}_{1} \big( 48 \vec{v}_{1}\cdot\vec{a}_{1} \vec{v}_{2}\cdot\vec{n} - 34 \vec{v}_{2}\cdot\vec{n} \vec{a}_{1}\cdot\vec{v}_{2} + 7 \vec{a}_{1}\cdot\vec{n} v_{2}^2 \nn\\ 
&& - 10 v_{1}^2 \vec{a}_{2}\cdot\vec{n} - 69 \vec{v}_{1}\cdot\vec{v}_{2} \vec{a}_{2}\cdot\vec{n} + 88 v_{2}^2 \vec{a}_{2}\cdot\vec{n} + 34 \vec{v}_{1}\cdot\vec{n} \vec{v}_{1}\cdot\vec{a}_{2} - 137 \vec{v}_{2}\cdot\vec{n} \vec{v}_{1}\cdot\vec{a}_{2} \nn\\ 
&& - 49 \vec{v}_{1}\cdot\vec{n} \vec{v}_{2}\cdot\vec{a}_{2} + 224 \vec{v}_{2}\cdot\vec{n} \vec{v}_{2}\cdot\vec{a}_{2} + 17 \vec{v}_{1}\cdot\vec{n} \vec{v}_{2}\cdot\vec{n} \vec{a}_{2}\cdot\vec{n} + 7 \vec{a}_{2}\cdot\vec{n} ( \vec{v}_{1}\cdot\vec{n})^{2} \nn\\ 
&& - 7 \vec{a}_{1}\cdot\vec{n} ( \vec{v}_{2}\cdot\vec{n})^{2} - 60 \vec{a}_{2}\cdot\vec{n} ( \vec{v}_{2}\cdot\vec{n})^{2} \big) - 2 \vec{S}_{1}\times\vec{a}_{1} \big( 24 v_{1}^2 \vec{v}_{2}\cdot\vec{n} - 102 \vec{v}_{2}\cdot\vec{n} \vec{v}_{1}\cdot\vec{v}_{2} \nn\\ 
&& + 21 \vec{v}_{1}\cdot\vec{n} v_{2}^2 + 132 \vec{v}_{2}\cdot\vec{n} v_{2}^2 -21 \vec{v}_{1}\cdot\vec{n} ( \vec{v}_{2}\cdot\vec{n})^{2} - 28 ( \vec{v}_{2}\cdot\vec{n})^{3} \big) \nn\\ 
&& - 3 \vec{S}_{1}\times\vec{v}_{2} \big( 14 \vec{v}_{1}\cdot\vec{a}_{1} \vec{v}_{2}\cdot\vec{n} + 8 \vec{v}_{2}\cdot\vec{n} \vec{a}_{1}\cdot\vec{v}_{2} - 12 \vec{a}_{1}\cdot\vec{n} v_{2}^2 + 37 v_{1}^2 \vec{a}_{2}\cdot\vec{n} \nn\\ 
&& + 60 \vec{v}_{1}\cdot\vec{v}_{2} \vec{a}_{2}\cdot\vec{n} - 134 v_{2}^2 \vec{a}_{2}\cdot\vec{n} - 24 \vec{v}_{1}\cdot\vec{n} \vec{v}_{1}\cdot\vec{a}_{2} + 108 \vec{v}_{2}\cdot\vec{n} \vec{v}_{1}\cdot\vec{a}_{2} + 92 \vec{v}_{1}\cdot\vec{n} \vec{v}_{2}\cdot\vec{a}_{2} \nn\\ 
&& - 348 \vec{v}_{2}\cdot\vec{n} \vec{v}_{2}\cdot\vec{a}_{2} -12 \vec{v}_{1}\cdot\vec{n} \vec{v}_{2}\cdot\vec{n} \vec{a}_{2}\cdot\vec{n} - 12 \vec{a}_{2}\cdot\vec{n} ( \vec{v}_{1}\cdot\vec{n})^{2} + 12 \vec{a}_{1}\cdot\vec{n} ( \vec{v}_{2}\cdot\vec{n})^{2} \nn\\ 
&& + 96 \vec{a}_{2}\cdot\vec{n} ( \vec{v}_{2}\cdot\vec{n})^{2} \big) + \vec{S}_{1}\times\vec{a}_{2} \big( 81 \vec{v}_{1}\cdot\vec{n} v_{1}^2 - 201 v_{1}^2 \vec{v}_{2}\cdot\vec{n} - 24 \vec{v}_{1}\cdot\vec{n} \vec{v}_{1}\cdot\vec{v}_{2} \nn\\ 
&& - 480 \vec{v}_{2}\cdot\vec{n} \vec{v}_{1}\cdot\vec{v}_{2} - 108 \vec{v}_{1}\cdot\vec{n} v_{2}^2 + 1116 \vec{v}_{2}\cdot\vec{n} v_{2}^2 + 84 \vec{v}_{2}\cdot\vec{n} ( \vec{v}_{1}\cdot\vec{n})^{2} \nn\\ 
&& - 12 \vec{v}_{1}\cdot\vec{n} ( \vec{v}_{2}\cdot\vec{n})^{2} - 272 ( \vec{v}_{2}\cdot\vec{n})^{3} \big) \Big] - 	\frac{1}{96} G \Big[ 3 \vec{S}_{2}\times\vec{n}\cdot\vec{v}_{1} \big( 2 \vec{v}_{1}\cdot\vec{a}_{2} \vec{v}_{1} + 2 \vec{v}_{2}\cdot\vec{a}_{2} \vec{v}_{1} \nn\\ 
&& - 6 \vec{v}_{1}\cdot\vec{a}_{2} \vec{v}_{2} - 2 \vec{v}_{2}\cdot\vec{a}_{2} \vec{v}_{2} - 3 v_{1}^2 \vec{a}_{2} - 2 \vec{v}_{1}\cdot\vec{v}_{2} \vec{a}_{2} - 13 v_{2}^2 \vec{a}_{2} - v_{1}^2 \vec{a}_{2}\cdot\vec{n} \vec{n} \nn\\ 
&& + 2 \vec{v}_{1}\cdot\vec{v}_{2} \vec{a}_{2}\cdot\vec{n} \vec{n} - 7 v_{2}^2 \vec{a}_{2}\cdot\vec{n} \vec{n} + 6 \vec{v}_{2}\cdot\vec{n} \vec{v}_{1}\cdot\vec{a}_{2} \vec{n} - 2 \vec{v}_{1}\cdot\vec{n} \vec{v}_{2}\cdot\vec{a}_{2} \vec{n} - 14 \vec{v}_{2}\cdot\vec{n} \vec{v}_{2}\cdot\vec{a}_{2} \vec{n} \nn\\ 
&& - 2 \vec{v}_{1}\cdot\vec{n} \vec{a}_{2}\cdot\vec{n} \vec{v}_{1} - 2 \vec{v}_{2}\cdot\vec{n} \vec{a}_{2}\cdot\vec{n} \vec{v}_{1} - 2 \vec{v}_{1}\cdot\vec{n} \vec{a}_{2}\cdot\vec{n} \vec{v}_{2} + 8 \vec{v}_{2}\cdot\vec{n} \vec{a}_{2}\cdot\vec{n} \vec{v}_{2} \nn\\ 
&& - 2 \vec{v}_{1}\cdot\vec{n} \vec{v}_{2}\cdot\vec{n} \vec{a}_{2} + 16 ( \vec{v}_{2}\cdot\vec{n})^{2} \vec{a}_{2} + 6 \vec{v}_{1}\cdot\vec{n} \vec{v}_{2}\cdot\vec{n} \vec{a}_{2}\cdot\vec{n} \vec{n} + 24 \vec{a}_{2}\cdot\vec{n} ( \vec{v}_{2}\cdot\vec{n})^{2} \vec{n} \big) \nn\\ 
&& - 6 \vec{S}_{2}\times\vec{n}\cdot\vec{a}_{1} \big( v_{2}^2 \vec{v}_{2} - \vec{v}_{2}\cdot\vec{n} v_{2}^2 \vec{n} - ( \vec{v}_{2}\cdot\vec{n})^{2} \vec{v}_{2} + ( \vec{v}_{2}\cdot\vec{n})^{3} \vec{n} \big) \nn\\ 
&& + 3 \vec{S}_{2}\times\vec{n}\cdot\vec{v}_{2} \big( 24 \vec{v}_{1}\cdot\vec{a}_{1} \vec{v}_{1} - 8 \vec{a}_{1}\cdot\vec{v}_{2} \vec{v}_{1} + 4 \vec{v}_{1}\cdot\vec{a}_{2} \vec{v}_{1} + 6 \vec{v}_{2}\cdot\vec{a}_{2} \vec{v}_{1} + 21 v_{1}^2 \vec{a}_{1} \nn\\ 
&& - 14 \vec{v}_{1}\cdot\vec{v}_{2} \vec{a}_{1} - 7 v_{2}^2 \vec{a}_{1} - 38 \vec{v}_{1}\cdot\vec{a}_{1} \vec{v}_{2} + 22 \vec{a}_{1}\cdot\vec{v}_{2} \vec{v}_{2} + 4 \vec{v}_{1}\cdot\vec{a}_{2} \vec{v}_{2} - 18 \vec{v}_{2}\cdot\vec{a}_{2} \vec{v}_{2} \nn\\ 
&& - 2 \vec{v}_{1}\cdot\vec{v}_{2} \vec{a}_{2} -5 v_{1}^2 \vec{a}_{1}\cdot\vec{n} \vec{n} - 11 \vec{v}_{1}\cdot\vec{n} \vec{v}_{1}\cdot\vec{a}_{1} \vec{n} + 14 \vec{v}_{1}\cdot\vec{a}_{1} \vec{v}_{2}\cdot\vec{n} \vec{n} + 4 \vec{a}_{1}\cdot\vec{n} \vec{v}_{1}\cdot\vec{v}_{2} \vec{n} \nn\\ 
&& + 4 \vec{v}_{1}\cdot\vec{n} \vec{a}_{1}\cdot\vec{v}_{2} \vec{n} - 14 \vec{v}_{2}\cdot\vec{n} \vec{a}_{1}\cdot\vec{v}_{2} \vec{n} + \vec{a}_{1}\cdot\vec{n} v_{2}^2 \vec{n} - 2 v_{1}^2 \vec{a}_{2}\cdot\vec{n} \vec{n} + 2 \vec{v}_{1}\cdot\vec{v}_{2} \vec{a}_{2}\cdot\vec{n} \vec{n} \nn\\ 
&& + 4 v_{2}^2 \vec{a}_{2}\cdot\vec{n} \vec{n} + 14 \vec{v}_{1}\cdot\vec{n} \vec{v}_{1}\cdot\vec{a}_{2} \vec{n} + 4 \vec{v}_{2}\cdot\vec{n} \vec{v}_{1}\cdot\vec{a}_{2} \vec{n} - 12 \vec{v}_{1}\cdot\vec{n} \vec{v}_{2}\cdot\vec{a}_{2} \vec{n} \nn\\ 
&& + 2 \vec{v}_{2}\cdot\vec{n} \vec{v}_{2}\cdot\vec{a}_{2} \vec{n} - 23 \vec{v}_{1}\cdot\vec{n} \vec{a}_{1}\cdot\vec{n} \vec{v}_{1} + 8 \vec{a}_{1}\cdot\vec{n} \vec{v}_{2}\cdot\vec{n} \vec{v}_{1} - 8 \vec{v}_{1}\cdot\vec{n} \vec{a}_{2}\cdot\vec{n} \vec{v}_{1} \nn\\ 
&& + 6 \vec{v}_{2}\cdot\vec{n} \vec{a}_{2}\cdot\vec{n} \vec{v}_{1} + 14 \vec{v}_{1}\cdot\vec{n} \vec{v}_{2}\cdot\vec{n} \vec{a}_{1} - 21 ( \vec{v}_{1}\cdot\vec{n})^{2} \vec{a}_{1} + 7 ( \vec{v}_{2}\cdot\vec{n})^{2} \vec{a}_{1} \nn\\ 
&& + 28 \vec{v}_{1}\cdot\vec{n} \vec{a}_{1}\cdot\vec{n} \vec{v}_{2} - 6 \vec{a}_{1}\cdot\vec{n} \vec{v}_{2}\cdot\vec{n} \vec{v}_{2} + 8 \vec{v}_{1}\cdot\vec{n} \vec{a}_{2}\cdot\vec{n} \vec{v}_{2} + 16 \vec{v}_{1}\cdot\vec{n} \vec{v}_{2}\cdot\vec{n} \vec{a}_{2} \nn\\ 
&& + 7 ( \vec{v}_{1}\cdot\vec{n})^{2} \vec{a}_{2} - 6 ( \vec{v}_{2}\cdot\vec{n})^{2} \vec{a}_{2} -12 \vec{v}_{1}\cdot\vec{n} \vec{a}_{1}\cdot\vec{n} \vec{v}_{2}\cdot\vec{n} \vec{n} + 15 \vec{a}_{1}\cdot\vec{n} ( \vec{v}_{1}\cdot\vec{n})^{2} \vec{n} \nn\\ 
&& + 3 \vec{a}_{2}\cdot\vec{n} ( \vec{v}_{1}\cdot\vec{n})^{2} \vec{n} - 3 \vec{a}_{1}\cdot\vec{n} ( \vec{v}_{2}\cdot\vec{n})^{2} \vec{n} - 18 \vec{a}_{2}\cdot\vec{n} ( \vec{v}_{2}\cdot\vec{n})^{2} \vec{n} \big) \nn\\ 
&& - 3 \vec{S}_{2}\times\vec{v}_{1}\cdot\vec{v}_{2} \big( 21 \vec{v}_{1}\cdot\vec{a}_{1} \vec{n} - 4 \vec{a}_{1}\cdot\vec{v}_{2} \vec{n} - 8 \vec{v}_{1}\cdot\vec{a}_{2} \vec{n} + 12 \vec{v}_{2}\cdot\vec{a}_{2} \vec{n} - 23 \vec{a}_{1}\cdot\vec{n} \vec{v}_{1} \nn\\ 
&& - 10 \vec{a}_{2}\cdot\vec{n} \vec{v}_{1} - 18 \vec{v}_{1}\cdot\vec{n} \vec{a}_{1} - 10 \vec{v}_{2}\cdot\vec{n} \vec{a}_{1} - 12 \vec{a}_{1}\cdot\vec{n} \vec{v}_{2} + 2 \vec{a}_{2}\cdot\vec{n} \vec{v}_{2} - 16 \vec{v}_{1}\cdot\vec{n} \vec{a}_{2} \nn\\ 
&& + 2 \vec{v}_{2}\cdot\vec{n} \vec{a}_{2} -14 \vec{v}_{1}\cdot\vec{n} \vec{a}_{1}\cdot\vec{n} \vec{n} + 20 \vec{a}_{1}\cdot\vec{n} \vec{v}_{2}\cdot\vec{n} \vec{n} - 10 \vec{v}_{2}\cdot\vec{n} \vec{a}_{2}\cdot\vec{n} \vec{n} \big) \nn\\ 
&& - 3 \vec{S}_{2}\times\vec{a}_{1}\cdot\vec{v}_{2} \big( 19 v_{1}^2 \vec{n} - 28 \vec{v}_{1}\cdot\vec{v}_{2} \vec{n} - 9 v_{2}^2 \vec{n} + 17 \vec{v}_{1}\cdot\vec{n} \vec{v}_{1} - 40 \vec{v}_{2}\cdot\vec{n} \vec{v}_{1} - 52 \vec{v}_{1}\cdot\vec{n} \vec{v}_{2} \nn\\ 
&& + 22 \vec{v}_{2}\cdot\vec{n} \vec{v}_{2} + 44 \vec{v}_{1}\cdot\vec{n} \vec{v}_{2}\cdot\vec{n} \vec{n} - 19 ( \vec{v}_{1}\cdot\vec{n})^{2} \vec{n} + ( \vec{v}_{2}\cdot\vec{n})^{2} \vec{n} \big) - 6 \vec{S}_{2}\times\vec{n}\cdot\vec{a}_{2} \big( 5 v_{1}^2 \vec{v}_{1} \nn\\ 
&& - 6 \vec{v}_{1}\cdot\vec{v}_{2} \vec{v}_{1} - v_{1}^2 \vec{v}_{2} + 2 \vec{v}_{1}\cdot\vec{v}_{2} \vec{v}_{2} + 2 v_{2}^2 \vec{v}_{2} + 4 \vec{v}_{1}\cdot\vec{n} v_{1}^2 \vec{n} + 2 v_{1}^2 \vec{v}_{2}\cdot\vec{n} \vec{n} \nn\\ 
&& - 6 \vec{v}_{1}\cdot\vec{n} \vec{v}_{1}\cdot\vec{v}_{2} \vec{n} - 2 \vec{v}_{2}\cdot\vec{n} \vec{v}_{1}\cdot\vec{v}_{2} \vec{n} + 3 \vec{v}_{1}\cdot\vec{n} v_{2}^2 \vec{n} - 4 \vec{v}_{2}\cdot\vec{n} v_{2}^2 \vec{n} + 8 \vec{v}_{1}\cdot\vec{n} \vec{v}_{2}\cdot\vec{n} \vec{v}_{1} \nn\\ 
&& + 3 ( \vec{v}_{1}\cdot\vec{n})^{2} \vec{v}_{1} - 3 ( \vec{v}_{2}\cdot\vec{n})^{2} \vec{v}_{1} - 8 \vec{v}_{1}\cdot\vec{n} \vec{v}_{2}\cdot\vec{n} \vec{v}_{2} - 3 ( \vec{v}_{1}\cdot\vec{n})^{2} \vec{v}_{2} -2 ( \vec{v}_{1}\cdot\vec{n})^{3} \vec{n} \nn\\ 
&& - 3 \vec{v}_{2}\cdot\vec{n} ( \vec{v}_{1}\cdot\vec{n})^{2} \vec{n} + 6 ( \vec{v}_{2}\cdot\vec{n})^{3} \vec{n} \big) - 3 \vec{S}_{2}\times\vec{v}_{1}\cdot\vec{a}_{2} \big( 17 v_{1}^2 \vec{n} - 6 \vec{v}_{1}\cdot\vec{v}_{2} \vec{n} + 13 v_{2}^2 \vec{n} \nn\\ 
&& + 6 \vec{v}_{1}\cdot\vec{n} \vec{v}_{1} - 18 \vec{v}_{2}\cdot\vec{n} \vec{v}_{1} - 2 \vec{v}_{1}\cdot\vec{n} \vec{v}_{2} + 8 \vec{v}_{2}\cdot\vec{n} \vec{v}_{2} -14 \vec{v}_{1}\cdot\vec{n} \vec{v}_{2}\cdot\vec{n} \vec{n} - 20 ( \vec{v}_{1}\cdot\vec{n})^{2} \vec{n} \nn\\ 
&& - 4 ( \vec{v}_{2}\cdot\vec{n})^{2} \vec{n} \big) + 3 \vec{S}_{2}\times\vec{v}_{2}\cdot\vec{a}_{2} \big( v_{1}^2 \vec{n} - 6 \vec{v}_{1}\cdot\vec{v}_{2} \vec{n} + 9 v_{2}^2 \vec{n} + 2 \vec{v}_{1}\cdot\vec{n} \vec{v}_{1} - 2 \vec{v}_{2}\cdot\vec{n} \vec{v}_{1} \nn\\ 
&& + 4 \vec{v}_{1}\cdot\vec{n} \vec{v}_{2} + 4 \vec{v}_{2}\cdot\vec{n} \vec{v}_{2} -12 \vec{v}_{1}\cdot\vec{n} \vec{v}_{2}\cdot\vec{n} \vec{n} - ( \vec{v}_{1}\cdot\vec{n})^{2} \vec{n} + 4 ( \vec{v}_{2}\cdot\vec{n})^{2} \vec{n} \big) \nn\\ 
&& + 6 \vec{S}_{2}\times\vec{n} \big( \vec{a}_{1}\cdot\vec{v}_{2} v_{2}^2 - v_{1}^2 \vec{v}_{1}\cdot\vec{a}_{2} - 2 \vec{v}_{1}\cdot\vec{v}_{2} \vec{v}_{1}\cdot\vec{a}_{2} - 7 v_{2}^2 \vec{v}_{1}\cdot\vec{a}_{2} - v_{1}^2 \vec{v}_{2}\cdot\vec{a}_{2} \nn\\ 
&& - 2 \vec{v}_{1}\cdot\vec{v}_{2} \vec{v}_{2}\cdot\vec{a}_{2} + 20 v_{2}^2 \vec{v}_{2}\cdot\vec{a}_{2} - \vec{a}_{1}\cdot\vec{n} \vec{v}_{2}\cdot\vec{n} v_{2}^2 - \vec{v}_{1}\cdot\vec{n} v_{1}^2 \vec{a}_{2}\cdot\vec{n} + v_{1}^2 \vec{v}_{2}\cdot\vec{n} \vec{a}_{2}\cdot\vec{n} \nn\\ 
&& + 2 \vec{v}_{1}\cdot\vec{n} \vec{v}_{1}\cdot\vec{v}_{2} \vec{a}_{2}\cdot\vec{n} + 4 \vec{v}_{2}\cdot\vec{n} \vec{v}_{1}\cdot\vec{v}_{2} \vec{a}_{2}\cdot\vec{n} - 3 \vec{v}_{1}\cdot\vec{n} v_{2}^2 \vec{a}_{2}\cdot\vec{n} - 20 \vec{v}_{2}\cdot\vec{n} v_{2}^2 \vec{a}_{2}\cdot\vec{n} \nn\\ 
&& + 2 \vec{v}_{1}\cdot\vec{n} \vec{v}_{2}\cdot\vec{n} \vec{v}_{1}\cdot\vec{a}_{2} - 6 \vec{v}_{1}\cdot\vec{n} \vec{v}_{2}\cdot\vec{n} \vec{v}_{2}\cdot\vec{a}_{2} + \vec{v}_{2}\cdot\vec{a}_{2} ( \vec{v}_{1}\cdot\vec{n})^{2} - \vec{a}_{1}\cdot\vec{v}_{2} ( \vec{v}_{2}\cdot\vec{n})^{2} \nn\\ 
&& + 8 \vec{v}_{1}\cdot\vec{a}_{2} ( \vec{v}_{2}\cdot\vec{n})^{2} - 20 \vec{v}_{2}\cdot\vec{a}_{2} ( \vec{v}_{2}\cdot\vec{n})^{2} -3 \vec{v}_{2}\cdot\vec{n} \vec{a}_{2}\cdot\vec{n} ( \vec{v}_{1}\cdot\vec{n})^{2} + \vec{a}_{1}\cdot\vec{n} ( \vec{v}_{2}\cdot\vec{n})^{3} \nn\\ 
&& + 12 \vec{v}_{1}\cdot\vec{n} \vec{a}_{2}\cdot\vec{n} ( \vec{v}_{2}\cdot\vec{n})^{2} + 20 \vec{a}_{2}\cdot\vec{n} ( \vec{v}_{2}\cdot\vec{n})^{3} \big) + 3 \vec{S}_{2}\times\vec{v}_{1} \big( v_{1}^2 \vec{a}_{2}\cdot\vec{n} - 2 \vec{v}_{1}\cdot\vec{v}_{2} \vec{a}_{2}\cdot\vec{n} \nn\\ 
&& - v_{2}^2 \vec{a}_{2}\cdot\vec{n} + 2 \vec{v}_{2}\cdot\vec{n} \vec{v}_{1}\cdot\vec{a}_{2} - 6 \vec{v}_{1}\cdot\vec{n} \vec{v}_{2}\cdot\vec{a}_{2} - 2 \vec{v}_{2}\cdot\vec{n} \vec{v}_{2}\cdot\vec{a}_{2} + 6 \vec{v}_{1}\cdot\vec{n} \vec{v}_{2}\cdot\vec{n} \vec{a}_{2}\cdot\vec{n} \big) \nn\\ 
&& + 4 \vec{S}_{2}\times\vec{a}_{1} \big( 3 \vec{v}_{2}\cdot\vec{n} v_{2}^2 - ( \vec{v}_{2}\cdot\vec{n})^{3} \big) + 3 \vec{S}_{2}\times\vec{v}_{2} \big( 67 v_{1}^2 \vec{a}_{1}\cdot\vec{n} + 157 \vec{v}_{1}\cdot\vec{n} \vec{v}_{1}\cdot\vec{a}_{1} \nn\\ 
&& - 34 \vec{v}_{1}\cdot\vec{a}_{1} \vec{v}_{2}\cdot\vec{n} - 60 \vec{a}_{1}\cdot\vec{n} \vec{v}_{1}\cdot\vec{v}_{2} - 76 \vec{v}_{1}\cdot\vec{n} \vec{a}_{1}\cdot\vec{v}_{2} + 22 \vec{v}_{2}\cdot\vec{n} \vec{a}_{1}\cdot\vec{v}_{2} + 11 \vec{a}_{1}\cdot\vec{n} v_{2}^2 \nn\\ 
&& - 6 v_{1}^2 \vec{a}_{2}\cdot\vec{n} + 2 \vec{v}_{1}\cdot\vec{v}_{2} \vec{a}_{2}\cdot\vec{n} + 20 v_{2}^2 \vec{a}_{2}\cdot\vec{n} - 22 \vec{v}_{1}\cdot\vec{n} \vec{v}_{1}\cdot\vec{a}_{2} - 4 \vec{v}_{2}\cdot\vec{n} \vec{v}_{1}\cdot\vec{a}_{2} \nn\\ 
&& + 32 \vec{v}_{1}\cdot\vec{n} \vec{v}_{2}\cdot\vec{a}_{2} + 22 \vec{v}_{2}\cdot\vec{n} \vec{v}_{2}\cdot\vec{a}_{2} + 44 \vec{v}_{1}\cdot\vec{n} \vec{a}_{1}\cdot\vec{n} \vec{v}_{2}\cdot\vec{n} - 16 \vec{v}_{1}\cdot\vec{n} \vec{v}_{2}\cdot\vec{n} \vec{a}_{2}\cdot\vec{n} \nn\\ 
&& - 67 \vec{a}_{1}\cdot\vec{n} ( \vec{v}_{1}\cdot\vec{n})^{2} + 3 \vec{a}_{2}\cdot\vec{n} ( \vec{v}_{1}\cdot\vec{n})^{2} - 3 \vec{a}_{1}\cdot\vec{n} ( \vec{v}_{2}\cdot\vec{n})^{2} - 22 \vec{a}_{2}\cdot\vec{n} ( \vec{v}_{2}\cdot\vec{n})^{2} \big) \nn\\ 
&& + 2 \vec{S}_{2}\times\vec{a}_{2} \big( 27 \vec{v}_{1}\cdot\vec{n} v_{1}^2 - 12 v_{1}^2 \vec{v}_{2}\cdot\vec{n} - 12 \vec{v}_{1}\cdot\vec{n} \vec{v}_{1}\cdot\vec{v}_{2} + 30 \vec{v}_{2}\cdot\vec{n} \vec{v}_{1}\cdot\vec{v}_{2} + 36 \vec{v}_{1}\cdot\vec{n} v_{2}^2 \nn\\ 
&& - 6 \vec{v}_{2}\cdot\vec{n} v_{2}^2 -10 ( \vec{v}_{1}\cdot\vec{n})^{3} - 6 \vec{v}_{2}\cdot\vec{n} ( \vec{v}_{1}\cdot\vec{n})^{2} - 24 \vec{v}_{1}\cdot\vec{n} ( \vec{v}_{2}\cdot\vec{n})^{2} + 4 ( \vec{v}_{2}\cdot\vec{n})^{3} \big) \Big] \nn\\ 
&& - 	\frac{G m_{2}}{64 m_{1}} \Big[ 2 \dot{\vec{S}}_{1}\times\vec{n}\cdot\vec{v}_{1} \big( v_{2}^2 \vec{v}_{1} + 4 v_{2}^2 \vec{v}_{2} + 8 \vec{v}_{2}\cdot\vec{n} v_{2}^2 \vec{n} - ( \vec{v}_{2}\cdot\vec{n})^{2} \vec{v}_{1} \nn\\ 
&& - 4 ( \vec{v}_{2}\cdot\vec{n})^{2} \vec{v}_{2} -8 ( \vec{v}_{2}\cdot\vec{n})^{3} \vec{n} \big) - 16 \dot{\vec{S}}_{1}\times\vec{n}\cdot\vec{v}_{2} \big( \vec{v}_{2}\cdot\vec{n} v_{2}^2 \vec{n} - ( \vec{v}_{2}\cdot\vec{n})^{3} \vec{n} \big) \nn\\ 
&& - 4 \dot{\vec{S}}_{1}\times\vec{v}_{1}\cdot\vec{v}_{2} \big( 6 v_{2}^2 \vec{n} - 7 \vec{v}_{2}\cdot\vec{n} \vec{v}_{1} -6 ( \vec{v}_{2}\cdot\vec{n})^{2} \vec{n} \big) + \dot{\vec{S}}_{1}\times\vec{n} \big( v_{1}^2 v_{2}^2 + 8 \vec{v}_{1}\cdot\vec{v}_{2} v_{2}^2 \nn\\ 
&& - 16 v_{2}^{4} + 16 \vec{v}_{1}\cdot\vec{n} \vec{v}_{2}\cdot\vec{n} v_{2}^2 - v_{1}^2 ( \vec{v}_{2}\cdot\vec{n})^{2} - 8 \vec{v}_{1}\cdot\vec{v}_{2} ( \vec{v}_{2}\cdot\vec{n})^{2} \nn\\ 
&& + 24 v_{2}^2 ( \vec{v}_{2}\cdot\vec{n})^{2} -16 \vec{v}_{1}\cdot\vec{n} ( \vec{v}_{2}\cdot\vec{n})^{3} - 8 ( \vec{v}_{2}\cdot\vec{n})^{4} \big) + 12 \dot{\vec{S}}_{1}\times\vec{v}_{1} v_{1}^2 \vec{v}_{2}\cdot\vec{n} \nn\\ 
&& - 2 \dot{\vec{S}}_{1}\times\vec{v}_{2} \big( 7 v_{1}^2 \vec{v}_{2}\cdot\vec{n} - 12 \vec{v}_{1}\cdot\vec{n} v_{2}^2 - 48 \vec{v}_{2}\cdot\vec{n} v_{2}^2 + 12 \vec{v}_{1}\cdot\vec{n} ( \vec{v}_{2}\cdot\vec{n})^{2} + 8 ( \vec{v}_{2}\cdot\vec{n})^{3} \big) \Big] \nn\\ 
&& - 	\frac{1}{48} G \Big[ 12 \dot{\vec{S}}_{2}\times\vec{n}\cdot\vec{v}_{1} \big( 2 \vec{v}_{1}\cdot\vec{v}_{2} \vec{v}_{1} + v_{2}^2 \vec{v}_{1} - 2 \vec{v}_{1}\cdot\vec{v}_{2} \vec{v}_{2} - v_{2}^2 \vec{v}_{2} - \vec{v}_{1}\cdot\vec{n} v_{2}^2 \vec{n} \nn\\ 
&& - 2 \vec{v}_{2}\cdot\vec{n} v_{2}^2 \vec{n} + 2 \vec{v}_{1}\cdot\vec{n} \vec{v}_{2}\cdot\vec{n} \vec{v}_{1} - ( \vec{v}_{2}\cdot\vec{n})^{2} \vec{v}_{1} + ( \vec{v}_{2}\cdot\vec{n})^{2} \vec{v}_{2} + 3 \vec{v}_{1}\cdot\vec{n} ( \vec{v}_{2}\cdot\vec{n})^{2} \vec{n} \nn\\ 
&& + 2 ( \vec{v}_{2}\cdot\vec{n})^{3} \vec{n} \big) + 3 \dot{\vec{S}}_{2}\times\vec{n}\cdot\vec{v}_{2} \big( v_{1}^2 \vec{v}_{1} - 2 \vec{v}_{1}\cdot\vec{v}_{2} \vec{v}_{1} - 9 v_{2}^2 \vec{v}_{1} - v_{1}^2 \vec{v}_{2} + 2 \vec{v}_{1}\cdot\vec{v}_{2} \vec{v}_{2} \nn\\ 
&& + 5 v_{2}^2 \vec{v}_{2} -4 \vec{v}_{1}\cdot\vec{n} v_{1}^2 \vec{n} - 3 v_{1}^2 \vec{v}_{2}\cdot\vec{n} \vec{n} + 6 \vec{v}_{1}\cdot\vec{n} \vec{v}_{1}\cdot\vec{v}_{2} \vec{n} + 2 \vec{v}_{2}\cdot\vec{n} \vec{v}_{1}\cdot\vec{v}_{2} \vec{n} + 7 \vec{v}_{2}\cdot\vec{n} v_{2}^2 \vec{n} \nn\\ 
&& - 6 \vec{v}_{1}\cdot\vec{n} \vec{v}_{2}\cdot\vec{n} \vec{v}_{1} - 3 ( \vec{v}_{1}\cdot\vec{n})^{2} \vec{v}_{1} + 6 ( \vec{v}_{2}\cdot\vec{n})^{2} \vec{v}_{1} + 2 \vec{v}_{1}\cdot\vec{n} \vec{v}_{2}\cdot\vec{n} \vec{v}_{2} + 3 ( \vec{v}_{1}\cdot\vec{n})^{2} \vec{v}_{2} \nn\\ 
&& + 2 ( \vec{v}_{1}\cdot\vec{n})^{3} \vec{n} + 3 \vec{v}_{2}\cdot\vec{n} ( \vec{v}_{1}\cdot\vec{n})^{2} \vec{n} - 9 \vec{v}_{1}\cdot\vec{n} ( \vec{v}_{2}\cdot\vec{n})^{2} \vec{n} - 6 ( \vec{v}_{2}\cdot\vec{n})^{3} \vec{n} \big) \nn\\ 
&& - 3 \dot{\vec{S}}_{2}\times\vec{v}_{1}\cdot\vec{v}_{2} \big( 8 v_{1}^2 \vec{n} - 2 \vec{v}_{1}\cdot\vec{v}_{2} \vec{n} + 8 v_{2}^2 \vec{n} - 6 \vec{v}_{1}\cdot\vec{n} \vec{v}_{1} + 10 \vec{v}_{2}\cdot\vec{n} \vec{v}_{1} - 2 \vec{v}_{1}\cdot\vec{n} \vec{v}_{2} \nn\\ 
&& - 6 \vec{v}_{2}\cdot\vec{n} \vec{v}_{2} -14 \vec{v}_{1}\cdot\vec{n} \vec{v}_{2}\cdot\vec{n} \vec{n} - 10 ( \vec{v}_{1}\cdot\vec{n})^{2} \vec{n} - 3 ( \vec{v}_{2}\cdot\vec{n})^{2} \vec{n} \big) + 6 \dot{\vec{S}}_{2}\times\vec{n} \big( v_{1}^2 v_{2}^2 \nn\\ 
&& - 2 \vec{v}_{1}\cdot\vec{v}_{2} v_{2}^2 - 2 ( \vec{v}_{1}\cdot\vec{v}_{2})^{2} + 5 v_{2}^{4} -4 \vec{v}_{1}\cdot\vec{n} \vec{v}_{2}\cdot\vec{n} v_{2}^2 - v_{2}^2 ( \vec{v}_{1}\cdot\vec{n})^{2} - v_{1}^2 ( \vec{v}_{2}\cdot\vec{n})^{2} \nn\\ 
&& + 2 \vec{v}_{1}\cdot\vec{v}_{2} ( \vec{v}_{2}\cdot\vec{n})^{2} - 10 v_{2}^2 ( \vec{v}_{2}\cdot\vec{n})^{2} + 3 ( \vec{v}_{1}\cdot\vec{n})^{2} ( \vec{v}_{2}\cdot\vec{n})^{2} + 4 \vec{v}_{1}\cdot\vec{n} ( \vec{v}_{2}\cdot\vec{n})^{3} \nn\\ 
&& + 5 ( \vec{v}_{2}\cdot\vec{n})^{4} \big) + \dot{\vec{S}}_{2}\times\vec{v}_{2} \big( 24 \vec{v}_{1}\cdot\vec{n} v_{1}^2 + 15 v_{1}^2 \vec{v}_{2}\cdot\vec{n} - 6 \vec{v}_{1}\cdot\vec{n} \vec{v}_{1}\cdot\vec{v}_{2} - 18 \vec{v}_{2}\cdot\vec{n} \vec{v}_{1}\cdot\vec{v}_{2} \nn\\ 
&& + 24 \vec{v}_{1}\cdot\vec{n} v_{2}^2 + 21 \vec{v}_{2}\cdot\vec{n} v_{2}^2 -10 ( \vec{v}_{1}\cdot\vec{n})^{3} - 21 \vec{v}_{2}\cdot\vec{n} ( \vec{v}_{1}\cdot\vec{n})^{2} - 9 \vec{v}_{1}\cdot\vec{n} ( \vec{v}_{2}\cdot\vec{n})^{2} \nn\\ 
&& - 16 ( \vec{v}_{2}\cdot\vec{n})^{3} \big) \Big]\nn\\ && +  	\frac{G^2 m_{2}{}^2}{384 m_{1} r} \Big[ \vec{S}_{1}\times\vec{n}\cdot\vec{v}_{1} \big( 283 \vec{a}_{2} + 1063 \vec{a}_{2}\cdot\vec{n} \vec{n} \big) - 16 \vec{S}_{1}\times\vec{n}\cdot\vec{a}_{1} \big( 5 \vec{v}_{2} -4 \vec{v}_{2}\cdot\vec{n} \vec{n} \big) \nn\\ 
&& - 6 \vec{S}_{1}\times\vec{n}\cdot\vec{v}_{2} \big( 184 \vec{a}_{2} + 273 \vec{a}_{2}\cdot\vec{n} \vec{n} \big) + 80 \vec{S}_{1}\times\vec{a}_{1}\cdot\vec{v}_{2} \vec{n} + \vec{S}_{1}\times\vec{n}\cdot\vec{a}_{2} \big( 1037 \vec{v}_{1} \nn\\ 
&& - 2092 \vec{v}_{2} + 32 \vec{v}_{1}\cdot\vec{n} \vec{n} - 874 \vec{v}_{2}\cdot\vec{n} \vec{n} \big) + 56 \vec{S}_{1}\times\vec{v}_{1}\cdot\vec{a}_{2} \vec{n} + 1292 \vec{S}_{1}\times\vec{v}_{2}\cdot\vec{a}_{2} \vec{n} \nn\\ 
&& + \vec{S}_{1}\times\vec{n} \big( 80 \vec{a}_{1}\cdot\vec{v}_{2} + 650 \vec{v}_{1}\cdot\vec{a}_{2} - 1054 \vec{v}_{2}\cdot\vec{a}_{2} -64 \vec{a}_{1}\cdot\vec{n} \vec{v}_{2}\cdot\vec{n} - 55 \vec{v}_{1}\cdot\vec{n} \vec{a}_{2}\cdot\vec{n} \nn\\ 
&& + 344 \vec{v}_{2}\cdot\vec{n} \vec{a}_{2}\cdot\vec{n} \big) + 730 \vec{S}_{1}\times\vec{v}_{1} \vec{a}_{2}\cdot\vec{n} + 256 \vec{S}_{1}\times\vec{a}_{1} \vec{v}_{2}\cdot\vec{n} + 4 \vec{S}_{1}\times\vec{v}_{2} \big( 20 \vec{a}_{1}\cdot\vec{n} \nn\\ 
&& + 107 \vec{a}_{2}\cdot\vec{n} \big) - 3 \vec{S}_{1}\times\vec{a}_{2} \big( 607 \vec{v}_{1}\cdot\vec{n} - 1322 \vec{v}_{2}\cdot\vec{n} \big) \Big] - 	\frac{G^2 m_{2}}{384 r} \Big[ 2 \vec{S}_{1}\times\vec{n}\cdot\vec{v}_{1} \big( 296 \vec{a}_{1} \nn\\ 
&& + 394 \vec{a}_{2} + 54 \vec{a}_{1}\cdot\vec{n} \vec{n} + 305 \vec{a}_{2}\cdot\vec{n} \vec{n} \big) + 2 \vec{S}_{1}\times\vec{n}\cdot\vec{a}_{1} \big( 217 \vec{v}_{1} + 537 \vec{v}_{2} -176 \vec{v}_{1}\cdot\vec{n} \vec{n} \nn\\ 
&& - 375 \vec{v}_{2}\cdot\vec{n} \vec{n} \big) + 2677 \vec{S}_{1}\times\vec{v}_{1}\cdot\vec{a}_{1} \vec{n} + 8 \vec{S}_{1}\times\vec{n}\cdot\vec{v}_{2} \big( 172 \vec{a}_{1} + 97 \vec{a}_{2} -26 \vec{a}_{1}\cdot\vec{n} \vec{n} \nn\\ 
&& + 94 \vec{a}_{2}\cdot\vec{n} \vec{n} \big) + 2294 \vec{S}_{1}\times\vec{a}_{1}\cdot\vec{v}_{2} \vec{n} + 8 \vec{S}_{1}\times\vec{n}\cdot\vec{a}_{2} \big( 60 \vec{v}_{1} + 34 \vec{v}_{2} -15 \vec{v}_{1}\cdot\vec{n} \vec{n} \nn\\ 
&& + 175 \vec{v}_{2}\cdot\vec{n} \vec{n} \big) - 866 \vec{S}_{1}\times\vec{v}_{1}\cdot\vec{a}_{2} \vec{n} - 408 \vec{S}_{1}\times\vec{v}_{2}\cdot\vec{a}_{2} \vec{n} + 2 \vec{S}_{1}\times\vec{n} \big( 538 \vec{v}_{1}\cdot\vec{a}_{1} \nn\\ 
&& - 1502 \vec{a}_{1}\cdot\vec{v}_{2} - 1012 \vec{v}_{1}\cdot\vec{a}_{2} + 2128 \vec{v}_{2}\cdot\vec{a}_{2} -458 \vec{v}_{1}\cdot\vec{n} \vec{a}_{1}\cdot\vec{n} + 1771 \vec{a}_{1}\cdot\vec{n} \vec{v}_{2}\cdot\vec{n} \nn\\ 
&& + 3488 \vec{v}_{1}\cdot\vec{n} \vec{a}_{2}\cdot\vec{n} - 4088 \vec{v}_{2}\cdot\vec{n} \vec{a}_{2}\cdot\vec{n} \big) - \vec{S}_{1}\times\vec{v}_{1} \big( 3665 \vec{a}_{1}\cdot\vec{n} + 3508 \vec{a}_{2}\cdot\vec{n} \big) \nn\\ 
&& - 2 \vec{S}_{1}\times\vec{a}_{1} \big( 2931 \vec{v}_{1}\cdot\vec{n} + 1823 \vec{v}_{2}\cdot\vec{n} \big) + 6 \vec{S}_{1}\times\vec{v}_{2} \big( 607 \vec{a}_{1}\cdot\vec{n} + 152 \vec{a}_{2}\cdot\vec{n} \big) \nn\\ 
&& + 16 \vec{S}_{1}\times\vec{a}_{2} \big( 427 \vec{v}_{1}\cdot\vec{n} - 324 \vec{v}_{2}\cdot\vec{n} \big) \Big] + 	\frac{G^2 m_{2}}{192 r} \Big[ 8 \vec{S}_{2}\times\vec{n}\cdot\vec{v}_{1} \big( 41 \vec{a}_{2} + 98 \vec{a}_{2}\cdot\vec{n} \vec{n} \big) \nn\\ 
&& + \vec{S}_{2}\times\vec{n}\cdot\vec{v}_{2} \big( 60 \vec{a}_{1} - 503 \vec{a}_{2} -84 \vec{a}_{1}\cdot\vec{n} \vec{n} - 1069 \vec{a}_{2}\cdot\vec{n} \vec{n} \big) - 543 \vec{S}_{2}\times\vec{a}_{1}\cdot\vec{v}_{2} \vec{n} \nn\\ 
&& - 4 \vec{S}_{2}\times\vec{n}\cdot\vec{a}_{2} \big( 174 \vec{v}_{1} - 95 \vec{v}_{2} -126 \vec{v}_{1}\cdot\vec{n} \vec{n} + 133 \vec{v}_{2}\cdot\vec{n} \vec{n} \big) + 632 \vec{S}_{2}\times\vec{v}_{1}\cdot\vec{a}_{2} \vec{n} \nn\\ 
&& + 79 \vec{S}_{2}\times\vec{v}_{2}\cdot\vec{a}_{2} \vec{n} + 4 \vec{S}_{2}\times\vec{n} \big( 151 \vec{v}_{1}\cdot\vec{a}_{2} - 206 \vec{v}_{2}\cdot\vec{a}_{2} + 13 \vec{v}_{1}\cdot\vec{n} \vec{a}_{2}\cdot\vec{n} \nn\\ 
&& + 256 \vec{v}_{2}\cdot\vec{n} \vec{a}_{2}\cdot\vec{n} \big) + 324 \vec{S}_{2}\times\vec{v}_{1} \vec{a}_{2}\cdot\vec{n} - 3 \vec{S}_{2}\times\vec{v}_{2} \big( 469 \vec{a}_{1}\cdot\vec{n} - 229 \vec{a}_{2}\cdot\vec{n} \big) \nn\\ 
&& - 8 \vec{S}_{2}\times\vec{a}_{2} \big( 103 \vec{v}_{1}\cdot\vec{n} - 280 \vec{v}_{2}\cdot\vec{n} \big) \Big] + 	\frac{G^2 m_{1}}{384 r} \Big[ 2 \vec{S}_{2}\times\vec{n}\cdot\vec{v}_{1} \big( 2180 \vec{a}_{1} + 584 \vec{a}_{2} \nn\\ 
&& + 17 \vec{a}_{1}\cdot\vec{n} \vec{n} - 67 \vec{a}_{2}\cdot\vec{n} \vec{n} \big) - \vec{S}_{2}\times\vec{n}\cdot\vec{a}_{1} \big( 1005 \vec{v}_{1} - 1771 \vec{v}_{2} -3504 \vec{v}_{1}\cdot\vec{n} \vec{n} \nn\\ 
&& + 452 \vec{v}_{2}\cdot\vec{n} \vec{n} \big) + 845 \vec{S}_{2}\times\vec{v}_{1}\cdot\vec{a}_{1} \vec{n} - \vec{S}_{2}\times\vec{n}\cdot\vec{v}_{2} \big( 4628 \vec{a}_{1} - 1291 \vec{a}_{2} + 82 \vec{a}_{1}\cdot\vec{n} \vec{n} \nn\\ 
&& + 95 \vec{a}_{2}\cdot\vec{n} \vec{n} \big) + 978 \vec{S}_{2}\times\vec{a}_{1}\cdot\vec{v}_{2} \vec{n} - \vec{S}_{2}\times\vec{n}\cdot\vec{a}_{2} \big( 2044 \vec{v}_{1} - 1520 \vec{v}_{2} + 1696 \vec{v}_{1}\cdot\vec{n} \vec{n} \nn\\ 
&& - 23 \vec{v}_{2}\cdot\vec{n} \vec{n} \big) + 3040 \vec{S}_{2}\times\vec{v}_{1}\cdot\vec{a}_{2} \vec{n} - 38 \vec{S}_{2}\times\vec{v}_{2}\cdot\vec{a}_{2} \vec{n} + \vec{S}_{2}\times\vec{n} \big( 79 \vec{v}_{1}\cdot\vec{a}_{1} \nn\\ 
&& + 1119 \vec{a}_{1}\cdot\vec{v}_{2} + 2086 \vec{v}_{1}\cdot\vec{a}_{2} - 3135 \vec{v}_{2}\cdot\vec{a}_{2} -4328 \vec{v}_{1}\cdot\vec{n} \vec{a}_{1}\cdot\vec{n} + 1404 \vec{a}_{1}\cdot\vec{n} \vec{v}_{2}\cdot\vec{n} \nn\\ 
&& - 206 \vec{v}_{1}\cdot\vec{n} \vec{a}_{2}\cdot\vec{n} + 3432 \vec{v}_{2}\cdot\vec{n} \vec{a}_{2}\cdot\vec{n} \big) - 3 \vec{S}_{2}\times\vec{v}_{1} \big( 77 \vec{a}_{1}\cdot\vec{n} - 346 \vec{a}_{2}\cdot\vec{n} \big) \nn\\ 
&& - 28 \vec{S}_{2}\times\vec{a}_{1} \big( 35 \vec{v}_{1}\cdot\vec{n} - 59 \vec{v}_{2}\cdot\vec{n} \big) - 8 \vec{S}_{2}\times\vec{v}_{2} \big( 233 \vec{a}_{1}\cdot\vec{n} + 76 \vec{a}_{2}\cdot\vec{n} \big) \nn\\ 
&& - 25 \vec{S}_{2}\times\vec{a}_{2} \big( 88 \vec{v}_{1}\cdot\vec{n} + 23 \vec{v}_{2}\cdot\vec{n} \big) \Big] - 	\frac{G^2 m_{2}}{384 r} \Big[ 4 \dot{\vec{S}}_{1}\times\vec{n}\cdot\vec{v}_{1} \big( 31 \vec{v}_{1} + 97 \vec{v}_{2} \nn\\ 
&& + 742 \vec{v}_{1}\cdot\vec{n} \vec{n} - 629 \vec{v}_{2}\cdot\vec{n} \vec{n} \big) + 8 \dot{\vec{S}}_{1}\times\vec{n}\cdot\vec{v}_{2} \big( 64 \vec{v}_{1} + 175 \vec{v}_{2} -80 \vec{v}_{1}\cdot\vec{n} \vec{n} \nn\\ 
&& + 133 \vec{v}_{2}\cdot\vec{n} \vec{n} \big) + 1606 \dot{\vec{S}}_{1}\times\vec{v}_{1}\cdot\vec{v}_{2} \vec{n} - \dot{\vec{S}}_{1}\times\vec{n} \big( 435 v_{1}^2 - 2104 \vec{v}_{1}\cdot\vec{v}_{2} \nn\\ 
&& + 1852 v_{2}^2 -2368 \vec{v}_{1}\cdot\vec{n} \vec{v}_{2}\cdot\vec{n} + 504 ( \vec{v}_{1}\cdot\vec{n})^{2} + 760 ( \vec{v}_{2}\cdot\vec{n})^{2} \big) + 2 \dot{\vec{S}}_{1}\times\vec{v}_{1} \big( 532 \vec{v}_{1}\cdot\vec{n} \nn\\ 
&& - 771 \vec{v}_{2}\cdot\vec{n} \big) - 8 \dot{\vec{S}}_{1}\times\vec{v}_{2} \big( 782 \vec{v}_{1}\cdot\vec{n} - 565 \vec{v}_{2}\cdot\vec{n} \big) \Big] \nn\\ 
&& - 	\frac{G^2 m_{2}{}^2}{24 m_{1} r} \Big[ 3 \dot{\vec{S}}_{1}\times\vec{n}\cdot\vec{v}_{1} \big( 5 \vec{v}_{2} -4 \vec{v}_{2}\cdot\vec{n} \vec{n} \big) - 2 \dot{\vec{S}}_{1}\times\vec{n}\cdot\vec{v}_{2} \big( \vec{v}_{2} + 4 \vec{v}_{2}\cdot\vec{n} \vec{n} \big) \nn\\ 
&& - 15 \dot{\vec{S}}_{1}\times\vec{v}_{1}\cdot\vec{v}_{2} \vec{n} + 3 \dot{\vec{S}}_{1}\times\vec{n} \big( 5 \vec{v}_{1}\cdot\vec{v}_{2} - 6 v_{2}^2 -4 \vec{v}_{1}\cdot\vec{n} \vec{v}_{2}\cdot\vec{n} - 5 ( \vec{v}_{2}\cdot\vec{n})^{2} \big) \nn\\ 
&& + 18 \dot{\vec{S}}_{1}\times\vec{v}_{1} \vec{v}_{2}\cdot\vec{n} + \dot{\vec{S}}_{1}\times\vec{v}_{2} \big( 15 \vec{v}_{1}\cdot\vec{n} + 221 \vec{v}_{2}\cdot\vec{n} \big) \Big] + 	\frac{G^2 m_{2}}{48 r} \Big[ 24 \dot{\vec{S}}_{2}\times\vec{n}\cdot\vec{v}_{1} \big( 4 \vec{v}_{2} \nn\\ 
&& + 11 \vec{v}_{2}\cdot\vec{n} \vec{n} \big) - \dot{\vec{S}}_{2}\times\vec{n}\cdot\vec{v}_{2} \big( 125 \vec{v}_{1} - 165 \vec{v}_{2} -142 \vec{v}_{1}\cdot\vec{n} \vec{n} + 420 \vec{v}_{2}\cdot\vec{n} \vec{n} \big) \nn\\ 
&& + 229 \dot{\vec{S}}_{2}\times\vec{v}_{1}\cdot\vec{v}_{2} \vec{n} + \dot{\vec{S}}_{2}\times\vec{n} \big( 96 \vec{v}_{1}\cdot\vec{v}_{2} - 29 v_{2}^2 + 264 \vec{v}_{1}\cdot\vec{n} \vec{v}_{2}\cdot\vec{n} - 149 ( \vec{v}_{2}\cdot\vec{n})^{2} \big) \nn\\ 
&& - \dot{\vec{S}}_{2}\times\vec{v}_{2} \big( 340 \vec{v}_{1}\cdot\vec{n} - 593 \vec{v}_{2}\cdot\vec{n} \big) \Big] + 	\frac{G^2 m_{1}}{768 r} \Big[ 4 \dot{\vec{S}}_{2}\times\vec{n}\cdot\vec{v}_{1} \big( 719 \vec{v}_{1} \nn\\ 
&& + 865 \vec{v}_{2} -3148 \vec{v}_{1}\cdot\vec{n} \vec{n} + 2137 \vec{v}_{2}\cdot\vec{n} \vec{n} \big) - 2 \dot{\vec{S}}_{2}\times\vec{n}\cdot\vec{v}_{2} \big( 1126 \vec{v}_{1} \nn\\ 
&& - 435 \vec{v}_{2} -1898 \vec{v}_{1}\cdot\vec{n} \vec{n} + 3519 \vec{v}_{2}\cdot\vec{n} \vec{n} \big) + 12180 \dot{\vec{S}}_{2}\times\vec{v}_{1}\cdot\vec{v}_{2} \vec{n} - \dot{\vec{S}}_{2}\times\vec{n} \big( 1966 v_{1}^2 \nn\\ 
&& - 9236 \vec{v}_{1}\cdot\vec{v}_{2} + 6903 v_{2}^2 -4292 \vec{v}_{1}\cdot\vec{n} \vec{v}_{2}\cdot\vec{n} + 11728 ( \vec{v}_{1}\cdot\vec{n})^{2} - 7884 ( \vec{v}_{2}\cdot\vec{n})^{2} \big) \nn\\ 
&& + 8 \dot{\vec{S}}_{2}\times\vec{v}_{1} \big( 835 \vec{v}_{1}\cdot\vec{n} + 92 \vec{v}_{2}\cdot\vec{n} \big) - 4 \dot{\vec{S}}_{2}\times\vec{v}_{2} \big( 3575 \vec{v}_{1}\cdot\vec{n} + 354 \vec{v}_{2}\cdot\vec{n} \big) \Big]\nn\\ && - 	\frac{225 G^3 m_{1} m_{2}}{16 r{}^2} \dot{\vec{S}}_{1}\times\vec{n} + 	\frac{5 G^3 m_{2}{}^3}{3 m_{1} r{}^2} \dot{\vec{S}}_{1}\times\vec{n} - 	\frac{17 G^3 m_{1} m_{2}}{3 r{}^2} {\Big( \frac{1}{\epsilon} - 3\log \frac{r}{R_0} \Big)} \dot{\vec{S}}_{1}\times\vec{n} \nn\\ 
&& - 	\frac{29 G^3 m_{2}{}^2}{6 r{}^2} {\Big( \frac{1}{\epsilon} - 3\log \frac{r}{R_0} \Big)} \dot{\vec{S}}_{1}\times\vec{n} + 	\frac{2 G^3 m_{2}{}^3}{3 m_{1} r{}^2} {\Big( \frac{1}{\epsilon} - 3\log \frac{r}{R_0} \Big)} \dot{\vec{S}}_{1}\times\vec{n} \nn\\ 
&& - 	\frac{G^3 m_{2}{}^2}{576 r{}^2} {(2767 - 108 \pi^2)} \dot{\vec{S}}_{1}\times\vec{n} + 	\frac{3325 G^3 m_{1}{}^2}{288 r{}^2} \dot{\vec{S}}_{2}\times\vec{n} + 	\frac{23 G^3 m_{2}{}^2}{3 r{}^2} \dot{\vec{S}}_{2}\times\vec{n} \nn\\ 
&& + 	\frac{37 G^3 m_{1}{}^2}{6 r{}^2} {\Big( \frac{1}{\epsilon} - 3\log \frac{r}{R_0} \Big)} \dot{\vec{S}}_{2}\times\vec{n} + 	\frac{29 G^3 m_{1} m_{2}}{6 r{}^2} {\Big( \frac{1}{\epsilon} - 3\log \frac{r}{R_0} \Big)} \dot{\vec{S}}_{2}\times\vec{n} \nn\\ 
&& + 	\frac{8 G^3 m_{2}{}^2}{3 r{}^2} {\Big( \frac{1}{\epsilon} - 3\log \frac{r}{R_0} \Big)} \dot{\vec{S}}_{2}\times\vec{n} + 	\frac{G^3 m_{1} m_{2}}{288 r{}^2} {(2441 - 99 \pi^2)} \dot{\vec{S}}_{2}\times\vec{n}, 
\eea
\bea
\stackrel{(2)}{ \Delta \vec{x}_1}_{(3,1)}&=& - 	\frac{G m_{2} r}{768 m_{1}} \Big[ (4 \vec{S}_{1}\times\vec{n}\cdot\vec{v}_{1} \big( 4 \vec{v}_{1}\cdot\dot{\vec{a}}_{2} \vec{n} - 28 \vec{v}_{2}\cdot\dot{\vec{a}}_{2} \vec{n} + 7 \dot{\vec{a}}_{2}\cdot\vec{n} \vec{v}_{1} + 44 \dot{\vec{a}}_{2}\cdot\vec{n} \vec{v}_{2} \nn\\ 
&& - 32 \vec{v}_{1}\cdot\vec{n} \dot{\vec{a}}_{2} + 152 \vec{v}_{2}\cdot\vec{n} \dot{\vec{a}}_{2} -4 \vec{v}_{1}\cdot\vec{n} \dot{\vec{a}}_{2}\cdot\vec{n} \vec{n} + 28 \vec{v}_{2}\cdot\vec{n} \dot{\vec{a}}_{2}\cdot\vec{n} \vec{n} \big) \nn\\ 
&& - 16 \vec{S}_{1}\times\vec{n}\cdot\vec{v}_{2} \big( \vec{v}_{1}\cdot\dot{\vec{a}}_{2} \vec{n} - 7 \vec{v}_{2}\cdot\dot{\vec{a}}_{2} \vec{n} + \dot{\vec{a}}_{2}\cdot\vec{n} \vec{v}_{1} + 5 \dot{\vec{a}}_{2}\cdot\vec{n} \vec{v}_{2} - 5 \vec{v}_{1}\cdot\vec{n} \dot{\vec{a}}_{2} \nn\\ 
&& + 23 \vec{v}_{2}\cdot\vec{n} \dot{\vec{a}}_{2} - \vec{v}_{1}\cdot\vec{n} \dot{\vec{a}}_{2}\cdot\vec{n} \vec{n} + 7 \vec{v}_{2}\cdot\vec{n} \dot{\vec{a}}_{2}\cdot\vec{n} \vec{n} \big) + 144 \vec{S}_{1}\times\vec{v}_{1}\cdot\vec{v}_{2} \dot{\vec{a}}_{2} \nn\\ 
&& - 48 \vec{S}_{1}\times\vec{n}\cdot\dot{\vec{a}}_{2} \big( \vec{v}_{1}\cdot\vec{v}_{2} \vec{n} - 3 v_{2}^2 \vec{n} + \vec{v}_{2}\cdot\vec{n} \vec{v}_{1} - \vec{v}_{1}\cdot\vec{n} \vec{v}_{2} + 4 \vec{v}_{2}\cdot\vec{n} \vec{v}_{2} - \vec{v}_{1}\cdot\vec{n} \vec{v}_{2}\cdot\vec{n} \vec{n} \nn\\ 
&& + 3 ( \vec{v}_{2}\cdot\vec{n})^{2} \vec{n} \big) + 4 \vec{S}_{1}\times\vec{v}_{1}\cdot\dot{\vec{a}}_{2} \big( 53 \vec{v}_{1} + 100 \vec{v}_{2} + 8 \vec{v}_{1}\cdot\vec{n} \vec{n} - 44 \vec{v}_{2}\cdot\vec{n} \vec{n} \big) \nn\\ 
&& - 16 \vec{S}_{1}\times\vec{v}_{2}\cdot\dot{\vec{a}}_{2} \big( 2 \vec{v}_{1} + 7 \vec{v}_{2} + 2 \vec{v}_{1}\cdot\vec{n} \vec{n} - 11 \vec{v}_{2}\cdot\vec{n} \vec{n} \big) + 4 \vec{S}_{1}\times\vec{n} \big( 7 v_{1}^2 \dot{\vec{a}}_{2}\cdot\vec{n} \nn\\ 
&& + 52 \vec{v}_{1}\cdot\vec{v}_{2} \dot{\vec{a}}_{2}\cdot\vec{n} - 104 v_{2}^2 \dot{\vec{a}}_{2}\cdot\vec{n} - 28 \vec{v}_{1}\cdot\vec{n} \vec{v}_{1}\cdot\dot{\vec{a}}_{2} + 160 \vec{v}_{2}\cdot\vec{n} \vec{v}_{1}\cdot\dot{\vec{a}}_{2} + 4 \vec{v}_{1}\cdot\vec{n} \vec{v}_{2}\cdot\dot{\vec{a}}_{2} \nn\\ 
&& - 280 \vec{v}_{2}\cdot\vec{n} \vec{v}_{2}\cdot\dot{\vec{a}}_{2} + 20 \vec{v}_{1}\cdot\vec{n} \vec{v}_{2}\cdot\vec{n} \dot{\vec{a}}_{2}\cdot\vec{n} - 4 \dot{\vec{a}}_{2}\cdot\vec{n} ( \vec{v}_{1}\cdot\vec{n})^{2} + 80 \dot{\vec{a}}_{2}\cdot\vec{n} ( \vec{v}_{2}\cdot\vec{n})^{2} \big) \nn\\ 
&& + 4 \vec{S}_{1}\times\vec{v}_{1} \big( 307 \vec{v}_{1}\cdot\dot{\vec{a}}_{2} - 352 \vec{v}_{2}\cdot\dot{\vec{a}}_{2} + 7 \vec{v}_{1}\cdot\vec{n} \dot{\vec{a}}_{2}\cdot\vec{n} - 184 \vec{v}_{2}\cdot\vec{n} \dot{\vec{a}}_{2}\cdot\vec{n} \big) \nn\\ 
&& - 16 \vec{S}_{1}\times\vec{v}_{2} \big( 49 \vec{v}_{1}\cdot\dot{\vec{a}}_{2} - 115 \vec{v}_{2}\cdot\dot{\vec{a}}_{2} -2 \vec{v}_{1}\cdot\vec{n} \dot{\vec{a}}_{2}\cdot\vec{n} - 49 \vec{v}_{2}\cdot\vec{n} \dot{\vec{a}}_{2}\cdot\vec{n} \big) \nn\\ 
&& - 4 \vec{S}_{1}\times\dot{\vec{a}}_{2} \big( 53 v_{1}^2 + 356 \vec{v}_{1}\cdot\vec{v}_{2} - 628 v_{2}^2 + 44 \vec{v}_{1}\cdot\vec{n} \vec{v}_{2}\cdot\vec{n} + 8 ( \vec{v}_{1}\cdot\vec{n})^{2} \nn\\ 
&& - 460 ( \vec{v}_{2}\cdot\vec{n})^{2} \big)) - (32 \ddot{\vec{S}}_{1}\times\vec{n} \big( 3 \vec{v}_{2}\cdot\vec{n} v_{2}^2 - ( \vec{v}_{2}\cdot\vec{n})^{3} \big) - 288 \ddot{\vec{S}}_{1}\times\vec{v}_{2} \big( v_{2}^2 + ( \vec{v}_{2}\cdot\vec{n})^{2} \big)) \nn\\ 
&& - (4 \vec{S}_{1}\times\vec{n}\cdot\vec{v}_{1} \big( 40 a_{2}^2 \vec{n} - \vec{a}_{2}\cdot\vec{n} \vec{a}_{1} - 168 \vec{a}_{2}\cdot\vec{n} \vec{a}_{2} \big) - 96 \vec{S}_{1}\times\vec{n}\cdot\vec{a}_{1} \big( \vec{a}_{2}\cdot\vec{n} \vec{v}_{2} \nn\\ 
&& + 2 \vec{v}_{2}\cdot\vec{n} \vec{a}_{2} \big) + 4 \vec{S}_{1}\times\vec{v}_{1}\cdot\vec{a}_{1} \big( 97 \vec{a}_{2} -15 \vec{a}_{2}\cdot\vec{n} \vec{n} \big) - 96 \vec{S}_{1}\times\vec{n}\cdot\vec{v}_{2} \big( a_{2}^2 \vec{n} - 4 \vec{a}_{2}\cdot\vec{n} \vec{a}_{2} \big) \nn\\ 
&& + 96 \vec{S}_{1}\times\vec{a}_{1}\cdot\vec{v}_{2} \big( \vec{a}_{2} - \vec{a}_{2}\cdot\vec{n} \vec{n} \big) + 64 \vec{S}_{1}\times\vec{n}\cdot\vec{a}_{2} \big( 2 \vec{v}_{1}\cdot\vec{a}_{2} \vec{n} - 6 \vec{v}_{2}\cdot\vec{a}_{2} \vec{n} + 3 \vec{a}_{2}\cdot\vec{n} \vec{v}_{2} \nn\\ 
&& - 2 \vec{v}_{1}\cdot\vec{n} \vec{a}_{2} + 12 \vec{v}_{2}\cdot\vec{n} \vec{a}_{2} + 6 \vec{v}_{2}\cdot\vec{n} \vec{a}_{2}\cdot\vec{n} \vec{n} \big) - 4 \vec{S}_{1}\times\vec{v}_{1}\cdot\vec{a}_{2} \big( 15 \vec{a}_{1} \nn\\ 
&& + 304 \vec{a}_{2} -24 \vec{a}_{2}\cdot\vec{n} \vec{n} \big) - 96 \vec{S}_{1}\times\vec{a}_{1}\cdot\vec{a}_{2} \big( \vec{v}_{2} + 2 \vec{v}_{2}\cdot\vec{n} \vec{n} \big) + 192 \vec{S}_{1}\times\vec{v}_{2}\cdot\vec{a}_{2} \big( 2 \vec{a}_{2} \nn\\ 
&& - \vec{a}_{2}\cdot\vec{n} \vec{n} \big) - 4 \vec{S}_{1}\times\vec{n} \big( 3 \vec{v}_{1}\cdot\vec{a}_{1} \vec{a}_{2}\cdot\vec{n} + 168 \vec{a}_{2}\cdot\vec{n} \vec{v}_{1}\cdot\vec{a}_{2} + 24 \vec{v}_{2}\cdot\vec{n} \vec{a}_{1}\cdot\vec{a}_{2} \nn\\ 
&& - 24 \vec{a}_{1}\cdot\vec{n} \vec{v}_{2}\cdot\vec{a}_{2} - 384 \vec{a}_{2}\cdot\vec{n} \vec{v}_{2}\cdot\vec{a}_{2} - 8 \vec{v}_{1}\cdot\vec{n} a_{2}^2 - 264 \vec{v}_{2}\cdot\vec{n} a_{2}^2 \nn\\ 
&& + 24 \vec{a}_{1}\cdot\vec{n} \vec{v}_{2}\cdot\vec{n} \vec{a}_{2}\cdot\vec{n} \big) - 8 \vec{S}_{1}\times\vec{v}_{1} \big( 97 \vec{a}_{1}\cdot\vec{a}_{2} - 126 a_{2}^2 -15 \vec{a}_{1}\cdot\vec{n} \vec{a}_{2}\cdot\vec{n} \big) \nn\\ 
&& - 12 \vec{S}_{1}\times\vec{a}_{1} \big( 97 \vec{v}_{1}\cdot\vec{a}_{2} - 120 \vec{v}_{2}\cdot\vec{a}_{2} -15 \vec{v}_{1}\cdot\vec{n} \vec{a}_{2}\cdot\vec{n} - 56 \vec{v}_{2}\cdot\vec{n} \vec{a}_{2}\cdot\vec{n} \big) \nn\\ 
&& + 192 \vec{S}_{1}\times\vec{v}_{2} \big( 2 \vec{a}_{1}\cdot\vec{a}_{2} - 8 a_{2}^2 - \vec{a}_{1}\cdot\vec{n} \vec{a}_{2}\cdot\vec{n} \big) + 4 \vec{S}_{1}\times\vec{a}_{2} \big( 45 \vec{v}_{1}\cdot\vec{a}_{1} + 48 \vec{a}_{1}\cdot\vec{v}_{2} \nn\\ 
&& + 496 \vec{v}_{1}\cdot\vec{a}_{2} - 1440 \vec{v}_{2}\cdot\vec{a}_{2} -120 \vec{a}_{1}\cdot\vec{n} \vec{v}_{2}\cdot\vec{n} - 24 \vec{v}_{1}\cdot\vec{n} \vec{a}_{2}\cdot\vec{n} - 816 \vec{v}_{2}\cdot\vec{n} \vec{a}_{2}\cdot\vec{n} \big)) \nn\\ 
&& + (48 \dot{\vec{S}}_{1}\times\vec{n}\cdot\vec{v}_{1} \big( \vec{v}_{2}\cdot\vec{a}_{2} \vec{n} + 2 \vec{a}_{2}\cdot\vec{n} \vec{v}_{1} + \vec{a}_{2}\cdot\vec{n} \vec{v}_{2} + \vec{v}_{2}\cdot\vec{n} \vec{a}_{2} - \vec{v}_{2}\cdot\vec{n} \vec{a}_{2}\cdot\vec{n} \vec{n} \big) \nn\\ 
&& - 48 \dot{\vec{S}}_{1}\times\vec{n}\cdot\vec{v}_{2} \big( \vec{v}_{2}\cdot\vec{a}_{2} \vec{n} + \vec{a}_{2}\cdot\vec{n} \vec{v}_{2} + \vec{v}_{2}\cdot\vec{n} \vec{a}_{2} - \vec{v}_{2}\cdot\vec{n} \vec{a}_{2}\cdot\vec{n} \vec{n} \big) \nn\\ 
&& - 48 \dot{\vec{S}}_{1}\times\vec{v}_{1}\cdot\vec{v}_{2} \big( \vec{a}_{2} -3 \vec{a}_{2}\cdot\vec{n} \vec{n} \big) - 48 \dot{\vec{S}}_{1}\times\vec{n}\cdot\vec{a}_{2} \big( v_{2}^2 \vec{n} + 2 \vec{v}_{2}\cdot\vec{n} \vec{v}_{2} - ( \vec{v}_{2}\cdot\vec{n})^{2} \vec{n} \big) \nn\\ 
&& + 48 \dot{\vec{S}}_{1}\times\vec{v}_{1}\cdot\vec{a}_{2} \big( 14 \vec{v}_{1} + 3 \vec{v}_{2} + 7 \vec{v}_{2}\cdot\vec{n} \vec{n} \big) - 48 \dot{\vec{S}}_{1}\times\vec{v}_{2}\cdot\vec{a}_{2} \big( \vec{v}_{2} + \vec{v}_{2}\cdot\vec{n} \vec{n} \big) \nn\\ 
&& + 3 \dot{\vec{S}}_{1}\times\vec{n} \big( 17 v_{1}^2 \vec{a}_{2}\cdot\vec{n} - 32 \vec{v}_{1}\cdot\vec{v}_{2} \vec{a}_{2}\cdot\vec{n} + 48 v_{2}^2 \vec{a}_{2}\cdot\vec{n} - 32 \vec{v}_{2}\cdot\vec{n} \vec{v}_{1}\cdot\vec{a}_{2} \nn\\ 
&& - 32 \vec{v}_{1}\cdot\vec{n} \vec{v}_{2}\cdot\vec{a}_{2} + 96 \vec{v}_{2}\cdot\vec{n} \vec{v}_{2}\cdot\vec{a}_{2} + 32 \vec{v}_{1}\cdot\vec{n} \vec{v}_{2}\cdot\vec{n} \vec{a}_{2}\cdot\vec{n} - 64 \vec{a}_{2}\cdot\vec{n} ( \vec{v}_{2}\cdot\vec{n})^{2} \big) \nn\\ 
&& + 6 \dot{\vec{S}}_{1}\times\vec{v}_{1} \big( 81 \vec{v}_{1}\cdot\vec{a}_{2} - 88 \vec{v}_{2}\cdot\vec{a}_{2} -31 \vec{v}_{1}\cdot\vec{n} \vec{a}_{2}\cdot\vec{n} - 24 \vec{v}_{2}\cdot\vec{n} \vec{a}_{2}\cdot\vec{n} \big) \nn\\ 
&& + 192 \dot{\vec{S}}_{1}\times\vec{a}_{1} \big( v_{2}^2 + ( \vec{v}_{2}\cdot\vec{n})^{2} \big) + 48 \dot{\vec{S}}_{1}\times\vec{v}_{2} \big( 2 \vec{v}_{1}\cdot\vec{a}_{2} - \vec{v}_{2}\cdot\vec{a}_{2} + 6 \vec{v}_{1}\cdot\vec{n} \vec{a}_{2}\cdot\vec{n} \nn\\ 
&& - 15 \vec{v}_{2}\cdot\vec{n} \vec{a}_{2}\cdot\vec{n} \big) - 3 \dot{\vec{S}}_{1}\times\vec{a}_{2} \big( 127 v_{1}^2 - 160 \vec{v}_{1}\cdot\vec{v}_{2} + 192 v_{2}^2 -224 \vec{v}_{1}\cdot\vec{n} \vec{v}_{2}\cdot\vec{n} \nn\\ 
&& + 304 ( \vec{v}_{2}\cdot\vec{n})^{2} \big)) \Big] - 	\frac{1}{768} G r \Big[ (16 \vec{S}_{2}\times\vec{n}\cdot\vec{v}_{1} \big( \vec{v}_{1}\cdot\dot{\vec{a}}_{2} \vec{n} - 7 \vec{v}_{2}\cdot\dot{\vec{a}}_{2} \vec{n} + \dot{\vec{a}}_{2}\cdot\vec{n} \vec{v}_{1} \nn\\ 
&& + 5 \dot{\vec{a}}_{2}\cdot\vec{n} \vec{v}_{2} - 5 \vec{v}_{1}\cdot\vec{n} \dot{\vec{a}}_{2} + 23 \vec{v}_{2}\cdot\vec{n} \dot{\vec{a}}_{2} - \vec{v}_{1}\cdot\vec{n} \dot{\vec{a}}_{2}\cdot\vec{n} \vec{n} + 7 \vec{v}_{2}\cdot\vec{n} \dot{\vec{a}}_{2}\cdot\vec{n} \vec{n} \big) \nn\\ 
&& + 16 \vec{S}_{2}\times\vec{n}\cdot\vec{v}_{2} \big( 36 \vec{v}_{1}\cdot\dot{\vec{a}}_{1} \vec{n} - 33 \dot{\vec{a}}_{1}\cdot\vec{v}_{2} \vec{n} + 10 \vec{v}_{1}\cdot\dot{\vec{a}}_{2} \vec{n} - 4 \vec{v}_{2}\cdot\dot{\vec{a}}_{2} \vec{n} + 36 \dot{\vec{a}}_{1}\cdot\vec{n} \vec{v}_{1} \nn\\ 
&& + \dot{\vec{a}}_{2}\cdot\vec{n} \vec{v}_{1} + 63 \vec{v}_{1}\cdot\vec{n} \dot{\vec{a}}_{1} + 21 \vec{v}_{2}\cdot\vec{n} \dot{\vec{a}}_{1} - 33 \dot{\vec{a}}_{1}\cdot\vec{n} \vec{v}_{2} - \dot{\vec{a}}_{2}\cdot\vec{n} \vec{v}_{2} + 19 \vec{v}_{1}\cdot\vec{n} \dot{\vec{a}}_{2} \nn\\ 
&& - 4 \vec{v}_{2}\cdot\vec{n} \dot{\vec{a}}_{2} -15 \vec{v}_{1}\cdot\vec{n} \dot{\vec{a}}_{1}\cdot\vec{n} \vec{n} - 3 \dot{\vec{a}}_{1}\cdot\vec{n} \vec{v}_{2}\cdot\vec{n} \vec{n} + 2 \vec{v}_{1}\cdot\vec{n} \dot{\vec{a}}_{2}\cdot\vec{n} \vec{n} - 5 \vec{v}_{2}\cdot\vec{n} \dot{\vec{a}}_{2}\cdot\vec{n} \vec{n} \big) \nn\\ 
&& - 16 \vec{S}_{2}\times\vec{v}_{1}\cdot\vec{v}_{2} \big( 9 \dot{\vec{a}}_{1} + 4 \dot{\vec{a}}_{2} + 57 \dot{\vec{a}}_{1}\cdot\vec{n} \vec{n} - 5 \dot{\vec{a}}_{2}\cdot\vec{n} \vec{n} \big) - 48 \vec{S}_{2}\times\dot{\vec{a}}_{1}\cdot\vec{v}_{2} \big( 40 \vec{v}_{1} - 25 \vec{v}_{2} \nn\\ 
&& + 47 \vec{v}_{1}\cdot\vec{n} \vec{n} - 5 \vec{v}_{2}\cdot\vec{n} \vec{n} \big) - 24 \vec{S}_{2}\times\vec{n}\cdot\dot{\vec{a}}_{2} \big( 3 v_{1}^2 \vec{n} - 4 \vec{v}_{1}\cdot\vec{v}_{2} \vec{n} - v_{2}^2 \vec{n} + 10 \vec{v}_{1}\cdot\vec{n} \vec{v}_{1} \nn\\ 
&& + 2 \vec{v}_{2}\cdot\vec{n} \vec{v}_{1} - 10 \vec{v}_{1}\cdot\vec{n} \vec{v}_{2} - 2 \vec{v}_{2}\cdot\vec{n} \vec{v}_{2} -4 \vec{v}_{1}\cdot\vec{n} \vec{v}_{2}\cdot\vec{n} \vec{n} - ( \vec{v}_{1}\cdot\vec{n})^{2} \vec{n} + 4 ( \vec{v}_{2}\cdot\vec{n})^{2} \vec{n} \big) \nn\\ 
&& + 16 \vec{S}_{2}\times\vec{v}_{1}\cdot\dot{\vec{a}}_{2} \big( 8 \vec{v}_{1} - 14 \vec{v}_{2} + 20 \vec{v}_{1}\cdot\vec{n} \vec{n} + 25 \vec{v}_{2}\cdot\vec{n} \vec{n} \big) - 16 \vec{S}_{2}\times\vec{v}_{2}\cdot\dot{\vec{a}}_{2} \big( 34 \vec{v}_{1} \nn\\ 
&& - 34 \vec{v}_{2} + 22 \vec{v}_{1}\cdot\vec{n} \vec{n} + 11 \vec{v}_{2}\cdot\vec{n} \vec{n} \big) + 16 \vec{S}_{2}\times\vec{n} \big( v_{1}^2 \dot{\vec{a}}_{2}\cdot\vec{n} + 7 \vec{v}_{1}\cdot\vec{v}_{2} \dot{\vec{a}}_{2}\cdot\vec{n} - 20 v_{2}^2 \dot{\vec{a}}_{2}\cdot\vec{n} \nn\\ 
&& - 4 \vec{v}_{1}\cdot\vec{n} \vec{v}_{1}\cdot\dot{\vec{a}}_{2} + 25 \vec{v}_{2}\cdot\vec{n} \vec{v}_{1}\cdot\dot{\vec{a}}_{2} - 5 \vec{v}_{1}\cdot\vec{n} \vec{v}_{2}\cdot\dot{\vec{a}}_{2} - 40 \vec{v}_{2}\cdot\vec{n} \vec{v}_{2}\cdot\dot{\vec{a}}_{2} \nn\\ 
&& + 5 \vec{v}_{1}\cdot\vec{n} \vec{v}_{2}\cdot\vec{n} \dot{\vec{a}}_{2}\cdot\vec{n} - \dot{\vec{a}}_{2}\cdot\vec{n} ( \vec{v}_{1}\cdot\vec{n})^{2} + 20 \dot{\vec{a}}_{2}\cdot\vec{n} ( \vec{v}_{2}\cdot\vec{n})^{2} \big) - 16 \vec{S}_{2}\times\vec{v}_{1} \big( 5 \vec{v}_{1}\cdot\dot{\vec{a}}_{2} \nn\\ 
&& - 2 \vec{v}_{2}\cdot\dot{\vec{a}}_{2} - \vec{v}_{1}\cdot\vec{n} \dot{\vec{a}}_{2}\cdot\vec{n} - 2 \vec{v}_{2}\cdot\vec{n} \dot{\vec{a}}_{2}\cdot\vec{n} \big) + 48 \vec{S}_{2}\times\vec{v}_{2} \big( 32 \vec{v}_{1}\cdot\dot{\vec{a}}_{1} - 17 \dot{\vec{a}}_{1}\cdot\vec{v}_{2} \nn\\ 
&& + 4 \vec{v}_{1}\cdot\dot{\vec{a}}_{2} - 5 \vec{v}_{2}\cdot\dot{\vec{a}}_{2} + 39 \vec{v}_{1}\cdot\vec{n} \dot{\vec{a}}_{1}\cdot\vec{n} + 3 \dot{\vec{a}}_{1}\cdot\vec{n} \vec{v}_{2}\cdot\vec{n} - 2 \vec{v}_{1}\cdot\vec{n} \dot{\vec{a}}_{2}\cdot\vec{n} - 8 \vec{v}_{2}\cdot\vec{n} \dot{\vec{a}}_{2}\cdot\vec{n} \big) \nn\\ 
&& - 8 \vec{S}_{2}\times\dot{\vec{a}}_{2} \big( 13 v_{1}^2 + 4 \vec{v}_{1}\cdot\vec{v}_{2} + 37 v_{2}^2 + 100 \vec{v}_{1}\cdot\vec{n} \vec{v}_{2}\cdot\vec{n} + 19 ( \vec{v}_{1}\cdot\vec{n})^{2} - 2 ( \vec{v}_{2}\cdot\vec{n})^{2} \big)) \nn\\ 
&& - (192 \ddot{\vec{S}}_{2}\times\vec{n}\cdot\vec{v}_{1} \big( v_{2}^2 \vec{n} - \vec{v}_{1}\cdot\vec{n} \vec{v}_{1} + \vec{v}_{2}\cdot\vec{n} \vec{v}_{1} - \vec{v}_{2}\cdot\vec{n} \vec{v}_{2} - \vec{v}_{1}\cdot\vec{n} \vec{v}_{2}\cdot\vec{n} \vec{n} \nn\\ 
&& - ( \vec{v}_{2}\cdot\vec{n})^{2} \vec{n} \big) + 6 \ddot{\vec{S}}_{2}\times\vec{n}\cdot\vec{v}_{2} \big( 12 v_{1}^2 \vec{n} - 8 \vec{v}_{1}\cdot\vec{v}_{2} \vec{n} - 27 v_{2}^2 \vec{n} + 24 \vec{v}_{1}\cdot\vec{n} \vec{v}_{1} - 48 \vec{v}_{2}\cdot\vec{n} \vec{v}_{1} \nn\\ 
&& - 8 \vec{v}_{1}\cdot\vec{n} \vec{v}_{2} + 24 \vec{v}_{1}\cdot\vec{n} \vec{v}_{2}\cdot\vec{n} \vec{n} - 4 ( \vec{v}_{1}\cdot\vec{n})^{2} \vec{n} + 24 ( \vec{v}_{2}\cdot\vec{n})^{2} \vec{n} \big) - 48 \ddot{\vec{S}}_{2}\times\vec{v}_{1}\cdot\vec{v}_{2} \big( 9 \vec{v}_{1} \nn\\ 
&& - 5 \vec{v}_{2} + 11 \vec{v}_{1}\cdot\vec{n} \vec{n} + \vec{v}_{2}\cdot\vec{n} \vec{n} \big) + 32 \ddot{\vec{S}}_{2}\times\vec{n} \big( 3 v_{1}^2 \vec{v}_{2}\cdot\vec{n} - 6 \vec{v}_{2}\cdot\vec{n} \vec{v}_{1}\cdot\vec{v}_{2} + 6 \vec{v}_{1}\cdot\vec{n} v_{2}^2 \nn\\ 
&& + 30 \vec{v}_{2}\cdot\vec{n} v_{2}^2 -3 \vec{v}_{2}\cdot\vec{n} ( \vec{v}_{1}\cdot\vec{n})^{2} - 6 \vec{v}_{1}\cdot\vec{n} ( \vec{v}_{2}\cdot\vec{n})^{2} - 10 ( \vec{v}_{2}\cdot\vec{n})^{3} \big) + 6 \ddot{\vec{S}}_{2}\times\vec{v}_{2} \big( 36 v_{1}^2 \nn\\ 
&& - 40 \vec{v}_{1}\cdot\vec{v}_{2} - 5 v_{2}^2 + 8 \vec{v}_{1}\cdot\vec{n} \vec{v}_{2}\cdot\vec{n} + 44 ( \vec{v}_{1}\cdot\vec{n})^{2} + 40 ( \vec{v}_{2}\cdot\vec{n})^{2} \big)) \nn\\ 
&& - (96 \vec{S}_{2}\times\vec{n}\cdot\vec{v}_{1} \big( a_{2}^2 \vec{n} - 4 \vec{a}_{2}\cdot\vec{n} \vec{a}_{2} \big) - 64 \vec{S}_{2}\times\vec{n}\cdot\vec{a}_{1} \big( \vec{a}_{2}\cdot\vec{n} \vec{v}_{2} + 2 \vec{v}_{2}\cdot\vec{n} \vec{a}_{2} \big) \nn\\ 
&& - 4 \vec{S}_{2}\times\vec{n}\cdot\vec{v}_{2} \big( 54 a_{1}^2 \vec{n} - 90 \vec{a}_{1}\cdot\vec{a}_{2} \vec{n} - 19 a_{2}^2 \vec{n} + 216 \vec{a}_{1}\cdot\vec{n} \vec{a}_{1} + 58 \vec{a}_{2}\cdot\vec{n} \vec{a}_{1} \nn\\ 
&& - 42 \vec{a}_{1}\cdot\vec{n} \vec{a}_{2} - 36 \vec{a}_{2}\cdot\vec{n} \vec{a}_{2} -6 \vec{a}_{1}\cdot\vec{n} \vec{a}_{2}\cdot\vec{n} \vec{n} \big) + 8 \vec{S}_{2}\times\vec{a}_{1}\cdot\vec{v}_{2} \big( 84 \vec{a}_{1} - 91 \vec{a}_{2} \nn\\ 
&& + 162 \vec{a}_{1}\cdot\vec{n} \vec{n} - 3 \vec{a}_{2}\cdot\vec{n} \vec{n} \big) - 4 \vec{S}_{2}\times\vec{n}\cdot\vec{a}_{2} \big( 84 \vec{v}_{1}\cdot\vec{a}_{1} \vec{n} - 84 \vec{a}_{1}\cdot\vec{v}_{2} \vec{n} + 36 \vec{v}_{1}\cdot\vec{a}_{2} \vec{n} \nn\\ 
&& - 3 \vec{v}_{2}\cdot\vec{a}_{2} \vec{n} + 48 \vec{a}_{1}\cdot\vec{n} \vec{v}_{1} + 24 \vec{a}_{2}\cdot\vec{n} \vec{v}_{1} + 84 \vec{v}_{1}\cdot\vec{n} \vec{a}_{1} + 116 \vec{v}_{2}\cdot\vec{n} \vec{a}_{1} - 36 \vec{a}_{1}\cdot\vec{n} \vec{v}_{2} \nn\\ 
&& + 132 \vec{v}_{1}\cdot\vec{n} \vec{a}_{2} - 72 \vec{v}_{2}\cdot\vec{n} \vec{a}_{2} -24 \vec{v}_{1}\cdot\vec{n} \vec{a}_{1}\cdot\vec{n} \vec{n} - 12 \vec{a}_{1}\cdot\vec{n} \vec{v}_{2}\cdot\vec{n} \vec{n} + 12 \vec{v}_{1}\cdot\vec{n} \vec{a}_{2}\cdot\vec{n} \vec{n} \nn\\ 
&& - 72 \vec{v}_{2}\cdot\vec{n} \vec{a}_{2}\cdot\vec{n} \vec{n} \big) + 48 \vec{S}_{2}\times\vec{v}_{1}\cdot\vec{a}_{2} \big( 5 \vec{a}_{1} - 3 \vec{a}_{2} + 10 \vec{a}_{1}\cdot\vec{n} \vec{n} - 3 \vec{a}_{2}\cdot\vec{n} \vec{n} \big) \nn\\ 
&& + 16 \vec{S}_{2}\times\vec{a}_{1}\cdot\vec{a}_{2} \big( 60 \vec{v}_{1} - 11 \vec{v}_{2} + 66 \vec{v}_{1}\cdot\vec{n} \vec{n} + 21 \vec{v}_{2}\cdot\vec{n} \vec{n} \big) - 4 \vec{S}_{2}\times\vec{v}_{2}\cdot\vec{a}_{2} \big( 6 \vec{a}_{1} \nn\\ 
&& + 23 \vec{a}_{2} -30 \vec{a}_{1}\cdot\vec{n} \vec{n} + 15 \vec{a}_{2}\cdot\vec{n} \vec{n} \big) + 32 \vec{S}_{2}\times\vec{n} \big( \vec{a}_{1}\cdot\vec{v}_{2} \vec{a}_{2}\cdot\vec{n} - 12 \vec{a}_{2}\cdot\vec{n} \vec{v}_{1}\cdot\vec{a}_{2} \nn\\ 
&& - \vec{v}_{2}\cdot\vec{n} \vec{a}_{1}\cdot\vec{a}_{2} + 3 \vec{a}_{1}\cdot\vec{n} \vec{v}_{2}\cdot\vec{a}_{2} + 30 \vec{a}_{2}\cdot\vec{n} \vec{v}_{2}\cdot\vec{a}_{2} + 3 \vec{v}_{1}\cdot\vec{n} a_{2}^2 \nn\\ 
&& + 15 \vec{v}_{2}\cdot\vec{n} a_{2}^2 -3 \vec{a}_{1}\cdot\vec{n} \vec{v}_{2}\cdot\vec{n} \vec{a}_{2}\cdot\vec{n} \big) + 96 \vec{S}_{2}\times\vec{a}_{1} \big( \vec{v}_{2}\cdot\vec{a}_{2} + \vec{v}_{2}\cdot\vec{n} \vec{a}_{2}\cdot\vec{n} \big) \nn\\ 
&& - 4 \vec{S}_{2}\times\vec{v}_{2} \big( 282 a_{1}^2 - 158 \vec{a}_{1}\cdot\vec{a}_{2} - 45 a_{2}^2 -30 \vec{a}_{1}\cdot\vec{n} \vec{a}_{2}\cdot\vec{n} \big) - 4 \vec{S}_{2}\times\vec{a}_{2} \big( 60 \vec{v}_{1}\cdot\vec{a}_{1} \nn\\ 
&& - 68 \vec{a}_{1}\cdot\vec{v}_{2} - 36 \vec{v}_{1}\cdot\vec{a}_{2} - 117 \vec{v}_{2}\cdot\vec{a}_{2} + 120 \vec{v}_{1}\cdot\vec{n} \vec{a}_{1}\cdot\vec{n} + 12 \vec{a}_{1}\cdot\vec{n} \vec{v}_{2}\cdot\vec{n} \nn\\ 
&& - 132 \vec{v}_{1}\cdot\vec{n} \vec{a}_{2}\cdot\vec{n} + 48 \vec{v}_{2}\cdot\vec{n} \vec{a}_{2}\cdot\vec{n} \big)) + (48 \dot{\vec{S}}_{2}\times\vec{n}\cdot\vec{v}_{1} \big( 3 \vec{v}_{1}\cdot\vec{a}_{2} \vec{n} - 8 \vec{v}_{2}\cdot\vec{a}_{2} \vec{n} \nn\\ 
&& - \vec{a}_{2}\cdot\vec{n} \vec{v}_{1} + 4 \vec{a}_{2}\cdot\vec{n} \vec{v}_{2} - \vec{v}_{1}\cdot\vec{n} \vec{a}_{2} + 16 \vec{v}_{2}\cdot\vec{n} \vec{a}_{2} + \vec{v}_{1}\cdot\vec{n} \vec{a}_{2}\cdot\vec{n} \vec{n} + 8 \vec{v}_{2}\cdot\vec{n} \vec{a}_{2}\cdot\vec{n} \vec{n} \big) \nn\\ 
&& + 48 \dot{\vec{S}}_{2}\times\vec{n}\cdot\vec{a}_{1} \big( v_{2}^2 \vec{n} + 2 \vec{v}_{2}\cdot\vec{n} \vec{v}_{2} - ( \vec{v}_{2}\cdot\vec{n})^{2} \vec{n} \big) + 12 \dot{\vec{S}}_{2}\times\vec{n}\cdot\vec{v}_{2} \big( 28 \vec{v}_{1}\cdot\vec{a}_{1} \vec{n} \nn\\ 
&& - 28 \vec{a}_{1}\cdot\vec{v}_{2} \vec{n} + 8 \vec{v}_{1}\cdot\vec{a}_{2} \vec{n} + 7 \vec{v}_{2}\cdot\vec{a}_{2} \vec{n} + 16 \vec{a}_{1}\cdot\vec{n} \vec{v}_{1} + 12 \vec{a}_{2}\cdot\vec{n} \vec{v}_{1} + 28 \vec{v}_{1}\cdot\vec{n} \vec{a}_{1} \nn\\ 
&& + 28 \vec{v}_{2}\cdot\vec{n} \vec{a}_{1} - 12 \vec{a}_{1}\cdot\vec{n} \vec{v}_{2} + 32 \vec{v}_{1}\cdot\vec{n} \vec{a}_{2} - 24 \vec{v}_{2}\cdot\vec{n} \vec{a}_{2} -8 \vec{v}_{1}\cdot\vec{n} \vec{a}_{1}\cdot\vec{n} \vec{n} \nn\\ 
&& - 4 \vec{a}_{1}\cdot\vec{n} \vec{v}_{2}\cdot\vec{n} \vec{n} - 24 \vec{v}_{2}\cdot\vec{n} \vec{a}_{2}\cdot\vec{n} \vec{n} \big) - 48 \dot{\vec{S}}_{2}\times\vec{v}_{1}\cdot\vec{v}_{2} \big( 5 \vec{a}_{1} + 6 \vec{a}_{2} + 10 \vec{a}_{1}\cdot\vec{n} \vec{n} \nn\\ 
&& - 6 \vec{a}_{2}\cdot\vec{n} \vec{n} \big) - 48 \dot{\vec{S}}_{2}\times\vec{a}_{1}\cdot\vec{v}_{2} \big( 20 \vec{v}_{1} - 3 \vec{v}_{2} + 22 \vec{v}_{1}\cdot\vec{n} \vec{n} + 9 \vec{v}_{2}\cdot\vec{n} \vec{n} \big) \nn\\ 
&& - 3 \dot{\vec{S}}_{2}\times\vec{n}\cdot\vec{a}_{2} \big( 32 v_{1}^2 \vec{n} - 32 \vec{v}_{1}\cdot\vec{v}_{2} \vec{n} - 61 v_{2}^2 \vec{n} + 128 \vec{v}_{1}\cdot\vec{n} \vec{v}_{1} - 96 \vec{v}_{2}\cdot\vec{n} \vec{v}_{1} \nn\\ 
&& - 128 \vec{v}_{1}\cdot\vec{n} \vec{v}_{2} -16 ( \vec{v}_{1}\cdot\vec{n})^{2} \vec{n} + 96 ( \vec{v}_{2}\cdot\vec{n})^{2} \vec{n} \big) - 48 \dot{\vec{S}}_{2}\times\vec{v}_{1}\cdot\vec{a}_{2} \big( 7 \vec{v}_{1} \nn\\ 
&& + 2 \vec{v}_{2} -15 \vec{v}_{1}\cdot\vec{n} \vec{n} - 6 \vec{v}_{2}\cdot\vec{n} \vec{n} \big) + 6 \dot{\vec{S}}_{2}\times\vec{v}_{2}\cdot\vec{a}_{2} \big( 8 \vec{v}_{1} - \vec{v}_{2} -48 \vec{v}_{1}\cdot\vec{n} \vec{n} + 31 \vec{v}_{2}\cdot\vec{n} \vec{n} \big) \nn\\ 
&& - 48 \dot{\vec{S}}_{2}\times\vec{n} \big( 2 \vec{v}_{2}\cdot\vec{n} \vec{a}_{1}\cdot\vec{v}_{2} + \vec{a}_{1}\cdot\vec{n} v_{2}^2 - v_{1}^2 \vec{a}_{2}\cdot\vec{n} - 4 \vec{v}_{1}\cdot\vec{v}_{2} \vec{a}_{2}\cdot\vec{n} + 20 v_{2}^2 \vec{a}_{2}\cdot\vec{n} \nn\\ 
&& - 2 \vec{v}_{1}\cdot\vec{n} \vec{v}_{1}\cdot\vec{a}_{2} - 16 \vec{v}_{2}\cdot\vec{n} \vec{v}_{1}\cdot\vec{a}_{2} + 8 \vec{v}_{1}\cdot\vec{n} \vec{v}_{2}\cdot\vec{a}_{2} \nn\\ 
&& + 40 \vec{v}_{2}\cdot\vec{n} \vec{v}_{2}\cdot\vec{a}_{2} -8 \vec{v}_{1}\cdot\vec{n} \vec{v}_{2}\cdot\vec{n} \vec{a}_{2}\cdot\vec{n} + \vec{a}_{2}\cdot\vec{n} ( \vec{v}_{1}\cdot\vec{n})^{2} - \vec{a}_{1}\cdot\vec{n} ( \vec{v}_{2}\cdot\vec{n})^{2} \nn\\ 
&& - 20 \vec{a}_{2}\cdot\vec{n} ( \vec{v}_{2}\cdot\vec{n})^{2} \big) - 48 \dot{\vec{S}}_{2}\times\vec{v}_{1} \big( \vec{v}_{1}\cdot\vec{a}_{2} -3 \vec{v}_{1}\cdot\vec{n} \vec{a}_{2}\cdot\vec{n} \big) - 96 \dot{\vec{S}}_{2}\times\vec{a}_{1} \big( v_{2}^2 \nn\\ 
&& + ( \vec{v}_{2}\cdot\vec{n})^{2} \big) + 12 \dot{\vec{S}}_{2}\times\vec{v}_{2} \big( 20 \vec{v}_{1}\cdot\vec{a}_{1} - 28 \vec{a}_{1}\cdot\vec{v}_{2} + 24 \vec{v}_{1}\cdot\vec{a}_{2} - 63 \vec{v}_{2}\cdot\vec{a}_{2} \nn\\ 
&& + 40 \vec{v}_{1}\cdot\vec{n} \vec{a}_{1}\cdot\vec{n} + 4 \vec{a}_{1}\cdot\vec{n} \vec{v}_{2}\cdot\vec{n} - 48 \vec{v}_{1}\cdot\vec{n} \vec{a}_{2}\cdot\vec{n} - 64 \vec{v}_{2}\cdot\vec{n} \vec{a}_{2}\cdot\vec{n} \big) + 3 \dot{\vec{S}}_{2}\times\vec{a}_{2} \big( 48 v_{1}^2 \nn\\ 
&& - 160 \vec{v}_{1}\cdot\vec{v}_{2} + 19 v_{2}^2 -384 \vec{v}_{1}\cdot\vec{n} \vec{v}_{2}\cdot\vec{n} - 96 ( \vec{v}_{1}\cdot\vec{n})^{2} + 160 ( \vec{v}_{2}\cdot\vec{n})^{2} \big)) \Big] \nn\\ 
&& - 	\frac{G m_{2} r}{16 m_{1}} \Big[ 2 \vec{S}_{1}\times\vec{n}\cdot\vec{v}_{1} ( \vec{a}_{2}\cdot\vec{n})^{2} \vec{n} - 2 \vec{S}_{1}\times\vec{n}\cdot\vec{v}_{2} ( \vec{a}_{2}\cdot\vec{n})^{2} \vec{n} + 2 \vec{S}_{1}\times\vec{n} \big( \vec{v}_{1}\cdot\vec{n} ( \vec{a}_{2}\cdot\vec{n})^{2} \nn\\ 
&& + 5 \vec{v}_{2}\cdot\vec{n} ( \vec{a}_{2}\cdot\vec{n})^{2} \big) - 9 \vec{S}_{1}\times\vec{v}_{1} ( \vec{a}_{2}\cdot\vec{n})^{2} + 12 \vec{S}_{1}\times\vec{v}_{2} ( \vec{a}_{2}\cdot\vec{n})^{2} \Big] \nn\\ 
&& - 	\frac{1}{32} G r \Big[ 4 \vec{S}_{2}\times\vec{n}\cdot\vec{v}_{1} ( \vec{a}_{2}\cdot\vec{n})^{2} \vec{n} - \vec{S}_{2}\times\vec{n}\cdot\vec{v}_{2} \big( 5 ( \vec{a}_{1}\cdot\vec{n})^{2} \vec{n} + 3 ( \vec{a}_{2}\cdot\vec{n})^{2} \vec{n} \big) \nn\\ 
&& + 4 \vec{S}_{2}\times\vec{n} \big( \vec{v}_{1}\cdot\vec{n} ( \vec{a}_{2}\cdot\vec{n})^{2} + 5 \vec{v}_{2}\cdot\vec{n} ( \vec{a}_{2}\cdot\vec{n})^{2} \big) + \vec{S}_{2}\times\vec{v}_{2} \big( 51 ( \vec{a}_{1}\cdot\vec{n})^{2} - 11 ( \vec{a}_{2}\cdot\vec{n})^{2} \big) \Big]\nn\\ && +  	\frac{G^2 m_{2}{}^2}{576 m_{1}} \Big[ (48 \vec{S}_{1}\times\vec{n}\cdot\dot{\vec{a}}_{2} \vec{n} + 1404 \vec{S}_{1}\times\vec{n} \dot{\vec{a}}_{2}\cdot\vec{n} + 4952 \vec{S}_{1}\times\dot{\vec{a}}_{2}) - (333 \dot{\vec{S}}_{1}\times\vec{n} \vec{a}_{2}\cdot\vec{n} \nn\\ 
&& - 1647 \dot{\vec{S}}_{1}\times\vec{a}_{2}) \Big] + 	\frac{1}{6} G^2 m_{2} {\Big( \frac{1}{\epsilon} - 2\log \frac{r}{R_0} \Big)} \Big[ (10 \vec{S}_{1}\times\dot{\vec{a}}_{1} + 40 \vec{S}_{1}\times\dot{\vec{a}}_{2}) - (6 \ddot{\vec{S}}_{1}\times\vec{v}_{1} \nn\\ 
&& - 8 \ddot{\vec{S}}_{1}\times\vec{v}_{2}) - (28 \dot{\vec{S}}_{1}\times\vec{a}_{1} - 5 \dot{\vec{S}}_{1}\times\vec{a}_{2}) \Big] + 	\frac{1}{576} G^2 m_{2} \Big[ (51 \vec{S}_{1}\times\vec{n}\cdot\dot{\vec{a}}_{1} \vec{n} \nn\\ 
&& + 288 \vec{S}_{1}\times\vec{n}\cdot\dot{\vec{a}}_{2} \vec{n} - 3 \vec{S}_{1}\times\vec{n} \big( 383 \dot{\vec{a}}_{1}\cdot\vec{n} - 812 \dot{\vec{a}}_{2}\cdot\vec{n} \big) + 3720 \vec{S}_{1}\times\dot{\vec{a}}_{1} \nn\\ 
&& + 2608 \vec{S}_{1}\times\dot{\vec{a}}_{2}) + (756 \ddot{\vec{S}}_{1}\times\vec{n}\cdot\vec{v}_{1} \vec{n} - 588 \ddot{\vec{S}}_{1}\times\vec{n}\cdot\vec{v}_{2} \vec{n} - 36 \ddot{\vec{S}}_{1}\times\vec{n} \big( 20 \vec{v}_{1}\cdot\vec{n} \nn\\ 
&& - 89 \vec{v}_{2}\cdot\vec{n} \big) + 52 \ddot{\vec{S}}_{1}\times\vec{v}_{1} + 576 \ddot{\vec{S}}_{1}\times\vec{v}_{2}) - (117 \dot{\vec{S}}_{1}\times\vec{n}\cdot\vec{a}_{1} \vec{n} + 360 \dot{\vec{S}}_{1}\times\vec{n}\cdot\vec{a}_{2} \vec{n} \nn\\ 
&& + 1068 \dot{\vec{S}}_{1}\times\vec{n} \big( \vec{a}_{1}\cdot\vec{n} - 2 \vec{a}_{2}\cdot\vec{n} \big) - 47 \dot{\vec{S}}_{1}\times\vec{a}_{1} + 1368 \dot{\vec{S}}_{1}\times\vec{a}_{2}) \Big] \nn\\ 
&& + 	\frac{1}{576} G^2 m_{1} \Big[ (624 \vec{S}_{2}\times\vec{n}\cdot\dot{\vec{a}}_{1} \vec{n} + 750 \vec{S}_{2}\times\vec{n}\cdot\dot{\vec{a}}_{2} \vec{n} + 15 \vec{S}_{2}\times\vec{n} \big( 32 \dot{\vec{a}}_{1}\cdot\vec{n} + 71 \dot{\vec{a}}_{2}\cdot\vec{n} \big) \nn\\ 
&& - 1320 \vec{S}_{2}\times\dot{\vec{a}}_{1} - 2174 \vec{S}_{2}\times\dot{\vec{a}}_{2}) + (2124 \ddot{\vec{S}}_{2}\times\vec{n}\cdot\vec{v}_{1} \vec{n} - 1479 \ddot{\vec{S}}_{2}\times\vec{n}\cdot\vec{v}_{2} \vec{n} \nn\\ 
&& + 6 \ddot{\vec{S}}_{2}\times\vec{n} \big( 44 \vec{v}_{1}\cdot\vec{n} + 625 \vec{v}_{2}\cdot\vec{n} \big) + 1649 \ddot{\vec{S}}_{2}\times\vec{v}_{1} - 1590 \ddot{\vec{S}}_{2}\times\vec{v}_{2}) \nn\\ 
&& + (273 \dot{\vec{S}}_{2}\times\vec{n}\cdot\vec{a}_{1} \vec{n} - 216 \dot{\vec{S}}_{2}\times\vec{n}\cdot\vec{a}_{2} \vec{n} + 3 \dot{\vec{S}}_{2}\times\vec{n} \big( 95 \vec{a}_{1}\cdot\vec{n} + 1093 \vec{a}_{2}\cdot\vec{n} \big) \nn\\ 
&& - 322 \dot{\vec{S}}_{2}\times\vec{a}_{1} - 1558 \dot{\vec{S}}_{2}\times\vec{a}_{2}) \Big] - 	\frac{1}{6} G^2 m_{1} {\Big( \frac{1}{\epsilon} - 2\log \frac{r}{R_0} \Big)} \Big[ 30 \vec{S}_{2}\times\dot{\vec{a}}_{2} - (8 \ddot{\vec{S}}_{2}\times\vec{v}_{1} \nn\\ 
&& - 6 \ddot{\vec{S}}_{2}\times\vec{v}_{2}) - (5 \dot{\vec{S}}_{2}\times\vec{a}_{1} - 28 \dot{\vec{S}}_{2}\times\vec{a}_{2}) \Big] - 	\frac{1}{6} G^2 m_{2} {\Big( \frac{1}{\epsilon} - 2\log \frac{r}{R_0} \Big)} \Big[ 30 \vec{S}_{2}\times\dot{\vec{a}}_{2} \nn\\ 
&& - 2 \ddot{\vec{S}}_{2}\times\vec{v}_{2} + 23 \dot{\vec{S}}_{2}\times\vec{a}_{2} \Big] + 	\frac{1}{288} G^2 m_{2} \Big[ (243 \vec{S}_{2}\times\vec{n}\cdot\dot{\vec{a}}_{2} \vec{n} + 936 \vec{S}_{2}\times\vec{n} \dot{\vec{a}}_{2}\cdot\vec{n} \nn\\ 
&& - 2063 \vec{S}_{2}\times\dot{\vec{a}}_{2}) + (480 \ddot{\vec{S}}_{2}\times\vec{n}\cdot\vec{v}_{1} \vec{n} - 315 \ddot{\vec{S}}_{2}\times\vec{n}\cdot\vec{v}_{2} \vec{n} + 24 \ddot{\vec{S}}_{2}\times\vec{n} \big( 20 \vec{v}_{1}\cdot\vec{n} \nn\\ 
&& + 13 \vec{v}_{2}\cdot\vec{n} \big) + 87 \ddot{\vec{S}}_{2}\times\vec{v}_{2}) - (318 \dot{\vec{S}}_{2}\times\vec{n}\cdot\vec{a}_{2} \vec{n} - 1674 \dot{\vec{S}}_{2}\times\vec{n} \vec{a}_{2}\cdot\vec{n} + 2058 \dot{\vec{S}}_{2}\times\vec{a}_{2}) \Big] \nn\\ 
&& + 	\frac{20 G^2 m_{2}{}^2}{3 m_{1}} {\Big( \frac{1}{\epsilon} - 2\log \frac{r}{R_0} \Big)} \vec{S}_{1}\times\dot{\vec{a}}_{2} ,
\eea
\bea
\stackrel{(3)}{ \Delta \vec{x}_1}_{(3,1)}&=& 	\frac{G m_{2} r{}^2}{48 m_{1}} \Big[ ( \vec{S}_{1}\times\vec{n}\cdot\vec{v}_{1} \big( 8 \ddot{\vec{a}}_{2} - \ddot{\vec{a}}_{2}\cdot\vec{n} \vec{n} \big) - \vec{S}_{1}\times\vec{n}\cdot\vec{v}_{2} \big( 5 \ddot{\vec{a}}_{2} - \ddot{\vec{a}}_{2}\cdot\vec{n} \vec{n} \big) \nn\\ 
&& - 3 \vec{S}_{1}\times\vec{n}\cdot\ddot{\vec{a}}_{2} \big( \vec{v}_{2} - \vec{v}_{2}\cdot\vec{n} \vec{n} \big) - 2 \vec{S}_{1}\times\vec{v}_{1}\cdot\ddot{\vec{a}}_{2} \vec{n} + 2 \vec{S}_{1}\times\vec{v}_{2}\cdot\ddot{\vec{a}}_{2} \vec{n} + \vec{S}_{1}\times\vec{n} \big( 10 \vec{v}_{1}\cdot\ddot{\vec{a}}_{2} \nn\\ 
&& - 16 \vec{v}_{2}\cdot\ddot{\vec{a}}_{2} + \vec{v}_{1}\cdot\vec{n} \ddot{\vec{a}}_{2}\cdot\vec{n} - 10 \vec{v}_{2}\cdot\vec{n} \ddot{\vec{a}}_{2}\cdot\vec{n} \big) - 10 \vec{S}_{1}\times\vec{v}_{1} \ddot{\vec{a}}_{2}\cdot\vec{n} + 13 \vec{S}_{1}\times\vec{v}_{2} \ddot{\vec{a}}_{2}\cdot\vec{n} \nn\\ 
&& - \vec{S}_{1}\times\ddot{\vec{a}}_{2} \big( 20 \vec{v}_{1}\cdot\vec{n} - 83 \vec{v}_{2}\cdot\vec{n} \big)) + ( \vec{S}_{1}\times\vec{n}\cdot\vec{a}_{1} \big( 8 \dot{\vec{a}}_{2} - \dot{\vec{a}}_{2}\cdot\vec{n} \vec{n} \big) \nn\\ 
&& - \vec{S}_{1}\times\vec{n}\cdot\vec{a}_{2} \big( 23 \dot{\vec{a}}_{2} -7 \dot{\vec{a}}_{2}\cdot\vec{n} \vec{n} \big) - 3 \vec{S}_{1}\times\vec{n}\cdot\dot{\vec{a}}_{2} \big( 7 \vec{a}_{2} -3 \vec{a}_{2}\cdot\vec{n} \vec{n} \big) - 2 \vec{S}_{1}\times\vec{a}_{1}\cdot\dot{\vec{a}}_{2} \vec{n} \nn\\ 
&& + 2 \vec{S}_{1}\times\vec{a}_{2}\cdot\dot{\vec{a}}_{2} \vec{n} + \vec{S}_{1}\times\vec{n} \big( 8 \vec{a}_{1}\cdot\dot{\vec{a}}_{2} - 50 \vec{a}_{2}\cdot\dot{\vec{a}}_{2} - \vec{a}_{1}\cdot\vec{n} \dot{\vec{a}}_{2}\cdot\vec{n} - 20 \vec{a}_{2}\cdot\vec{n} \dot{\vec{a}}_{2}\cdot\vec{n} \big) \nn\\ 
&& - 12 \vec{S}_{1}\times\vec{a}_{1} \dot{\vec{a}}_{2}\cdot\vec{n} + 69 \vec{S}_{1}\times\vec{a}_{2} \dot{\vec{a}}_{2}\cdot\vec{n} + \vec{S}_{1}\times\dot{\vec{a}}_{2} \big( 2 \vec{a}_{1}\cdot\vec{n} + 115 \vec{a}_{2}\cdot\vec{n} \big)) \nn\\ 
&& + 6 \dot{\vec{S}}_{1}\times\vec{n} ( \vec{a}_{2}\cdot\vec{n})^{2} + ( \dot{\vec{S}}_{1}\times\vec{n}\cdot\vec{v}_{1} \big( 8 \dot{\vec{a}}_{2} - \dot{\vec{a}}_{2}\cdot\vec{n} \vec{n} \big) - \dot{\vec{S}}_{1}\times\vec{n}\cdot\vec{v}_{2} \big( 5 \dot{\vec{a}}_{2} - \dot{\vec{a}}_{2}\cdot\vec{n} \vec{n} \big) \nn\\ 
&& - 3 \dot{\vec{S}}_{1}\times\vec{n}\cdot\dot{\vec{a}}_{2} \big( \vec{v}_{2} - \vec{v}_{2}\cdot\vec{n} \vec{n} \big) - 2 \dot{\vec{S}}_{1}\times\vec{v}_{1}\cdot\dot{\vec{a}}_{2} \vec{n} + 2 \dot{\vec{S}}_{1}\times\vec{v}_{2}\cdot\dot{\vec{a}}_{2} \vec{n} + \dot{\vec{S}}_{1}\times\vec{n} \big( 7 \vec{v}_{1}\cdot\dot{\vec{a}}_{2} \nn\\ 
&& - \vec{v}_{2}\cdot\dot{\vec{a}}_{2} -2 \vec{v}_{1}\cdot\vec{n} \dot{\vec{a}}_{2}\cdot\vec{n} + 5 \vec{v}_{2}\cdot\vec{n} \dot{\vec{a}}_{2}\cdot\vec{n} \big) - \dot{\vec{S}}_{1}\times\vec{v}_{1} \dot{\vec{a}}_{2}\cdot\vec{n} - 20 \dot{\vec{S}}_{1}\times\vec{v}_{2} \dot{\vec{a}}_{2}\cdot\vec{n} \nn\\ 
&& + \dot{\vec{S}}_{1}\times\dot{\vec{a}}_{2} \big( 13 \vec{v}_{1}\cdot\vec{n} - 34 \vec{v}_{2}\cdot\vec{n} \big)) + (3 \ddot{\vec{S}}_{1}\times\vec{n} \big( \vec{v}_{2}\cdot\vec{a}_{2} + \vec{v}_{2}\cdot\vec{n} \vec{a}_{2}\cdot\vec{n} \big) \nn\\ 
&& + 6 \ddot{\vec{S}}_{1}\times\vec{v}_{1} \vec{a}_{2}\cdot\vec{n} - 9 \ddot{\vec{S}}_{1}\times\vec{v}_{2} \vec{a}_{2}\cdot\vec{n} - 21 \ddot{\vec{S}}_{1}\times\vec{a}_{2} \vec{v}_{2}\cdot\vec{n}) - (8 \dot{\vec{S}}_{1}\times\vec{n}\cdot\vec{a}_{2} \vec{a}_{2} \nn\\ 
&& - 2 \dot{\vec{S}}_{1}\times\vec{n} a_{2}^2 - 12 \dot{\vec{S}}_{1}\times\vec{a}_{1} \vec{a}_{2}\cdot\vec{n} + 60 \dot{\vec{S}}_{1}\times\vec{a}_{2} \vec{a}_{2}\cdot\vec{n}) \Big] + 	\frac{1}{96} G r{}^2 \Big[ (2 \vec{S}_{2}\times\vec{n}\cdot\vec{v}_{1} \big( 5 \ddot{\vec{a}}_{2} \nn\\ 
&& - \ddot{\vec{a}}_{2}\cdot\vec{n} \vec{n} \big) + \vec{S}_{2}\times\vec{n}\cdot\vec{v}_{2} \big( 7 \ddot{\vec{a}}_{2} + \ddot{\vec{a}}_{2}\cdot\vec{n} \vec{n} \big) + 24 \vec{S}_{2}\times\ddot{\vec{a}}_{1}\cdot\vec{v}_{2} \vec{n} - 2 \vec{S}_{2}\times\vec{n}\cdot\ddot{\vec{a}}_{2} \big( 4 \vec{v}_{1} \nn\\ 
&& - 4 \vec{v}_{2} + \vec{v}_{1}\cdot\vec{n} \vec{n} - \vec{v}_{2}\cdot\vec{n} \vec{n} \big) + 24 \vec{S}_{2}\times\vec{v}_{1}\cdot\ddot{\vec{a}}_{2} \vec{n} - 17 \vec{S}_{2}\times\vec{v}_{2}\cdot\ddot{\vec{a}}_{2} \vec{n} + 2 \vec{S}_{2}\times\vec{n} \big( 7 \vec{v}_{1}\cdot\ddot{\vec{a}}_{2} \nn\\ 
&& - 10 \vec{v}_{2}\cdot\ddot{\vec{a}}_{2} + \vec{v}_{1}\cdot\vec{n} \ddot{\vec{a}}_{2}\cdot\vec{n} - 10 \vec{v}_{2}\cdot\vec{n} \ddot{\vec{a}}_{2}\cdot\vec{n} \big) + 4 \vec{S}_{2}\times\vec{v}_{1} \ddot{\vec{a}}_{2}\cdot\vec{n} + 3 \vec{S}_{2}\times\vec{v}_{2} \big( 8 \ddot{\vec{a}}_{1}\cdot\vec{n} \nn\\ 
&& - 5 \ddot{\vec{a}}_{2}\cdot\vec{n} \big) - 20 \vec{S}_{2}\times\ddot{\vec{a}}_{2} \big( \vec{v}_{1}\cdot\vec{n} + 2 \vec{v}_{2}\cdot\vec{n} \big)) - (8 \dddot{\vec{S}}_{2}\times\vec{n}\cdot\vec{v}_{1} \big( \vec{v}_{1} - \vec{v}_{2} + \vec{v}_{1}\cdot\vec{n} \vec{n} \nn\\ 
&& + 2 \vec{v}_{2}\cdot\vec{n} \vec{n} \big) - 6 \dddot{\vec{S}}_{2}\times\vec{n}\cdot\vec{v}_{2} \big( 2 \vec{v}_{1} + \vec{v}_{1}\cdot\vec{n} \vec{n} + 2 \vec{v}_{2}\cdot\vec{n} \vec{n} \big) + 2 \dddot{\vec{S}}_{2}\times\vec{v}_{1}\cdot\vec{v}_{2} \vec{n} \nn\\ 
&& + 4 \dddot{\vec{S}}_{2}\times\vec{n} \big( v_{1}^2 - 2 \vec{v}_{1}\cdot\vec{v}_{2} + 10 v_{2}^2 + 4 \vec{v}_{1}\cdot\vec{n} \vec{v}_{2}\cdot\vec{n} + ( \vec{v}_{1}\cdot\vec{n})^{2} + 10 ( \vec{v}_{2}\cdot\vec{n})^{2} \big) \nn\\ 
&& - 2 \dddot{\vec{S}}_{2}\times\vec{v}_{2} \big( \vec{v}_{1}\cdot\vec{n} - 4 \vec{v}_{2}\cdot\vec{n} \big)) + (2 \vec{S}_{2}\times\vec{n}\cdot\vec{a}_{1} \big( 5 \dot{\vec{a}}_{2} - \dot{\vec{a}}_{2}\cdot\vec{n} \vec{n} \big) + 2 \vec{S}_{2}\times\vec{n}\cdot\vec{a}_{2} \big( 21 \dot{\vec{a}}_{1} \nn\\ 
&& - 4 \dot{\vec{a}}_{2} + 3 \dot{\vec{a}}_{1}\cdot\vec{n} \vec{n} + 5 \dot{\vec{a}}_{2}\cdot\vec{n} \vec{n} \big) - 42 \vec{S}_{2}\times\dot{\vec{a}}_{1}\cdot\vec{a}_{2} \vec{n} + 3 \vec{S}_{2}\times\vec{n}\cdot\dot{\vec{a}}_{2} \big( 11 \vec{a}_{1} - 2 \vec{a}_{2} \nn\\ 
&& + \vec{a}_{1}\cdot\vec{n} \vec{n} + 4 \vec{a}_{2}\cdot\vec{n} \vec{n} \big) - 43 \vec{S}_{2}\times\vec{a}_{1}\cdot\dot{\vec{a}}_{2} \vec{n} - 16 \vec{S}_{2}\times\vec{a}_{2}\cdot\dot{\vec{a}}_{2} \vec{n} + 2 \vec{S}_{2}\times\vec{n} \big( 5 \vec{a}_{1}\cdot\dot{\vec{a}}_{2} \nn\\ 
&& - 20 \vec{a}_{2}\cdot\dot{\vec{a}}_{2} - \vec{a}_{1}\cdot\vec{n} \dot{\vec{a}}_{2}\cdot\vec{n} - 20 \vec{a}_{2}\cdot\vec{n} \dot{\vec{a}}_{2}\cdot\vec{n} \big) + 2 \vec{S}_{2}\times\vec{a}_{2} \big( 45 \dot{\vec{a}}_{1}\cdot\vec{n} + 2 \dot{\vec{a}}_{2}\cdot\vec{n} \big) \nn\\ 
&& + \vec{S}_{2}\times\dot{\vec{a}}_{2} \big( 19 \vec{a}_{1}\cdot\vec{n} + 2 \vec{a}_{2}\cdot\vec{n} \big)) - 60 \dot{\vec{S}}_{2}\times\vec{n} ( \vec{a}_{2}\cdot\vec{n})^{2} \nn\\ 
&& + (2 \dot{\vec{S}}_{2}\times\vec{n}\cdot\vec{v}_{1} \big( 23 \dot{\vec{a}}_{2} -7 \dot{\vec{a}}_{2}\cdot\vec{n} \vec{n} \big) + 2 \dot{\vec{S}}_{2}\times\vec{n}\cdot\vec{v}_{2} \big( 21 \dot{\vec{a}}_{1} - 4 \dot{\vec{a}}_{2} + 3 \dot{\vec{a}}_{1}\cdot\vec{n} \vec{n} \nn\\ 
&& + 5 \dot{\vec{a}}_{2}\cdot\vec{n} \vec{n} \big) - 42 \dot{\vec{S}}_{2}\times\dot{\vec{a}}_{1}\cdot\vec{v}_{2} \vec{n} - 6 \dot{\vec{S}}_{2}\times\vec{n}\cdot\dot{\vec{a}}_{2} \big( \vec{v}_{1} - \vec{v}_{2} + 2 \vec{v}_{1}\cdot\vec{n} \vec{n} - 4 \vec{v}_{2}\cdot\vec{n} \vec{n} \big) \nn\\ 
&& + 74 \dot{\vec{S}}_{2}\times\vec{v}_{1}\cdot\dot{\vec{a}}_{2} \vec{n} - 22 \dot{\vec{S}}_{2}\times\vec{v}_{2}\cdot\dot{\vec{a}}_{2} \vec{n} + 10 \dot{\vec{S}}_{2}\times\vec{n} \big( 5 \vec{v}_{1}\cdot\dot{\vec{a}}_{2} - 8 \vec{v}_{2}\cdot\dot{\vec{a}}_{2} - \vec{v}_{1}\cdot\vec{n} \dot{\vec{a}}_{2}\cdot\vec{n} \nn\\ 
&& - 8 \vec{v}_{2}\cdot\vec{n} \dot{\vec{a}}_{2}\cdot\vec{n} \big) + 4 \dot{\vec{S}}_{2}\times\vec{v}_{1} \dot{\vec{a}}_{2}\cdot\vec{n} + 18 \dot{\vec{S}}_{2}\times\vec{v}_{2} \big( 5 \dot{\vec{a}}_{1}\cdot\vec{n} - 2 \dot{\vec{a}}_{2}\cdot\vec{n} \big) \nn\\ 
&& - 2 \dot{\vec{S}}_{2}\times\dot{\vec{a}}_{2} \big( 71 \vec{v}_{1}\cdot\vec{n} - 11 \vec{v}_{2}\cdot\vec{n} \big)) + (24 \ddot{\vec{S}}_{2}\times\vec{n}\cdot\vec{v}_{1} \big( 2 \vec{a}_{2} - \vec{a}_{2}\cdot\vec{n} \vec{n} \big) \nn\\ 
&& + 6 \ddot{\vec{S}}_{2}\times\vec{n}\cdot\vec{a}_{1} \big( \vec{v}_{2} + \vec{v}_{2}\cdot\vec{n} \vec{n} \big) + 3 \ddot{\vec{S}}_{2}\times\vec{n}\cdot\vec{v}_{2} \big( 7 \vec{a}_{1} - 6 \vec{a}_{2} + \vec{a}_{1}\cdot\vec{n} \vec{n} + 6 \vec{a}_{2}\cdot\vec{n} \vec{n} \big) \nn\\ 
&& - 51 \ddot{\vec{S}}_{2}\times\vec{a}_{1}\cdot\vec{v}_{2} \vec{n} + 18 \ddot{\vec{S}}_{2}\times\vec{n}\cdot\vec{a}_{2} \big( \vec{v}_{1} + 2 \vec{v}_{2}\cdot\vec{n} \vec{n} \big) + 24 \ddot{\vec{S}}_{2}\times\vec{v}_{1}\cdot\vec{a}_{2} \vec{n} \nn\\ 
&& + 12 \ddot{\vec{S}}_{2}\times\vec{v}_{2}\cdot\vec{a}_{2} \vec{n} - 6 \ddot{\vec{S}}_{2}\times\vec{n} \big( \vec{a}_{1}\cdot\vec{v}_{2} - 8 \vec{v}_{1}\cdot\vec{a}_{2} + 20 \vec{v}_{2}\cdot\vec{a}_{2} + \vec{a}_{1}\cdot\vec{n} \vec{v}_{2}\cdot\vec{n} \nn\\ 
&& + 4 \vec{v}_{1}\cdot\vec{n} \vec{a}_{2}\cdot\vec{n} + 20 \vec{v}_{2}\cdot\vec{n} \vec{a}_{2}\cdot\vec{n} \big) - 12 \ddot{\vec{S}}_{2}\times\vec{a}_{1} \vec{v}_{2}\cdot\vec{n} + 15 \ddot{\vec{S}}_{2}\times\vec{v}_{2} \big( \vec{a}_{1}\cdot\vec{n} - 2 \vec{a}_{2}\cdot\vec{n} \big) \nn\\ 
&& - 96 \ddot{\vec{S}}_{2}\times\vec{a}_{2} \big( \vec{v}_{1}\cdot\vec{n} - \vec{v}_{2}\cdot\vec{n} \big)) + (16 \dot{\vec{S}}_{2}\times\vec{n}\cdot\vec{a}_{1} \vec{a}_{2} + 2 \dot{\vec{S}}_{2}\times\vec{n}\cdot\vec{a}_{2} \big( 29 \vec{a}_{1} - 18 \vec{a}_{2} \nn\\ 
&& + 3 \vec{a}_{1}\cdot\vec{n} \vec{n} + 18 \vec{a}_{2}\cdot\vec{n} \vec{n} \big) - 90 \dot{\vec{S}}_{2}\times\vec{a}_{1}\cdot\vec{a}_{2} \vec{n} + 4 \dot{\vec{S}}_{2}\times\vec{n} \big( \vec{a}_{1}\cdot\vec{a}_{2} \nn\\ 
&& - 15 a_{2}^2 -3 \vec{a}_{1}\cdot\vec{n} \vec{a}_{2}\cdot\vec{n} \big) - 12 \dot{\vec{S}}_{2}\times\vec{a}_{1} \vec{a}_{2}\cdot\vec{n} + 30 \dot{\vec{S}}_{2}\times\vec{a}_{2} \big( \vec{a}_{1}\cdot\vec{n} + 2 \vec{a}_{2}\cdot\vec{n} \big)) \Big]\nn\\ && +  	\frac{559}{576} G^2 m_{1} r \dddot{\vec{S}}_{2}\times\vec{n} + 	\frac{7}{6} G^2 m_{2} r \dddot{\vec{S}}_{1}\times\vec{n} + 	\frac{77}{48} G^2 m_{2} r \dddot{\vec{S}}_{2}\times\vec{n} \nn\\ 
&& + 	\frac{4}{3} G^2 m_{1} r {\Big( \frac{1}{\epsilon} - 2\log \frac{r}{R_0} \Big)} \dddot{\vec{S}}_{2}\times\vec{n} + 	\frac{4}{3} G^2 m_{2} r {\Big( \frac{1}{\epsilon} - 2\log \frac{r}{R_0} \Big)} \dddot{\vec{S}}_{2}\times\vec{n} ,
\eea
\bea
\stackrel{(4)}{ \Delta \vec{x}_1}_{(3,1)}&=&  	\frac{G m_{2} r{}^3}{144 m_{1}} \Big[ (6 \vec{S}_{1}\times\vec{n} \dddot{\vec{a}}_{2}\cdot\vec{n} - 22 \vec{S}_{1}\times\dddot{\vec{a}}_{2}) + (3 \dot{\vec{S}}_{1}\times\vec{n} \ddot{\vec{a}}_{2}\cdot\vec{n} - 11 \dot{\vec{S}}_{1}\times\ddot{\vec{a}}_{2}) \nn\\ 
&& - (3 \ddot{\vec{S}}_{1}\times\vec{n} \dot{\vec{a}}_{2}\cdot\vec{n} - 11 \ddot{\vec{S}}_{1}\times\dot{\vec{a}}_{2}) \Big] + 	\frac{1}{288} G r{}^3 \Big[ (3 \vec{S}_{2}\times\vec{n}\cdot\dddot{\vec{a}}_{2} \vec{n} + 12 \vec{S}_{2}\times\vec{n} \dddot{\vec{a}}_{2}\cdot\vec{n} \nn\\ 
&& + 27 \vec{S}_{2}\times\dddot{\vec{a}}_{2}) + (12 \ddddot{\vec{S}}_{2}\times\vec{n}\cdot\vec{v}_{1} \vec{n} - 9 \ddddot{\vec{S}}_{2}\times\vec{n}\cdot\vec{v}_{2} \vec{n} + 12 \ddddot{\vec{S}}_{2}\times\vec{n} \big( \vec{v}_{1}\cdot\vec{n} + 5 \vec{v}_{2}\cdot\vec{n} \big) \nn\\ 
&& - \ddddot{\vec{S}}_{2}\times\vec{v}_{2}) - (6 \dot{\vec{S}}_{2}\times\vec{n}\cdot\ddot{\vec{a}}_{2} \vec{n} - 60 \dot{\vec{S}}_{2}\times\vec{n} \ddot{\vec{a}}_{2}\cdot\vec{n} - 46 \dot{\vec{S}}_{2}\times\ddot{\vec{a}}_{2}) - (36 \ddot{\vec{S}}_{2}\times\vec{n}\cdot\dot{\vec{a}}_{2} \vec{n} \nn\\ 
&& - 120 \ddot{\vec{S}}_{2}\times\vec{n} \dot{\vec{a}}_{2}\cdot\vec{n} + 20 \ddot{\vec{S}}_{2}\times\dot{\vec{a}}_{2}) - (6 \dddot{\vec{S}}_{2}\times\vec{n}\cdot\vec{a}_{1} \vec{n} + 36 \dddot{\vec{S}}_{2}\times\vec{n}\cdot\vec{a}_{2} \vec{n} \nn\\ 
&& - 6 \dddot{\vec{S}}_{2}\times\vec{n} \big( \vec{a}_{1}\cdot\vec{n} + 20 \vec{a}_{2}\cdot\vec{n} \big) - 4 \dddot{\vec{S}}_{2}\times\vec{a}_{1} + 44 \dddot{\vec{S}}_{2}\times\vec{a}_{2}) \Big],
\eea
\bea
\stackrel{(5)}{ \Delta \vec{x}_1}_{(3,1)}&=& - 	\frac{1}{72} G r{}^4 \stackrel{(5)}{\vec{S}_{2}}\times\vec{n} ,
\eea
and
\bea
\stackrel{(0)}{ \omega^{ij}_1}_{(3,1)}&=& - 	\frac{G m_{2}}{64 r} \Big[ 50 v_{1}^2 \vec{v}_{1}\cdot\vec{v}_{2} v_{1}^i v_{2}^j - 75 v_{1}^2 v_{2}^2 v_{1}^i v_{2}^j + 80 \vec{v}_{1}\cdot\vec{v}_{2} v_{2}^2 v_{1}^i v_{2}^j + 16 ( \vec{v}_{1}\cdot\vec{v}_{2})^{2} v_{1}^i v_{2}^j \nn\\ 
&& - 72 v_{1}^{4} v_{1}^i v_{2}^j - 152 v_{2}^{4} v_{1}^i v_{2}^j + 38 v_{1}^2 \vec{v}_{2}\cdot\vec{n} \vec{v}_{1}\cdot\vec{v}_{2} v_{1}^i n^j - 9 \vec{v}_{1}\cdot\vec{n} v_{1}^2 v_{2}^2 v_{1}^i n^j \nn\\ 
&& - 45 v_{1}^2 \vec{v}_{2}\cdot\vec{n} v_{2}^2 v_{1}^i n^j + 8 \vec{v}_{1}\cdot\vec{n} \vec{v}_{1}\cdot\vec{v}_{2} v_{2}^2 v_{1}^i n^j + 72 \vec{v}_{2}\cdot\vec{n} \vec{v}_{1}\cdot\vec{v}_{2} v_{2}^2 v_{1}^i n^j \nn\\ 
&& - 24 \vec{v}_{2}\cdot\vec{n} v_{1}^{4} v_{1}^i n^j - 16 \vec{v}_{1}\cdot\vec{n} v_{2}^{4} v_{1}^i n^j - 88 \vec{v}_{2}\cdot\vec{n} v_{2}^{4} v_{1}^i n^j + 8 v_{1}^2 \vec{v}_{2}\cdot\vec{n} v_{2}^2 v_{2}^i n^j \nn\\ 
&& - 32 \vec{v}_{2}\cdot\vec{n} \vec{v}_{1}\cdot\vec{v}_{2} v_{2}^2 v_{2}^i n^j - 16 \vec{v}_{2}\cdot\vec{n} ( \vec{v}_{1}\cdot\vec{v}_{2})^{2} v_{2}^i n^j + 16 \vec{v}_{1}\cdot\vec{n} v_{2}^{4} v_{2}^i n^j \nn\\ 
&& + 88 \vec{v}_{2}\cdot\vec{n} v_{2}^{4} v_{2}^i n^j + 6 \vec{v}_{1}\cdot\vec{n} v_{1}^2 \vec{v}_{2}\cdot\vec{n} v_{1}^i v_{2}^j + 16 \vec{v}_{1}\cdot\vec{n} \vec{v}_{2}\cdot\vec{n} \vec{v}_{1}\cdot\vec{v}_{2} v_{1}^i v_{2}^j \nn\\ 
&& - 24 \vec{v}_{1}\cdot\vec{n} \vec{v}_{2}\cdot\vec{n} v_{2}^2 v_{1}^i v_{2}^j + 27 v_{1}^2 ( \vec{v}_{2}\cdot\vec{n})^{2} v_{1}^i v_{2}^j - 32 \vec{v}_{1}\cdot\vec{v}_{2} ( \vec{v}_{2}\cdot\vec{n})^{2} v_{1}^i v_{2}^j \nn\\ 
&& + 128 v_{2}^2 ( \vec{v}_{2}\cdot\vec{n})^{2} v_{1}^i v_{2}^j + 24 \vec{v}_{2}\cdot\vec{n} v_{2}^2 ( \vec{v}_{1}\cdot\vec{n})^{2} v_{1}^i n^j + 27 \vec{v}_{1}\cdot\vec{n} v_{1}^2 ( \vec{v}_{2}\cdot\vec{n})^{2} v_{1}^i n^j \nn\\ 
&& + 29 v_{1}^2 ( \vec{v}_{2}\cdot\vec{n})^{3} v_{1}^i n^j - 24 \vec{v}_{1}\cdot\vec{n} \vec{v}_{1}\cdot\vec{v}_{2} ( \vec{v}_{2}\cdot\vec{n})^{2} v_{1}^i n^j - 40 \vec{v}_{1}\cdot\vec{v}_{2} ( \vec{v}_{2}\cdot\vec{n})^{3} v_{1}^i n^j \nn\\ 
&& + 72 \vec{v}_{1}\cdot\vec{n} v_{2}^2 ( \vec{v}_{2}\cdot\vec{n})^{2} v_{1}^i n^j + 104 v_{2}^2 ( \vec{v}_{2}\cdot\vec{n})^{3} v_{1}^i n^j - 24 \vec{v}_{2}\cdot\vec{n} v_{2}^2 ( \vec{v}_{1}\cdot\vec{n})^{2} v_{2}^i n^j \nn\\ 
&& - 8 v_{1}^2 ( \vec{v}_{2}\cdot\vec{n})^{3} v_{2}^i n^j + 16 \vec{v}_{1}\cdot\vec{v}_{2} ( \vec{v}_{2}\cdot\vec{n})^{3} v_{2}^i n^j - 72 \vec{v}_{1}\cdot\vec{n} v_{2}^2 ( \vec{v}_{2}\cdot\vec{n})^{2} v_{2}^i n^j \nn\\ 
&& - 104 v_{2}^2 ( \vec{v}_{2}\cdot\vec{n})^{3} v_{2}^i n^j + 24 \vec{v}_{1}\cdot\vec{n} ( \vec{v}_{2}\cdot\vec{n})^{3} v_{1}^i v_{2}^j \nn\\ 
&& - 48 ( \vec{v}_{2}\cdot\vec{n})^{4} v_{1}^i v_{2}^j -40 ( \vec{v}_{1}\cdot\vec{n})^{2} ( \vec{v}_{2}\cdot\vec{n})^{3} v_{1}^i n^j - 40 \vec{v}_{1}\cdot\vec{n} ( \vec{v}_{2}\cdot\vec{n})^{4} v_{1}^i n^j \nn\\ 
&& - 40 ( \vec{v}_{2}\cdot\vec{n})^{5} v_{1}^i n^j + 40 ( \vec{v}_{1}\cdot\vec{n})^{2} ( \vec{v}_{2}\cdot\vec{n})^{3} v_{2}^i n^j + 40 \vec{v}_{1}\cdot\vec{n} ( \vec{v}_{2}\cdot\vec{n})^{4} v_{2}^i n^j \nn\\ 
&& + 40 ( \vec{v}_{2}\cdot\vec{n})^{5} v_{2}^i n^j \Big]\nn\\ && +  	\frac{G^2 m_{2}{}^2}{96 r{}^2} \Big[ 172 v_{1}^2 v_{1}^i v_{2}^j + 988 \vec{v}_{1}\cdot\vec{v}_{2} v_{1}^i v_{2}^j - 570 v_{2}^2 v_{1}^i v_{2}^j -42 \vec{v}_{1}\cdot\vec{n} v_{1}^2 v_{1}^i n^j \nn\\ 
&& + 214 v_{1}^2 \vec{v}_{2}\cdot\vec{n} v_{1}^i n^j - 136 \vec{v}_{1}\cdot\vec{n} \vec{v}_{1}\cdot\vec{v}_{2} v_{1}^i n^j - 1146 \vec{v}_{2}\cdot\vec{n} \vec{v}_{1}\cdot\vec{v}_{2} v_{1}^i n^j + 51 \vec{v}_{1}\cdot\vec{n} v_{2}^2 v_{1}^i n^j \nn\\ 
&& + 3366 \vec{v}_{2}\cdot\vec{n} v_{2}^2 v_{1}^i n^j + 127 v_{1}^2 \vec{v}_{2}\cdot\vec{n} v_{2}^i n^j - 16 \vec{v}_{1}\cdot\vec{n} \vec{v}_{1}\cdot\vec{v}_{2} v_{2}^i n^j \nn\\ 
&& - 2868 \vec{v}_{2}\cdot\vec{n} \vec{v}_{1}\cdot\vec{v}_{2} v_{2}^i n^j - 267 \vec{v}_{1}\cdot\vec{n} v_{2}^2 v_{2}^i n^j - 252 \vec{v}_{2}\cdot\vec{n} v_{2}^2 v_{2}^i n^j - 548 \vec{v}_{1}\cdot\vec{n} \vec{v}_{2}\cdot\vec{n} v_{1}^i v_{2}^j \nn\\ 
&& - 56 ( \vec{v}_{1}\cdot\vec{n})^{2} v_{1}^i v_{2}^j + 222 ( \vec{v}_{2}\cdot\vec{n})^{2} v_{1}^i v_{2}^j + 128 \vec{v}_{2}\cdot\vec{n} ( \vec{v}_{1}\cdot\vec{n})^{2} v_{1}^i n^j \nn\\ 
&& + 129 \vec{v}_{1}\cdot\vec{n} ( \vec{v}_{2}\cdot\vec{n})^{2} v_{1}^i n^j + 394 ( \vec{v}_{2}\cdot\vec{n})^{3} v_{1}^i n^j + 128 \vec{v}_{2}\cdot\vec{n} ( \vec{v}_{1}\cdot\vec{n})^{2} v_{2}^i n^j \nn\\ 
&& + 1665 \vec{v}_{1}\cdot\vec{n} ( \vec{v}_{2}\cdot\vec{n})^{2} v_{2}^i n^j - 1408 ( \vec{v}_{2}\cdot\vec{n})^{3} v_{2}^i n^j \Big] + 	\frac{G^2 m_{1} m_{2}}{96 r{}^2} \Big[ 2043 v_{1}^2 v_{1}^i v_{2}^j \nn\\ 
&& - 2364 \vec{v}_{1}\cdot\vec{v}_{2} v_{1}^i v_{2}^j + 1620 v_{2}^2 v_{1}^i v_{2}^j -2295 \vec{v}_{1}\cdot\vec{n} v_{1}^2 v_{1}^i n^j - 318 v_{1}^2 \vec{v}_{2}\cdot\vec{n} v_{1}^i n^j \nn\\ 
&& + 3060 \vec{v}_{1}\cdot\vec{n} \vec{v}_{1}\cdot\vec{v}_{2} v_{1}^i n^j - 1300 \vec{v}_{2}\cdot\vec{n} \vec{v}_{1}\cdot\vec{v}_{2} v_{1}^i n^j - 2780 \vec{v}_{1}\cdot\vec{n} v_{2}^2 v_{1}^i n^j \nn\\ 
&& + 1536 \vec{v}_{2}\cdot\vec{n} v_{2}^2 v_{1}^i n^j + 192 \vec{v}_{1}\cdot\vec{n} v_{1}^2 v_{2}^i n^j + 64 v_{1}^2 \vec{v}_{2}\cdot\vec{n} v_{2}^i n^j + 488 \vec{v}_{1}\cdot\vec{n} \vec{v}_{1}\cdot\vec{v}_{2} v_{2}^i n^j \nn\\ 
&& + 936 \vec{v}_{2}\cdot\vec{n} \vec{v}_{1}\cdot\vec{v}_{2} v_{2}^i n^j + 672 \vec{v}_{1}\cdot\vec{n} v_{2}^2 v_{2}^i n^j - 1416 \vec{v}_{2}\cdot\vec{n} v_{2}^2 v_{2}^i n^j \nn\\ 
&& + 2772 \vec{v}_{1}\cdot\vec{n} \vec{v}_{2}\cdot\vec{n} v_{1}^i v_{2}^j - 2508 ( \vec{v}_{1}\cdot\vec{n})^{2} v_{1}^i v_{2}^j - 1668 ( \vec{v}_{2}\cdot\vec{n})^{2} v_{1}^i v_{2}^j -729 ( \vec{v}_{1}\cdot\vec{n})^{3} v_{1}^i n^j \nn\\ 
&& + 2364 \vec{v}_{2}\cdot\vec{n} ( \vec{v}_{1}\cdot\vec{n})^{2} v_{1}^i n^j + 1808 \vec{v}_{1}\cdot\vec{n} ( \vec{v}_{2}\cdot\vec{n})^{2} v_{1}^i n^j - 1296 ( \vec{v}_{2}\cdot\vec{n})^{3} v_{1}^i n^j \nn\\ 
&& + 328 ( \vec{v}_{1}\cdot\vec{n})^{3} v_{2}^i n^j - 1552 \vec{v}_{2}\cdot\vec{n} ( \vec{v}_{1}\cdot\vec{n})^{2} v_{2}^i n^j - 384 \vec{v}_{1}\cdot\vec{n} ( \vec{v}_{2}\cdot\vec{n})^{2} v_{2}^i n^j \nn\\ 
&& + 1056 ( \vec{v}_{2}\cdot\vec{n})^{3} v_{2}^i n^j \Big]\nn\\ && - 	\frac{19 G^3 m_{1}{}^2 m_{2}}{30 r{}^3} {\Big( \frac{1}{\epsilon} - 3\log \frac{r}{R_0} \Big)} \Big[ v_{1}^i v_{2}^j -9 \vec{v}_{1}\cdot\vec{n} v_{1}^i n^j + 9 \vec{v}_{2}\cdot\vec{n} v_{1}^i n^j + 12 \vec{v}_{1}\cdot\vec{n} v_{2}^i n^j \nn\\ 
&& - 12 \vec{v}_{2}\cdot\vec{n} v_{2}^i n^j \Big] + 	\frac{G^3 m_{1}{}^2 m_{2}}{1800 r{}^3} \Big[ 20299 v_{1}^i v_{2}^j + 107253 \vec{v}_{1}\cdot\vec{n} v_{1}^i n^j - 55251 \vec{v}_{2}\cdot\vec{n} v_{1}^i n^j \nn\\ 
&& - 42594 \vec{v}_{1}\cdot\vec{n} v_{2}^i n^j + 34434 \vec{v}_{2}\cdot\vec{n} v_{2}^i n^j \Big] + 	\frac{G^3 m_{2}{}^3}{1800 r{}^3} \Big[ 20525 v_{1}^i v_{2}^j + 12600 \vec{v}_{1}\cdot\vec{n} v_{1}^i n^j \nn\\ 
&& - 11001 \vec{v}_{2}\cdot\vec{n} v_{1}^i n^j + 14799 \vec{v}_{1}\cdot\vec{n} v_{2}^i n^j + 85548 \vec{v}_{2}\cdot\vec{n} v_{2}^i n^j \Big] \nn\\ 
&& + 	\frac{G^3 m_{1} m_{2}{}^2}{288 r{}^3} \Big[ {(11024 + 468 \pi^2)} v_{1}^i v_{2}^j + {(12063 - 1512 \pi^2)} \vec{v}_{1}\cdot\vec{n} v_{1}^i n^j \nn\\ 
&& - {(11112 + 459 \pi^2)} \vec{v}_{2}\cdot\vec{n} v_{1}^i n^j - {(13689 - 945 \pi^2)} \vec{v}_{1}\cdot\vec{n} v_{2}^i n^j \nn\\ 
&&  + {(16194 - 2862 \pi^2)} \vec{v}_{2}\cdot\vec{n} v_{2}^i n^j \Big],
\eea
\bea
\stackrel{(1)}{ \omega^{ij}_1}_{(3,1)}&=& \frac{1}{192} G m_{2} \Big[ 42 \vec{v}_{1}\cdot\vec{a}_{1} v_{2}^2 v_{1}^i n^j - 24 \vec{a}_{1}\cdot\vec{v}_{2} v_{2}^2 v_{1}^i n^j - 48 v_{2}^2 \vec{v}_{1}\cdot\vec{a}_{2} v_{1}^i n^j + 42 v_{1}^2 \vec{v}_{2}\cdot\vec{a}_{2} v_{1}^i n^j \nn\\ 
&& - 96 \vec{v}_{1}\cdot\vec{v}_{2} \vec{v}_{2}\cdot\vec{a}_{2} v_{1}^i n^j + 264 v_{2}^2 \vec{v}_{2}\cdot\vec{a}_{2} v_{1}^i n^j + 21 v_{1}^2 v_{2}^2 a_{1}^i n^j - 48 \vec{v}_{1}\cdot\vec{v}_{2} v_{2}^2 a_{1}^i n^j \nn\\ 
&& + 72 v_{2}^{4} a_{1}^i n^j + 24 v_{2}^2 \vec{v}_{1}\cdot\vec{a}_{2} v_{2}^i n^j + 48 \vec{v}_{1}\cdot\vec{v}_{2} \vec{v}_{2}\cdot\vec{a}_{2} v_{2}^i n^j - 264 v_{2}^2 \vec{v}_{2}\cdot\vec{a}_{2} v_{2}^i n^j \nn\\ 
&& + 24 \vec{v}_{1}\cdot\vec{v}_{2} v_{2}^2 a_{2}^i n^j - 72 v_{2}^{4} a_{2}^i n^j + 24 \vec{v}_{2}\cdot\vec{n} v_{2}^2 v_{1}^i a_{1}^j - 204 \vec{v}_{1}\cdot\vec{a}_{1} \vec{v}_{2}\cdot\vec{n} v_{1}^i v_{2}^j \nn\\ 
&& - 72 \vec{a}_{1}\cdot\vec{n} v_{2}^2 v_{1}^i v_{2}^j - 102 v_{1}^2 \vec{a}_{2}\cdot\vec{n} v_{1}^i v_{2}^j + 144 \vec{v}_{1}\cdot\vec{v}_{2} \vec{a}_{2}\cdot\vec{n} v_{1}^i v_{2}^j - 432 v_{2}^2 \vec{a}_{2}\cdot\vec{n} v_{1}^i v_{2}^j \nn\\ 
&& + 144 \vec{v}_{2}\cdot\vec{n} \vec{v}_{1}\cdot\vec{a}_{2} v_{1}^i v_{2}^j - 864 \vec{v}_{2}\cdot\vec{n} \vec{v}_{2}\cdot\vec{a}_{2} v_{1}^i v_{2}^j - 102 v_{1}^2 \vec{v}_{2}\cdot\vec{n} a_{1}^i v_{2}^j \nn\\ 
&& + 144 \vec{v}_{2}\cdot\vec{n} \vec{v}_{1}\cdot\vec{v}_{2} a_{1}^i v_{2}^j - 480 \vec{v}_{2}\cdot\vec{n} v_{2}^2 a_{1}^i v_{2}^j - 102 v_{1}^2 \vec{v}_{2}\cdot\vec{n} v_{1}^i a_{2}^j + 144 \vec{v}_{2}\cdot\vec{n} \vec{v}_{1}\cdot\vec{v}_{2} v_{1}^i a_{2}^j \nn\\ 
&& - 528 \vec{v}_{2}\cdot\vec{n} v_{2}^2 v_{1}^i a_{2}^j + 24 \vec{v}_{2}\cdot\vec{n} v_{2}^2 v_{2}^i a_{2}^j -48 \vec{a}_{1}\cdot\vec{n} \vec{v}_{2}\cdot\vec{n} v_{2}^2 v_{1}^i n^j - 42 v_{1}^2 \vec{v}_{2}\cdot\vec{n} \vec{a}_{2}\cdot\vec{n} v_{1}^i n^j \nn\\ 
&& + 96 \vec{v}_{2}\cdot\vec{n} \vec{v}_{1}\cdot\vec{v}_{2} \vec{a}_{2}\cdot\vec{n} v_{1}^i n^j - 24 \vec{v}_{1}\cdot\vec{n} v_{2}^2 \vec{a}_{2}\cdot\vec{n} v_{1}^i n^j - 216 \vec{v}_{2}\cdot\vec{n} v_{2}^2 \vec{a}_{2}\cdot\vec{n} v_{1}^i n^j \nn\\ 
&& - 48 \vec{v}_{1}\cdot\vec{n} \vec{v}_{2}\cdot\vec{n} \vec{v}_{2}\cdot\vec{a}_{2} v_{1}^i n^j - 42 \vec{v}_{1}\cdot\vec{a}_{1} ( \vec{v}_{2}\cdot\vec{n})^{2} v_{1}^i n^j + 24 \vec{a}_{1}\cdot\vec{v}_{2} ( \vec{v}_{2}\cdot\vec{n})^{2} v_{1}^i n^j \nn\\ 
&& + 48 \vec{v}_{1}\cdot\vec{a}_{2} ( \vec{v}_{2}\cdot\vec{n})^{2} v_{1}^i n^j - 216 \vec{v}_{2}\cdot\vec{a}_{2} ( \vec{v}_{2}\cdot\vec{n})^{2} v_{1}^i n^j - 24 \vec{v}_{1}\cdot\vec{n} \vec{v}_{2}\cdot\vec{n} v_{2}^2 a_{1}^i n^j \nn\\ 
&& - 21 v_{1}^2 ( \vec{v}_{2}\cdot\vec{n})^{2} a_{1}^i n^j + 48 \vec{v}_{1}\cdot\vec{v}_{2} ( \vec{v}_{2}\cdot\vec{n})^{2} a_{1}^i n^j - 120 v_{2}^2 ( \vec{v}_{2}\cdot\vec{n})^{2} a_{1}^i n^j \nn\\ 
&& + 48 \vec{a}_{1}\cdot\vec{n} \vec{v}_{2}\cdot\vec{n} v_{2}^2 v_{2}^i n^j - 48 \vec{v}_{2}\cdot\vec{n} \vec{v}_{1}\cdot\vec{v}_{2} \vec{a}_{2}\cdot\vec{n} v_{2}^i n^j + 24 \vec{v}_{1}\cdot\vec{n} v_{2}^2 \vec{a}_{2}\cdot\vec{n} v_{2}^i n^j \nn\\ 
&& + 216 \vec{v}_{2}\cdot\vec{n} v_{2}^2 \vec{a}_{2}\cdot\vec{n} v_{2}^i n^j + 48 \vec{v}_{1}\cdot\vec{n} \vec{v}_{2}\cdot\vec{n} \vec{v}_{2}\cdot\vec{a}_{2} v_{2}^i n^j - 24 \vec{v}_{1}\cdot\vec{a}_{2} ( \vec{v}_{2}\cdot\vec{n})^{2} v_{2}^i n^j \nn\\ 
&& + 216 \vec{v}_{2}\cdot\vec{a}_{2} ( \vec{v}_{2}\cdot\vec{n})^{2} v_{2}^i n^j + 24 \vec{v}_{1}\cdot\vec{n} \vec{v}_{2}\cdot\vec{n} v_{2}^2 a_{2}^i n^j - 24 \vec{v}_{1}\cdot\vec{v}_{2} ( \vec{v}_{2}\cdot\vec{n})^{2} a_{2}^i n^j \nn\\ 
&& + 120 v_{2}^2 ( \vec{v}_{2}\cdot\vec{n})^{2} a_{2}^i n^j - 8 ( \vec{v}_{2}\cdot\vec{n})^{3} v_{1}^i a_{1}^j + 72 \vec{a}_{1}\cdot\vec{n} ( \vec{v}_{2}\cdot\vec{n})^{2} v_{1}^i v_{2}^j \nn\\ 
&& + 288 \vec{a}_{2}\cdot\vec{n} ( \vec{v}_{2}\cdot\vec{n})^{2} v_{1}^i v_{2}^j + 112 ( \vec{v}_{2}\cdot\vec{n})^{3} a_{1}^i v_{2}^j + 128 ( \vec{v}_{2}\cdot\vec{n})^{3} v_{1}^i a_{2}^j - 8 ( \vec{v}_{2}\cdot\vec{n})^{3} v_{2}^i a_{2}^j \nn\\ 
&& + 48 \vec{a}_{1}\cdot\vec{n} ( \vec{v}_{2}\cdot\vec{n})^{3} v_{1}^i n^j + 72 \vec{v}_{1}\cdot\vec{n} \vec{a}_{2}\cdot\vec{n} ( \vec{v}_{2}\cdot\vec{n})^{2} v_{1}^i n^j + 168 \vec{a}_{2}\cdot\vec{n} ( \vec{v}_{2}\cdot\vec{n})^{3} v_{1}^i n^j \nn\\ 
&& + 24 \vec{v}_{1}\cdot\vec{n} ( \vec{v}_{2}\cdot\vec{n})^{3} a_{1}^i n^j + 48 ( \vec{v}_{2}\cdot\vec{n})^{4} a_{1}^i n^j - 48 \vec{a}_{1}\cdot\vec{n} ( \vec{v}_{2}\cdot\vec{n})^{3} v_{2}^i n^j \nn\\ 
&& - 72 \vec{v}_{1}\cdot\vec{n} \vec{a}_{2}\cdot\vec{n} ( \vec{v}_{2}\cdot\vec{n})^{2} v_{2}^i n^j - 168 \vec{a}_{2}\cdot\vec{n} ( \vec{v}_{2}\cdot\vec{n})^{3} v_{2}^i n^j - 24 \vec{v}_{1}\cdot\vec{n} ( \vec{v}_{2}\cdot\vec{n})^{3} a_{2}^i n^j \nn\\ 
&& - 48 ( \vec{v}_{2}\cdot\vec{n})^{4} a_{2}^i n^j \Big]\nn\\ && - 	\frac{G^2 m_{1} m_{2}}{96 r} \Big[ 174 \vec{v}_{1}\cdot\vec{a}_{1} v_{1}^i n^j + 160 \vec{a}_{1}\cdot\vec{v}_{2} v_{1}^i n^j + 376 \vec{v}_{1}\cdot\vec{a}_{2} v_{1}^i n^j - 836 \vec{v}_{2}\cdot\vec{a}_{2} v_{1}^i n^j \nn\\ 
&& + 185 v_{1}^2 a_{1}^i n^j + 180 \vec{v}_{1}\cdot\vec{v}_{2} a_{1}^i n^j - 428 v_{2}^2 a_{1}^i n^j + 168 \vec{v}_{1}\cdot\vec{a}_{1} v_{2}^i n^j + 112 \vec{a}_{1}\cdot\vec{v}_{2} v_{2}^i n^j \nn\\ 
&& - 8 \vec{v}_{1}\cdot\vec{a}_{2} v_{2}^i n^j + 344 \vec{v}_{2}\cdot\vec{a}_{2} v_{2}^i n^j + 16 v_{1}^2 a_{2}^i n^j + 128 \vec{v}_{1}\cdot\vec{v}_{2} a_{2}^i n^j + 164 v_{2}^2 a_{2}^i n^j \nn\\ 
&& - 196 \vec{v}_{1}\cdot\vec{n} v_{1}^i a_{1}^j - 20 \vec{v}_{2}\cdot\vec{n} v_{1}^i a_{1}^j - 1364 \vec{a}_{1}\cdot\vec{n} v_{1}^i v_{2}^j + 996 \vec{a}_{2}\cdot\vec{n} v_{1}^i v_{2}^j - 1500 \vec{v}_{1}\cdot\vec{n} a_{1}^i v_{2}^j \nn\\ 
&& + 1012 \vec{v}_{2}\cdot\vec{n} a_{1}^i v_{2}^j - 1168 \vec{v}_{1}\cdot\vec{n} v_{1}^i a_{2}^j + 1016 \vec{v}_{2}\cdot\vec{n} v_{1}^i a_{2}^j - 136 \vec{v}_{1}\cdot\vec{n} v_{2}^i a_{2}^j \nn\\ 
&& + 16 \vec{v}_{2}\cdot\vec{n} v_{2}^i a_{2}^j + 444 \vec{v}_{1}\cdot\vec{n} \vec{a}_{1}\cdot\vec{n} v_{1}^i n^j - 452 \vec{a}_{1}\cdot\vec{n} \vec{v}_{2}\cdot\vec{n} v_{1}^i n^j - 884 \vec{v}_{1}\cdot\vec{n} \vec{a}_{2}\cdot\vec{n} v_{1}^i n^j \nn\\ 
&& + 688 \vec{v}_{2}\cdot\vec{n} \vec{a}_{2}\cdot\vec{n} v_{1}^i n^j - 492 \vec{v}_{1}\cdot\vec{n} \vec{v}_{2}\cdot\vec{n} a_{1}^i n^j + 26 ( \vec{v}_{1}\cdot\vec{n})^{2} a_{1}^i n^j + 364 ( \vec{v}_{2}\cdot\vec{n})^{2} a_{1}^i n^j \nn\\ 
&& - 168 \vec{v}_{1}\cdot\vec{n} \vec{a}_{1}\cdot\vec{n} v_{2}^i n^j + 184 \vec{a}_{1}\cdot\vec{n} \vec{v}_{2}\cdot\vec{n} v_{2}^i n^j + 424 \vec{v}_{1}\cdot\vec{n} \vec{a}_{2}\cdot\vec{n} v_{2}^i n^j \nn\\ 
&& - 496 \vec{v}_{2}\cdot\vec{n} \vec{a}_{2}\cdot\vec{n} v_{2}^i n^j + 152 \vec{v}_{1}\cdot\vec{n} \vec{v}_{2}\cdot\vec{n} a_{2}^i n^j + 52 ( \vec{v}_{1}\cdot\vec{n})^{2} a_{2}^i n^j - 232 ( \vec{v}_{2}\cdot\vec{n})^{2} a_{2}^i n^j \Big] \nn\\ 
&& - 	\frac{G^2 m_{2}{}^2}{96 r} \Big[ 60 \vec{a}_{1}\cdot\vec{v}_{2} v_{1}^i n^j + 60 \vec{v}_{1}\cdot\vec{a}_{2} v_{1}^i n^j - 336 \vec{v}_{2}\cdot\vec{a}_{2} v_{1}^i n^j + 60 \vec{v}_{1}\cdot\vec{v}_{2} a_{1}^i n^j \nn\\ 
&& - 168 v_{2}^2 a_{1}^i n^j - 8 \vec{a}_{1}\cdot\vec{v}_{2} v_{2}^i n^j - 8 \vec{v}_{1}\cdot\vec{a}_{2} v_{2}^i n^j + 902 \vec{v}_{2}\cdot\vec{a}_{2} v_{2}^i n^j - 8 \vec{v}_{1}\cdot\vec{v}_{2} a_{2}^i n^j \nn\\ 
&& + 451 v_{2}^2 a_{2}^i n^j + 60 \vec{a}_{1}\cdot\vec{n} v_{1}^i v_{2}^j + 1316 \vec{a}_{2}\cdot\vec{n} v_{1}^i v_{2}^j + 60 \vec{v}_{1}\cdot\vec{n} a_{1}^i v_{2}^j + 1316 \vec{v}_{2}\cdot\vec{n} a_{1}^i v_{2}^j \nn\\ 
&& + 60 \vec{v}_{1}\cdot\vec{n} v_{1}^i a_{2}^j + 1316 \vec{v}_{2}\cdot\vec{n} v_{1}^i a_{2}^j -48 \vec{a}_{1}\cdot\vec{n} \vec{v}_{2}\cdot\vec{n} v_{1}^i n^j - 48 \vec{v}_{1}\cdot\vec{n} \vec{a}_{2}\cdot\vec{n} v_{1}^i n^j \nn\\ 
&& - 24 \vec{v}_{2}\cdot\vec{n} \vec{a}_{2}\cdot\vec{n} v_{1}^i n^j - 48 \vec{v}_{1}\cdot\vec{n} \vec{v}_{2}\cdot\vec{n} a_{1}^i n^j - 12 ( \vec{v}_{2}\cdot\vec{n})^{2} a_{1}^i n^j - 32 \vec{a}_{1}\cdot\vec{n} \vec{v}_{2}\cdot\vec{n} v_{2}^i n^j \nn\\ 
&& - 32 \vec{v}_{1}\cdot\vec{n} \vec{a}_{2}\cdot\vec{n} v_{2}^i n^j - 634 \vec{v}_{2}\cdot\vec{n} \vec{a}_{2}\cdot\vec{n} v_{2}^i n^j - 32 \vec{v}_{1}\cdot\vec{n} \vec{v}_{2}\cdot\vec{n} a_{2}^i n^j \nn\\ 
&& - 317 ( \vec{v}_{2}\cdot\vec{n})^{2} a_{2}^i n^j \Big]\nn\\ && - 	\frac{G^3 m_{1}{}^2 m_{2}}{72 r{}^2} \Big[ 829 a_{1}^i n^j - 102 a_{2}^i n^j \Big] + 	\frac{2 G^3 m_{2}{}^3}{3 r{}^2} {\Big( \frac{1}{\epsilon} - 3\log \frac{r}{R_0} \Big)} \Big[ a_{1}^i n^j + a_{2}^i n^j \Big] \nn\\ 
&& + 	\frac{G^3 m_{2}{}^3}{9 r{}^2} \Big[ 15 a_{1}^i n^j + 11 a_{2}^i n^j \Big] + 	\frac{G^3 m_{1} m_{2}{}^2}{288 r{}^2} \Big[ {(5813 - 450 \pi^2)} a_{1}^i n^j \nn\\ 
&& + {(358 + 63 \pi^2)} a_{2}^i n^j \Big] - 	\frac{5 G^3 m_{1}{}^2 m_{2}}{2 r{}^2} {\Big( \frac{1}{\epsilon} - 3\log \frac{r}{R_0} \Big)} a_{1}^i n^j,
\eea
\bea
\stackrel{(2)}{ \omega^{ij}_1}_{(3,1)}&=& \frac{1}{24} G m_{2} r \Big[ \big( 18 \vec{v}_{2}\cdot\dot{\vec{a}}_{2} v_{1}^i v_{2}^j + 9 v_{2}^2 \dot{a}_{1}^i v_{2}^j + 9 v_{2}^2 v_{1}^i \dot{a}_{2}^j -3 v_{2}^2 \dot{\vec{a}}_{2}\cdot\vec{n} v_{1}^i n^j - 6 \vec{v}_{2}\cdot\vec{n} \vec{v}_{2}\cdot\dot{\vec{a}}_{2} v_{1}^i n^j \nn\\ 
&& - 3 \vec{v}_{2}\cdot\vec{n} v_{2}^2 \dot{a}_{1}^i n^j + 3 v_{2}^2 \dot{\vec{a}}_{2}\cdot\vec{n} v_{2}^i n^j + 6 \vec{v}_{2}\cdot\vec{n} \vec{v}_{2}\cdot\dot{\vec{a}}_{2} v_{2}^i n^j + 3 \vec{v}_{2}\cdot\vec{n} v_{2}^2 \dot{a}_{2}^i n^j \nn\\ 
&& + 18 \vec{v}_{2}\cdot\vec{n} \dot{\vec{a}}_{2}\cdot\vec{n} v_{1}^i v_{2}^j + 9 ( \vec{v}_{2}\cdot\vec{n})^{2} \dot{a}_{1}^i v_{2}^j + 9 ( \vec{v}_{2}\cdot\vec{n})^{2} v_{1}^i \dot{a}_{2}^j + 3 \dot{\vec{a}}_{2}\cdot\vec{n} ( \vec{v}_{2}\cdot\vec{n})^{2} v_{1}^i n^j \nn\\ 
&& + ( \vec{v}_{2}\cdot\vec{n})^{3} \dot{a}_{1}^i n^j - 3 \dot{\vec{a}}_{2}\cdot\vec{n} ( \vec{v}_{2}\cdot\vec{n})^{2} v_{2}^i n^j - ( \vec{v}_{2}\cdot\vec{n})^{3} \dot{a}_{2}^i n^j \big) + 6 \big( 3 a_{2}^2 v_{1}^i v_{2}^j \nn\\ 
&& + 6 \vec{v}_{2}\cdot\vec{a}_{2} a_{1}^i v_{2}^j + 6 \vec{v}_{2}\cdot\vec{a}_{2} v_{1}^i a_{2}^j + 3 v_{2}^2 a_{1}^i a_{2}^j -2 \vec{a}_{2}\cdot\vec{n} \vec{v}_{2}\cdot\vec{a}_{2} v_{1}^i n^j - \vec{v}_{2}\cdot\vec{n} a_{2}^2 v_{1}^i n^j \nn\\ 
&& - v_{2}^2 \vec{a}_{2}\cdot\vec{n} a_{1}^i n^j - 2 \vec{v}_{2}\cdot\vec{n} \vec{v}_{2}\cdot\vec{a}_{2} a_{1}^i n^j + 2 \vec{a}_{2}\cdot\vec{n} \vec{v}_{2}\cdot\vec{a}_{2} v_{2}^i n^j + \vec{v}_{2}\cdot\vec{n} a_{2}^2 v_{2}^i n^j \nn\\ 
&& + v_{2}^2 \vec{a}_{2}\cdot\vec{n} a_{2}^i n^j + 2 \vec{v}_{2}\cdot\vec{n} \vec{v}_{2}\cdot\vec{a}_{2} a_{2}^i n^j + 6 \vec{v}_{2}\cdot\vec{n} \vec{a}_{2}\cdot\vec{n} a_{1}^i v_{2}^j + 6 \vec{v}_{2}\cdot\vec{n} \vec{a}_{2}\cdot\vec{n} v_{1}^i a_{2}^j \nn\\ 
&& + 3 ( \vec{v}_{2}\cdot\vec{n})^{2} a_{1}^i a_{2}^j + \vec{a}_{2}\cdot\vec{n} ( \vec{v}_{2}\cdot\vec{n})^{2} a_{1}^i n^j - \vec{a}_{2}\cdot\vec{n} ( \vec{v}_{2}\cdot\vec{n})^{2} a_{2}^i n^j \big) \Big] \nn\\ 
&& + 	\frac{1}{4} G m_{2} r \Big[ 3 ( \vec{a}_{2}\cdot\vec{n})^{2} v_{1}^i v_{2}^j + \vec{v}_{2}\cdot\vec{n} ( \vec{a}_{2}\cdot\vec{n})^{2} v_{1}^i n^j - \vec{v}_{2}\cdot\vec{n} ( \vec{a}_{2}\cdot\vec{n})^{2} v_{2}^i n^j \Big]\nn\\ && - 	\frac{1}{24} G^2 m_{1} m_{2} \Big[ \big( 12 \dot{a}_{1}^i v_{2}^j + 12 v_{1}^i \dot{a}_{2}^j + 49 \dot{\vec{a}}_{1}\cdot\vec{n} v_{1}^i n^j + 5 \dot{\vec{a}}_{2}\cdot\vec{n} v_{1}^i n^j + 49 \vec{v}_{1}\cdot\vec{n} \dot{a}_{1}^i n^j \nn\\ 
&& + 5 \vec{v}_{2}\cdot\vec{n} \dot{a}_{1}^i n^j - 34 \dot{\vec{a}}_{1}\cdot\vec{n} v_{2}^i n^j + 4 \dot{\vec{a}}_{2}\cdot\vec{n} v_{2}^i n^j - 34 \vec{v}_{1}\cdot\vec{n} \dot{a}_{2}^i n^j + 4 \vec{v}_{2}\cdot\vec{n} \dot{a}_{2}^i n^j \big) \nn\\ 
&& + 2 \big( 12 a_{1}^i a_{2}^j + 49 \vec{a}_{1}\cdot\vec{n} a_{1}^i n^j + 5 \vec{a}_{2}\cdot\vec{n} a_{1}^i n^j - 34 \vec{a}_{1}\cdot\vec{n} a_{2}^i n^j + 4 \vec{a}_{2}\cdot\vec{n} a_{2}^i n^j \big) \Big].
\eea

\section{Generic Action}
\label{explicitstdaction}

As noted in section \ref{finalaction}, after our reduction process, we get the following generic 
action: 
\bea
\label{reducedwopole}
V_{\text{N}^3\text{LO}}^{\text{SO}} &=&
\frac{G m_{2}}{128 r{}^2} \Big[ \vec{S}_{1}\times\vec{n}\cdot\vec{v}_{1} \big( 100 v_{1}^2 \vec{v}_{1}\cdot\vec{v}_{2} v_{2}^2 - 76 v_{1}^2 ( \vec{v}_{1}\cdot\vec{v}_{2})^{2} - 128 v_{2}^2 ( \vec{v}_{1}\cdot\vec{v}_{2})^{2} + 5 v_{1}^{6} \nn\\ 
&& + 32 \vec{v}_{1}\cdot\vec{v}_{2} v_{1}^{4} - 14 v_{2}^2 v_{1}^{4} - 26 v_{1}^2 v_{2}^{4} + 160 \vec{v}_{1}\cdot\vec{v}_{2} v_{2}^{4} - 48 v_{2}^{6} + 120 \vec{v}_{1}\cdot\vec{n} v_{1}^2 \vec{v}_{2}\cdot\vec{n} \vec{v}_{1}\cdot\vec{v}_{2} \nn\\ 
&& - 204 \vec{v}_{1}\cdot\vec{n} v_{1}^2 \vec{v}_{2}\cdot\vec{n} v_{2}^2 + 96 \vec{v}_{1}\cdot\vec{n} \vec{v}_{2}\cdot\vec{n} \vec{v}_{1}\cdot\vec{v}_{2} v_{2}^2 - 54 v_{1}^2 v_{2}^2 ( \vec{v}_{1}\cdot\vec{n})^{2} \nn\\ 
&& - 252 v_{1}^2 \vec{v}_{1}\cdot\vec{v}_{2} ( \vec{v}_{2}\cdot\vec{n})^{2} + 96 \vec{v}_{1}\cdot\vec{n} \vec{v}_{2}\cdot\vec{n} ( \vec{v}_{1}\cdot\vec{v}_{2})^{2} + 192 ( \vec{v}_{2}\cdot\vec{n})^{2} ( \vec{v}_{1}\cdot\vec{v}_{2})^{2} \nn\\ 
&& + 228 v_{1}^2 v_{2}^2 ( \vec{v}_{2}\cdot\vec{n})^{2} - 768 \vec{v}_{1}\cdot\vec{v}_{2} v_{2}^2 ( \vec{v}_{2}\cdot\vec{n})^{2} - 144 \vec{v}_{1}\cdot\vec{n} \vec{v}_{2}\cdot\vec{n} v_{1}^{4} + 78 ( \vec{v}_{2}\cdot\vec{n})^{2} v_{1}^{4} \nn\\ 
&& - 48 ( \vec{v}_{1}\cdot\vec{n})^{2} v_{2}^{4} - 144 \vec{v}_{1}\cdot\vec{n} \vec{v}_{2}\cdot\vec{n} v_{2}^{4} + 480 ( \vec{v}_{2}\cdot\vec{n})^{2} v_{2}^{4} + 240 \vec{v}_{2}\cdot\vec{n} v_{2}^2 ( \vec{v}_{1}\cdot\vec{n})^{3} \nn\\ 
&& + 270 v_{1}^2 ( \vec{v}_{1}\cdot\vec{n})^{2} ( \vec{v}_{2}\cdot\vec{n})^{2} + 180 \vec{v}_{1}\cdot\vec{n} v_{1}^2 ( \vec{v}_{2}\cdot\vec{n})^{3} - 180 v_{1}^2 ( \vec{v}_{2}\cdot\vec{n})^{4} \nn\\ 
&& + 480 \vec{v}_{1}\cdot\vec{v}_{2} ( \vec{v}_{2}\cdot\vec{n})^{4} + 240 v_{2}^2 ( \vec{v}_{1}\cdot\vec{n})^{2} ( \vec{v}_{2}\cdot\vec{n})^{2} \nn\\ 
&& - 480 v_{2}^2 ( \vec{v}_{2}\cdot\vec{n})^{4} -560 ( \vec{v}_{1}\cdot\vec{n})^{3} ( \vec{v}_{2}\cdot\vec{n})^{3} + 140 ( \vec{v}_{2}\cdot\vec{n})^{6} \big) \nn\\ 
&& - 16 \vec{S}_{1}\times\vec{n}\cdot\vec{v}_{2} \big( v_{1}^2 \vec{v}_{1}\cdot\vec{v}_{2} v_{2}^2 - 2 ( \vec{v}_{1}\cdot\vec{v}_{2})^{3} - 2 v_{2}^2 ( \vec{v}_{1}\cdot\vec{v}_{2})^{2} + v_{1}^2 v_{2}^{4} - 3 \vec{v}_{1}\cdot\vec{v}_{2} v_{2}^{4} \nn\\ 
&& + 5 v_{2}^{6} -9 \vec{v}_{1}\cdot\vec{n} v_{1}^2 \vec{v}_{2}\cdot\vec{n} v_{2}^2 - 3 \vec{v}_{1}\cdot\vec{v}_{2} v_{2}^2 ( \vec{v}_{1}\cdot\vec{n})^{2} - 3 v_{1}^2 \vec{v}_{1}\cdot\vec{v}_{2} ( \vec{v}_{2}\cdot\vec{n})^{2} \nn\\ 
&& + 6 \vec{v}_{1}\cdot\vec{n} \vec{v}_{2}\cdot\vec{n} ( \vec{v}_{1}\cdot\vec{v}_{2})^{2} - 3 v_{1}^2 v_{2}^2 ( \vec{v}_{2}\cdot\vec{n})^{2} - 3 ( \vec{v}_{1}\cdot\vec{n})^{2} v_{2}^{4} - 9 \vec{v}_{1}\cdot\vec{n} \vec{v}_{2}\cdot\vec{n} v_{2}^{4} \nn\\ 
&& + 15 \vec{v}_{2}\cdot\vec{n} v_{2}^2 ( \vec{v}_{1}\cdot\vec{n})^{3} + 15 \vec{v}_{1}\cdot\vec{n} v_{1}^2 ( \vec{v}_{2}\cdot\vec{n})^{3} + 15 \vec{v}_{1}\cdot\vec{v}_{2} ( \vec{v}_{1}\cdot\vec{n})^{2} ( \vec{v}_{2}\cdot\vec{n})^{2} \nn\\ 
&& + 15 v_{2}^2 ( \vec{v}_{1}\cdot\vec{n})^{2} ( \vec{v}_{2}\cdot\vec{n})^{2} -35 ( \vec{v}_{1}\cdot\vec{n})^{3} ( \vec{v}_{2}\cdot\vec{n})^{3} \big) - 4 \vec{S}_{1}\times\vec{v}_{1}\cdot\vec{v}_{2} \big( 22 \vec{v}_{1}\cdot\vec{n} v_{1}^2 \vec{v}_{1}\cdot\vec{v}_{2} \nn\\ 
&& - 30 v_{1}^2 \vec{v}_{2}\cdot\vec{n} \vec{v}_{1}\cdot\vec{v}_{2} - 31 \vec{v}_{1}\cdot\vec{n} v_{1}^2 v_{2}^2 + 38 v_{1}^2 \vec{v}_{2}\cdot\vec{n} v_{2}^2 + 64 \vec{v}_{1}\cdot\vec{n} \vec{v}_{1}\cdot\vec{v}_{2} v_{2}^2 \nn\\ 
&& - 128 \vec{v}_{2}\cdot\vec{n} \vec{v}_{1}\cdot\vec{v}_{2} v_{2}^2 + 24 \vec{v}_{2}\cdot\vec{n} ( \vec{v}_{1}\cdot\vec{v}_{2})^{2} - 32 \vec{v}_{1}\cdot\vec{n} v_{1}^{4} + 15 \vec{v}_{2}\cdot\vec{n} v_{1}^{4} - 56 \vec{v}_{1}\cdot\vec{n} v_{2}^{4} \nn\\ 
&& + 124 \vec{v}_{2}\cdot\vec{n} v_{2}^{4} + 9 v_{1}^2 \vec{v}_{2}\cdot\vec{n} ( \vec{v}_{1}\cdot\vec{n})^{2} + 24 \vec{v}_{2}\cdot\vec{n} \vec{v}_{1}\cdot\vec{v}_{2} ( \vec{v}_{1}\cdot\vec{n})^{2} - 36 \vec{v}_{2}\cdot\vec{n} v_{2}^2 ( \vec{v}_{1}\cdot\vec{n})^{2} \nn\\ 
&& + 33 \vec{v}_{1}\cdot\vec{n} v_{1}^2 ( \vec{v}_{2}\cdot\vec{n})^{2} - 24 v_{1}^2 ( \vec{v}_{2}\cdot\vec{n})^{3} - 120 \vec{v}_{1}\cdot\vec{n} \vec{v}_{1}\cdot\vec{v}_{2} ( \vec{v}_{2}\cdot\vec{n})^{2} + 96 \vec{v}_{1}\cdot\vec{v}_{2} ( \vec{v}_{2}\cdot\vec{n})^{3} \nn\\ 
&& + 168 \vec{v}_{1}\cdot\vec{n} v_{2}^2 ( \vec{v}_{2}\cdot\vec{n})^{2} - 192 v_{2}^2 ( \vec{v}_{2}\cdot\vec{n})^{3} + 60 ( \vec{v}_{1}\cdot\vec{n})^{2} ( \vec{v}_{2}\cdot\vec{n})^{3} - 120 \vec{v}_{1}\cdot\vec{n} ( \vec{v}_{2}\cdot\vec{n})^{4} \nn\\ 
&& + 90 ( \vec{v}_{2}\cdot\vec{n})^{5} \big) \Big]\nn\\ && +  	\frac{G^2 m_{2}{}^2}{96 r{}^3} \Big[ \vec{S}_{1}\times\vec{n}\cdot\vec{v}_{1} \big( 66 v_{1}^2 \vec{v}_{1}\cdot\vec{v}_{2} - 136 v_{1}^2 v_{2}^2 + 5331 \vec{v}_{1}\cdot\vec{v}_{2} v_{2}^2 - 1196 ( \vec{v}_{1}\cdot\vec{v}_{2})^{2} + 6 v_{1}^{4} \nn\\ 
&& - 4110 v_{2}^{4} -228 \vec{v}_{1}\cdot\vec{n} v_{1}^2 \vec{v}_{2}\cdot\vec{n} + 3154 \vec{v}_{1}\cdot\vec{n} \vec{v}_{2}\cdot\vec{n} \vec{v}_{1}\cdot\vec{v}_{2} - 13254 \vec{v}_{1}\cdot\vec{n} \vec{v}_{2}\cdot\vec{n} v_{2}^2 \nn\\ 
&& + 168 v_{1}^2 ( \vec{v}_{1}\cdot\vec{n})^{2} + 672 \vec{v}_{1}\cdot\vec{v}_{2} ( \vec{v}_{1}\cdot\vec{n})^{2} - 614 v_{2}^2 ( \vec{v}_{1}\cdot\vec{n})^{2} + 697 v_{1}^2 ( \vec{v}_{2}\cdot\vec{n})^{2} \nn\\ 
&& - 4725 \vec{v}_{1}\cdot\vec{v}_{2} ( \vec{v}_{2}\cdot\vec{n})^{2} + 14760 v_{2}^2 ( \vec{v}_{2}\cdot\vec{n})^{2} -768 \vec{v}_{2}\cdot\vec{n} ( \vec{v}_{1}\cdot\vec{n})^{3} \nn\\ 
&& - 258 ( \vec{v}_{1}\cdot\vec{n})^{2} ( \vec{v}_{2}\cdot\vec{n})^{2} + 186 \vec{v}_{1}\cdot\vec{n} ( \vec{v}_{2}\cdot\vec{n})^{3} - 42 ( \vec{v}_{2}\cdot\vec{n})^{4} \big) \nn\\ 
&& - \vec{S}_{1}\times\vec{n}\cdot\vec{v}_{2} \big( 192 v_{1}^2 \vec{v}_{1}\cdot\vec{v}_{2} + 319 v_{1}^2 v_{2}^2 - 69 \vec{v}_{1}\cdot\vec{v}_{2} v_{2}^2 + 1568 ( \vec{v}_{1}\cdot\vec{v}_{2})^{2} \nn\\ 
&& - 2010 v_{2}^{4} -492 \vec{v}_{1}\cdot\vec{n} v_{1}^2 \vec{v}_{2}\cdot\vec{n} - 11362 \vec{v}_{1}\cdot\vec{n} \vec{v}_{2}\cdot\vec{n} \vec{v}_{1}\cdot\vec{v}_{2} + 1626 \vec{v}_{1}\cdot\vec{n} \vec{v}_{2}\cdot\vec{n} v_{2}^2 \nn\\ 
&& - 192 \vec{v}_{1}\cdot\vec{v}_{2} ( \vec{v}_{1}\cdot\vec{n})^{2} - 1894 v_{2}^2 ( \vec{v}_{1}\cdot\vec{n})^{2} - 1105 v_{1}^2 ( \vec{v}_{2}\cdot\vec{n})^{2} + 9927 \vec{v}_{1}\cdot\vec{v}_{2} ( \vec{v}_{2}\cdot\vec{n})^{2} \nn\\ 
&& + 4566 v_{2}^2 ( \vec{v}_{2}\cdot\vec{n})^{2} + 768 \vec{v}_{2}\cdot\vec{n} ( \vec{v}_{1}\cdot\vec{n})^{3} + 8988 ( \vec{v}_{1}\cdot\vec{n})^{2} ( \vec{v}_{2}\cdot\vec{n})^{2} \nn\\ 
&& - 14166 \vec{v}_{1}\cdot\vec{n} ( \vec{v}_{2}\cdot\vec{n})^{3} + 2520 ( \vec{v}_{2}\cdot\vec{n})^{4} \big) - \vec{S}_{1}\times\vec{v}_{1}\cdot\vec{v}_{2} \big( 522 \vec{v}_{1}\cdot\vec{n} v_{1}^2 - 1025 v_{1}^2 \vec{v}_{2}\cdot\vec{n} \nn\\ 
&& - 3760 \vec{v}_{1}\cdot\vec{n} \vec{v}_{1}\cdot\vec{v}_{2} + 7756 \vec{v}_{2}\cdot\vec{n} \vec{v}_{1}\cdot\vec{v}_{2} + 3716 \vec{v}_{1}\cdot\vec{n} v_{2}^2 - 5970 \vec{v}_{2}\cdot\vec{n} v_{2}^2 \nn\\ 
&& + 224 ( \vec{v}_{1}\cdot\vec{n})^{3} + 1994 \vec{v}_{2}\cdot\vec{n} ( \vec{v}_{1}\cdot\vec{n})^{2} - 6572 \vec{v}_{1}\cdot\vec{n} ( \vec{v}_{2}\cdot\vec{n})^{2} + 2590 ( \vec{v}_{2}\cdot\vec{n})^{3} \big) \Big] \nn\\ 
&& + 	\frac{G^2 m_{1} m_{2}}{96 r{}^3} \Big[ \vec{S}_{1}\times\vec{n}\cdot\vec{v}_{1} \big( 3180 v_{1}^2 \vec{v}_{1}\cdot\vec{v}_{2} - 1142 v_{1}^2 v_{2}^2 + 1776 \vec{v}_{1}\cdot\vec{v}_{2} v_{2}^2 \nn\\ 
&& - 2668 ( \vec{v}_{1}\cdot\vec{v}_{2})^{2} - 1461 v_{1}^{4} + 270 v_{2}^{4} -6510 \vec{v}_{1}\cdot\vec{n} v_{1}^2 \vec{v}_{2}\cdot\vec{n} + 17264 \vec{v}_{1}\cdot\vec{n} \vec{v}_{2}\cdot\vec{n} \vec{v}_{1}\cdot\vec{v}_{2} \nn\\ 
&& - 14232 \vec{v}_{1}\cdot\vec{n} \vec{v}_{2}\cdot\vec{n} v_{2}^2 + 8025 v_{1}^2 ( \vec{v}_{1}\cdot\vec{n})^{2} - 9633 \vec{v}_{1}\cdot\vec{v}_{2} ( \vec{v}_{1}\cdot\vec{n})^{2} + 7256 v_{2}^2 ( \vec{v}_{1}\cdot\vec{n})^{2} \nn\\ 
&& + 410 v_{1}^2 ( \vec{v}_{2}\cdot\vec{n})^{2} - 5472 \vec{v}_{1}\cdot\vec{v}_{2} ( \vec{v}_{2}\cdot\vec{n})^{2} + 2724 v_{2}^2 ( \vec{v}_{2}\cdot\vec{n})^{2} + 1110 ( \vec{v}_{1}\cdot\vec{n})^{4} \nn\\ 
&& - 7518 \vec{v}_{2}\cdot\vec{n} ( \vec{v}_{1}\cdot\vec{n})^{3} - 2802 ( \vec{v}_{1}\cdot\vec{n})^{2} ( \vec{v}_{2}\cdot\vec{n})^{2} + 12372 \vec{v}_{1}\cdot\vec{n} ( \vec{v}_{2}\cdot\vec{n})^{3} \nn\\ 
&& - 2730 ( \vec{v}_{2}\cdot\vec{n})^{4} \big) + 8 \vec{S}_{1}\times\vec{n}\cdot\vec{v}_{2} \big( 126 v_{1}^2 \vec{v}_{1}\cdot\vec{v}_{2} - 62 v_{1}^2 v_{2}^2 + 120 \vec{v}_{1}\cdot\vec{v}_{2} v_{2}^2 \nn\\ 
&& - 112 ( \vec{v}_{1}\cdot\vec{v}_{2})^{2} - 72 v_{2}^{4} -336 \vec{v}_{1}\cdot\vec{n} v_{1}^2 \vec{v}_{2}\cdot\vec{n} + 398 \vec{v}_{1}\cdot\vec{n} \vec{v}_{2}\cdot\vec{n} \vec{v}_{1}\cdot\vec{v}_{2} \nn\\ 
&& + 138 \vec{v}_{1}\cdot\vec{n} \vec{v}_{2}\cdot\vec{n} v_{2}^2 + 15 v_{1}^2 ( \vec{v}_{1}\cdot\vec{n})^{2} - 426 \vec{v}_{1}\cdot\vec{v}_{2} ( \vec{v}_{1}\cdot\vec{n})^{2} + 5 v_{2}^2 ( \vec{v}_{1}\cdot\vec{n})^{2} \nn\\ 
&& + 158 v_{1}^2 ( \vec{v}_{2}\cdot\vec{n})^{2} -102 ( \vec{v}_{1}\cdot\vec{n})^{4} + 798 \vec{v}_{2}\cdot\vec{n} ( \vec{v}_{1}\cdot\vec{n})^{3} - 588 ( \vec{v}_{1}\cdot\vec{n})^{2} ( \vec{v}_{2}\cdot\vec{n})^{2} \big) \nn\\ 
&& + \vec{S}_{1}\times\vec{v}_{1}\cdot\vec{v}_{2} \big( 1386 \vec{v}_{1}\cdot\vec{n} v_{1}^2 - 1550 v_{1}^2 \vec{v}_{2}\cdot\vec{n} + 620 \vec{v}_{1}\cdot\vec{n} \vec{v}_{1}\cdot\vec{v}_{2} - 292 \vec{v}_{2}\cdot\vec{n} \vec{v}_{1}\cdot\vec{v}_{2} \nn\\ 
&& - 890 \vec{v}_{1}\cdot\vec{n} v_{2}^2 - 30 \vec{v}_{2}\cdot\vec{n} v_{2}^2 -4489 ( \vec{v}_{1}\cdot\vec{n})^{3} + 6686 \vec{v}_{2}\cdot\vec{n} ( \vec{v}_{1}\cdot\vec{n})^{2} \nn\\ 
&& - 976 \vec{v}_{1}\cdot\vec{n} ( \vec{v}_{2}\cdot\vec{n})^{2} - 380 ( \vec{v}_{2}\cdot\vec{n})^{3} \big) \Big]\nn\\ && +  	\frac{G^3 m_{1}{}^2 m_{2}}{1200 r{}^4} \Big[ \vec{S}_{1}\times\vec{n}\cdot\vec{v}_{1} \big( 68911 v_{1}^2 - 97654 \vec{v}_{1}\cdot\vec{v}_{2} + 36343 v_{2}^2 + 445520 \vec{v}_{1}\cdot\vec{n} \vec{v}_{2}\cdot\vec{n} \nn\\ 
&& - 248930 ( \vec{v}_{1}\cdot\vec{n})^{2} - 113690 ( \vec{v}_{2}\cdot\vec{n})^{2} \big) - 8 \vec{S}_{1}\times\vec{n}\cdot\vec{v}_{2} \big( 2296 v_{1}^2 - 3587 \vec{v}_{1}\cdot\vec{v}_{2} \nn\\ 
&& + 1116 v_{2}^2 + 9985 \vec{v}_{1}\cdot\vec{n} \vec{v}_{2}\cdot\vec{n} - 4355 ( \vec{v}_{1}\cdot\vec{n})^{2} + 195 ( \vec{v}_{2}\cdot\vec{n})^{2} \big) \nn\\ 
&& + 2 \vec{S}_{1}\times\vec{v}_{1}\cdot\vec{v}_{2} \big( 41559 \vec{v}_{1}\cdot\vec{n} - 21695 \vec{v}_{2}\cdot\vec{n} \big) \Big] + 	\frac{G^3 m_{2}{}^3}{2400 r{}^4} \Big[ 2 \vec{S}_{1}\times\vec{n}\cdot\vec{v}_{1} \big( 10950 v_{1}^2 \nn\\ 
&& - 16134 \vec{v}_{1}\cdot\vec{v}_{2} + 9659 v_{2}^2 + 101520 \vec{v}_{1}\cdot\vec{n} \vec{v}_{2}\cdot\vec{n} - 48000 ( \vec{v}_{1}\cdot\vec{n})^{2} - 1945 ( \vec{v}_{2}\cdot\vec{n})^{2} \big) \nn\\ 
&& + \vec{S}_{1}\times\vec{n}\cdot\vec{v}_{2} \big( 1832 v_{1}^2 + 92232 \vec{v}_{1}\cdot\vec{v}_{2} - 94289 v_{2}^2 -361860 \vec{v}_{1}\cdot\vec{n} \vec{v}_{2}\cdot\vec{n} \nn\\ 
&& - 105160 ( \vec{v}_{1}\cdot\vec{n})^{2} + 401995 ( \vec{v}_{2}\cdot\vec{n})^{2} \big) + 4 \vec{S}_{1}\times\vec{v}_{1}\cdot\vec{v}_{2} \big( 15733 \vec{v}_{1}\cdot\vec{n} + 30874 \vec{v}_{2}\cdot\vec{n} \big) \Big] \nn\\ 
&& + 	\frac{G^3 m_{1} m_{2}{}^2}{384 r{}^4} \Big[ \vec{S}_{1}\times\vec{n}\cdot\vec{v}_{1} \big( {(25844 - 2313 \pi^2)} v_{1}^2 - {(45404 - 1998 \pi^2)} \vec{v}_{1}\cdot\vec{v}_{2} \nn\\ 
&& + {(35144 + 315 \pi^2)} v_{2}^2 + {(172012 - 9990 \pi^2)} \vec{v}_{1}\cdot\vec{n} \vec{v}_{2}\cdot\vec{n} \nn\\ 
&& - {(69688 - 11565 \pi^2)} ( \vec{v}_{1}\cdot\vec{n})^{2} - {(44056 + 1575 \pi^2)} ( \vec{v}_{2}\cdot\vec{n})^{2} \big) \nn\\ 
&& - 2 \vec{S}_{1}\times\vec{n}\cdot\vec{v}_{2} \big( {(8386 - 747 \pi^2)} v_{1}^2 - {(12010 - 2772 \pi^2)} \vec{v}_{1}\cdot\vec{v}_{2} + {(9392 - 2025 \pi^2)} v_{2}^2 \nn\\ 
&& + {(61394 - 13860 \pi^2)} \vec{v}_{1}\cdot\vec{n} \vec{v}_{2}\cdot\vec{n} - {(19286 - 3735 \pi^2)} ( \vec{v}_{1}\cdot\vec{n})^{2} \nn\\ 
&& - {(25072 - 10125 \pi^2)} ( \vec{v}_{2}\cdot\vec{n})^{2} \big) + 2 \vec{S}_{1}\times\vec{v}_{1}\cdot\vec{v}_{2} \big( {(16640 + 495 \pi^2)} \vec{v}_{1}\cdot\vec{n} \nn\\ 
&& - {(2140 + 3087 \pi^2)} \vec{v}_{2}\cdot\vec{n} \big) \Big]\nn\\ && - 	\frac{G^4 m_{2}{}^4}{1800 r{}^5} \Big[ 23175 \vec{S}_{1}\times\vec{n}\cdot\vec{v}_{1} - 14801 \vec{S}_{1}\times\vec{n}\cdot\vec{v}_{2} \Big] + 	\frac{G^4 m_{1}{}^3 m_{2}}{1800 r{}^5} \Big[ 6773 \vec{S}_{1}\times\vec{n}\cdot\vec{v}_{1} \nn\\ 
&& + 5892 \vec{S}_{1}\times\vec{n}\cdot\vec{v}_{2} \Big] - 	\frac{G^4 m_{1} m_{2}{}^3}{7200 r{}^5} \Big[ 3 {(222643 - 22425 \pi^2)} \vec{S}_{1}\times\vec{n}\cdot\vec{v}_{1} \nn\\ 
&& - {(1489367 - 59625 \pi^2)} \vec{S}_{1}\times\vec{n}\cdot\vec{v}_{2} \Big] - 	\frac{G^4 m_{1}{}^2 m_{2}{}^2}{3600 r{}^5} \Big[ {(297683 - 7500 \pi^2)} \vec{S}_{1}\times\vec{n}\cdot\vec{v}_{1} \nn\\ 
&& - 3 {(179518 - 17425 \pi^2)} \vec{S}_{1}\times\vec{n}\cdot\vec{v}_{2} \Big].
\eea

\section{General Hamiltonian}
\label{explicitgenham}

As noted in section \ref{hamilton}, we obtain the following general Hamiltonian for the present sector:
\bea
H_{\text{N}^3\text{LO}}^{\text{SO}}&=&  	\frac{G}{64 m_{1}{}^5 m_{2} r{}^2} \Big[ \vec{S}_{1}\times\vec{n}\cdot\vec{p}_{1} \big( 58 p_{1}^2 ( \vec{p}_{1}\cdot\vec{p}_{2})^{2} - 7 p_{2}^2 p_{1}^{4} -84 \vec{p}_{1}\cdot\vec{n} p_{1}^2 \vec{p}_{2}\cdot\vec{n} \vec{p}_{1}\cdot\vec{p}_{2} \nn\\ 
&& + 45 p_{1}^2 p_{2}^2 ( \vec{p}_{1}\cdot\vec{n})^{2} + 3 ( \vec{p}_{2}\cdot\vec{n})^{2} p_{1}^{4} -225 p_{1}^2 ( \vec{p}_{1}\cdot\vec{n})^{2} ( \vec{p}_{2}\cdot\vec{n})^{2} \big) \nn\\ 
&& + 24 \vec{S}_{1}\times\vec{n}\cdot\vec{p}_{2} \big( \vec{p}_{1}\cdot\vec{p}_{2} p_{1}^{4} + 3 \vec{p}_{1}\cdot\vec{n} \vec{p}_{2}\cdot\vec{n} p_{1}^{4} \big) + 2 \vec{S}_{1}\times\vec{p}_{1}\cdot\vec{p}_{2} \big( 50 \vec{p}_{1}\cdot\vec{n} p_{1}^2 \vec{p}_{1}\cdot\vec{p}_{2} \nn\\ 
&& - 15 \vec{p}_{2}\cdot\vec{n} p_{1}^{4} -9 p_{1}^2 \vec{p}_{2}\cdot\vec{n} ( \vec{p}_{1}\cdot\vec{n})^{2} \big) \Big] - 	\frac{G}{32 m_{1}{}^4 m_{2}{}^2 r{}^2} \Big[ \vec{S}_{1}\times\vec{n}\cdot\vec{p}_{1} \big( 11 p_{1}^2 \vec{p}_{1}\cdot\vec{p}_{2} p_{2}^2 \nn\\ 
&& + 27 \vec{p}_{1}\cdot\vec{n} p_{1}^2 \vec{p}_{2}\cdot\vec{n} p_{2}^2 - 33 p_{1}^2 \vec{p}_{1}\cdot\vec{p}_{2} ( \vec{p}_{2}\cdot\vec{n})^{2} \nn\\ 
&& - 24 \vec{p}_{1}\cdot\vec{n} \vec{p}_{2}\cdot\vec{n} ( \vec{p}_{1}\cdot\vec{p}_{2})^{2} -60 \vec{p}_{2}\cdot\vec{n} p_{2}^2 ( \vec{p}_{1}\cdot\vec{n})^{3} - 45 \vec{p}_{1}\cdot\vec{n} p_{1}^2 ( \vec{p}_{2}\cdot\vec{n})^{3} \nn\\ 
&& + 140 ( \vec{p}_{1}\cdot\vec{n})^{3} ( \vec{p}_{2}\cdot\vec{n})^{3} \big) + 8 \vec{S}_{1}\times\vec{n}\cdot\vec{p}_{2} \big( 2 p_{1}^2 ( \vec{p}_{1}\cdot\vec{p}_{2})^{2} - p_{2}^2 p_{1}^{4} + 3 p_{1}^2 p_{2}^2 ( \vec{p}_{1}\cdot\vec{n})^{2} \nn\\ 
&& + 3 ( \vec{p}_{2}\cdot\vec{n})^{2} p_{1}^{4} -15 p_{1}^2 ( \vec{p}_{1}\cdot\vec{n})^{2} ( \vec{p}_{2}\cdot\vec{n})^{2} \big) + \vec{S}_{1}\times\vec{p}_{1}\cdot\vec{p}_{2} \big( 34 p_{1}^2 \vec{p}_{2}\cdot\vec{n} \vec{p}_{1}\cdot\vec{p}_{2} \nn\\ 
&& + 5 \vec{p}_{1}\cdot\vec{n} p_{1}^2 p_{2}^2 + 24 \vec{p}_{2}\cdot\vec{n} \vec{p}_{1}\cdot\vec{p}_{2} ( \vec{p}_{1}\cdot\vec{n})^{2} - 15 \vec{p}_{1}\cdot\vec{n} p_{1}^2 ( \vec{p}_{2}\cdot\vec{n})^{2} \big) \Big] \nn\\ 
&& - 	\frac{G}{64 m_{1}{}^3 m_{2}{}^3 r{}^2} \Big[ \vec{S}_{1}\times\vec{n}\cdot\vec{p}_{1} \big( 5 p_{1}^2 p_{2}^{4} + 48 \vec{p}_{1}\cdot\vec{n} \vec{p}_{2}\cdot\vec{n} \vec{p}_{1}\cdot\vec{p}_{2} p_{2}^2 - 96 ( \vec{p}_{2}\cdot\vec{n})^{2} ( \vec{p}_{1}\cdot\vec{p}_{2})^{2} \nn\\ 
&& - 30 p_{1}^2 p_{2}^2 ( \vec{p}_{2}\cdot\vec{n})^{2} - 24 ( \vec{p}_{1}\cdot\vec{n})^{2} p_{2}^{4} + 60 p_{1}^2 ( \vec{p}_{2}\cdot\vec{n})^{4} + 120 p_{2}^2 ( \vec{p}_{1}\cdot\vec{n})^{2} ( \vec{p}_{2}\cdot\vec{n})^{2} \big) \nn\\ 
&& - 8 \vec{S}_{1}\times\vec{n}\cdot\vec{p}_{2} \big( p_{1}^2 \vec{p}_{1}\cdot\vec{p}_{2} p_{2}^2 + 2 ( \vec{p}_{1}\cdot\vec{p}_{2})^{3} + 15 \vec{p}_{1}\cdot\vec{n} p_{1}^2 \vec{p}_{2}\cdot\vec{n} p_{2}^2 + 3 \vec{p}_{1}\cdot\vec{p}_{2} p_{2}^2 ( \vec{p}_{1}\cdot\vec{n})^{2} \nn\\ 
&& + 3 p_{1}^2 \vec{p}_{1}\cdot\vec{p}_{2} ( \vec{p}_{2}\cdot\vec{n})^{2} - 6 \vec{p}_{1}\cdot\vec{n} \vec{p}_{2}\cdot\vec{n} ( \vec{p}_{1}\cdot\vec{p}_{2})^{2} -15 \vec{p}_{2}\cdot\vec{n} p_{2}^2 ( \vec{p}_{1}\cdot\vec{n})^{3} \nn\\ 
&& - 15 \vec{p}_{1}\cdot\vec{n} p_{1}^2 ( \vec{p}_{2}\cdot\vec{n})^{3} - 15 \vec{p}_{1}\cdot\vec{p}_{2} ( \vec{p}_{1}\cdot\vec{n})^{2} ( \vec{p}_{2}\cdot\vec{n})^{2} + 35 ( \vec{p}_{1}\cdot\vec{n})^{3} ( \vec{p}_{2}\cdot\vec{n})^{3} \big) \nn\\ 
&& + 4 \vec{S}_{1}\times\vec{p}_{1}\cdot\vec{p}_{2} \big( 5 p_{1}^2 \vec{p}_{2}\cdot\vec{n} p_{2}^2 + 8 \vec{p}_{1}\cdot\vec{n} \vec{p}_{1}\cdot\vec{p}_{2} p_{2}^2 \nn\\ 
&& + 12 \vec{p}_{2}\cdot\vec{n} ( \vec{p}_{1}\cdot\vec{p}_{2})^{2} -18 \vec{p}_{2}\cdot\vec{n} p_{2}^2 ( \vec{p}_{1}\cdot\vec{n})^{2} + 6 p_{1}^2 ( \vec{p}_{2}\cdot\vec{n})^{3} - 60 \vec{p}_{1}\cdot\vec{n} \vec{p}_{1}\cdot\vec{p}_{2} ( \vec{p}_{2}\cdot\vec{n})^{2} \nn\\ 
&& + 30 ( \vec{p}_{1}\cdot\vec{n})^{2} ( \vec{p}_{2}\cdot\vec{n})^{3} \big) \Big] - 	\frac{G}{4 m_{1}{}^2 m_{2}{}^4 r{}^2} \Big[ \vec{S}_{1}\times\vec{n}\cdot\vec{p}_{1} \big( \vec{p}_{1}\cdot\vec{p}_{2} p_{2}^{4} \nn\\ 
&& + 6 \vec{p}_{1}\cdot\vec{p}_{2} p_{2}^2 ( \vec{p}_{2}\cdot\vec{n})^{2} -15 \vec{p}_{1}\cdot\vec{p}_{2} ( \vec{p}_{2}\cdot\vec{n})^{4} \big) + \vec{S}_{1}\times\vec{n}\cdot\vec{p}_{2} \big( 2 p_{2}^2 ( \vec{p}_{1}\cdot\vec{p}_{2})^{2} - p_{1}^2 p_{2}^{4} \nn\\ 
&& + 3 p_{1}^2 p_{2}^2 ( \vec{p}_{2}\cdot\vec{n})^{2} + 3 ( \vec{p}_{1}\cdot\vec{n})^{2} p_{2}^{4} -15 p_{2}^2 ( \vec{p}_{1}\cdot\vec{n})^{2} ( \vec{p}_{2}\cdot\vec{n})^{2} \big) \nn\\ 
&& - \vec{S}_{1}\times\vec{p}_{1}\cdot\vec{p}_{2} \big( 4 \vec{p}_{2}\cdot\vec{n} \vec{p}_{1}\cdot\vec{p}_{2} p_{2}^2 + \vec{p}_{1}\cdot\vec{n} p_{2}^{4} -12 \vec{p}_{1}\cdot\vec{p}_{2} ( \vec{p}_{2}\cdot\vec{n})^{3} - 12 \vec{p}_{1}\cdot\vec{n} p_{2}^2 ( \vec{p}_{2}\cdot\vec{n})^{2} \nn\\ 
&& + 15 \vec{p}_{1}\cdot\vec{n} ( \vec{p}_{2}\cdot\vec{n})^{4} \big) \Big] + 	\frac{G}{32 m_{1} m_{2}{}^5 r{}^2} \Big[ \vec{S}_{1}\times\vec{n}\cdot\vec{p}_{1} \big( 4 p_{2}^{6} -60 p_{2}^2 ( \vec{p}_{2}\cdot\vec{n})^{4} \nn\\ 
&& + 35 ( \vec{p}_{2}\cdot\vec{n})^{6} \big) + 12 \vec{S}_{1}\times\vec{n}\cdot\vec{p}_{2} \big( \vec{p}_{1}\cdot\vec{p}_{2} p_{2}^{4} + 3 \vec{p}_{1}\cdot\vec{n} \vec{p}_{2}\cdot\vec{n} p_{2}^{4} \big) \nn\\ 
&& - 6 \vec{S}_{1}\times\vec{p}_{1}\cdot\vec{p}_{2} \big( 2 \vec{p}_{2}\cdot\vec{n} p_{2}^{4} -8 p_{2}^2 ( \vec{p}_{2}\cdot\vec{n})^{3} + 15 ( \vec{p}_{2}\cdot\vec{n})^{5} \big) \Big] \nn\\ 
&& - 	\frac{G}{8 m_{1}{}^6 r{}^2} \Big[ \vec{S}_{1}\times\vec{n}\cdot\vec{p}_{1} \big( 2 \vec{p}_{1}\cdot\vec{p}_{2} p_{1}^{4} + 9 \vec{p}_{1}\cdot\vec{n} \vec{p}_{2}\cdot\vec{n} p_{1}^{4} \big) - 4 \vec{S}_{1}\times\vec{p}_{1}\cdot\vec{p}_{2} \vec{p}_{1}\cdot\vec{n} p_{1}^{4} \Big] \nn\\ 
&& - 	\frac{45 G m_{2}}{128 m_{1}{}^7 r{}^2} \vec{S}_{1}\times\vec{n}\cdot\vec{p}_{1} p_{1}^{6} \nn\\ && - 	\frac{G^2 m_{2}{}^2}{16 m_{1}{}^5 r{}^3} \vec{S}_{1}\times\vec{n}\cdot\vec{p}_{1} \big( 41 p_{1}^{4} + 44 p_{1}^2 ( \vec{p}_{1}\cdot\vec{n})^{2} \big) - 	\frac{G^2 m_{2}}{96 m_{1}{}^4 r{}^3} \Big[ 3 \vec{S}_{1}\times\vec{n}\cdot\vec{p}_{1} \big( 34 p_{1}^2 \vec{p}_{1}\cdot\vec{p}_{2} \nn\\ 
&& + 335 p_{1}^{4} + 380 \vec{p}_{1}\cdot\vec{n} p_{1}^2 \vec{p}_{2}\cdot\vec{n} - 1027 p_{1}^2 ( \vec{p}_{1}\cdot\vec{n})^{2} - 224 \vec{p}_{1}\cdot\vec{p}_{2} ( \vec{p}_{1}\cdot\vec{n})^{2} -370 ( \vec{p}_{1}\cdot\vec{n})^{4} \nn\\ 
&& + 256 \vec{p}_{2}\cdot\vec{n} ( \vec{p}_{1}\cdot\vec{n})^{3} \big) - 3 \vec{S}_{1}\times\vec{n}\cdot\vec{p}_{2} \big( 31 p_{1}^{4} -64 p_{1}^2 ( \vec{p}_{1}\cdot\vec{n})^{2} \big) \nn\\ 
&& + 2 \vec{S}_{1}\times\vec{p}_{1}\cdot\vec{p}_{2} \big( 417 \vec{p}_{1}\cdot\vec{n} p_{1}^2 + 112 ( \vec{p}_{1}\cdot\vec{n})^{3} \big) \Big] \nn\\ 
&& + 	\frac{G^2}{96 m_{1}{}^3 r{}^3} \Big[ \vec{S}_{1}\times\vec{n}\cdot\vec{p}_{1} \big( 2964 p_{1}^2 \vec{p}_{1}\cdot\vec{p}_{2} - 34 p_{1}^2 p_{2}^2 \nn\\ 
&& - 212 ( \vec{p}_{1}\cdot\vec{p}_{2})^{2} -3228 \vec{p}_{1}\cdot\vec{n} p_{1}^2 \vec{p}_{2}\cdot\vec{n} + 1738 \vec{p}_{1}\cdot\vec{n} \vec{p}_{2}\cdot\vec{n} \vec{p}_{1}\cdot\vec{p}_{2} \nn\\ 
&& - 8721 \vec{p}_{1}\cdot\vec{p}_{2} ( \vec{p}_{1}\cdot\vec{n})^{2} + 58 p_{2}^2 ( \vec{p}_{1}\cdot\vec{n})^{2} + 469 p_{1}^2 ( \vec{p}_{2}\cdot\vec{n})^{2} -3918 \vec{p}_{2}\cdot\vec{n} ( \vec{p}_{1}\cdot\vec{n})^{3} \nn\\ 
&& - 4290 ( \vec{p}_{1}\cdot\vec{n})^{2} ( \vec{p}_{2}\cdot\vec{n})^{2} \big) + 6 \vec{S}_{1}\times\vec{n}\cdot\vec{p}_{2} \big( 99 p_{1}^2 \vec{p}_{1}\cdot\vec{p}_{2} - 36 p_{1}^{4} + 114 \vec{p}_{1}\cdot\vec{n} p_{1}^2 \vec{p}_{2}\cdot\vec{n} \nn\\ 
&& + 180 p_{1}^2 ( \vec{p}_{1}\cdot\vec{n})^{2} + 32 \vec{p}_{1}\cdot\vec{p}_{2} ( \vec{p}_{1}\cdot\vec{n})^{2} -136 ( \vec{p}_{1}\cdot\vec{n})^{4} - 128 \vec{p}_{2}\cdot\vec{n} ( \vec{p}_{1}\cdot\vec{n})^{3} \big) \nn\\ 
&& + \vec{S}_{1}\times\vec{p}_{1}\cdot\vec{p}_{2} \big( 1638 \vec{p}_{1}\cdot\vec{n} p_{1}^2 + 1733 p_{1}^2 \vec{p}_{2}\cdot\vec{n} + 5512 \vec{p}_{1}\cdot\vec{n} \vec{p}_{1}\cdot\vec{p}_{2} -4513 ( \vec{p}_{1}\cdot\vec{n})^{3} \nn\\ 
&& - 1490 \vec{p}_{2}\cdot\vec{n} ( \vec{p}_{1}\cdot\vec{n})^{2} \big) \Big] - 	\frac{G^2}{96 m_{2}{}^3 r{}^3} \Big[ 6 \vec{S}_{1}\times\vec{n}\cdot\vec{p}_{1} \big( 31 p_{2}^{4} -178 p_{2}^2 ( \vec{p}_{2}\cdot\vec{n})^{2} \nn\\ 
&& + 275 ( \vec{p}_{2}\cdot\vec{n})^{4} \big) - 3 \vec{S}_{1}\times\vec{n}\cdot\vec{p}_{2} \big( 144 \vec{p}_{1}\cdot\vec{p}_{2} p_{2}^2 + 81 p_{2}^{4} + 128 \vec{p}_{1}\cdot\vec{n} \vec{p}_{2}\cdot\vec{n} p_{2}^2 \nn\\ 
&& - 77 p_{2}^2 ( \vec{p}_{2}\cdot\vec{n})^{2} -909 ( \vec{p}_{2}\cdot\vec{n})^{4} \big) - 2 \vec{S}_{1}\times\vec{p}_{1}\cdot\vec{p}_{2} \big( 759 \vec{p}_{2}\cdot\vec{n} p_{2}^2 -1486 ( \vec{p}_{2}\cdot\vec{n})^{3} \big) \Big] \nn\\ 
&& - 	\frac{G^2}{96 m_{1}{}^2 m_{2} r{}^3} \Big[ \vec{S}_{1}\times\vec{n}\cdot\vec{p}_{1} \big( 1388 p_{1}^2 p_{2}^2 - 4245 \vec{p}_{1}\cdot\vec{p}_{2} p_{2}^2 \nn\\ 
&& + 1228 ( \vec{p}_{1}\cdot\vec{p}_{2})^{2} -12680 \vec{p}_{1}\cdot\vec{n} \vec{p}_{2}\cdot\vec{n} \vec{p}_{1}\cdot\vec{p}_{2} + 13092 \vec{p}_{1}\cdot\vec{n} \vec{p}_{2}\cdot\vec{n} p_{2}^2 - 7304 p_{2}^2 ( \vec{p}_{1}\cdot\vec{n})^{2} \nn\\ 
&& + 226 p_{1}^2 ( \vec{p}_{2}\cdot\vec{n})^{2} + 1725 \vec{p}_{1}\cdot\vec{p}_{2} ( \vec{p}_{2}\cdot\vec{n})^{2} + 7806 ( \vec{p}_{1}\cdot\vec{n})^{2} ( \vec{p}_{2}\cdot\vec{n})^{2} \nn\\ 
&& - 3714 \vec{p}_{1}\cdot\vec{n} ( \vec{p}_{2}\cdot\vec{n})^{3} \big) - \vec{S}_{1}\times\vec{n}\cdot\vec{p}_{2} \big( 864 p_{1}^2 \vec{p}_{1}\cdot\vec{p}_{2} - 88 p_{1}^2 p_{2}^2 \nn\\ 
&& - 2342 ( \vec{p}_{1}\cdot\vec{p}_{2})^{2} -2352 \vec{p}_{1}\cdot\vec{n} p_{1}^2 \vec{p}_{2}\cdot\vec{n} + 11278 \vec{p}_{1}\cdot\vec{n} \vec{p}_{2}\cdot\vec{n} \vec{p}_{1}\cdot\vec{p}_{2} \nn\\ 
&& - 3408 \vec{p}_{1}\cdot\vec{p}_{2} ( \vec{p}_{1}\cdot\vec{n})^{2} + 1117 p_{2}^2 ( \vec{p}_{1}\cdot\vec{n})^{2} + 679 p_{1}^2 ( \vec{p}_{2}\cdot\vec{n})^{2} + 3072 \vec{p}_{2}\cdot\vec{n} ( \vec{p}_{1}\cdot\vec{n})^{3} \nn\\ 
&& - 4857 ( \vec{p}_{1}\cdot\vec{n})^{2} ( \vec{p}_{2}\cdot\vec{n})^{2} \big) + 2 \vec{S}_{1}\times\vec{p}_{1}\cdot\vec{p}_{2} \big( 466 p_{1}^2 \vec{p}_{2}\cdot\vec{n} - 1774 \vec{p}_{1}\cdot\vec{n} \vec{p}_{1}\cdot\vec{p}_{2} \nn\\ 
&& + 5024 \vec{p}_{2}\cdot\vec{n} \vec{p}_{1}\cdot\vec{p}_{2} + 2476 \vec{p}_{1}\cdot\vec{n} p_{2}^2 -3193 \vec{p}_{2}\cdot\vec{n} ( \vec{p}_{1}\cdot\vec{n})^{2} - 3364 \vec{p}_{1}\cdot\vec{n} ( \vec{p}_{2}\cdot\vec{n})^{2} \big) \Big] \nn\\ 
&& + 	\frac{G^2}{96 m_{1} m_{2}{}^2 r{}^3} \Big[ 6 \vec{S}_{1}\times\vec{n}\cdot\vec{p}_{1} \big( 162 \vec{p}_{1}\cdot\vec{p}_{2} p_{2}^2 - 584 p_{2}^{4} -1736 \vec{p}_{1}\cdot\vec{n} \vec{p}_{2}\cdot\vec{n} p_{2}^2 \nn\\ 
&& - 252 \vec{p}_{1}\cdot\vec{p}_{2} ( \vec{p}_{2}\cdot\vec{n})^{2} + 2229 p_{2}^2 ( \vec{p}_{2}\cdot\vec{n})^{2} + 1864 \vec{p}_{1}\cdot\vec{n} ( \vec{p}_{2}\cdot\vec{n})^{3} + 17 ( \vec{p}_{2}\cdot\vec{n})^{4} \big) \nn\\ 
&& - \vec{S}_{1}\times\vec{n}\cdot\vec{p}_{2} \big( 40 p_{1}^2 p_{2}^2 - 1569 \vec{p}_{1}\cdot\vec{p}_{2} p_{2}^2 + 1040 ( \vec{p}_{1}\cdot\vec{p}_{2})^{2} -3952 \vec{p}_{1}\cdot\vec{n} \vec{p}_{2}\cdot\vec{n} \vec{p}_{1}\cdot\vec{p}_{2} \nn\\ 
&& + 2406 \vec{p}_{1}\cdot\vec{n} \vec{p}_{2}\cdot\vec{n} p_{2}^2 - 88 p_{2}^2 ( \vec{p}_{1}\cdot\vec{n})^{2} - 544 p_{1}^2 ( \vec{p}_{2}\cdot\vec{n})^{2} + 10659 \vec{p}_{1}\cdot\vec{p}_{2} ( \vec{p}_{2}\cdot\vec{n})^{2} \nn\\ 
&& + 528 ( \vec{p}_{1}\cdot\vec{n})^{2} ( \vec{p}_{2}\cdot\vec{n})^{2} - 10710 \vec{p}_{1}\cdot\vec{n} ( \vec{p}_{2}\cdot\vec{n})^{3} \big) - 2 \vec{S}_{1}\times\vec{p}_{1}\cdot\vec{p}_{2} \big( 1778 \vec{p}_{2}\cdot\vec{n} \vec{p}_{1}\cdot\vec{p}_{2} \nn\\ 
&& + 1078 \vec{p}_{1}\cdot\vec{n} p_{2}^2 - 4251 \vec{p}_{2}\cdot\vec{n} p_{2}^2 -1072 \vec{p}_{1}\cdot\vec{n} ( \vec{p}_{2}\cdot\vec{n})^{2} + 1889 ( \vec{p}_{2}\cdot\vec{n})^{3} \big) \Big]\nn\\ && +  	\frac{7 G^3 m_{2}{}^3}{16 m_{1}{}^3 r{}^4} \vec{S}_{1}\times\vec{n}\cdot\vec{p}_{1} \big( 9 p_{1}^2 -152 ( \vec{p}_{1}\cdot\vec{n})^{2} \big) + 	\frac{3 G^3 m_{1}{}^2}{50 m_{2}{}^2 r{}^4} \vec{S}_{1}\times\vec{n}\cdot\vec{p}_{2} \big( p_{2}^2 -5 ( \vec{p}_{2}\cdot\vec{n})^{2} \big) \nn\\ 
&& + 	\frac{G^3 m_{2}{}^2}{9600 m_{1}{}^2 r{}^4} \Big[ \vec{S}_{1}\times\vec{n}\cdot\vec{p}_{1} \big( {(1292900 - 57825 \pi^2)} p_{1}^2 - 165072 \vec{p}_{1}\cdot\vec{p}_{2} \nn\\ 
&& + 279360 \vec{p}_{1}\cdot\vec{n} \vec{p}_{2}\cdot\vec{n} - {(4348600 - 289125 \pi^2)} ( \vec{p}_{1}\cdot\vec{n})^{2} \big) \nn\\ 
&& + 16 \vec{S}_{1}\times\vec{n}\cdot\vec{p}_{2} \big( 3383 p_{1}^2 -30190 ( \vec{p}_{1}\cdot\vec{n})^{2} \big) + 371728 \vec{S}_{1}\times\vec{p}_{1}\cdot\vec{p}_{2} \vec{p}_{1}\cdot\vec{n} \Big] \nn\\ 
&& - 	\frac{G^3 m_{1}}{4800 m_{2} r{}^4} \Big[ 4 \vec{S}_{1}\times\vec{n}\cdot\vec{p}_{1} \big( 14957 p_{2}^2 + 68690 ( \vec{p}_{2}\cdot\vec{n})^{2} \big) - \vec{S}_{1}\times\vec{n}\cdot\vec{p}_{2} \big( 153184 \vec{p}_{1}\cdot\vec{p}_{2} \nn\\ 
&& - {(940700 - 50625 \pi^2)} p_{2}^2 -319520 \vec{p}_{1}\cdot\vec{n} \vec{p}_{2}\cdot\vec{n} + {(1920400 - 253125 \pi^2)} ( \vec{p}_{2}\cdot\vec{n})^{2} \big) \nn\\ 
&& + 397960 \vec{S}_{1}\times\vec{p}_{1}\cdot\vec{p}_{2} \vec{p}_{2}\cdot\vec{n} \Big] + 	\frac{G^3 m_{2}}{4800 m_{1} r{}^4} \Big[ \vec{S}_{1}\times\vec{n}\cdot\vec{p}_{1} \big( 382144 p_{1}^2 \nn\\ 
&& - {(1143250 - 24975 \pi^2)} \vec{p}_{1}\cdot\vec{p}_{2} + 114686 p_{2}^2 + {(3907550 - 124875 \pi^2)} \vec{p}_{1}\cdot\vec{n} \vec{p}_{2}\cdot\vec{n} \nn\\ 
&& - 1735220 ( \vec{p}_{1}\cdot\vec{n})^{2} + 339170 ( \vec{p}_{2}\cdot\vec{n})^{2} \big) - \vec{S}_{1}\times\vec{n}\cdot\vec{p}_{2} \big( {(662950 - 18675 \pi^2)} p_{1}^2 \nn\\ 
&& - 455964 \vec{p}_{1}\cdot\vec{p}_{2} + 595020 \vec{p}_{1}\cdot\vec{n} \vec{p}_{2}\cdot\vec{n} - {(1479350 - 93375 \pi^2)} ( \vec{p}_{1}\cdot\vec{n})^{2} \big) \nn\\ 
&& + \vec{S}_{1}\times\vec{p}_{1}\cdot\vec{p}_{2} \big( {(747200 + 12375 \pi^2)} \vec{p}_{1}\cdot\vec{n} + 281792 \vec{p}_{2}\cdot\vec{n} \big) \Big] \nn\\ 
&& - 	\frac{G^3}{9600 r{}^4} \Big[ \vec{S}_{1}\times\vec{n}\cdot\vec{p}_{1} \big( 492032 \vec{p}_{1}\cdot\vec{p}_{2} \nn\\ 
&& - {(1377500 + 7875 \pi^2)} p_{2}^2 -3978760 \vec{p}_{1}\cdot\vec{n} \vec{p}_{2}\cdot\vec{n} + {(2523700 + 39375 \pi^2)} ( \vec{p}_{2}\cdot\vec{n})^{2} \big) \nn\\ 
&& + 4 \vec{S}_{1}\times\vec{n}\cdot\vec{p}_{2} \big( 85336 p_{1}^2 - {(747575 - 34650 \pi^2)} \vec{p}_{1}\cdot\vec{p}_{2} + 265364 p_{2}^2 \nn\\ 
&& + {(1851325 - 173250 \pi^2)} \vec{p}_{1}\cdot\vec{n} \vec{p}_{2}\cdot\vec{n} - 149480 ( \vec{p}_{1}\cdot\vec{n})^{2} - 441820 ( \vec{p}_{2}\cdot\vec{n})^{2} \big) \nn\\ 
&& - 2 \vec{S}_{1}\times\vec{p}_{1}\cdot\vec{p}_{2} \big( 677472 \vec{p}_{1}\cdot\vec{n} - {(294700 + 77175 \pi^2)} \vec{p}_{2}\cdot\vec{n} \big) \Big]\nn\\ && - 	\frac{303 G^4 m_{2}{}^4}{8 m_{1} r{}^5} \vec{S}_{1}\times\vec{n}\cdot\vec{p}_{1} - 	\frac{G^4 m_{2}{}^3}{14400 r{}^5} \Big[ {(3373058 - 193275 \pi^2)} \vec{S}_{1}\times\vec{n}\cdot\vec{p}_{1} \nn\\ 
&& - 560008 \vec{S}_{1}\times\vec{n}\cdot\vec{p}_{2} \Big] + 	\frac{1497 G^4 m_{1}{}^3}{50 r{}^5} \vec{S}_{1}\times\vec{n}\cdot\vec{p}_{2} \nn\\ 
&& - 	\frac{G^4 m_{1} m_{2}{}^2}{14400 r{}^5} \Big[ 4 {(927908 - 27075 \pi^2)} \vec{S}_{1}\times\vec{n}\cdot\vec{p}_{1} \nn\\ 
&& - {(5053634 - 177975 \pi^2)} \vec{S}_{1}\times\vec{n}\cdot\vec{p}_{2} \Big] - 	\frac{G^4 m_{1}{}^2 m_{2}}{3600 r{}^5} \Big[ 104254 \vec{S}_{1}\times\vec{n}\cdot\vec{p}_{1} \nn\\ 
&& - 3 {(369443 - 23950 \pi^2)} \vec{S}_{1}\times\vec{n}\cdot\vec{p}_{2} \Big].
\eea

\bibliographystyle{jhep}
\bibliography{gwbibtex}

\end{document}